\documentclass[11pt,epsfig]{article} 
\usepackage[utf8]{inputenc}
\usepackage[top=1in, left=0.95in, bottom=1.1in, right=0.95in]{geometry}
\usepackage[colorlinks=true,
linkcolor=blue,
urlcolor=red,
citecolor=red]{hyperref}
\usepackage[vcentermath,enableskew]{youngtab}
\usepackage{booktabs}
\usepackage{ytableau}
\usepackage{tikz}
\usepackage{young}
\usepackage{tabularx}
\usepackage{diagbox}
\usepackage[toc]{appendix}
\usepackage{longtable}
\usepackage{enumerate}
\usepackage{float}
\usepackage{subfig}
\usepackage{xcolor}

\usepackage{cite}

\let\counterwithin\relax
\usepackage{chngcntr}
\usepackage{amssymb, amsmath,mathrsfs}
\usepackage{multicol,multirow}

\usepackage{graphics}
\usepackage{graphicx}
\usepackage{epsf}
\usepackage{epsfig}
\usepackage{float}
\usepackage{makecell}

\usepackage{color}
\usepackage{xcolor}
\usepackage{simplewick}
\usepackage{amsmath}
\usepackage{amsfonts}
\usepackage{makeidx} 
\usepackage[section]{placeins}
\newcommand{\bea}{\begin{eqnarray}}
\newcommand{\eea}{\end{eqnarray}}
\newcommand{\dd}[1]{(N^{\dagger} #1 N^{\dagger})}
\newcommand{\nn}[1]{(N #1 N)}
\newcommand{\dn}[1]{(N^{\dagger} #1 N)}
\newcommand{\rsd}{\Vec{\sigma}\cdot\overrightarrow{\nabla}}
\newcommand{\lsd}{\Vec{\sigma}\cdot\overleftarrow{\nabla}}
\newcommand{\rd}[1]{\overrightarrow{\nabla}^{#1}}
\newcommand{\ld}[1]{\overleftarrow{\nabla}^{#1}}

\def\baselinestretch{1.16}

\newcommand{\ykl}[1]{{\color{red}{\{YK: #1\}}}}

\begin{document}
\begin{center}


{\Large \textbf  {Systematic Operator Construction for Non-relativistic Effective Field Theories: Hilbert Series versus Young Tensor}}\\[10mm]

Yong-Kang Li$^{a, b}$\footnote{liyongkang@itp.ac.cn}, Yi-Ning Wang$^{a, b}$\footnote{wangyining@itp.ac.cn}, Jiang-Hao Yu$^{a, b, c, d}$\footnote{jhyu@itp.ac.cn}\\[10mm]

\noindent 
$^a${\em \small Institute of Theoretical Physics, Chinese Academy of Sciences,  Beijing 100190, P. R. China}  \\
$^b${\em \small School of Physical Sciences, University of Chinese Academy of Sciences,  Beijing 100049, P.R. China}   \\
$^c${\em \small School of Fundamental Physics and Mathematical Sciences, Hangzhou Institute for Advanced Study, UCAS, Hangzhou 310024, China} \\
$^d${\em \small International Centre for Theoretical Physics Asia-Pacific, Beijing/Hangzhou, China}\\[10mm]

\date{\today}   
          
\end{center}
\begin{abstract}
This work establishes a systematic framework for operator construction in the non-relativistic effective field theory, incorporating both the three-dimensional Euclidean symmetry and the internal symmetries. By employing double cover of the rotation group, we extend the Hilbert series to the non-relativistic systems, and eliminates redundancies introduced by the spin operator. We also generalize the Young tensor method to the non-relativistic cases through the $SU(2)$ semi‑standard Young tableaux, which allows for the construction of operator bases with repeated fields
at any given mass dimension. Utilizing the Young tensor technique and Hibert series as cross-check, we obtain the complete operator bases for the following cases: heavy particle (and also heavy quark) effective theory operators up to mass dimension 9; pion-less effective theory operators, including nucleon‑nucleon contact interactions up to $\mathcal{O}(Q^4)$ and three‑nucleon interactions at $\mathcal{O}(Q^2)$; and finally the spin‑1/2 dark matter-nucleon operators up to $\mathcal{O}(v^4)$.

\end{abstract}


\newpage
\setcounter{tocdepth}{3}
\setcounter{secnumdepth}{3}

\tableofcontents

\setcounter{footnote}{0}

\def\baselinestretch{1.5}
\counterwithin{equation}{section}


\section{Introduction}

%
In quantum field theory, two kinds of spacetime symmetry, the Poincaré group and the Galilean group, are applied to describe relativistic and non-relativistic quantum system, respectively. 
Both groups contain rotations, spacetime translations, and boosts, but differ in the commutation relations of their boost generators due to the different velocity transformation laws between the inertial frames. These groups are nonetheless related by the Inönü–Wigner contraction~\cite{Inonu:1953sp}, whereby the Galilean group emerges from the Poincaré group in the limit that the speed of light $c$ tends to infinity $c \to \infty$. 
When the boost transformation is irrelevant or spontaneously broken, the residue symmetry of both spacetime groups share the same sub-group $\mathbb{R}^{3,1}\rtimes SO(3) = \mathbb{R}^{1} \times \text{E}(3)$. Here, the three‑dimensional Euclidean group is defined as the semi‑direct product of spatial translations and rotations, $\text{E}(3)=\mathbb{R}^{3}\rtimes SO(3)$, while $\mathbb{R}^{1}$ corresponds to time translation, describing the Hamiltonian of the quantum system. 

Within the specific spacetime and internal symmetry, the power of effective field theory (EFT) lies in providing a systematic framework to describe the low-energy dynamics by leveraging a hierarchy of scales.  The construction of a non-relativistic effective field theory (NREFT), applicable when the typical energy is much smaller than the particle mass, is also guided by the three fundamental elements: the relevant degrees of freedom, the underlying symmetries, and a consistent power-counting scheme.
{First, particle-antiparticle production is kinematically forbidden at low energies, allowing antiparticle degrees of freedom to be excluded from the effective Lagrangian. Thus the dynamics can be described by a two-component heavy spinor field. Second, for the particle degree of freedom, the NR field transforms non-linearly under the Lorentz boost transformation, but linearly under the Galilean boost one. On the other hand, the NR field transforms linearly under the three-dimensional Euclidean group $\text{E}(3)=\mathbb{R}^{3} \rtimes SO(3)$. Thus the underlying symmetry is taken to be Euclidean group, while the boost transformation can be further applied to constrain the Wilson coefficients between effective operators. }
Finally, to organize the infinite series of operators in the Lagrangian, a well-defined power-counting rule must be established based on the relevant energy scales~\cite{Weinberg:1978kz}.

The applications of these general principles to various EFTs are well illustrated in several specific contexts. In this work, we discuss the following representative examples.
Heavy quark effective theory (HQET)~\cite{Isgur:1989vq, Eichten:1989zv, Georgi:1990um, Luke:1992cs,Alberte:2020eil,
Caswell:1985ui,Grinstein:1990mj,Eichten:1990vp,Manohar:1997qy,Manohar:2000dt
} describes the dynamics of a single heavy quark (e.g., the bottom quark with mass $m_b$) inside a hadron, possess the non-linearized Lorentz symmetry, but explicit $\mathbb{R}^{3,1}\rtimes SO(3)$ symmetry, the Euclidean symmetry with additional time translation. The theory involves two widely separated scales: the large heavy quark mass $m$ and the small momentum fluctuations of order $\Lambda_{\text{QCD}} \ll m_b$. The effective Lagrangian, which excludes anti-particle degrees of freedom, admits a simple power counting in $\Lambda_{\text{QCD}}/m_b$, denoted as $\mathcal{O}(1/m)$. The Lagrangian was first constructed up to $\mathcal{O}(1/m^3)$ (mass dimension 7) in~\cite{Manohar:1997qy}. Subsequent work~\cite{Gunawardana:2017zix} developed a manual method to construct the operator basis up to dimension 8, while the Hilbert series was later introduced to systematically count operators~\cite{Kobach:2017xkw, Kobach:2018nmt}.
Pionless effective field theory~\cite{Weinberg:1990rz,Weinberg:1991um,Weinberg:1992yk,Ordonez:1995rz,vanKolck:1999mw,Chen:1999tn,Bedaque:2002mn,Epelbaum:2005pn,Epelbaum:2008ga,Entem:2003ft,Epelbaum:2004fk,Kaplan:1998we,
Girlanda:2010ya, Girlanda:2011fh, Xiao:2018jot,Epelbaum:2002vt
} describes nuclear forces at momentum scales $|\vec{k}| \ll m_\pi$, where pions are integrated out and nucleons interact via contact operators. Although this effective theory possess the Galilean symmetry, it can be more generally described by the explicit $\mathbb{R}^{1}\rtimes \text{E}(3)$ symmetry, which include both the center-of-mass and relative motion between nucleons.  Under the $\text{E}(3)$ symmetry, a complete basis of these operators allows for a systematic parameterization of short-range nuclear interactions and potential contributions from new heavy particles through their Wilson coefficients. For instance, the basis for nucleon-nucleon (N-N) contact operators at $\mathcal{O}(Q^2)$ in a general frame was given in~\cite{Pastore:2009is,Girlanda:2010ya}. This was extended to $\mathcal{O}(Q^4)$ in the center-of-mass frame~\cite{Xiao:2018jot} and in momentum space for a general frame~\cite{Filandri:2023qio}. Furthermore, three-nucleon (3N) contact operators up to $\mathcal{O}(Q^2)$ have been constructed~\cite{Girlanda:2011fh}.
Dark matter (DM) direct detection~\cite{Fan:2010gt,Fitzpatrick:2012ix,Cirelli:2013ufw,Bishara:2016hek,Roszkowski:2017nbc,DelNobile:2021wmp,Hill:2011be,Bishara:2017pfq} provides another  application for the framework of the NREFTs. The scattering of cold, heavy, and non-relativistic dark matter  particles on nucleons is described by an effective Lagrangian of the contact operators. For the relevant $2\to 2$ scattering process, the Galilean boost invariance is often imposed on the $\text{E}(3)$ symmetry to reduce the number of independent structures and simplify the basis. One set of momentum-space non-relativistic operators for DM-nucleon interactions have been constructed in~\cite{Fitzpatrick:2012ix, DelNobile:2018dfg}.

The conventional construction of the operator bases in these NREFTs has often proceeded through case-specific methodologies, a practice that can lead to a lack of uniformity and introduce potential redundancies. As an illustration, the resulting challenges can be categorized as follows:
First, for operator counting in the HQET and the heavy particle effective theory (HPET), Refs.~\cite{Kobach:2017xkw,Kobach:2018nmt} employ the Hilbert series method~\cite{Lehman:2015via,Henning:2015alf,Henning:2015daa,Marinissen:2020jmb,Graf:2020yxt}. However, the approach in Ref.~\cite{Kobach:2017xkw} formulates building blocks within the homogeneous $SO(3) \times SU(2)$ symmetry framework—where the $SO(3)$ corresponds to the spatial rotations and the $SU(2)$ to the spin rotations—by introducing explicit spin operators. Due to the redundancy of the spin operator, this construction fails to generalize to the multi-fermion sectors, such as those containing four-fermion operators.
Second, to explicitly construct operators,
conventional methods for establishing a complete and independent operator basis face growing challenges at higher orders. These difficulties arise from the need to systematically eliminate redundancies due to the equations of motion (EOM), integration by parts (IBP), the Fierz identities, and the permutation symmetries of identical fields. For instance, in the HQET and the NRQCD, Ref.~\cite{Gunawardana:2017zix} manually enumerates all independent combinations of building blocks in matrix elements to remove the EOM and IBP redundancies for the operators. Similarly, in the context of the pionless effective field theory, Ref.~\cite{Girlanda:2011fh} employs three specific Fierz identities Eqs.~\eqref{eq:Fierz}, \eqref{eq:Fierz2}, and \eqref{eq:Fierz3} to eliminate redundancies related to the rearrangement of $SO(3)$ indices. A more systematic treatment of these redundancies is required.
Third, the manual imposition of the discrete symmetry transformations—such as the spatial inversion ($P$) and the time reversal ($T$)—on operators, as done in Ref.~\cite{Kobach:2017xkw}, calls for a more efficient and reliable  approach.
Fourth, in the context of the dark matter–nucleon scattering~\cite{Fitzpatrick:2012ix,DelNobile:2018dfg}, the Galilean boost invariance is frequently adopted as a simplifying assumption. However, the relativistic origin of the underlying effective field theory implies the existence of the physically valid non-relativistic operators that explicitly break this Galilean invariance, a complication that must be properly accounted for.

In this work, we address these challenges by systematically construct the operator basis using the group theoretic technique. First, apart from the translations, we formulate the building blocks to transform under a single rotation group, $SO(3)$, with the understanding that the $SU(2)$ spin group acts as its double cover. Consequently, the spin operator becomes redundant and is excluded from the set of building blocks. This approach allows us to derive the correct Hilbert series for a range of theories, including  the HPET, the HQET, the pionless effective field theories, and the dark matter-nucleon contact interactions.
Second, we implement an alternative formulation in which building blocks are expressed in terms of $SU(2)$ indices. In this representation, the spin operator is implicitly encoded, and the multiple Fierz identities reduce to a single Schouten identity, given in Eq.~\eqref{eq:schouten0}. This formulation facilitates the application of the Young tensor method~\cite{Li:2020gnx,Li:2022tec} for explicit operator construction, which is also free from the Schouten identity Eq.~\eqref{eq:schouten0}. By leveraging the correspondence between semi-standard Young tableaux (SSYT) and operators, we can automatically generate a complete, non-redundant operator basis to any desired order. The operators are symmetrized by the Young operators, such that the redundancies arising from the repeated fields can be removed directly.
Third, we systematically classify operators according to their  $P$ and  $T$ eigenvalues, a procedure that applies across those NREFTs under consideration. Building on recent advances in the treatment of the discrete symmetries within the Hilbert series formalism~\cite{Sun:2022aag, Song:2024fae, Li:2024ghg}, we employ this method to count invariants with specified $P$ and $T$ properties and consistently incorporate these symmetries in the Young tensor method. Fourth, for the specific case of the spin-1/2 dark matter-nucleon contact interactions, we adopt the building blocks that transform under the irreducible representations of the rotational symmetry, thereby obtaining the most general operator basis.

This paper is organized as follows. Section \ref{review} discusses the properties of spin and spin groups in non-relativistic field theory, explaining why the explicit introduction of spin operators is unnecessary, and provides a brief review of several representative non-relativistic effective field theories. In sections \ref{NRop} and \ref{ytm}, we employ the Hilbert series and Young tensor method to systematically construct the non-relativistic operator bases. Finally, the explicit complete operator bases  are presented in the appendices: HPET operators in appendix~\ref{app:HQETop}, HQET operators in appendix~\ref{ap:HQET}, nucleon-nucleon contact operators in appendix~\ref{ap:NN}, three-nucleon contact operators in appendix~\ref{ap:3N}, and spin-1/2 dark matter-nucleon contact operators in appendix~\ref{ap:DMN}.

\section{Symmetry for NR Fields}\label{review}
In Section~\ref{sec:symmetry3}, we briefly review the Poincar\'{e} group, Galilean group and Euclidean group, where we concentrate on the rotation group to construct the non-relativistic operator bases.
In Section~\ref{symmetry}, we analyze the rotational symmetry and clarify a common misconception: it is sometimes mistakenly believed that the spin angular momentum $\boldsymbol{S}$ and the orbital angular momentum $\boldsymbol{L}$ generate independent rotation groups. In reality, \emph{only one} rotation group governs the system, as evidenced physically by spin-orbit coupling and formally by the redundancy of the spin operators.

We then introduce the spin group in Section~\ref{spingroup} and classify the EFT building blocks according to its irreducible representations. This classification simplifies the  subsequent basis construction. Finally, in Section~\ref{eftreview}, we briefly review the heavy particle effective theory (HPET), the pionless EFT, and the dark matter-nucleon contact interactions, illustrating how the symmetry principles are implemented in specific contexts.


\subsection{Poincar\'{e}, Galilean and Euclidean Symmetries}\label{sec:symmetry3}

\paragraph{Poincar\'{e} symmetry}
The Poincar\'{e} group, defined as the semi-direct product $\mathbb{R}^{3,1}\rtimes SO(3,1)$, encodes the fundamental spacetime symmetries of relativistic systems. Its generators correspond to the spacetime translations, the spatial rotations, and the Lorentz boosts.
Under a Poincar\'{e} transformation specified by a translation $b \in \mathbb{R}^{3,1}$ and a Lorentz transformation $\Lambda \in SO(3,1)$, a one-particle state created by the operator $a_{p,\sigma}^{\dagger}$ transforms according to the unitary induced  representation as
\begin{eqnarray}
     U(\Lambda,b) a^\dagger_{p,\sigma} U^{-1}(\Lambda,b) = e^{\mathbf{i}\Lambda p\cdot b}\sum_{\sigma'} D^{(s)}_{\sigma'\sigma}(W(\Lambda,p)) a^\dagger_{\Lambda p,\sigma'},
\end{eqnarray}
where $W(\Lambda,p)$ denotes the Wigner rotation associated with the Lorentz transformation $\Lambda$ and the momentum $p^{\mu}$. In contrast, the corresponding relativistic 
quantum field $\phi_{\alpha}(x)$ that creates such state, transforms linearly under the Lorentz group $SO(3,1)$ as 
\begin{eqnarray}
U(\Lambda)^{-1}\phi_{\alpha}(x)U(\Lambda) = D_{\alpha}^{\beta}(\Lambda) \phi_{\beta}(\Lambda^{-1}x),
\end{eqnarray}
with $D(\Lambda)$ representing a finite-dimensional representation of the Lorentz group.

However,  the non-relativistic fields transform nonlinearly under the Lorentz boost due to the hidden Lorentz symmetry. As an illustrative example, consider the non-relativistic spinor field $N_I(x)$, which carries $SU(2)$ spin index $i$ and lacks anti-particle degrees of freedom~\cite{Li:2025ejk}. It is defined by
\begin{eqnarray}\label{eq:Ndef}
    N_{i}(x)=\int \frac{d^3\vec{k}}{(2\pi)^3\sqrt{2E}}\sum_{\sigma}u_{i}^{\sigma}(k)\,a_{v,k}^{\sigma}\,e^{-\mathbf{i} k\cdot x},
\end{eqnarray}
where $E = \sqrt{m^2 + \vec{k}^2}$ and $u_i^\sigma(k)$ are the non-relativistic spinor wavefunctions. Under an infinitesimal Lorentz boost $\Lambda = B(q) = \exp\!\bigl[\mathbf{i}  J_{\alpha\beta} q^{\alpha} v^{\beta} \theta\bigr]$, with $J_{\alpha\beta}$ the Lorentz generators, $v^\mu$ a timelike unit vector ($v^2 = 1$), and $\theta = \sinh^{-1}\!\bigl(\sqrt{[(v - q)\cdot v]^2 - 1}\,\bigr) / \sqrt{[(v - q)\cdot v]^2 - 1}$ for infinitesimal boost parameter $q^\alpha$, the field transforms nonlinearly as
\begin{eqnarray}
    U(B(q))^{-1} N_i(x) U(B(q)) = e^{\mathbf{i} \vec{q}\cdot\vec{x}} \left(1 + \frac{\vec{q}}{m} \cdot \vec{\mathcal{K}} \right)_{i}^{\;j} N_j(x'),
\end{eqnarray}
where $\vec{\mathcal{K}} = -\mathbf{i} \,\frac{\vec{\sigma}}{2}\, f(\nabla)$. The operator-valued function $f(\nabla)$ encodes the influence of the eliminated anti-particle modes and involve spatial derivatives as well as terms proportional to the gauge field strength, reflecting possible interactions with external gauge fields.

\paragraph{Galilean symmetry}
The Galilean group $\text{Gal}(3)$ is utilized in describing the spacetime symmetry of the non-relativistic quantum mechanics, where the velocities of particles are infinitesimal comparing to the speed of light. This group preserves the Newtonian absolute time and Euclidean space. Its Lie algebra is given by:
\begin{equation}\label{eq:galileanalgebra}
    \begin{aligned}
        [\hat J^I, \hat J^J] &= \mathbf{i} \epsilon^{IJK} \hat J^K, &
        [\hat J^I, \hat P^J] &= \mathbf{i} \epsilon^{IJK} \hat P^K, &
        [\hat P^I, \hat P^J] &= [\hat P^I, \hat H] = [\hat J^I, \hat H] = 0, \\[4pt]
        [\hat J^I, \hat B^J] &= \mathbf{i} \epsilon^{IJK} \hat B^K, &
        [\hat B^I, \hat P^J] &= \mathbf{i} m \delta^{IJ}, &
        [\hat B^I, \hat H] &= \mathbf{i} \hat P^I, \quad [\hat B^I, \hat B^J] = 0,
    \end{aligned}
\end{equation}
where $\hat J^I$, $\hat B^I$, $\hat P^I$, and $\hat H$ are the generators of rotations, Galilean boosts, spatial translations, and time translation, respectively. The central charge $m$, obtained from the central extension~\cite{LEVYLEBLOND1971221}, commutes with all other generators. Defining the coordinate operator as $\hat X^I \equiv \frac{1}{m} \hat B^I$, we recover the canonical commutation relation:
\begin{equation}
    [\hat X^I, \hat P^J] = \mathbf{i} \delta^{IJ}.
\end{equation}

The algebra in Eq.~\eqref{eq:galileanalgebra} can be obtained from the Poincaré algebra via Inönü–Wigner contraction~\cite{Inonu:1953sp}. In this construction, a contracted group $G'$ arises from an original group $G$ by keeping a subgroup $H$ unchanged while scaling the remaining generators so that they form an Abelian invariant subgroup $V$ in the limit. This is implemented as a one-parameter transformation labelled by $\epsilon$.  
For the contraction from the Poincaré group to the Galilean group, the parameter $\epsilon$ corresponds to reinstating the speed of light $c$ in the Poincaré algebra:
\begin{equation}\label{eq:poincarealgebra}
\begin{array}{llll}
     \lbrack\hat J^I,\hat J^J\rbrack=\mathbf{i}\epsilon^{IJK}J^K,& 
     \lbrack\hat J^I,\hat P^J\rbrack=\mathbf{i}\epsilon^{IJK}\hat P^K,&[\hat P^I,\hat P^J]=[\hat P^I,\hat P^0]=[\hat J^I,\hat P^0]=0,
      \\
     \\  
        \lbrack\hat J^I,\hat K^J\rbrack=\mathbf{i}\epsilon^{IJK}K^K,  &  
          \lbrack\hat K^I,\hat  P^J\rbrack=\frac{1}{c}\mathbf{i}\hat P^0\delta^{IJ},&\lbrack\hat K^I,\hat P^0c\rbrack=\mathbf{i}\hat P^I,\quad \lbrack\hat K^I,\hat K^J\rbrack=-\frac{1}{c^2}\mathbf{i}\epsilon^{IJK}J^I,
\end{array}    
\end{equation}
where $\hat P^0 c \equiv E = m c^2 + \hat H$. Taking the limit $c \to \infty$ yields
\begin{align}\label{eq:contraction1}
\bigl[\hat K^I,\hat P^J\bigr] &= \frac{1}{c}\,\mathbf{i}\hat P^0\,\delta^{IJ}
\;\longrightarrow\; \bigl[\hat B^I,\hat P^J\bigr] = \mathbf{i} m \delta^{IJ}, \nonumber \\
\bigl[\hat K^I,\hat K^J\bigr] &= -\frac{1}{c^2}\,\mathbf{i}\epsilon^{IJK} \hat J^I
\;\longrightarrow\; \bigl[\hat B^I,\hat B^J\bigr] = 0. 
\end{align}
Thus the Poincaré algebra Eq.~\eqref{eq:poincarealgebra} reduces to the Galilean algebra Eq.~\eqref{eq:galileanalgebra}. In this contraction the subgroup $H$ is the rotation group $SO(3)$, while the Abelian invariant subgroup $V$ is generated by the Galilean-boost operators $\hat B^I$, which satisfy $\bigl[\hat B^I,\hat B^J\bigr]=0$.

Physically, the contraction shows that the Galilean spacetime emerges as the strict non‑relativistic limit ($c\to\infty$) of the Minkowski spacetime. In this limit the Poincar\'{e} group translate into the Galilean group, whose algebra encodes the familiar symmetries of the non‑relativistic quantum mechanics.
The same contraction can also be understood from the power-counting perspective of the non-relativistic velocity in the literature~\cite{Weinberg:1995mt}. Consider a particle of mass $m$ and velocity $|\vec v| \ll 1$. Its momentum scales as $|\vec P| \sim m|\vec v|$, while the canonical commutation relation $|\vec P||\vec X|\sim|\vec J| \sim 1$ implies $|\vec X| \sim 1/(m|\vec v|)$, and the Lorentz‑boost generator scales as $|\vec K| \sim m|\vec X| \sim 1/|\vec v|$. The total energy is $P^0 \sim m + m|\vec v|^2$.
Keeping terms up to $\mathcal{O}(|\vec v|)$, the Poincar\'{e} commutation relations for the boosts reduce to
\begin{align}\label{eq:contraction2}
\bigl[\hat K^I,\hat P^J\bigr] &= \mathbf{i}\hat P^0\delta^{IJ}
\;\longrightarrow\; \mathbf{i}m\delta^{IJ}, \nonumber\\
\bigl[\hat K^I,\hat K^J\bigr] &= -\mathbf{i}\epsilon^{IJK}\hat J^I
\;\longrightarrow\; 0 .
\end{align}
Thus, in the non‑relativistic limit the Lorentz boosts $\hat K^I$ contract to the Galilean boosts $\hat B^I$, reproducing the algebra of Eq.~\eqref{eq:galileanalgebra}. The two descriptions in Eq.~\eqref{eq:contraction1} and Eq.~\eqref{eq:contraction2} coincide, as both employ the non-relativistic expansion  $|\vec{v}|/c \ll 1$. This condition is satisfied in two equivalent ways: either $|\vec{v}|$ is fixed while $c \to \infty$, or $c = 1$ is kept constant while $|\vec{v}| \ll 1$.

The non-relativistic field $N(x)$ defined in Eq.~\eqref{eq:Ndef} transforms linearly under Galilean boosts generated by $e^{-\mathbf{i} \boldsymbol{v} \cdot \hat{\boldsymbol{B}}}$, where $\boldsymbol{v}$ denotes the boost velocity. Specifically, its transformation law reads
\begin{eqnarray}\label{eq:galileanboostofN}
    N(x) \longrightarrow e^{-\mathbf{i}  f(x')} N(x'),
\end{eqnarray}
with the phase factor given by $f(x) = m\, \boldsymbol{v} \cdot \boldsymbol{x} - \frac{1}{2} m\, \boldsymbol{v}^2 t$, and the spacetime coordinate denoted as $x = (\boldsymbol{x}, t)$. Under the full Galilean group, this field creates particle states in configuration space, denoted $|\boldsymbol{x}, t, \sigma\rangle$~\cite{10.1088/978-1-6270-5624-3,Dawson2009QuantumMF}.
In particular, for a multi-particle scattering state, the  Galilean boost depends on the center-of-mass (COM) coordinate,
\begin{eqnarray}
    \boldsymbol{X}_{\text{COM}} \equiv \frac{1}{M} \sum_i m_i \boldsymbol{x}_i,
\end{eqnarray}
where $M = \sum_i m_i$ is the total mass. The incoming (or outgoing) state transforms as
\begin{eqnarray}
    |\Psi_{\text{in}}\rangle = \bigotimes_i |\boldsymbol{x}_i, t, \sigma_i\rangle 
    \;\longrightarrow\;
    |\Psi'_{\text{in}}\rangle = e^{-\mathbf{i}  \left( M \boldsymbol{v} \cdot \boldsymbol{X}_{\text{COM}} + \frac{1}{2} M \boldsymbol{v}^2 t \right)} \bigotimes_i |\boldsymbol{x}'_i, t, \sigma'_i\rangle.
\end{eqnarray}
For interactions and scattering amplitudes that respect Galilean invariance, the matrix element must remain unchanged under such a boost. This requirement imposes the constraint
\begin{eqnarray}\label{eq:galileanboostconstrait}
    \langle\Psi_{\text{out}}|\mathcal{O}|\Psi_{\text{in}}\rangle 
    &\longrightarrow& 
    \langle\Psi_{\text{out}}| e^{\mathbf{i}  \left( \boldsymbol{v} \cdot \boldsymbol{X}_{\text{COM}} + \frac{1}{2} M\boldsymbol{v}^2 t \right)} \,
    \mathcal{O} \,
    e^{-\mathbf{i}  \left( M \boldsymbol{v} \cdot \boldsymbol{X}_{\text{COM}} + \frac{1}{2} M \boldsymbol{v}^2 t \right)} 
    |\Psi_{\text{in}}\rangle \nonumber\\
    &=& \langle\Psi_{\text{out}}|\mathcal{O}|\Psi_{\text{in}}\rangle.
\end{eqnarray}
This equality holds only when the interaction operator $\mathcal{O}$  is independent of the COM momentum $\boldsymbol{P}_{\text{COM}}$ in momentum space.
Note that, the momentum of each particle can be decomposed as
\begin{eqnarray}\label{eq:ppcmpr}
    \boldsymbol{p}_i = \boldsymbol{P}_{\text{COM}} + \boldsymbol{p}_{\text{r},i},
\end{eqnarray}
where $\boldsymbol{p}_{\text{r},i}$ denotes the relative momentum of particle $i$ with respect to the COM frame, satisfying $\sum_i \boldsymbol{p}_{\text{r},i} = 0$. Therefore, due to Eq.~\eqref{eq:galileanboostconstrait} and Eq.~\eqref{eq:ppcmpr}, any Galilean-invariant scattering amplitude can depend only on the relative momenta $\boldsymbol{p}_{\text{r}}$, and not on the overall motion of the system.

\paragraph{Euclidean Symmetry}
The construction of an EFT begins with identifying the underlying physical symmetries relevant at the energy scale of interest. We consider the three-dimensional Euclidean group $\text{E}(3)$, which is the common subgroup of the Poincar\'{e} group and the Galilean group. For quantum systems, the existence of a Hamiltonian implies the time translation symmetry, described by the group $\mathbb{R}^1$, which corresponds to the energy conservation. In addition, the spacetime symmetries are also  described by the Euclidean group $\mathrm{E}(3)=\mathbb{R}^{3}\rtimes SO(3)$ together with an appropriate realization of boosts. For spacetime coordinates $(\boldsymbol{x},t)$, the transformations of the Euclidean group are represented by the matrices:
\begin{eqnarray}
    \text{E}(3)=\left\{T=\left\lbrack\begin{array}{cc}
         R& \boldsymbol{r} \\
        0 & 1
    \end{array}\right\rbrack\in\mathbb{R}^{4\times4}|R\in SO(3),\boldsymbol{r}\in \mathbb{R}^3\right\}
\end{eqnarray}
The non-relativistic $N(x)$ in Eq.~\eqref{eq:Ndef} also transforms linearly under the $\text{E}(3)$ group. Under the rotation $R\in SO(3)$, we have the transformation
\begin{eqnarray}
    U(R)^{-1}N(x)U(R)=D(R)N(R^{-1}x).
\end{eqnarray}
The non-relativistic field $N(x)$ is also selected as the relevant degree of freedom.

In the non-relativistic effective theory, the momentum of each particle can be decomposed into the COM and relative components,
\begin{eqnarray}\label{eq:ppCMpr}
\underbrace{\boldsymbol{p}}_{\text{Euclidean Invariant}} = \boldsymbol{P}_{\text{COM}} + \underbrace{\boldsymbol{p}_{\text{r}}}_{\text{Galilean Invariant}}.
\end{eqnarray}
To retain the COM momentum dependence $\boldsymbol{P}_{\text{COM}}$ arising from relativistic corrections, the construction of the interaction operator $\mathcal{O}$ must exclude covariance under the Galilean boosts Eq.~\eqref{eq:galileanboostofN}. Subsequently, the relevant symmetry reduces to the three-dimensional Euclidean group $\text{E}(3)$ together with time translations $\mathbb{R}^1$,
\begin{eqnarray}
\text{Gal}(3) \xrightarrow{\text{without Galilean boosts}} \mathbb{R}^1 \times \text{E}(3).
\end{eqnarray}
On the contrary, an effective theory formulated under the $\mathbb{R}^1 \times \text{E}(3)$ symmetry restores full Galilean invariance $\text{Gal}(3)$ in the COM frame,
\begin{eqnarray}
\boldsymbol{P}_{\text{COM}} = 0,
\end{eqnarray}
where the COM coordinate and momentum dependence in Eq.~\eqref{eq:ppCMpr} vanishes, leaving only relative motion that is invariant under Galilean boosts Eq.~\eqref{eq:galileanboostofN} and satisfies the constraints imposed by the boost transformation Eq.~\eqref{eq:galileanboostconstrait}.

The Euclidean group itself consists of rotations and spatial translations, while the boost transformations are required to relate reference frames moving with different velocities.  
When supplemented with Lorentz boosts, which act as
\begin{eqnarray}\label{eq:lorentzboost}
    (\boldsymbol{x},t)\mapsto \left(\boldsymbol{x}+\left[\frac{\boldsymbol{\beta}\cdot\boldsymbol{x}}{|\boldsymbol{\beta}|^2}(\gamma-1)-\gamma t\right]\boldsymbol{\beta}, \gamma\left[ t-\boldsymbol{\beta}\cdot \boldsymbol{x}\right]\right),
\end{eqnarray}
where $\boldsymbol{\beta}=\boldsymbol{v}/c$ and $\gamma=1/\sqrt{1-\boldsymbol{\beta}^2}$, the full symmetry group becomes the Poincar\'{e} group, governing relativistic systems. Alternatively, when supplemented with Galilean boosts, which act as
\begin{eqnarray}\label{eq:galileanboost}
(\boldsymbol{x},t)\mapsto(\boldsymbol{x}-\boldsymbol{v}t,t),
\end{eqnarray}
the symmetry group becomes the Galilean group, which describes the non‑relativistic quantum mechanics. Note that the Lorentz boost Eq.~\eqref{eq:lorentzboost} reduces to the Galilean boost Eq.~\eqref{eq:galileanboost} in the limit $c\rightarrow\infty$. The Euclidean group $\mathrm{E}(3)$ constitutes the common subgroup shared by both the Poincar\'{e} and Galilean groups when their distinct boost transformations Eqs.~\eqref{eq:lorentzboost} and \eqref{eq:galileanboost} are disregarded.

To construct the non‑relativistic operators, 
the symmetry $\mathrm{E}(3)$ (or $\mathbb{R}^{1}\times\mathrm{E}(3)$) can be used directly, as the boost symmetry is nonlinearly realized in the low‑energy description. The hidden boost symmetry  only impose relations among Wilson coefficients of different operators, and then the Euclidean subgroup alone is sufficient for organizing the operator basis. 
Furthermore, at the level of local operators, translations are implemented as shifts in the spacetime coordinates. Spacetime translations $\mathbb{R}^{3,1}=\mathbb{R}^{3}\times\mathbb{R}^{1}$, with parameters $\boldsymbol{r}\in\mathbb{R}^{3}$ and $\tau\in\mathbb{R}^{1}$, act as
\begin{eqnarray}
    (\boldsymbol{x},t) &\longmapsto& (\boldsymbol{x}+\boldsymbol{r},t),\nonumber\\
    (\boldsymbol{x},t) &\longmapsto& (\boldsymbol{x},t+\tau).
\end{eqnarray}
The invariance under translations manifests as delta functions enforcing the conservation of energy and momentum,
\begin{eqnarray}
    \mathbb{R}^1\times \text{E}(3)\rightarrow SO(3) + \text{energy and momentum conservation}. 
\end{eqnarray}
Consequently, when constructing the operator basis, one only needs to consider the homogeneous part of the symmetry group.
For example, in the relativistic quantum field theory with Poincaré symmetry $\mathbb{R}^{3,1}\rtimes SO(3,1)$, local fields transform under the Lorentz group $SO(3,1)$. Similarly, in the non‑relativistic case with Euclidean symmetry $\text{E}(3)=\mathbb{R}^3\rtimes SO(3)$, we focus on the homogeneous rotational symmetry, where the indices of local fields transform according to the irreducible representations of the rotation group $SO(3)$. Rotations are represented by the set of three‑dimensional special orthogonal matrices
\begin{eqnarray}
    SO(3)=\left\{R\in\mathbb{R}^{3\times3}\;\big|\;RR^{T}=\mathbf{I}_{3\times3},\;\det(R)=1 \right\},
\end{eqnarray}
so that the coordinate transforms as
\begin{eqnarray}
    (\boldsymbol{x},t)\;\longmapsto\;(R\boldsymbol{x},t).
\end{eqnarray}

\begin{figure}
    \centering
    \includegraphics[width=13cm, height=8.5cm]{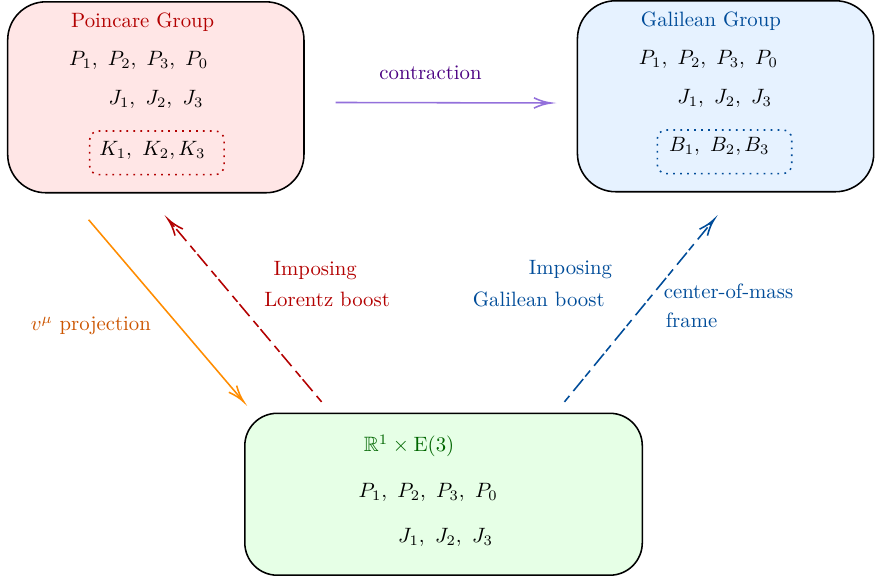}
    \caption{This figure depicts the relationships among the Poincar\'{e} group, the Galilean group, and the group $\mathbb{R}^1\times \text{E}(3)$. The Galilean group is obtained from the Poincar\'{e} group via an In\"{o}n\"{u}-Wigner contraction, while the subgroup $\mathbb{R}^1\times \text{E}(3)$ arises from the Poincar\'{e} group through velocity projection. By imposing the corresponding boost invariance on the rotation group, the full Poincar\'{e} or Galilean symmetry can be restored.}
    \label{fig:3group}
\end{figure}

In summary, the Euclidean group is a subgroup common to both the Poincar\'{e} group and the Galilean group. 
Through the projection along the velocity vector $v^\mu$, the Poincar\'{e} group is spontaneously broken to the Euclidean group  $\mathbb{R}^1\times\text{E}(3)$.\footnote{In what follows, for brevity we shall refer to the symmetry group $\mathbb{R}^1 \times \mathrm{E}(3)$ simply as the Euclidean symmetry, by neglecting the time translation, which is always applied as the Hamiltonian in quantum system. }
By imposing either a Galilean boost or a Lorentz boost, one can restore the corresponding spacetime symmetry: the Galilean symmetry or the Lorentz symmetry, respectively. 
Furthermore, imposing the Galilean boost invariance is equivalent to adopting the COM frame. We summarize the relations between these three group in Fig.~\ref{fig:3group}. We construct the non-relativistic effective operator bases under the $SO(3)$ rotation group.

\subsection{Rotation and its Double Cover}\label{symmetry}

As it is discussed in the previous subsection, to construct the non-relativistic effective operator bases, we begin by examining the homogeneous symmetry group—the rotation group, without knowing the explicit boost transformation.

The Stern–Gerlach experiment reveals a spin-dependent coupling of the form
\begin{equation}\label{eq:sdcxiB}
\Delta \mathcal{L} \sim \xi^{\dagger} \xi  \boldsymbol{B},
\end{equation}
where $\xi$ and $\xi^{\dagger}$ represent the non-relativistic incoming and outgoing fermion fields, respectively, and $\boldsymbol{B}$ denotes the magnetic field. The boldface notation denotes that $\boldsymbol{B}$ transforms as a three-vector under the spatial rotations. Explicitly, in terms of the $SO(3)$ index $I$, the components are written as $B^I$.

To incorporate such an interaction in a rotationally invariant Lagrangian, one can naively consider two distinct symmetry groups: the spatial rotation group $SO(3)$, or a direct product $SO(3) \times SU(2)$ accounting separately for spatial and spin rotations. However, analysis shows that only a single rotation group is physically relevant. This result originates from the fact that the spin degrees of freedom themselves transform under a spinor representation of the rotation group, whose double cover is the spin group $Spin(3)=SU(2)$, and that the $SO(3)$ vector $\boldsymbol{B}$ can be converted into the representations of the $SU(2)$. Thus, the apparent distinction between spin and spatial rotations is resolved within the $SU(2)$ spin group description, which fully captures both types of transformations.

\paragraph{1) $SO(3)$ Group}

Under this symmetry, the magnetic field $\boldsymbol{B}$ and the spinor bilinear $\xi^{\dagger}\xi$ both transform as three-dimensional representations
\begin{equation}\label{oneso3}
\begin{array}{ll}
 \boldsymbol{B} & \in (\mathbf{3}), \\
 \xi^{\dagger}\xi & \in (\mathbf{3}),
\end{array}
\end{equation}
but the exact contraction of the indices in Eq.~\eqref{eq:sdcxiB} remains undetermined.

The Hilbert space of a non-relativistic particle decomposes as $\mathcal{H}_{|\boldsymbol{k}|} \otimes \mathbf{C}^{2s+1}$, where $\mathcal{H}_{|\boldsymbol{k}|}$ contains states with fixed momentum magnitude and $\mathbf{C}^{2s+1}$ is the $(2s+1)$-dimensional spin space. The total angular momentum operator is correspondingly expressed as
\begin{equation}\label{eq:totalj=l+s}
\hat{\boldsymbol{J}} = \hat{\boldsymbol{L}} + \hat{\boldsymbol{S}} = \hat{\boldsymbol{L}} \otimes \mathbf{1} + \mathbf{1} \otimes \hat{\boldsymbol{S}},
\end{equation}
with $\hat{\boldsymbol{L}}$ and $\hat{\boldsymbol{S}}$ acting on the momentum and spin subspaces respectively.
Although the orbital and spin angular momentum operators commute, $[\hat{\boldsymbol{L}}, \hat{\boldsymbol{S}}] = 0$, physical rotations are generated by the total angular momentum $\hat{\boldsymbol{J}}$. Under a rotation by angle $\boldsymbol{\vartheta}$, the fields transform as
\begin{equation}\label{eq:Bxirotation}
\begin{cases}
    \boldsymbol{B} \xrightarrow{e^{-i\hat{\boldsymbol{J}}\cdot\boldsymbol{\vartheta}}} R(\boldsymbol{\vartheta})\boldsymbol{B}, \\[8pt]
    \xi \xrightarrow{e^{-i\hat{\boldsymbol{J}}\cdot\boldsymbol{\vartheta}}} u(\boldsymbol{\vartheta})\xi,
\end{cases}
\end{equation}
where $R(\boldsymbol{\vartheta})$ and $u(\boldsymbol{\vartheta})$ are the vector and spinor representations of the same rotation.

To construct a rotationally invariant interaction from $\xi^{\dagger}\xi$ and $\boldsymbol{B}$, we require an object to couple the two representations into a scalar. This is achieved through the Pauli matrices
\begin{equation}
    \boldsymbol{\Sigma} \equiv \frac{\boldsymbol{\sigma}}{2},
\end{equation}
which satisfy the following relation under the rotation
\begin{equation}\label{eq:uRrelation}
u^{-1}(\boldsymbol{\vartheta})~\boldsymbol{\Sigma}~u(\boldsymbol{\vartheta})=R(\boldsymbol{\vartheta})~\boldsymbol{\Sigma}.
\end{equation}
The Pauli matrices serve as the Clebsch-Gordan coefficients (CGC) for the reduction $(\mathbf{3}) \otimes (\mathbf{3}) \rightarrow (\mathbf{1})$:
\begin{equation}
    \underbrace{\left( \xi^{\dagger}\xi \otimes \boldsymbol{B} \right)}_{(\mathbf{3}) \otimes (\mathbf{3})} \times \underbrace{\boldsymbol{\Sigma}}_{\text{CGC}} \rightarrow \underbrace{\xi^{\dagger} \boldsymbol{\Sigma} \xi \cdot \boldsymbol{B}}_{(\mathbf{1})}\ .
\end{equation}
Utilizing the transformation in Eq.~\eqref{eq:Bxirotation} and the relation Eq.~\eqref{eq:uRrelation}, the above result becomes a singlet under the rotation generated by the total angular momentum $\hat{\boldsymbol{J}} $, using a common rotation angle $\boldsymbol{\vartheta}$,
\begin{equation}
    \xi^{\dagger}\boldsymbol{\Sigma}\xi\cdot\boldsymbol{B}\quad\xrightarrow{e^{-i\hat{\boldsymbol{J}}\cdot\boldsymbol{\vartheta}}}\quad \left(R(\boldsymbol{\vartheta})\xi^{\dagger}\boldsymbol{\Sigma}\xi\right)\cdot \left(R(\boldsymbol{\vartheta})\boldsymbol{B}\right) = \xi^{\dagger}\boldsymbol{\Sigma}\xi\cdot\boldsymbol{B}.
\end{equation}
As a consequence, we obtain the rotationally invariant interaction term
\begin{equation}\label{eq:deltaLSB}
    \Delta\mathcal{L} = \frac{c}{m}~\xi^{\dagger} \boldsymbol{\Sigma} \xi \cdot \boldsymbol{B},
\end{equation}
where $m$ is the mass of the fermion $\xi$, and $c$ is an undetermined dimensionless constant. This formulation demonstrates that a single rotation group suffices to describe both spatial and spin transformations. In practice, we adopt the spin group $Spin(3) = SU(2)$ in the following subsection to derive the invariants.

\paragraph{2) Direct Product Group $SO(3)\times SU(2)$}

Suppose the symmetry of the system is given by $SO(3) \times SU(2)$, where the $SO(3)$ corresponds to the spatial rotations and the $SU(2)$ to the intrinsic spin rotations, thus we denote them as the $SO(3)_{\text{spatial}}$ and the $SU(2)_{\text{spin}}$ group. This symmetry structure emerges  in systems where the spin and orbital degrees of freedom decouple in the infinite mass limit $m\rightarrow\infty$.
In the context of heavy quark physics, the intrinsic spin rotation group $SU(2)_{\text{spin}}$ is known as the heavy quark spin symmetry~\cite{Isgur:1990yhj}. This symmetry manifests when the quark mass becomes infinitely large, and then the spin-dependent interactions vanish.  In the infinite mass limit, the $SU(2)_{\text{spin}}$ symmetry implies that rotating the spin state of a heavy quark, such as the bottom quark $b$, leaves the physics invariant. Consequently, the bottom mesons with different total angular momenta---the pseudoscalar $B$ meson ($J=0$) and the vector $B^{*}$ meson ($J=1$)---can be transformed into one another under such rotations, resulting in the degenerate masses.

However, the spin-dependent interaction term explicitly breaks the full $SO(3)_{\text{spatial}} \times SU(2)_{\text{spin}}$ symmetry. At order $1/m$, the Lagrangian including a spin-dependent interaction takes the form
\begin{equation}\label{eq:Lso3xsu2}
    \mathcal{L} = \underbrace{\xi^{\dagger}\mathbf{i}\partial_t\xi + \frac{1}{2m}\xi^{\dagger}\nabla^2\xi}_{\text{spin-independent}} + \frac{c}{m}\xi^{\dagger}\xi\boldsymbol{B}.
\end{equation}
Here $c$ is the undetermined dimensionless constant for the magnetic moment coupling. This breaking occurs at order $1/m$ and introduces mass splittings between otherwise degenerate states, such as the $B$--$B^*$ mass difference. The presence of the spin-dependent term reduces the symmetry from the direct product group to the diagonal subgroup where spatial and spin rotations are identified:
    \begin{equation}
    SO(3)_{\text{spatial}} \times SU(2)_{\text{spin}}\longrightarrow SU(2)_{\text{spin}}.
\end{equation}
Here the $SU(2)_{\text{spin}}$ becomes the double cover of the $SO(3)_{\text{spatial}}$. This symmetry-breaking pattern reveals the emergence of spin-orbit coupling. In the infinite mass limit, the Hamiltonian commutes with both the spin and orbital angular momentum operators, $[H,\boldsymbol{S}]=0$ and $[H,\boldsymbol{L}]=0$, separately. However, once the magnetic moment interaction in Eq.~\eqref{eq:Lso3xsu2} is introduced, neither $\boldsymbol{S}$ nor $\boldsymbol{L}$ is conserved, as $[H,\boldsymbol{S}] \neq 0$ and $[H,\boldsymbol{L}] \neq 0$. The conserved quantity becomes the total angular momentum $\boldsymbol{J}$, with $[H,\boldsymbol{J}]=0$. The symmetry  thus breaks to the diagonal subgroup, under which the spin and orbital degrees of freedom transform simultaneously.

To  analyze the symmetry breaking pattern, we classify  the building blocks according to their transformation properties under the direct product group $SO(3)_{\text{spatial}} \times SU(2)_{\text{spin}}$, denoting representations as $(2l+1, 2s+1)$:
\begin{equation}
    \begin{array}{lll}
        \partial_t & \in & (1,1), \\
        \boldsymbol{\nabla} & \in & (3,1), \\
        \boldsymbol{B} & \in & (3,1), \\
        \xi & \in & (1,2), \\
        \xi^{\dagger} & \in & (1,2).
    \end{array}
\end{equation}
Under this classification, the kinetic term in Eq.~\eqref{eq:Lso3xsu2} transforms as a singlet
\begin{equation}
    \mathcal{L}^{(1,1)} \equiv \xi^{\dagger}\mathbf{i}\partial_t\xi + \frac{1}{2m}\xi^{\dagger}\nabla^2\xi \subset SO(3)_{\text{spatial}} \times SU(2)_{\text{spin}},
\end{equation}
whereas the magnetic moment interaction term does not form a singlet representation
\begin{equation}\label{eq:deltaL33}
    \mathcal{L}^{(3,3)} \equiv \frac{1}{m}\xi^{\dagger}\xi\boldsymbol{B} \subset \!\!\!\!\!/~\quad SO(3)_{\text{spatial}} \times SU(2)_{\text{spin}},
\end{equation}
which manifests the explicit symmetry breaking induced by spin-dependent term.

To maintain the original symmetry, a spurion in the representation $(3, 3)$ must be introduced such that Eq.~\eqref{eq:deltaL33} are converted into a singlet.  The spurion technique allows us to deal with the symmetry-breaking effects systematically while restoring the manifest invariance under the full symmetry group. Thus we define the spurion---spin operator $\boldsymbol{\Sigma}$ as
\begin{equation}
    \boldsymbol{\Sigma} \in (3,3).
\end{equation}
Referring to Tab.~\ref{tab:spuriontable}, the singlet interaction is constructed by
\begin{equation}
    (\Sigma^I)_{ij} \otimes \mathcal{L}^{(3,3)} \subset SO(3)_{\text{spatial}} \times SU(2)_{\text{spin}},
\end{equation}
where $I$ denotes the $SO(3)_{\text{spatial}}$ index and $i,j$ are the $SU(2)_{\text{spin}}$ indices. The full Lagrangian respecting $SO(3)_{\text{spatial}} \times SU(2)_{\text{spin}}$ symmetry then becomes:
\begin{equation}
    \mathcal{L}^{\prime} = \mathcal{L}^{(1,1)} + (\Sigma^I)_{ij} \otimes \mathcal{L}^{(3,3)},
\end{equation}
and the interaction term is written by
\begin{equation}\label{eq:DLso3su2}
    \Delta\mathcal{L}=(\Sigma^I)_{ij} \otimes \mathcal{L}^{(3,3)}=\frac{c}{m}~\xi^{\dagger} \boldsymbol{\Sigma} \xi \cdot \boldsymbol{B},
\end{equation}
which is the same as Eq.~\eqref{eq:deltaLSB} in the one $SO(3)$ group case.

\begin{table}[H]
    \centering
    \footnotesize
    \begin{tabular}{|c|c|c|}
    \hline
         Building blocks & $SO(3)_{\text{spatial}}\times SU(2)_{\text{spin}}$ & $U(1)_{\text{EM}}$ \\
         \hline
         $\partial_t$ & $(1,1)$ & 0 \\
         \hline
         $\nabla^I$ & $(3,1)$ & 0 \\
         \hline
         $B^I$ & $(3,1)$ & 0 \\
         \hline
         $\xi^{\dagger}_i$ & $(1,2)$ & $-1$ \\
         \hline
         $\xi_i$ & $(1,2)$ & $1$ \\
         \hline
         $(\Sigma^I)_{ij}$ & $(3,3)$ & 0 \\
         \hline
    \end{tabular}
    \caption{Representations of the building blocks under the direct product group $SO(3)_{\text{spatial}}\times SU(2)_{\text{spin}}$. Here the $\Sigma$ is the spurion. The last column shows the electromagnetic charge under $U(1)_{\text{EM}}$.}
    \label{tab:spuriontable}
\end{table}

However, the spurion $\boldsymbol{\Sigma}$ is not a dynamic variable and does not transform under the rotation $SO(3)_{\text{spatial}}\times SU(2)_{\text{spin}}$. In the following we examine how the invariance of the interaction term in Eq.~\eqref{eq:DLso3su2} is restored.

\begin{figure}[ht]
    \centering
    \includegraphics[width=0.5\linewidth]{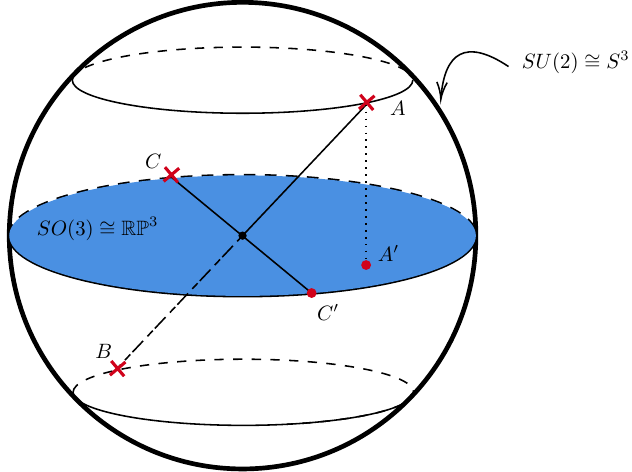}
    \caption{Schematic illustration of the manifold structures of $SU(2)$ and $SO(3)$ groups using their two-dimensional analogues. The $SO(3)$ group manifold is the real projective space $\mathbb{RP}^3$, obtained from the $S^3$ manifold of $SU(2)$ by identifying antipodal points. Although the actual manifolds are three-dimensional, this two-dimensional representation using $S^2$ and $\mathbb{RP}^2$ symbolically demonstrates the antipodal identification process: the blue region represents the resulting projective space after identifying opposite points on the sphere. Points $A$ and $B$ (antipodal on the sphere) are identified and projected to $A'$ in the projective region; similarly, $C$ and $C'$ are identified as $C'$ in the projective region.}
    \label{fig:sphere}
\end{figure}

A rotation $R(\vec n_{\phi},\phi)\in SO(3)_{\text{spin}}$ is induced by the element $u(\vec n_{\phi},\phi)\in SU(2)_{\text{spin}}$ through the adjoint action

\begin{eqnarray}
     &u^{-1}(\vec n_{\phi},\phi)~\sigma^{I'} ~u(\vec n_{\phi},\phi)&=R^{I'J'}(\vec n_{\phi},\phi)~\sigma^{J'},
\end{eqnarray}
and then we obtain the rotation from the double cover map
\begin{eqnarray}\label{eq:su2spinso3spin}
    R^{I'J'}(\vec n_{\phi},\phi)=\frac{1}{2}\text{tr}\left[u^{-1}(\vec n_{\phi},\phi)~\sigma^{I'} ~u(\vec n_{\phi},\phi)\sigma^{J'}\right].
\end{eqnarray}
Thus, we introduce the new group $SO(3)_{\text{spin}}$, and the above equation establishes a 2:1 homomorphism between $SU(2)_{\text{spin}}$ and $SO(3)_{\text{spin}}$, with the kernel $\mathbb{Z}_2=\{\pm 1\}$ corresponding to the identification of antipodal points in the group manifolds.
The group manifold of the group manifold of $SU(2)_{\text{spin}}$ is the three-dimensional sphere $S^3$,  while the group manifold of the $SO(3)_{\text{spin}}$ is the real projective space $\mathbb{RP}^3$, which can be visualized as a three-dimensional sphere $S^3$ with antipodal points identified.  As in Fig.~\ref{fig:sphere}, the identification $\mathbb{RP}^3 \cong S^3/\mathbb{Z}_2$ describes this 2:1 homomorphism, where each rotation in physical space corresponds to two distinct elements in the spinor representation.

With the group elements $u(\vec n_{\phi},\phi)\in SU(2)_{\text{spin}}$ and $R(\vec n_{\theta},\theta)\in SO(3)_{\text{spatial}}$, and utilizing the relation Eq.~\eqref{eq:su2spinso3spin}, we obtain the $SO(3)_{\text{spatial}}\times SU(2)_{\text{spin}}$ transformation:
\begin{equation}\label{eq:so3su2trans}
    \begin{cases}
    \Sigma^I&\longrightarrow\Sigma^I,\\
    \xi^{\dagger}\sigma^{I'}\xi&\longrightarrow R^{I'J'}(\vec n_{\phi},\phi)~\xi^{\dagger}\sigma^{J'}\xi,\\
    B^I&\longrightarrow R^{IJ}(\vec n_{\theta},\theta)B^J.
    \end{cases}
\end{equation}
Under these transformations, the invariance of the magnetic moment coupling in Eq.~\eqref{eq:DLso3su2} imposes the constraint
\begin{equation}
    R^{I'J'}(\vec n_{\phi},\phi)R^{KL}(\vec n_{\theta},\theta)\delta^{I'K}=\delta^{J'L}.
\end{equation}
Utilizing the orthogonality $R^{I'J'}=(R^{J'I'})^{-1}$, we deduce
\begin{equation}\label{eq:rso3=rso3}
   R(\vec n_{\phi},\phi)=R(\vec n_{\theta},\theta),\quad \Longrightarrow \quad SO(3)_{\text{spin}}=SO(3)_{\text{spatial}}.
\end{equation}
This demonstrates the identification of the spatial and spin rotation groups in maintaining the symmetry.

Furthermore, for $SO(3)$ rotations, noting the identity $R(\vec n_{\theta},\theta)=R(-\vec n_{\theta},2\pi-\theta)$, we find two distinct solutions for Eq.~\eqref{eq:rso3=rso3} for the rotation parameters:
\begin{equation}
    \left\{\begin{array}{l}
         \vec n_{\phi}=\vec n_{\theta}  \\
        \phi=\theta 
    \end{array}\right.\quad \text{or}\quad \left\{\begin{array}{l}
         \vec n_{\phi}=-\vec n_{\theta}  \\
        \phi=2\pi-\theta 
    \end{array}\right. .
\end{equation}
In the $SU(2)_{\text{spin}}$ group, using the fundamental property $u(-\vec n_{\theta},2\pi-\theta)=-u(\vec n_{\phi},\phi)$, we then establish the 2:1 homomorphism:
\begin{equation}\label{eq:21homo}
     R(\vec n_{\theta},\theta)=R(\vec n_{\phi},\phi)=\left\{\begin{array}{r}
          u(\vec n_{\phi},\phi),  \\
          -u(\vec n_{\phi},\phi). 
     \end{array}\right. 
\end{equation}
The Pauli matrices in Eq.~\eqref{eq:so3su2trans} connect the rotation group and its double cover, playing the role of the spurion. As a result, the spin operator is identified with the Pauli matrices
\begin{equation}
    \boldsymbol{\Sigma}=\frac{\boldsymbol{\sigma}}{2},
\end{equation}
and Eq.~\eqref{eq:21homo} confirms the fundamental 2:1 homomorphism
\begin{equation}
SU(2)_{\text{spin}}=\text{double cover of }SO(3)_{\text{spatial}}.
\end{equation}
This analysis demonstrates how the spurion restores the rotational symmetry while revealing the  relationship between the spatial  and spin rotations.

\paragraph{Redundancy of the Spurion}
As a spurion, the spin operator $\boldsymbol{\Sigma}$ is in fact a constant matrix rather than a dynamical field depending on spacetime coordinates. It should therefore be normalized, and its products expanded in a suitable basis with constant coefficients.
Quantitatively, fundamental principles of quantum mechanics impose constraints on the spin operator. First, for any direction $\boldsymbol{n}$, the operator $\boldsymbol{n} \cdot \boldsymbol{\Sigma}$ has eigenvalues $\pm \frac{1}{2}$, leading to the normalization condition:
\begin{equation}\label{spineq1}
    (\boldsymbol{n} \cdot \boldsymbol{\Sigma})^2 = \frac{1}{4}.
\end{equation}
Second, the four operators $\{I_{2\times2}, \Sigma^I\}$ span the space of $2\times2$ matrices, implying that the product $\Sigma^I \Sigma^J$ can be expanded as
\begin{equation}\label{spineq2}
    \Sigma^I \Sigma^J = A^{IJ} + C^{IJK} \Sigma^K,
\end{equation}
where $A^{IJ}$ and $C^{IJK}$ are undetermined coefficients. 
Combining the conditions in Eqs.~\eqref{spineq1} and \eqref{spineq2}, we derive the redundancy relation for the spin operator in the tensor product group $SO(3) \times SU(2)$:
\begin{equation}\label{eq:redundacyeqspin}
    \Sigma^I \Sigma^J = \frac{1}{4} \delta^{IJ} + \frac{i}{2} \epsilon^{IJK} \Sigma^K.
\end{equation}
This relation uniquely determines the spin operator to be proportional to the Pauli matrices
\begin{equation}
    \boldsymbol{\Sigma} = \frac{\boldsymbol{\sigma}}{2}.
\end{equation}

The redundancy in Eq.~\eqref{eq:redundacyeqspin} impose a specific relation between the $SO(3)$ and the $SU(2)$ within the direct product group $SO(3) \times SU(2)$, revealing that only one rotation group is actually necessary to construct singlet operators. This conclusion aligns with the earlier perspective expressed in Eq.~\eqref{oneso3}. Since any self-contraction of $\boldsymbol{\Sigma}$ becomes redundant due to Eq.~\eqref{eq:redundacyeqspin}, all spin operators introduced in the Lagrangian must be totally symmetric. In terms of the $SU(2)$ tensor $T_{SU(2)}$ and the $SO(3)$ tensor $T_{SO(3)}$ the spin-$S$ operator can be explicitly written as
\begin{equation}\label{eq:sigmas2s+1}
   \boldsymbol{\Sigma}^S = T_{SU(2)}^{\{i_1\cdots i_{2S}\}} \otimes T_{SO(3)}^{\{I_1\cdots I_{S}\}}\in (2S+1, 2S+1), \quad S \in \mathbb{Z}^+.
\end{equation}
With the remaining part $\mathcal{L}^{(2n+1,2n+1)}$ constructed by other building blocks in Tab.~\ref{tab:spuriontable}, 
\begin{equation}\label{eq:l2s+12s+1}
    \mathcal{L}^{(2S+1,2S+1)} \equiv (T_A)_{SU(2)}^{i_1\cdots i_{2S}} \otimes (T_B)_{SO(3)}^{I_1\cdots I_{S}}\in (2S+1, 2S+1),
\end{equation}
the singlet interaction term is obtained by
\begin{equation}\label{eq:deltaLSL}
    \Delta\mathcal{L} = \boldsymbol{\Sigma}^S \otimes \mathcal{L}^{(2S+1,2S+1)} \subset SO(3) \times SU(2).
\end{equation}
Crucially, in the above equation, all $SO(3)$ indices in $T_B$ in Eq.~\eqref{eq:l2s+12s+1}  are effectively replaced with the totally symmetric $SU(2)$ indices,
\begin{equation}\label{eq:TbSU2}
   \boldsymbol{\Sigma}^S \otimes (T_B)_{SO(3)}^{I_1\cdots I_{S}}  ~\simeq~ (T_B)_{SU(2)}^{\{i_1\cdots i_{2S}\}}.
\end{equation}
Consequently, in Eq.~\eqref{eq:deltaLSL}, only the $SU(2)$ index contraction survives 
\begin{equation}\label{eq:deltalsl}
    \Delta\mathcal{L} = \boldsymbol{\Sigma}^{S} \otimes \mathcal{L}^{(2S+1,2S+1)} = (T_A)_{SU(2)}^{i_1\cdots i_{2S}} \otimes (T_B)_{SU(2)}^{\{i_1\cdots i_{2S}\}},
\end{equation}
demonstrating that a single $SU(2)$ (or equivalently $SO(3)$) group suffices, rather than the direct product $SO(3) \times SU(2)$.

This structure can also be understood diagrammatically through the Young tableaux. The Young tableau of the spin-$S$  operator $\boldsymbol{\Sigma}^S$ in Eq.~\eqref{eq:sigmas2s+1} corresponds to 
\begin{center}
\tikzset{every picture/.style={line width=0.75pt}} 

\begin{tikzpicture}[x=0.75pt,y=0.75pt,yscale=-1,xscale=1]

\draw  [color={rgb, 255:red, 0; green, 0; blue, 0 }  ,draw opacity=1 ] (100,115.83) -- (126.79,115.83) -- (126.79,140.58) -- (100,140.58) -- cycle ;
\draw  [color={rgb, 255:red, 0; green, 0; blue, 0 }  ,draw opacity=1 ] (126.79,115.83) -- (153.58,115.83) -- (153.58,140.58) -- (126.79,140.58) -- cycle ;
\draw  [color={rgb, 255:red, 0; green, 0; blue, 0 }  ,draw opacity=1 ] (169.66,115.83) -- (196.45,115.83) -- (196.45,140.58) -- (169.66,140.58) -- cycle ;
\draw  [color={rgb, 255:red, 0; green, 0; blue, 0 }  ,draw opacity=1 ] (196.45,115.83) -- (223.24,115.83) -- (223.24,140.58) -- (196.45,140.58) -- cycle ;
\draw  [color={rgb, 255:red, 0; green, 0; blue, 0 }  ,draw opacity=1 ][fill={rgb, 255:red, 74; green, 144; blue, 226 }  ,fill opacity=1 ] (239.74,115.83) -- (266.53,115.83) -- (266.53,140.58) -- (239.74,140.58) -- cycle ;
\draw  [color={rgb, 255:red, 0; green, 0; blue, 0 }  ,draw opacity=1 ][fill={rgb, 255:red, 74; green, 144; blue, 226 }  ,fill opacity=1 ] (282.61,115.83) -- (309.4,115.83) -- (309.4,140.58) -- (282.61,140.58) -- cycle ;
\draw   (100.21,151.13) .. controls (100.22,155.8) and (102.55,158.13) .. (107.22,158.12) -- (151.69,158.07) .. controls (158.36,158.06) and (161.69,160.39) .. (161.7,165.06) .. controls (161.69,160.39) and (165.02,158.06) .. (171.69,158.05)(168.69,158.06) -- (216.16,158.01) .. controls (220.83,158) and (223.16,155.67) .. (223.15,151) ;
\draw   (239.92,150.8) .. controls (239.9,155.47) and (242.22,157.81) .. (246.89,157.83) -- (264.43,157.91) .. controls (271.1,157.94) and (274.42,160.28) .. (274.4,164.95) .. controls (274.42,160.28) and (277.76,157.97) .. (284.43,158)(281.43,157.99) -- (301.73,158.08) .. controls (306.4,158.1) and (308.74,155.78) .. (308.76,151.11) ;

\draw (154.89,124.66) node [anchor=north west][inner sep=0.75pt]   [align=left] {...};
\draw (153.37,165.89) node [anchor=north west][inner sep=0.75pt]    {$2S$};
\draw (267.52,124.07) node [anchor=north west][inner sep=0.75pt]   [align=left] {...};
\draw (269.46,166.95) node [anchor=north west][inner sep=0.75pt]    {$S$};
\draw (141.69,91.49) node [anchor=north west][inner sep=0.75pt]    {$SU( 2)$};
\draw (258.63,91.26) node [anchor=north west][inner sep=0.75pt]    {$SO( 3)$};

\end{tikzpicture}.
\end{center}
For the $(2j+1)$-dimensional irreducible representation of the $SO(3)$ group, the tensor $(T_B)_{SO(3)}^{I_1\cdots I_j}$ in Eq.~\eqref{eq:l2s+12s+1} corresponds to
\begin{center}

\tikzset{every picture/.style={line width=0.75pt}} 

\begin{tikzpicture}[x=0.75pt,y=0.75pt,yscale=-1,xscale=1]

\draw  [fill={rgb, 255:red, 74; green, 144; blue, 226 }  ,fill opacity=1 ] (281,81.88) -- (308.38,81.88) -- (308.38,107.18) -- (281,107.18) -- cycle ;
\draw  [fill={rgb, 255:red, 74; green, 144; blue, 226 }  ,fill opacity=1 ] (324.62,81.88) -- (352,81.88) -- (352,107.18) -- (324.62,107.18) -- cycle ;
\draw   (288.12,114.27) .. controls (288.12,118.94) and (290.45,121.27) .. (295.12,121.27) -- (305.5,121.27) .. controls (312.17,121.27) and (315.5,123.6) .. (315.5,128.27) .. controls (315.5,123.6) and (318.83,121.27) .. (325.5,121.27)(322.5,121.27) -- (335.88,121.27) .. controls (340.55,121.27) and (342.88,118.94) .. (342.88,114.27) ;

\draw (308.82,82.4) node [anchor=north west][inner sep=0.75pt]  [font=\large]  {$...$};
\draw (296.26,59.3) node [anchor=north west][inner sep=0.75pt]    {$SO( 3)$};
\draw (310.93,131.87) node [anchor=north west][inner sep=0.75pt]    {$j$};

\end{tikzpicture}.

\end{center}
For $j=S$, the spin operator $\boldsymbol{\Sigma}^S$ converts $(T_B)_{SO(3)}^{I_1\cdots I_S}$ into a  totally symmetric  $SU(2)$ tensor of rank $2S$, denoted as $(T_B)_{SU(2)}^{\{i_1\cdots i_{2S}\}}$ in Eq.~\eqref{eq:TbSU2}, represented by the Young tableau
\begin{center}

\tikzset{every picture/.style={line width=0.75pt}} 

\begin{tikzpicture}[x=0.75pt,y=0.75pt,yscale=-1,xscale=1]

\draw  [color={rgb, 255:red, 0; green, 0; blue, 0 }  ,draw opacity=1 ] (130,155.83) -- (151.34,155.83) -- (151.34,176.48) -- (130,176.48) -- cycle ;
\draw  [color={rgb, 255:red, 0; green, 0; blue, 0 }  ,draw opacity=1 ] (151.34,155.83) -- (172.69,155.83) -- (172.69,176.48) -- (151.34,176.48) -- cycle ;
\draw  [color={rgb, 255:red, 0; green, 0; blue, 0 }  ,draw opacity=1 ] (185.49,155.83) -- (206.83,155.83) -- (206.83,176.48) -- (185.49,176.48) -- cycle ;
\draw  [color={rgb, 255:red, 0; green, 0; blue, 0 }  ,draw opacity=1 ] (206.83,155.83) -- (228.18,155.83) -- (228.18,176.48) -- (206.83,176.48) -- cycle ;
\draw  [color={rgb, 255:red, 0; green, 0; blue, 0 }  ,draw opacity=1 ][fill={rgb, 255:red, 74; green, 144; blue, 226 }  ,fill opacity=1 ] (241.32,155.83) -- (262.67,155.83) -- (262.67,176.48) -- (241.32,176.48) -- cycle ;
\draw  [color={rgb, 255:red, 0; green, 0; blue, 0 }  ,draw opacity=1 ][fill={rgb, 255:red, 74; green, 144; blue, 226 }  ,fill opacity=1 ] (275.47,155.83) -- (296.81,155.83) -- (296.81,176.48) -- (275.47,176.48) -- cycle ;
\draw   (130.17,185.28) .. controls (130.18,189.95) and (132.51,192.28) .. (137.18,192.27) -- (169.15,192.23) .. controls (175.82,192.22) and (179.15,194.55) .. (179.16,199.22) .. controls (179.15,194.55) and (182.48,192.22) .. (189.15,192.21)(186.15,192.21) -- (221.12,192.17) .. controls (225.79,192.16) and (228.12,189.83) .. (228.11,185.16) ;
\draw   (241.47,185.01) .. controls (241.44,189.68) and (243.76,192.02) .. (248.43,192.04) -- (258.93,192.09) .. controls (265.6,192.12) and (268.92,194.46) .. (268.89,199.13) .. controls (268.92,194.46) and (272.26,192.15) .. (278.93,192.18)(275.93,192.17) -- (289.27,192.23) .. controls (293.94,192.25) and (296.28,189.93) .. (296.3,185.26) ;
\draw   (100.29,166.17) .. controls (100.29,161.64) and (103.85,157.98) .. (108.23,157.98) .. controls (112.62,157.98) and (116.17,161.64) .. (116.17,166.17) .. controls (116.17,170.69) and (112.62,174.36) .. (108.23,174.36) .. controls (103.85,174.36) and (100.29,170.69) .. (100.29,166.17) -- cycle ; \draw   (102.62,160.38) -- (113.85,171.96) ; \draw   (113.85,160.38) -- (102.62,171.96) ;
\draw  [color={rgb, 255:red, 0; green, 0; blue, 0 }  ,draw opacity=1 ][fill={rgb, 255:red, 74; green, 144; blue, 226 }  ,fill opacity=1 ] (26.48,156.11) -- (47.82,156.11) -- (47.82,176.76) -- (26.48,176.76) -- cycle ;
\draw  [color={rgb, 255:red, 0; green, 0; blue, 0 }  ,draw opacity=1 ][fill={rgb, 255:red, 74; green, 144; blue, 226 }  ,fill opacity=1 ] (60.63,156.11) -- (81.97,156.11) -- (81.97,176.76) -- (60.63,176.76) -- cycle ;
\draw   (26.62,185.28) .. controls (26.6,189.95) and (28.92,192.29) .. (33.59,192.32) -- (44.08,192.37) .. controls (50.75,192.4) and (54.07,194.74) .. (54.05,199.41) .. controls (54.07,194.74) and (57.41,192.43) .. (64.08,192.46)(61.08,192.45) -- (74.42,192.51) .. controls (79.09,192.53) and (81.43,190.21) .. (81.46,185.54) ;
\draw  [color={rgb, 255:red, 0; green, 0; blue, 0 }  ,draw opacity=1 ] (352.67,155.83) -- (374.01,155.83) -- (374.01,176.48) -- (352.67,176.48) -- cycle ;
\draw  [color={rgb, 255:red, 0; green, 0; blue, 0 }  ,draw opacity=1 ] (374.01,155.83) -- (395.35,155.83) -- (395.35,176.48) -- (374.01,176.48) -- cycle ;
\draw  [color={rgb, 255:red, 0; green, 0; blue, 0 }  ,draw opacity=1 ] (408.16,155.83) -- (429.5,155.83) -- (429.5,176.48) -- (408.16,176.48) -- cycle ;
\draw  [color={rgb, 255:red, 0; green, 0; blue, 0 }  ,draw opacity=1 ] (429.5,155.83) -- (450.84,155.83) -- (450.84,176.48) -- (429.5,176.48) -- cycle ;
\draw  [color={rgb, 255:red, 0; green, 0; blue, 0 }  ,draw opacity=1 ][fill={rgb, 255:red, 74; green, 144; blue, 226 }  ,fill opacity=1 ] (463.99,155.83) -- (485.33,155.83) -- (485.33,176.48) -- (463.99,176.48) -- cycle ;
\draw  [color={rgb, 255:red, 0; green, 0; blue, 0 }  ,draw opacity=1 ][fill={rgb, 255:red, 74; green, 144; blue, 226 }  ,fill opacity=1 ] (498.14,155.83) -- (519.48,155.83) -- (519.48,176.48) -- (498.14,176.48) -- cycle ;
\draw   (352.84,185.28) .. controls (352.85,189.95) and (355.18,192.28) .. (359.85,192.27) -- (391.81,192.23) .. controls (398.48,192.22) and (401.81,194.55) .. (401.82,199.22) .. controls (401.81,194.55) and (405.14,192.22) .. (411.81,192.21)(408.81,192.21) -- (443.78,192.17) .. controls (448.45,192.16) and (450.78,189.83) .. (450.77,185.16) ;
\draw   (464.13,185.01) .. controls (464.11,189.68) and (466.43,192.02) .. (471.1,192.04) -- (481.59,192.09) .. controls (488.26,192.12) and (491.58,194.46) .. (491.56,199.13) .. controls (491.58,194.46) and (494.92,192.15) .. (501.59,192.18)(498.59,192.17) -- (511.94,192.23) .. controls (516.61,192.25) and (518.95,189.93) .. (518.97,185.26) ;
\draw  [color={rgb, 255:red, 0; green, 0; blue, 0 }  ,draw opacity=1 ][fill={rgb, 255:red, 74; green, 144; blue, 226 }  ,fill opacity=1 ] (464.1,134.11) -- (485.44,134.11) -- (485.44,154.76) -- (464.1,154.76) -- cycle ;
\draw  [color={rgb, 255:red, 0; green, 0; blue, 0 }  ,draw opacity=1 ][fill={rgb, 255:red, 74; green, 144; blue, 226 }  ,fill opacity=1 ] (498.25,134.11) -- (519.59,134.11) -- (519.59,154.76) -- (498.25,154.76) -- cycle ;
\draw   (540.67,166.2) .. controls (540.67,161.78) and (544.07,158.2) .. (548.27,158.2) .. controls (552.46,158.2) and (555.87,161.78) .. (555.87,166.2) .. controls (555.87,170.62) and (552.46,174.2) .. (548.27,174.2) .. controls (544.07,174.2) and (540.67,170.62) .. (540.67,166.2) -- cycle ; \draw   (540.67,166.2) -- (555.87,166.2) ; \draw   (548.27,158.2) -- (548.27,174.2) ;
\draw  [color={rgb, 255:red, 0; green, 0; blue, 0 }  ,draw opacity=1 ] (349.33,253.83) -- (370.68,253.83) -- (370.68,274.48) -- (349.33,274.48) -- cycle ;
\draw  [color={rgb, 255:red, 0; green, 0; blue, 0 }  ,draw opacity=1 ] (370.68,253.83) -- (392.02,253.83) -- (392.02,274.48) -- (370.68,274.48) -- cycle ;
\draw  [color={rgb, 255:red, 0; green, 0; blue, 0 }  ,draw opacity=1 ] (404.82,253.83) -- (426.17,253.83) -- (426.17,274.48) -- (404.82,274.48) -- cycle ;
\draw  [color={rgb, 255:red, 0; green, 0; blue, 0 }  ,draw opacity=1 ] (426.17,253.83) -- (447.51,253.83) -- (447.51,274.48) -- (426.17,274.48) -- cycle ;
\draw   (349.5,283.28) .. controls (349.51,287.95) and (351.84,290.28) .. (356.51,290.27) -- (388.48,290.23) .. controls (395.15,290.22) and (398.48,292.55) .. (398.49,297.22) .. controls (398.48,292.55) and (401.81,290.22) .. (408.48,290.21)(405.48,290.21) -- (440.45,290.17) .. controls (445.12,290.16) and (447.45,287.83) .. (447.44,283.16) ;

\draw (172.3,161.79) node [anchor=north west][inner sep=0.75pt]   [align=left] {...};
\draw (170.48,196.34) node [anchor=north west][inner sep=0.75pt]    {$2S$};
\draw (262.03,161.29) node [anchor=north west][inner sep=0.75pt]   [align=left] {...};
\draw (263.78,197.22) node [anchor=north west][inner sep=0.75pt]    {$S$};
\draw (47.18,161.57) node [anchor=north west][inner sep=0.75pt]   [align=left] {...};
\draw (46.29,197.5) node [anchor=north west][inner sep=0.75pt]    {$j=S$};
\draw (311.33,159.4) node [anchor=north west][inner sep=0.75pt]    {$=$};
\draw (394.97,161.79) node [anchor=north west][inner sep=0.75pt]   [align=left] {...};
\draw (393.15,196.34) node [anchor=north west][inner sep=0.75pt]    {$2S$};
\draw (484.69,161.29) node [anchor=north west][inner sep=0.75pt]   [align=left] {...};
\draw (486.45,197.22) node [anchor=north west][inner sep=0.75pt]    {$S$};
\draw (484.8,139.57) node [anchor=north west][inner sep=0.75pt]   [align=left] {...};
\draw (577.63,161.12) node [anchor=north west][inner sep=0.75pt]   [align=left] {...};
\draw (312.67,254.73) node [anchor=north west][inner sep=0.75pt]    {$\simeq $};
\draw (391.63,259.79) node [anchor=north west][inner sep=0.75pt]   [align=left] {...};
\draw (389.82,294.34) node [anchor=north west][inner sep=0.75pt]    {$2j=2S$};

\end{tikzpicture}
\end{center}
This diagrammatic perspective again makes manifest why only one rotational symmetry group is ultimately required in Eq.~\eqref{eq:deltalsl} in the Lagrangian construction.

\subsection{$SO(3)$ Spin Group}\label{spingroup}

In the previous subsection, we discuss the unification of the rotation in the spatial and the spin space.
The spin degree of freedom is essential in both relativistic and non-relativistic regimes. In practice, effective theories are commonly formulated using the appropriate spin group, while keeping the spin operator itself implicit.

The spacetime symmetry of a theory is reflected in the transformation properties of its coordinate or momentum variables. For example, the four-momentum $p^\mu$ transforms under the fundamental representation of $SO(3,1)$, while the three-momentum $\boldsymbol{p}$ transforms under $SO(3)$. Spacetime invariants are constructed as singlets under the corresponding orthogonal group. However, these orthogonal groups are not simply connected, which motivates the introduction of their universal covers.

The spin group $Spin(n,m)$ is defined as the simply connected double cover of $SO(n,m)$, with analogous relations holding for $Spin(n)$ and $SO(n)$. In particular, $Spin(3,1) = SL(2,\mathbb{C})$ covers the Lorentz group $SO(3,1)$, and $Spin(3) = SU(2)$ covers the rotation group $SO(3)$. Fields and operators in a quantum theory more naturally transform under representations of the spin group, which faithfully capture both integer and half-integer spin states, serving as the fundamental building blocks for constructing spacetime invariants.

\paragraph{Spin(3,1)}
In the relativistic theories, the Lorentz indices $\mu$ can be mapped to the spinor indices through the relation $p^{\mu} \rightarrow p_{\alpha\dot\alpha} = p_{\mu}\sigma^{\mu}_{\alpha\dot\alpha}$, enabling the classification of building blocks by their transformation properties under the $(j_l,j_r)$ representations of the $Spin(3,1)$. For instance, the left-handed Weyl spinor $(\psi_L)_{\alpha} \in (1/2,0)$ transforms as:
\begin{equation}
     (\psi_L)_{\alpha} \rightarrow (\psi_L)_{\alpha}^{\prime} = \left[e^{-\frac{i}{2}S_L^{\mu\nu}\omega_{\mu\nu}}\right]_{\alpha}^{~\beta} (\psi_L)_{\beta},
\end{equation}
while the right-handed Weyl spinor $(\psi_R)^{\dot\alpha} \in (0,1/2)$ transforms as:
\begin{equation}
     (\psi_R)^{\dot\alpha} \rightarrow (\psi_R)^{\dot\alpha\prime} = \left[e^{-\frac{i}{2}S_R^{\mu\nu}\omega_{\mu\nu}}\right]^{\dot\alpha}_{~\dot\beta} (\psi_R)^{\dot\beta}.
\end{equation}
Here $\omega_{\mu\nu}$ parameterizes the Lorentz transformation, and $(S_L^{\mu\nu})_{\alpha}^{~\beta}$, $(S_R^{\mu\nu})^{\dot\alpha}_{~\dot\beta}$ are the generators of $Spin(3,1)$.
The $(1/2,0)$ and $(0,1/2)$ representations describe the transformation of the left- and right-handed Weyl fermions, with rotations generated by the spin operator $\boldsymbol{\Sigma} = \frac{\boldsymbol{\sigma}}{2}$ and boosts generated by $\pm\mathbf{i}\frac{\boldsymbol{\sigma}}{2}$.
Then the above transformations with rotation angle $\boldsymbol{\theta}$ and boost parameter $\boldsymbol{\beta}$ are
\begin{equation}
\begin{array}{lll}
     \psi_L \rightarrow \psi_L^{\prime} = e^{-i\frac{\boldsymbol{\sigma}}{2}\cdot\boldsymbol{\theta} - \frac{\boldsymbol{\sigma}}{2}\cdot\boldsymbol{\beta}}~\psi_L,  \\
     \\
     \psi_R \rightarrow \psi_R^{\prime} = e^{-i\frac{\boldsymbol{\sigma}}{2}\cdot\boldsymbol{\theta} + \frac{\boldsymbol{\sigma}}{2}\cdot\boldsymbol{\beta}}~\psi_R. 
\end{array} 
\end{equation}

For quantum fields in the $(j_l,j_r)$ representation, the helicity of massless excitations is given by $h = -j_l + j_r$ \cite{Li:2020gnx}. The representative massless fields and their $Spin(3,1)$ representations include \cite{Li:2022tec}:
\begin{align}
     &\phi \in (0,0), \quad \psi_{\alpha} \in (1/2,0), \quad \psi_{\dot\alpha}^{\dagger} \in (0,1/2), \nonumber\\
     &F_{L\alpha\beta} = \frac{i}{2}F_{\mu\nu}\sigma_{\alpha\beta}^{\mu\nu} \in (1,0), \quad F_{R\dot\alpha\dot\beta} = -\frac{i}{2}F_{\mu\nu}\bar\sigma^{\mu\nu}_{\dot\alpha\dot\beta} \in (0,1), \nonumber\\
     &C_{L\alpha\beta\gamma\delta} = C_{\mu\nu\rho\lambda}\sigma^{\mu\nu}_{\alpha\beta}\sigma^{\rho\lambda}_{\gamma\delta} \in (2,0), \quad C_{R\dot\alpha\dot\beta\dot\gamma\dot\delta} = C_{\mu\nu\rho\lambda}\bar\sigma^{\mu\nu}_{\dot\alpha\dot\beta}\bar\sigma^{\rho\lambda}_{\dot\gamma\dot\delta} \in (0,2).
\end{align}
Acting with $\omega$ derivatives on a massless field $\Psi \in (j_l,j_r)$ yields a field in the representation $D^{\omega}\Psi \in (j_l+\omega, j_r+\omega)$.
For the massive spin-$S$ particles with $S = j_l + j_r$ \cite{Arkani-Hamed:2017jhn}, the spin operator takes the symmetrized form
\begin{equation}
   \boldsymbol{\Sigma}^S = \text{sym}\left\{ \underbrace{\frac{\boldsymbol{\sigma}}{2}\otimes\cdots\otimes\frac{\boldsymbol{\sigma}}{2}}_{2j_l} \otimes \underbrace{\frac{\boldsymbol{\sigma}}{2}\otimes\cdots\otimes\frac{\boldsymbol{\sigma}}{2}}_{2j_r} \right\}.
\end{equation}

\paragraph{Spin(3)}
Once fields are classified according to the irreducible representations of the spacetime symmetry, both the helicity (or spin) of the corresponding particles and their spin operators become well-defined.
In the relativistic theories, the spin group $Spin(3,1)$ alone suffices—there is no need to introduce both $Spin(3,1)$ and $SO(3,1)$. Similarly, in the non-relativistic theories, we employ $Spin(3) = SU(2)$, the double cover of $SO(3)$, as the fundamental symmetry group. The $2:1$ homomorphism between $SU(2)$ and $SO(3)$ plays a crucial role: while both groups describe the same physical rotations, the $SU(2)$ group accommodates the  half-integer spin representations.

The two-component spinor $\xi^{r}_i$ and its conjugate $\xi^{\dagger r,i}$ (with $i=1,2$) carry the fundamental indices of the $SU(2)$ and the little group index $r$ with $r=1,2$. For the canonical state, the quantized spin axis is aligned along the $z$-direction. The corresponding little group indices are defined via the eigenvalue equations:
\begin{equation}
\begin{aligned}
     (\sigma^3)_i^{~j}~\xi_j^{1} &= +\xi^1_i, \\
     (\sigma^3)_i^{~j}~\xi_j^{2} &= -\xi^2_i.
\end{aligned}
\end{equation}
The spinor bilinear can be derived by
\begin{equation}\label{eq:2spinorproduct}
    \xi^{\dagger}_i\otimes \xi_j = \frac{1}{2}\underbrace{\left(\xi^{\dagger}\xi\right)}_{\text{scalar}}\epsilon_{ij} + \frac{1}{2}\underbrace{\left(\xi^{\dagger}\sigma^I\xi\right)}_{\text{vector}}(\sigma^I)_{ij},
\end{equation}
where $\left(\xi^{\dagger}\xi\right)$ is a spatial scalar and $\left(\xi^{\dagger}\sigma^I\xi\right)$ is a spatial vector.

The two-component spinor $\xi^r_i$ and its conjugation transform under the (anti-)fundamental irreducible representation of $SU(2)$ as:
\begin{equation}
\left\{
\begin{array}{lll}
    \xi^{r}_i \rightarrow \xi^{\prime r}_i &= (u)_{i}^{~j} \, \xi^{r}_j, \\[4pt]
    \xi^{\dagger r,i} \rightarrow \xi^{\prime\dagger r,i} &= \xi^{\dagger r,j} \, (u^{-1})_j^{~i}.
\end{array}
\right.
\end{equation}
Similar to  Eq.~\eqref{eq:uRrelation}, the relation between $SU(2)$ and $SO(3)$ representations is established through the spin operator $\boldsymbol{\Sigma} = \boldsymbol{\sigma}/2$, which converts the $SU(2)$ adjoint representation into the $SO(3)$ rotation $R$:
\begin{equation}\label{Sigmasu2rotation}
    (u^{-1})_{k}^{~i} \, (\Sigma^{I})_k^{~l} \, (u)_{l}^{~j} = (R)^{IJ} \, (\Sigma^J)_i^{~j}.
\end{equation}
This transformation rule explicitly demonstrates how the $2:1$ homomorphism operates: each $SO(3)$ rotation $R$ corresponds to two $SU(2)$ elements $\pm u$. As a result, the vector part in Eq.~\eqref{eq:2spinorproduct} transforms under $SU(2)$ as a spatial vector under $SO(3)$ rotations:
\begin{equation}
    \xi^{\dagger} \Sigma^I \xi \; \rightarrow \; (R)^{IJ} \, \xi^{\dagger} \Sigma^J \xi.
\end{equation}

The fundamental advantage of working with the covering group $SU(2)$ becomes apparent when constructing general representations. The $(2S+1)$-dimensional irreducible representations of the $SU(2)$ are constructed from the totally symmetric rank-$2S$ spinors $\xi_{\{i_1}\cdots\xi_{i_{2S}\}}$, while the $(2n+1)$-dimensional irreducible traceless representations of $SO(3)$, denoted as $V^{\{I_1\cdots I_n\}}$, can be obtained through the mapping
\begin{equation}\label{eq:su2toso3}
   \xi^{\dagger\{i_1}\cdots\xi^{\dagger i_{n}\}}  \xi_{\{j_1}\cdots\xi_{j_{n}\}} \stackrel{\boldsymbol{\sigma}}{\longrightarrow} V^{\{I_1\cdots I_n\}} = (\xi^{\dagger}\sigma^{I_1}\xi)\cdots(\xi^{\dagger}\sigma^{I_n}\xi),
\end{equation}
where $n$ must be an integer. Conversely, any irreducible traceless representation of $SO(3)$ can be expressed in $SU(2)$ index notation as
\begin{equation}\label{eq:so3tosu2}
    V^{\{I_1\cdots I_n\}} \stackrel{\boldsymbol{\sigma}}{\longrightarrow} V_{\{i_1\cdots i_n\}}^{~\ \{j_1\cdots j_n\}} \equiv V^{\{I_1\cdots I_n\}} (\sigma^{I_1})_{i_1}^{~j_1}\cdots(\sigma^{I_n})_{i_n}^{~j_n}.
\end{equation}
These relations establish a complete dictionary for translating between the tensor representations of $SO(3)$ and the spinor representations of $SU(2)$, leveraging the $2:1$ covering property. The anti-fundamental indices can be lowered using the $SU(2)$ invariant tensor $\epsilon_{ij}$, yielding a totally symmetric tensor:
\begin{equation}
    V_{\{i_1 j_1 \cdots i_n j_n\}} = \epsilon_{j_1 k_1}\cdots\epsilon_{j_n k_n} V_{\{i_1\cdots i_n\}}^{~\ \{k_1\cdots k_n\}}.
\end{equation}
Thus, the $SO(3)$ tensor $V^{\{I_1 \cdots I_n\}}$ is equivalently represented as a fully symmetric $SU(2)$ tensor of rank $2n$.

In practical calculations, the spinor $SU(2)$ formulation offers significant advantages due to the covering group structure. For instance, the magnetic moment coupling $\boldsymbol{\sigma}\cdot\boldsymbol{B}$ takes the particularly simple form
\begin{equation}
    \epsilon_{jk}(\boldsymbol{\sigma})_i^{~k} \cdot \boldsymbol{B} = B_{\{ij\}}.
\end{equation}
Note that the two indices in the above equation are symmetric since the Pauli matrix is traceless
\begin{equation}
    \epsilon^{ij}B_{ij} = \delta_k^i(\boldsymbol{\sigma})_i^{~k} \cdot \boldsymbol{B} = \text{Tr}[\boldsymbol{\sigma}]\cdot\boldsymbol{B} = 0.
\end{equation}
By expressing all degrees of freedom exclusively through $SU(2)$-index building blocks—such as $\xi_{\{i_1}\cdots\xi_{i_{2S}\}}$ and $V_{\{i_1 j_1 \cdots i_n j_n\}}$—we achieve a unified description of spin-$S$ particles and their interactions within non-relativistic effective field theory. 
Non-relativistic heavy fields are naturally classified by their $Spin(3) = SU(2)$ representations:
\begin{equation}
\begin{array}{ll}
     \phi \in (\mathbf{1}), & N_i \equiv \xi_i \in (\mathbf{2}), \\
     \Sigma_{\{ij\}} \equiv \xi_{\{i}\xi_{j\}} \in (\mathbf{3}),&
     \Delta_{\{ijk\}} \equiv \xi_{\{i}\xi_j\xi_{k\}} \in (\mathbf{4}),\\
     \xi_{\{i_1}\cdots\xi_{i_{2S}\}} \in (\mathbf{2S+1}),
\end{array}    
\end{equation}
while soft relativistic vector fields are represented as:
\begin{equation}
    E_{ij} \in (\mathbf{3}), \quad B_{ij} \in (\mathbf{3}),
\end{equation}
with the equations of motion imposing additional constraints when derivatives act on the electromagnetic fields. The derivatives $\nabla_{\{ij\}}$ themselves carry symmetric $SU(2)$ indices.

This approach, rooted in the covering group property, eliminates the need for explicit spin operators and the corresponding redundancies.
It demonstrates  efficiency and convenience which applies to the Hilbert series and the Young tensor method for operator construction.

\subsection{Typical NR EFTs}\label{eftreview}
Having discussed the rotational symmetry, we now proceed to construct several NREFTs. The power counting scheme in each case depends on the specific physical scenario under consideration. Among the building blocks,  the non-relativistic fermion field $N(x)$, which transforms under the fundamental irreducible representation of the $SU(2)$ rotation group, can be written as~\cite{Li:2025ejk}
\begin{equation}\label{Nellmodeexp}
    N_{i}(x)=\int \frac{d^3\vec k}{(2\pi)^3\sqrt{2E}}\sum_{\sigma}u_{i}^{\sigma}(k)a_{v,k}^{\sigma}e^{-\mathbf{i}kx},
\end{equation}
where $i=1,2$ is the fundamental indices of the $SU(2)$ group. Here $v^{\mu}=(1,0,0,0)$ is a reference vector that is invariant under the spatial rotation, and $\vec k$ is the three-momentum of the particle. The two-component wave function  $u_{i}^{\sigma}(k)$  carries the spin index $\sigma$ of the particle. This expression applies to different contexts: In the HPET and the HQET, $N(x)$ denotes the heavy particle or heavy quark field, respectively; in the pionless EFT, $N(x)$ denotes the nucleon field. Depending on the specific scenario, additional internal symmetry structures can be incorporated as required.

The corresponding NR effective Lagrangian can be organized as
\begin{eqnarray}\label{eq:sch}
\mathcal{L} = N^{\dagger}  \mathbf{i}\partial_tN + \frac{1}{2m} N^{\dagger}\nabla^2  N + \delta\mathcal{L},
\end{eqnarray}
where $\delta\mathcal{L}$ denotes the higher-order bilinear terms as well as the effective interactions. In the gauge theories, ordinary derivatives are promoted to the covariant derivatives via the replacement $(\partial_t, \nabla) \to (D_t, \vec{D})$.
Power counting for the kinetic terms depends on the number of the heavy particles involved. For processes involving a single heavy particle, in which the underlying symmetry is the Poincare group, the leading-order term $N^{\dagger} \mathbf{i}\partial_t N$ alone determines the propagator, while the spatial derivative term belongs to the sub-leading term, indicating a non-linear Lorentz boost.  
When interactions involve two or more heavy particles, both kinetic terms—$N^{\dagger} \mathbf{i}\partial_t N$ and $\frac{1}{2m} N^{\dagger} \nabla^2 N$—become equally important for power counting and contribute to the propagator \cite{Bodwin:1994jh, Luke:1996hj, Manohar:1997qy, vanKolck:1999mw}. The combination $N^{\dagger}\left(\mathbf{i}\partial_t+\frac{1}{2m}\nabla^2\right)N$ is invariant under the Galilean group \cite{Bargmann:1954gh, LEVYLEBLOND1971221}, indicating the Galilean symmetry in such system. 
In this work, we would take the Euclidean group as the symmetry of the NR field, and utilize the Poincar\'{e} symmetry or the Galilean symmetry as the further constraint on the operators. 


In this subsection, we adopt the conventional notation and employ $SO(3)$ indices for clarity. It should be noted, however, that in our systematic operator construction in section~\ref{ytm}, we will continue to use the $SU(2)$ spinor index formalism for its technical advantages.

\paragraph{Heavy Particle Effective Field Theory}

Consider a system with one heavy fermion and external $U(1)$ gauge fields. Comparing to the heavy fermion mass $m$, the momentum transfer $|\vec q|$ and the residual momentum of the heavy fermion $|\vec k|$ , $|\vec q|\sim|\vec k|\ll m$ are so small, that the velocity $v^{\mu}$ (usually $v^{\mu}=(1,0,0,0)$) is invariant during the interaction and then the spacetime Lorentz symmetry  spontaneously breaks to the rotational symmetry.  Apart from the continuous symmetry, this system also preserves parity $P$ and time reversal $T$. The traditional heavy particle effective Lagrangian has building blocks including the two-component heavy spinor field $N$ and its conjugation, the derivative $D^{\mu}=\partial^{\mu}+\mathbf{i}gA^{\mu}=(D_t, -\Vec{D})$, as wella as the electromagnetic field $E^I=\frac{\mathbf{i}}{g}[D_t,D^I]$, and $B^I=\frac{\mathbf{i}}{2g}\epsilon^{IJK}[D^J,D^K]$. Organized by the  expansion $|\vec k|/m$ and in terms of the $SO(3)$ indices, the HPET Lagrangian  up to $1/m^3$ is  
\begin{equation}
\begin{array}{lll}
      \mathcal{L}&=&N^{\dagger}\{\mathbf{i}D_t+c_2\frac{ \vec{D}^2}{2m}+c_Fg\frac{\vec{\sigma}\cdot\vec{B}}{2m}+c_Dg\frac{[D_I,E^I]}{8m^2}+\mathbf{i}c_Sg\frac{\epsilon^{IJK}\sigma^I\{D_J,E^K\}}{8m^2}+c_4\frac{\vec{D}^4}{8m^3}\\
      \\
        & & +c_{W1}g\frac{\{\vec{D}^2,\vec{\sigma}\cdot\vec{B}\}}{8m^3}-c_{W2}g\frac{D^I\vec{\sigma}\cdot\vec{B}D^I}{4m^3}+c_{p^{\prime}p}g\frac{\{\vec{\sigma}\cdot\vec{D},\vec{B}\cdot\vec{D}\}}{8m^3}\\
        \\
        &&+\mathbf{i}c_Mg\frac{\{D_I,\epsilon^{IJK}[D_J,B^K]\}}{8m^3}+c_{A1}g^2\frac{B^2-E^2}{8m^3}-c_{A2}g^2\frac{E^2}{16m^3}+\mathcal{O}(1/m^4)\}N.
    \end{array}
\end{equation}
The NRQED Lagrangian is obtained by further expanding according to the order of velocity $\boldsymbol{v}$, since it has more energy scale and different power counting~\cite{Bodwin:1994jh,Luke:1996hj,Manohar:1997qy}.

Similarly, HQET and NRQCD are effective theory constructed in the same way, but with external $SU(3)$ gauge fields. The difference between HQET and NRQCD can be seen as follow: The HQET describes the heavy-light system with one large scale, the heavy quark mass $m_Q$, and one low energy scale $\Lambda_{\text{QCD}}$. The leading order term is of order $\Lambda_{\text{QCD}}$,
\begin{equation}
\mathcal{L}_0=N^{\dagger}\mathbf{i}D_tN.
\end{equation}
This term depicts a heavy source with constant velocity $v^{\mu}=(1,0,0,0)$ even in the limit $m_Q\rightarrow\infty$. The higher order terms are suppressed by $\Lambda_{\text{QCD}}/m_Q$.
On the other hand, NRQCD describes the heavy quarkonium, counted by small velocity $|\boldsymbol{v}|\ll 1$. There are multiple scales other than $\Lambda_{\text{QCD}}$ and $m_Q$. For example, the momentum of heavy quarks are at order $m_Q|\boldsymbol{v}|$, while  the movement between these two heavy quarks gives the nonzero kinematic energy $m_Q|\boldsymbol{v}|^2$. The leading order terms are instead of order $m_Q|\boldsymbol{v}|^2$,
\begin{equation}
\mathcal{L}'_0=N^{\dagger}(\mathbf{i}D_t+\frac{\vec D^2}{2m_Q})N.
\end{equation}

To systematically construct the complete and independent HPET operator basis, we utilize the Hilbert Series and the Young tensor method. The operator basis up to $1/m^5$ are listed in appendix~\ref{app:HQETop}.

\paragraph{Pionless Effective Field Theory}
For the very low energy far from the cut off $\Lambda\sim m_{\pi}$,  the momentum transfer between nucleons is at $|\vec k|\sim Q$, $Q\ll \Lambda$, thus all the heavy internal degree of freedom such as pion (with mass $m_{\pi}$) should be integrate out, and the real degree of freedom are the external neutrons and protons whose Lagrangian are captured by short range interaction.

The  manifest  symmetry of the effective Lagrangian are the rotational symmetry $SU(2)$,  $SU(2)$ isospin symmetry and the $U(1)$ charge conservation symmetry. The Lagrangian is also $P$ and $T$ even. We denote the nucleon field with only positive energy modes, as the two-component spinor field $N_{i,p}$, $N_{i,p}^{\dagger}$, where $i$ is the $SU(2)$ spinor fundamental index and $p$ is the $SU(2)$ isospin fundamental index. The little group indices for a massive fermion is suppressed in this work. The building blocks are the $N_{i,p}, N^{\dagger}_{i,p}$ and the spatial derivative $\nabla$.

The Lagrangian is organized according to powers of $Q/\Lambda$, which in configuration space corresponds to an expansion in $\nabla/\Lambda$.  Consequently, the effective Lagrangian can be written in a power-counting scheme as
\begin{eqnarray}\label{eq:pionlessL}
    \mathcal{L} = &&\underbrace{\left( N^{\dagger} \mathbf{i} \partial_t N + \frac{1}{2m} N^{\dagger} \vec{\nabla}^2 N + \cdots \right)}_{\text{kinetic}} \nonumber\\
    &&+ \underbrace{\left( \mathcal{L}_{NN}^{(0)} + \mathcal{L}_{NN}^{(2)} + \mathcal{L}_{NN}^{(4)} + \cdots \right)}_{\text{two‑nucleon contact}} + \underbrace{\left( \mathcal{L}_{NNN}^{(0)} + \mathcal{L}_{NNN}^{(2)} + \cdots \right)}_{\text{three‑nucleon contact}} + \cdots ,
\end{eqnarray}
where the first underbrace contains the kinetic terms, including the higher‑order relativistic corrections such as $ \frac{1}{8m^3} N^{\dagger} \nabla^4 N $ \cite{Luke:1997ys,Kaplan:1998we}. The well-known leading‑order two-nucleon contact interaction given by   Refs.~\cite{Weinberg:1990rz,Weinberg:1991um}, is expressed in terms of $SO(3)$ indices as
\begin{eqnarray}\label{eq:csct}
    \mathcal{L}_{NN}^{(0)} = -\frac{1}{2}C_S (N^{\dagger} N)(N^{\dagger} N) - \frac{1}{2}C_T (N^{\dagger} \vec{\sigma} N) \cdot (N^{\dagger} \vec{\sigma} N).
\end{eqnarray}

To depict the operator basis construction procedure,  we  consider the nucleon-nucleon (N-N) contact interaction and three-nucleon (3N) contact interaction in this work as the example, and the N-N operator basis up to $\mathcal{O}(Q^4)$ are listed in appendix~\ref{ap:NN}, while the 3N operator basis  up to $\mathcal{O}(Q^2)$ are listed in appendix~\ref{ap:3N}.

\paragraph{DM-nucleon Contact Interaction}
Consider the contact NR interaction of a heavy spin-1/2 DM $\xi$ with a nucleon $N$, where $\xi$ and $N$ are NR two-component spinors. The DM $\xi_{i}$ carries fundamental index of the rotation group, while the nucleon $N_{i,p}$ carries the rotation and the  $SU(2)$ isospin fundamental indices. The spatial derivatives $\nabla$ or the corresponding momentum involve the effective interaction, and they are denoted as $\vec p $ $(\vec p^{~\prime})$ and $\vec k$ $ (\vec k^{\prime})$ as the initial (final) momentum of the nucleon and the DM respectively.
\begin{figure}[H]
    \centering
    \includegraphics[width=16cm,height=6cm]{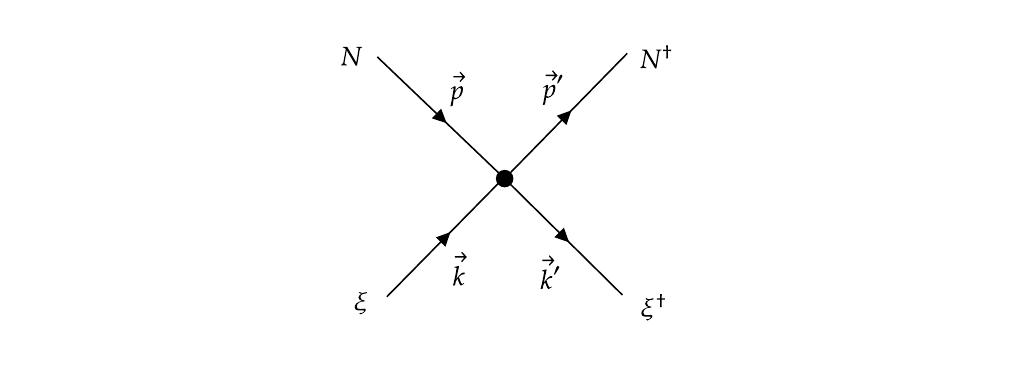}
    \caption{The DM-nucleon Contact Interaction.}
   
\end{figure}

According to the rotational symmetry, building blocks are determined as follows: In the general frame, there are three independent momentum due to the momentum conservation or integration by part (IBP), where we choose to be $\vec k$, $\vec k^{\prime}$ and  $\vec p^{~\prime}$, and in the configuration space they correspond to $-(N^{\dagger} N)(\xi^{\dagger}\mathbf{i}\overrightarrow{\nabla}\xi)$, $(N^{\dagger} N)(\xi^{\dagger}\mathbf{i}\overleftarrow{\nabla}\xi)$, and $-(N^{\dagger}\mathbf{i}\overrightarrow{\nabla} N)(\xi^{\dagger}\xi)$. This NR effective theory is counted by the velocity denoted by $\mathcal{O}(v)$, since
$m_N, m_{\xi}\gg|\vec p|\sim|\vec p^{\prime}|\sim|\vec k|\sim|\vec k^{\prime}|$. The time derivatives are always related to the higher order operators by EOM, and from the physical point of view, the energy for non-relativistic particles are mostly contained in their mass, i.e., $\frac{p_0}{m_N}\sim\frac{k_0}{m_{\xi}}\sim1$, thus time derivative is absent in the construction of effective operators. Therefore, the operator basis for the DM-nucleon contact interaction can be derived with the building blocks
$\xi_i$, $\xi^{\dagger,i}$, $N_{i,p}$, $N^{\dagger,i,p}$ as well as their spatial derivatives $\nabla_{ii}$.\footnote{ For simplicity, we use $\nabla/m_N$ to count the operators, similar treatment for four-fermion operators could also be found in Ref.~\cite{Hill:2012rh}.}

\begin{table}[H]
    \centering
    \begin{tabular}{c|c|c}
 
        \hline
        Ref.\cite{Fitzpatrick:2012ix,DelNobile:2018dfg} & Galilean Symmetry& $\vec{S}_{\xi},~\vec{S}_N, ~\vec{q}, ~\vec{v}$  \\
        \hline
        Our work& Rotational Symmetry & $N^{\dagger,i}, ~N_i,~\xi^{\dagger,i},~\xi_i$, $(\nabla_{N})_{ii},~(\nabla_{\xi})_{ii},~(\nabla_{\xi^{\dagger}})_{ii}$
        \\
        \hline
    \end{tabular}
    \caption{Building blocks in theories with different symmetry.}\label{bbsym}
\end{table}
For different descriptions of the contact interaction, we list their building blocks in Tab.~\ref{bbsym}.
In  COM frame~\cite{Fan:2010gt,Dobrescu:2006au} or owing to the consideration of Galilean invariance~\cite{Fitzpatrick:2012ix,DelNobile:2018dfg}, there are only two independent momentum. The momentum of COM, is either set to be zero or  absent in the theory.
Here $\vec q=\vec{k}-\vec{k}^{\prime}$, $~\vec v=\frac{\vec p^{~\prime}}{m_N}-\frac{\vec k^{\prime}}{m_{\xi}}$ are Galilean boost invariant, and the isospin index are suppressed. Since Galilean group is not a subgroup of the Poincare group, the Galilean boost invariance would break the Lorentz invariance at higher order. On the contrary, Lorentz covariance demands COM momentum (i.e., $\vec p+\vec k$) dependence in the non-relativistic operator.

As a result,  the non-relativistic operators based on the rotation group are needed, and the resulting effective operators could be constrained by Lorentz invariance up to any dimension order by order. The operator basis for DM-nucleon contact interaction is organized by the velocity $\mathcal{O}(v)$, and the operator bases up to $\mathcal{O}(v^4)$ are exhibited in appendix~\ref{ap:DMN}.

\subsection{Comparison with Galilean Group}

In this subsection, we clarify the distinction between the Galilean group and the Euclidean group in the context of non-relativistic EFT.

As an illustration,  we consider another example in the non-relativistic  nucleon contact interaction, to illustrate the difference between the Galilean invariance and the Euclidean invariance. The  bilinear and the two-nucleons sector in Eq.~\eqref{eq:pionlessL} can be written as~\cite{Epelbaum:2005pn}
\begin{eqnarray}
    \mathcal{L}=\underbrace{N^{\dagger}\left(  \mathbf{i} \partial_t  + \frac{1}{2m} \vec{\nabla}^2  \right)N-\frac{1}{2}C_S (N^{\dagger} N)(N^{\dagger} N) - \frac{1}{2}C_T (N^{\dagger} \vec{\sigma} N) \cdot (N^{\dagger} \vec{\sigma} N)}_{\text{Galilean invariant}}~+\cdots
\end{eqnarray}
The first term depicts the Schr\"odinger equation which is Galilean invariant. The another two leading-order nucleon contact interactions are introduced in Eq.~\eqref{eq:csct}
\begin{equation}
O_S=(N^{\dagger}N)(N^{\dagger}N),\quad O_T=(N^{\dagger}\vec\sigma N)\cdot(N^{\dagger}\vec \sigma N)
,
\end{equation}
these two terms are both rotational invariant and Galilean boost invariant. However, for the sub-leading contributions denoted as the type $\nabla^2 N^2 (N^\dagger)^2$,
the rotational-invariant basis    contains 12 independent operators obtained by the Young tensor method:
\begin{equation}
\begin{aligned}
\mathcal{O}^{p'(2)}_1 &= (N^\dagger \vec{\sigma} \cdot \overrightarrow{\nabla} \overrightarrow{\nabla}^i N^\dagger) (N \sigma^i N) I_1 + \text{h.c.}, &
\mathcal{O}^{p'(2)}_2 &= (N^\dagger \sigma^i N) (N^\dagger \overleftarrow{\nabla}^i \vec{\sigma} \cdot \overleftarrow{\nabla} N) I_1 + \text{h.c.}, \\[2pt]
\mathcal{O}^{p'(2)}_3 &= -(N^\dagger \overrightarrow{\nabla}^2 N^\dagger) (N N) I_1 + \text{h.c.}, &
\mathcal{O}^{p'(2)}_4 &= -(N^\dagger \overrightarrow{\nabla}^i N) (N^\dagger \sigma^i \vec{\sigma} \cdot \overleftarrow{\nabla} N) I_1 + \text{h.c.}, \\[2pt]
\mathcal{O}^{p'(2)}_5 &= -(N^\dagger \vec{\sigma} \cdot \overrightarrow{\nabla} \sigma^i N^\dagger) (N \overleftarrow{\nabla}^i N) I_1 + \text{h.c.}, &
\mathcal{O}^{p'(2)}_6 &= -(N^\dagger \sigma^i \vec{\sigma} \cdot \overrightarrow{\nabla} N) (N^\dagger \overleftarrow{\nabla}^i N) I_2 + \text{h.c.}, \\[2pt]
\mathcal{O}^{p'(2)}_7 &= -(N^\dagger \vec{\sigma} \cdot \overrightarrow{\nabla} N^\dagger) (N \vec{\sigma} \cdot \overleftarrow{\nabla} N) I_1 + \text{h.c.}, &
\mathcal{O}^{p'(2)}_8 &= 2 (N^\dagger \overrightarrow{\nabla}^i N^\dagger) (N \overleftarrow{\nabla}^i N) I_1 + \text{h.c.}, \\[2pt]
\mathcal{O}^{p'(2)}_9 &= -2 (N^\dagger N) (N^\dagger \overleftarrow{\nabla} \cdot \overrightarrow{\nabla} N) I_1 + \text{h.c.}, &
\mathcal{O}^{p'(2)}_{10} &= -(N^\dagger \sigma^i \sigma^j N) (N^\dagger \overleftarrow{\nabla}^i \overrightarrow{\nabla}^j N) I_1 + \text{h.c.}, \\[2pt]
\mathcal{O}^{p'(2)}_{11} &= (N^\dagger \sigma^i \sigma^j N) (N^\dagger \overleftarrow{\nabla}^i \overrightarrow{\nabla}^j N) I_2 + \text{h.c.}, &
\mathcal{O}^{p'(2)}_{12} &= (N^\dagger \vec{\sigma} \cdot \overrightarrow{\nabla} N^\dagger) (N \vec{\sigma} \cdot \overrightarrow{\nabla} N) I_1 + \text{h.c.}
\end{aligned}
\end{equation}
Here the isospin structures are defined by $I_1=\delta_{p_3}^{p_1}\delta_{p_4}^{p_2},
I_2=\delta_{p_3}^{p_2}\delta_{p_4}^{p_1}$, such that the isospin indices are contracted by
\begin{eqnarray}
(N^{\dagger}\cdots N)(N^{\dagger}\cdots N)I_1&=&(N^{\dagger,p_1}\cdots N_{p_1})(N^{\dagger,p_2}\cdots N_{p_2}),\nonumber\\
(N^{\dagger}\cdots N)(N^{\dagger}\cdots N)I_2&=&(N^{\dagger,p_1}\cdots N_{p_2})(N^{\dagger,p_2}\cdots N_{p_1}).
\end{eqnarray}

We can equivalently redefine a new operator basis as in Ref.~\cite{Filandri:2023qio}:
\begin{align}
   & O'_1 = \mathcal{O}^{p'(2)}_1 - \mathcal{O}^{p'(2)}_2 + \mathcal{O}^{p'(2)}_3 + \frac{1}{2}\mathcal{O}^{p'(2)}_9 ,\nonumber\\
    & O'_2 = \frac{1}{4}\mathcal{O}^{p'(2)}_1 - \frac{1}{4}\mathcal{O}^{p'(2)}_2 + \frac{1}{4}\mathcal{O}^{p'(2)}_3 + \frac{1}{8}\mathcal{O}^{p'(2)}_4 + \frac{1}{8}\mathcal{O}^{p'(2)}_5 - \frac{1}{8}\mathcal{O}^{p'(2)}_{11},\nonumber\\
    & O'_3 = -3\mathcal{O}^{p'(2)}_1 + 3\mathcal{O}^{p'(2)}_2 - \mathcal{O}^{p'(2)}_3 + \mathcal{O}^{p'(2)}_4 - \mathcal{O}^{p'(2)}_5 - 2\mathcal{O}^{p'(2)}_6 - \frac{1}{2}\mathcal{O}^{p'(2)}_9 + \mathcal{O}^{p'(2)}_{11} ,\nonumber\\
    & O'_4 = -\frac{3}{4}\mathcal{O}^{p'(2)}_1 + \frac{3}{4}\mathcal{O}^{p'(2)}_2 - \frac{1}{4}\mathcal{O}^{p'(2)}_3 - \frac{3}{8}\mathcal{O}^{p'(2)}_4 - \frac{3}{8}\mathcal{O}^{p'(2)}_5 + \frac{1}{4}\mathcal{O}^{p'(2)}_8 + \frac{3}{8}\mathcal{O}^{p'(2)}_{11} ,\nonumber\\
    & O'_5 = \frac{1}{4}\mathcal{O}^{p'(2)}_4 - \frac{1}{4}\mathcal{O}^{p'(2)}_5 + \frac{1}{4}\mathcal{O}^{p'(2)}_{11} ,\nonumber\\
    & O'_6 = -\mathcal{O}^{p'(2)}_2 + \mathcal{O}^{p'(2)}_6 - \mathcal{O}^{p'(2)}_{12} ,\nonumber\\
    & O'_7 = -\frac{1}{4}\mathcal{O}^{p'(2)}_2 + \frac{1}{4}\mathcal{O}^{p'(2)}_5 + \frac{1}{4}\mathcal{O}^{p'(2)}_7 - \frac{1}{8}\mathcal{O}^{p'(2)}_8 - \frac{1}{4}\mathcal{O}^{p'(2)}_{11},\nonumber \\
    & \textcolor{blue}{O^{'*}_1} = \frac{1}{2}\mathcal{O}^{p'(2)}_4 - \frac{1}{2}\mathcal{O}^{p'(2)}_5 + \frac{1}{2}\mathcal{O}^{p'(2)}_9 - \mathcal{O}^{p'(2)}_{10} + \frac{1}{2}\mathcal{O}^{p'(2)}_{11} ,\nonumber\\
    & \textcolor{blue}{O^{'*}_2} = \mathcal{O}^{p'(2)}_5 - \frac{1}{2}\mathcal{O}^{p'(2)}_8 ,\nonumber\\
    & \textcolor{blue}{O^{'*}_3} = -\frac{1}{2}\mathcal{O}^{p'(2)}_4 - \frac{1}{2}\mathcal{O}^{p'(2)}_5 - \frac{1}{2}\mathcal{O}^{p'(2)}_9 + \frac{1}{2}\mathcal{O}^{p'(2)}_{11} ,\nonumber\\
    & \textcolor{blue}{O^{'*}_4} = \frac{1}{2}\mathcal{O}^{p'(2)}_4 + \frac{5}{2}\mathcal{O}^{p'(2)}_5 + 2\mathcal{O}^{p'(2)}_6 - \mathcal{O}^{p'(2)}_8 + \frac{1}{2}\mathcal{O}^{p'(2)}_9 - \frac{5}{2}\mathcal{O}^{p'(2)}_{11} ,\nonumber\\
   &  \textcolor{blue}{O^{'*}_5} = -\mathcal{O}^{p'(2)}_5 - \mathcal{O}^{p'(2)}_6 - \mathcal{O}^{p'(2)}_7 + \frac{1}{2}\mathcal{O}^{p'(2)}_8 + \mathcal{O}^{p'(2)}_{11} + \mathcal{O}^{p'(2)}_{12},
\end{align}
where $O'_i$ denote the Galilean-invariant operators, and $O^{\prime*}_i$ are the ones that depend explicitly on the center-of-mass momentum. 
In the center-of-mass frame, the constraint
\begin{equation}\label{eq:CMconstraint}
(N_1^\dagger \cdots \overrightarrow{\nabla}^I \cdots N_3)(N_2^\dagger \cdots N_4)
= - (N_1^\dagger \cdots N_3)(N_2^\dagger \cdots \overrightarrow{\nabla}^I \cdots N_4),
\end{equation}
reduces the number of independent operators to the first seven ($O'_1$--$O'_7$)~\cite{Epelbaum:2004fk,Xiao:2018jot}.  These seven operators also form a basis for the Galilean group in the center-of-mass frame. In this frame, operators that depend on the center-of-mass momentum vanish. This is equivalent to imposing Galilean boost invariance, thereby restoring the Galilean group symmetry. In other words, consider the four-fermion sector of the Lagrangian at $\mathcal{O}(1/m^2)$ written as
\begin{equation}
    \mathcal{L} = \frac{1}{m^2} \left( \sum_{i=1}^{7} \mathcal{C}'_i \, O'_i \;+\; \sum_{j=1}^{5} \mathcal{C}^{\prime*}_j \, O^{\prime*}_j \right),
\end{equation}
Imposing the Galilean boost invariance or the Eq.~\eqref{eq:CMconstraint} forces the latter set to vanish, leading to the constraints of the Wilson coefficients
\begin{equation}
\mathcal{C}^{\prime*}_j = 0, \qquad j = 1,2,3,4,5.
\end{equation}
The relation between the three sets of operators is summarized in Tab.~\ref{tab:3basescompare}.

\begin{table}[ht]
\centering
\begin{tabular}{|c|c|c|}
\hline
\textbf{Euclidean-invariant basis} & 
$\begin{array}{l}
     \textbf{Euclidean-invariant}  \\
     \textbf{center-of-mass frame basis} 
\end{array}$
 & \textbf{Galilean-invariant basis} \\ \hline
$\begin{array}{ll}
O'_1 & O'_2 \\
O'_3 & O'_4 \\
O'_5 & O'_6 \\
O'_7 & \textcolor{blue}{O^{'*}_8} \\
\textcolor{blue}{O^{'*}_9} & \textcolor{blue}{O^{'*}_{10}} \\
\textcolor{blue}{O^{'*}_{11}} & \textcolor{blue}{O^{'*}_{12}}
\end{array}$ 
& 
$\begin{array}{ll}
O'_1 & O'_2 \\
O'_3 & O'_4 \\
O'_5 & O'_6 \\
O'_7 &
\end{array}$ 
& 
$\begin{array}{ll}
O'_1 & O'_2 \\
O'_3 & O'_4 \\
O'_5 & O'_6 \\
O'_7 &
\end{array}$ \\ \hline
\end{tabular}
\caption{Comparison of operator bases: full Euclidean-invariant set, the independent subset in the center-of-mass frame, and the general Galilean-invariant basis.}\label{tab:3basescompare}
\end{table}

When matching onto operators from the relativistic low-energy effective field theory, additional dependence on the center-of-mass momentum appears. For example, we can obtain a relativistic four-fermion operator built from the Dirac nucleon field $\Psi$ matches onto the non-relativistic operators as follows:
\begin{align}
(\bar\Psi\Psi)(\bar\Psi\Psi) &= (N^\dagger N)(N^\dagger N)I_1 
 + \frac{1}{4m^2} \Bigl(~
\underbrace{\overbrace{-4O_2'+2O_5'}^{\text{center-of-mass frame; Galilean}}
+ \textcolor{blue}{O_1^{'*}-O_3^{'*}}}_{\text{Euclidean-invariant set}} \Bigr)+\mathcal{O}(\frac{1}{m^4}),
\end{align}
which is the same as the results in Ref.~\cite{Filandri:2023qio}.
Neither the center-of-mass frame  basis nor the Galilean-invariant basis alone suffices to reproduce the relativistic result. Therefore, the underlying Lorentz invariance requires the full set of rotational-invariant operators for a consistent matching in the non-relativistic effective theories, and we adopt the rotational symmetry to write down the non-relativistic operator basis in this work.

\section{Hilbert Series for NR Theories}\label{NRop}

To systematically construct a complete and independent set of non-relativistic operators, we employ the Hilbert series method to enumerate operator bases. As discussed in the previous section, the symmetry underlying the non-relativistic action is the Euclidean group. 
Nevertheless, when implementing the Hilbert series method in terms of group characters to obtain results free of  EOM and IBP, we need to incorporate the scaling transformation. This scaling symmetry imposes constraints on the admissible structures of the spatial and time derivatives, thereby  determining the number of independent singlets. Consequently, we utilize the group characters of the Lifshitz symmetry within the Hilbert series to count the non-relativistic operator bases.

The Lifshitz algebra~\cite{Hornreich:1975zz,Grinstein:1981rbe,Kachru:2008yh,Ross:2009ar,Hoyos:2013eza} is generated  by the spatial rotations $\boldsymbol{J}$, the spatial translations $\boldsymbol{P}$, the time translation $H$, and the scaling transformation $\mathcal{D}$, satisfying the following commutation relations:
\begin{equation}
    \begin{array}{lll}
       [J^I,J^J]=\mathbf{i}\epsilon^{IJK}J^K,&[J^I,P^J]=\mathbf{i}\epsilon^{IJK}P^K,&[\mathcal{D},P^I]=\mathbf{i}P^I,\nonumber\\
\lbrack \mathcal{D},H\rbrack=2\mathbf{i}H,&[H,P^I]=0,&[P^I,P^J]=[H,J^I]=[\mathcal{D},J^I]=0.
    \end{array}
\end{equation}
The scaling transformation generated by $\mathcal{D}$ acts anisotropically on spacetime coordinates as
\begin{eqnarray}
(\boldsymbol{x},t) \longrightarrow (\lambda\boldsymbol{x},\lambda^z t),
\end{eqnarray}
where  $z$ implies the anisotropy between space and time. In the context of non-relativistic effective field theories describing heavy particles, this exponent is fixed to $z=2$, consistent with the leading-order equation of motion
\begin{eqnarray}
\mathbf{i}\partial_t N(x) = -\frac{1}{2m}\nabla^2 N(x).
\end{eqnarray}
The Lifshitz algebra constitutes a subalgebra of the Schr\"{o}dinger algebra. While the eigenstates and characters of the Schr\"{o}dinger group has been applied to the Hilbert series in Ref.~\cite{Kobach:2018nmt}, for heavy-particle states it suffices to utilize the eigenstates of the scaling generator $\mathcal{D}$ only, without explicit knowledge of the boost transformation.

The Hilbert series serves as a generating function that enumerates the number of independent group invariants constructed from derivatives and fields:
\begin{equation}
    H(\nabla,\{N\}) = \sum_{k=0}^{\infty} \sum_{r=0}^{\infty} c_{k,r} \nabla^k N^r,
\end{equation}
where $\nabla$ and $N$ are spurion variables representing spatial derivatives and fields respectively, and $c_{k,r}$ denotes the number of invariant operators composed of $k$ derivatives and $r$ fields. For the gauge interaction, we replace $\nabla\rightarrow D$, and the time derivative $D_t$ and the electromagnetic fields are included. This formulation provides both the operator types and their multiplicities at each order.

This generating function is computed through the integration
\begin{equation}\label{eq:int of the generating function}
    H(\nabla,\{N\}) = \int d\mu_{\text{st}}  d\mu_{\text{in}}  Z[N, \nabla, \chi],
\end{equation}
where $d\mu_{\text{st}}$ and $d\mu_{\text{in}}$ represent the Haar measures for the spacetime and internal (gauge) symmetry groups respectively. The function $Z[N, \nabla, \chi]$ sums characters of relevant representations with appropriate spurion weights $\nabla^k N^r$ for each operator type, analogous to a partition function in thermodynamic systems.
Equation~(\ref{eq:int of the generating function}) can be understood as a consequence of Schur's lemma, with $\chi$ denoting the character sum for the Single Particle Module (SPM)—the linear space spanned by fundamental building blocks like $\nabla^k N^1$. The totally (anti-)symmetric operator space is constructed via the plethystic exponential:
\begin{equation}\label{eq:PE}
    \mathrm{PE}[N \chi] = \exp\left( \sum_{n=1}^{\infty} \frac{1}{n} (\pm 1)^{n+1} N^n \chi(x_1^n, \dots, x_r^n) \right),
\end{equation}
where $x_1, \dots, x_r$ are group parameters, and the sign $(\pm 1)$ is $+$ for bosons and $-$ for fermions, consistent with the spin-statistics theorem.

Without additional constraints, the generating function takes the form $Z_0 = \mathrm{PE}[N \chi]$. However, physical operators must satisfy momentum conservation, requiring total derivatives to vanish. While incorporating this constraint is generally challenging, it has been established~\cite{Henning:2017fpj} that for operators with mass dimension exceeding the spacetime dimension, the generating function can be replaced by
\begin{equation}
    Z = \frac{Z_0}{P},
\end{equation}
where $P$ accounts for momentum conservation constraints.

To understand this factorization, consider that the complete operator space is generated by a set of primary operators $\{O_{\text{primary}}\}$ that cannot be expressed as total derivatives, while all other operators appear as descendants $\partial^{\mu_1}\cdots\partial^{\mu_n}O_{\text{primary}}$ of these primaries. In this framework, the desired character $Z$ corresponds specifically to the primary operator sector, leading to the approximate relation $Z_0 = P \cdot Z$. The factor $P$ represents the contribution from derivative operators and takes the explicit form
\begin{equation}
    P = \frac{1}{(1-x^2\nabla)(1-\nabla)(1-x^{-2}\nabla)}.
\end{equation}
This heuristic argument provides valuable intuition for the appearance of the $P$ factor in the generating function. The complete Hilbert series is therefore given by
\begin{equation}
    H(\nabla,\{N\}) = \int d\mu_{\text{st}}  d\mu_{\text{in}}  \frac{\mathrm{PE}[N \chi]}{P}.
\end{equation}

In this section, we clarify certain ambiguities regarding non-relativistic building blocks and their group characters in Section~\ref{rotation sym in QM}. In non-relativistic dynamics, only a single rotation group is required, as opposed to the direct product of the spinor rotation group $SU(2)$ and the spatial rotation group $SO(3)$. 
We further utilize this tool to enumerate operators that are even under charge conjugation ($C$) and space inversion ($P$). However, in non-relativistic contexts, the absence of antiparticles implies that $C$ and $P$ transformations do not necessarily preserve the operator type, preventing a direct application of the Hilbert series to identify invariants under these discrete symmetries. Instead, the anti-unitary time-reversal transformation ($T$) provides an effective approach for constructing Hermitian operators that are simultaneously $C$-even and $P$-even in subsection~\ref{cp+hs}. The explicit results for the HPET, the HQET, the pionless EFT and the DM-nucleon interaction are listed in subsection~\ref{sec:resultHS}.

\subsection{Redundancy of the Spin Operator}\label{rotation sym in QM}

When considering non-relativistic effective theories, one typically encounters two rotation groups: an $SU(2)$ group acting on particle spin indices $i,j,k$, and an $SO(3)$ group acting on vector quantities such as electromagnetic fields with indices $I,J,K$. These group indices are related through the spin operator $(s^I)_i^j$. For notational simplicity, we will suppress the fundamental spinor indices $i,j$ in subsequent discussions.

Taking Heavy Particle Effective Theory (HPET) as an example, the fundamental building blocks consist of $N$, $s^I$, $E$, and $B$, where $N$ represents the two-component non-relativistic spinor field. The Hilbert series for this model under the symmetry $SU(2)\times SO(3)$ was initially investigated in Refs.~\cite{Kobach:2017xkw,Kobach:2018nmt}. However, we have identified a significant issue in the operator counting for multi-fermion operators, particularly relevant to the pionless effective field theories. The problem arises from redundancies that occur when two spin operators associated with the same particle contract their indices. These redundancies are not adequately accounted for in the existing Hilbert series treatment, leading to an overcounting of independent operators in the multi-fermion sector.

In fact, it's equivalent to use only one single $SU(2)$ group, instead of the $SU(2)$ and $SO(3)$ both, once we take the commutation relation $[s^I,s^J]=\mathbf{i}\varepsilon^{IJK} s^K$ and $\{s^I,s^J\}=2\delta^{IJ}$ into account, and remove those redundancy. As a consequence, the spin operator having indices of two group should be dropped. Not only can the Hilbert series be simplified in HPET since only $N, E, B$ are needed now, but also the similar procedure can be used in counting the multi-fermion operators, even in spin-$S$ representation. We would proof this fact in the aspect of the Hilbert series, and it's nature to generalize it to the aspect of operators.

Indeed, the Hilbert Series are simplified by utilizing a single $SU(2)$ group rather than maintaining both $SU(2)$ and $SO(3)$ groups simultaneously. This simplification becomes possible when we account for the fundamental algebraic relations governing the spin operators: 
\begin{equation}
    [s^I,s^J]=\mathbf{i}\varepsilon^{IJK} s^K,\quad \{s^I,s^J\}=2\delta^{IJ}.
\end{equation}
These relations inherently encode the connection between spinor and vector representations, allowing us to eliminate redundant structures that arise from contracting multiple spin operators associated with the same particle.

Consequently, the spin operator $s^I$, which carries indices of both groups, can be eliminated from the set of fundamental building blocks. This reduction leads to two advantages. First, in the context of HPET, the Hilbert series construction is substantially simplified as only the fields $N$, $E$, and $B$ are required, with the spin degree of freedom now being fully incorporated through the algebraic properties of the $SU(2)$ group. Second, and more importantly, this approach provides a consistent framework for counting operators in multi-fermion systems, including those where fermions transform under arbitrary spin-$S$ representations.

We rigorously demonstrate this equivalence from the perspective of Hilbert series analysis, and it  naturally extends to the explicit operator construction, ensuring consistency between the counting of independent operators and their concrete realization in the effective field theory as in the end of subsection~\ref{symmetry}.

The complete Hilbert series  is denoted as
\begin{equation}\label{eq:Hdnseb}
    H(D,\{N,s,E,B\}) = \int d\mu_{SU(2)}(x) d\mu_{SO(3)}(y) d\mu_{\text{other}} Z[N,E,B,D] Z[s].
\end{equation}
In the group integration of the Hilbert series, we formally express the character $Z$ into two parts 
\begin{equation}
    Z = Z[s]Z[N,E,B,D],
\end{equation}
where $Z[s]$ is the generating function for the spin operators, and $Z[N,E,B,D]$ is for the remaining degree of freedom.
The character for a single spin operator $s^I$ is given by
\begin{equation}
    \chi_{s} = (x^2 + 1 + x^{-2})(y^2 + 1 + y^{-2}),
\end{equation}
where $x$ and $y$ are group parameters for $SU(2)$ and $SO(3)$ respectively. We emphasize that $Z[s] \neq \mathrm{PE}[s\chi_{s}]$ due to redundancies arising from spin operator contractions. Specifically, for two spin operators $s^I$, only the totally symmetric component survives since the $SU(2)$ indices cannot be contracted. More generally, we obtain the complete generating function for spin operators:
\begin{align}\label{eq:Zs}
    Z[s] & = \sum_n s^n \chi_{\mathbf{2n+1},SU(2)}(x) \chi_{\mathbf{2n+1},SO(3)}(y) \\
         & = \frac{(1-s)\left((1+s)^2 + s\left(x^2 + \frac{1}{x^2} + y^2 + \frac{1}{y^2}\right)\right)}{(1-sx^2/y^2)(1-sx^2y^2)(1-sy^2/x^2)(1-s/x^2y^2)}.
\end{align}
Besides, for the remaining degree of freedom, we expand $Z[N,E,B,D]$ in terms of group characters
\begin{equation}\label{eq:Znebd}
    Z[N,E,B,D] = \sum_{i,j} Z_{ij} \chi_{\mathbf{2i+1},SU(2)}(x) \chi_{\mathbf{2j+1},SO(3)}(y).
\end{equation}
Since $y$ is the group parameter of $SO(3)$, the index $j$ must be an integer. 

Using the orthogonality relation of $SO(3)$ characters
\begin{equation}
    \int d\mu_{SO(3)}(y) \chi_{\mathbf{2j+1},SO(3)}(y) \chi_{\mathbf{2n+1},SO(3)}(y) = \delta_{j,n},
\end{equation}
we plug the characters in Eq.~\eqref{eq:Zs} and Eq.~\eqref{eq:Znebd} into the Hilbert series in  Eq.~\eqref{eq:Hdnseb}, and integrate out the $SO(3)$ dependence, reducing the expression to a Hilbert series involving only the $SU(2)$ group
\begin{align}
   & H(D,\{N,s=1,E,B\}) \nonumber\\
&=\int d\mu_{SU(2)}(x) d\mu_{SO(3)}(y) d\mu_{\text{other}}
\nonumber\\
&\times\left(\sum_{i,j} Z_{ij} \chi_{\mathbf{2i+1},SU(2)}(x) \chi_{\mathbf{2j+1},SO(3)}(y)\right)\left( \chi_{\mathbf{2n+1},SU(2)}(x) \chi_{\mathbf{2n+1},SO(3)}(y)\right).
    \nonumber\\
    & = \int d\mu_{SU(2)}(x) d\mu_{\text{other}} \sum_{i,j} Z_{ij} \chi_{\mathbf{2i+1},SU(2)}(x) \chi_{\mathbf{2j+1},SU(2)}(x).
\end{align}
This reduction demonstrates that the essential step is to replace the $SO(3)$ group parameter $y$ in $Z$ with the $SU(2)$ parameter $x$, thereby identifying the two rotation groups. In the above equation, we see that the spin operator is eliminated.

\subsection{$C$, $P$, and $T$ Properties of the Hilbert Series}\label{cp+hs}

As established in Refs.~\cite{Graf:2020yxt,Sun:2022aag}, the Hilbert series can systematically count $C$-even and $P$-even operators when antiparticle degrees of freedom are explicitly included in the spectrum. However, in effective theories where antiparticles have been integrated out, direct analysis of $CP$ properties becomes challenging due to the complicated transformation rules of the combined $CP$ operation.

The $CPT$ theorem provides an alternative approach: instead of $CP$, we can analyze time-reversal symmetry $T$, which offers a more tractable framework in non-relativistic contexts. The transformation properties of the fundamental fields under $T$ are given by:
\begin{equation}
    N \xrightarrow{T} \mathbf{i}\sigma^2 N, \quad
    E \xrightarrow{T} E, \quad
    B \xrightarrow{T} -B.
\end{equation}
Our objective is to construct operators that are simultaneously $T$-symmetric and Hermitian. However, both $T$ and Hermitian conjugation are anti-linear operations, satisfying
\begin{equation}
    T(cN) = c^*T(N), \quad (cN)^{\dagger} = N^{\dagger}c^*,
\end{equation}
where $c$ is a complex coefficient. This demonstrates that the symmetry properties depend not only on the operator structure but also on the coefficients.
Since $T$ and $\dagger$ commute as real linear maps on the operator space, we can simultaneously diagonalize them
\begin{equation}
    T(O_{i,ab}) = a O_{i,ab}, \quad (O_{i,ab})^{\dagger} = b O_{i,ab}, \quad a,b = \pm 1,
\end{equation}
where $i$ indexes the remaining degrees of freedom. The operator space is then spanned by $O_{i,ab}$ with real coefficients. Straightforward analysis shows that $O_{i,++}$ and $\mathrm{i} O_{i,--}$ possess the desired symmetry properties.
This observation motivates defining a new transformation $T_0$ as
\begin{equation}
    T_0(O) \equiv T(O^{\dagger}),
\end{equation}
for any operator $O$. We therefore need to identify $T_0$-even operators using the Hilbert series method. This is achieved by extending the symmetry group to $\tilde{G} = G \rtimes \{1, T_0\}$, which decomposes into two disconnected components: $\tilde{G} = G \cup G'$.
The modified Hilbert series incorporating $T_0$ symmetry becomes:
\begin{equation}\label{eq:T_0}
    H(D,\{N\}) = \frac{1}{2} \int_{G} d\mu_{\text{st}} d\mu_{\text{in}} \frac{\mathrm{PE}[N \chi]}{P} 
                + \frac{1}{2} \int_{G'} d\mu_{\text{st}} d\mu_{\text{in}} \frac{\mathrm{PE}[N \chi]}{P}.
\end{equation}

To compute the character in the $G'$ sector, we note that the general character in Eq.~(\ref{eq:PE}) takes the form $\mathrm{tr}(g^n)$, where $g$ is the representation matrix. For a general group element $gT_0$, we have:
\begin{equation}
gT_0
\begin{pmatrix}
    N \\ N^{\dagger}
\end{pmatrix} 
 =
\begin{pmatrix}
    g_{N^{\dagger}} & 0 \\
    0 & g_{N}
\end{pmatrix}
\begin{pmatrix}
    0 & 1 \\
    1 & 0
\end{pmatrix}
\begin{pmatrix}
    N \\ N^{\dagger}
\end{pmatrix}.
\end{equation}
The character then evaluates to:
\begin{equation}
\mathrm{tr}((gT_0)^n) = 
\begin{cases}
(-1)^k \chi_{T_0}, & n = 2k, \\
0, & n = 2k-1.
\end{cases}
\end{equation}
The factor $(-1)^k$ arises because $T_0$ is an anti-homomorphism that reverses the order of multiplied fields:
\begin{equation}
    T_0(N N') = T_0(N') T_0(N) = -T_0(N) T_0(N'),
\end{equation}
where the minus sign originates from fermionic statistics. Apart from this statistical factor, the character matches that of the $CP$ transformation.
As concrete examples, the characters for $U(1)$ and $SU(2)$ representations transform as:
\begin{align}
    \chi_{q,U(1)}(x^{2k}) &= x^{2kq} + x^{-2kq} \rightarrow \chi_{T_0,q,U(1)}(x^{2k}) = 1 + 1, \\
    \chi_{\mathbf{2},SU(2)}(x^{2k}) &= x^{2k} + x^{-2k} \rightarrow \chi_{T_0,\mathbf{2},SU(2)}(x^{2k}) = (-1)^k (x^{2k} + x^{-2k}).
\end{align}
For HQET, where $N$ carries $U(1)$ charge $+1$ and transforms as a doublet under $SU(2)$, the character in the second term of Eq.~(\ref{eq:T_0}) becomes:
\begin{equation}
    \mathrm{PE}(N \chi_{N}) = \exp\left\{ \sum_{n=2k} \frac{1}{n} (-1) N^{2k} P_1(D^n,x^n) \left[(x^{2k} + x^{-2k}) \cdot 1 + (x^{2k} + x^{-2k}) \cdot 1 \right] \right\},
\end{equation}
while the Haar measures remain unchanged in this construction. Consequently, plugging  the above equation in Eq.~\eqref{eq:T_0}, we obtain the $T$ even Hilbert series.

\subsection{Invariants from Hilbert Series}\label{sec:resultHS}

In this subsection, we explicitly list the results obtained from the Hilbert series for several non-relativistic effective theories under the rotational group and the corresponding internal symmetry. We also exhibit the results with determined $P$ and $T$ transformation properties.

\paragraph{Hilber Series for HPET}\label{HSforHQET}
For the HPET, there are $U(1)$ group related to the electric charge and spinor rotation  group $SU(2)$, then the Hilbert series is given by
\begin{equation}\label{HPEThs}
     H(D,\{N,E,B\} )=\int d\mu_{U(1)}\int d\mu_{SU(2)}\frac{1}{P_0P_1}PE[N\chi_N]PE[N^{\dagger}\chi_{N^{\dagger}}]PE[E\chi_E]PE[B\chi_B].
\end{equation}
The group characters of the building blocks  are
\begin{equation}\label{chiHPET}
\left\{
    \begin{array}{lll}
         \chi_{N}&=&P_1[D,y]\chi_{U(1)}[1,x]\chi_{SU(2)\textbf{2}}[y],  \\
        \\
        \chi_{N^{\dagger}}&=&P_1[D,y]\chi_{U(1)}[-1,x]\chi_{SU(2)\textbf{2}}[y],
        \\
        \\
         \chi_{E}&=&P_1[D,y]P_0[D_t,y]\left(\chi_{SU(2)\textbf{3}}[y]-D\chi_{SU(2)\textbf{3}}[y]+D^2\right),
         \\
         \\
         \chi_{B}&=&P_1[D,y]P_0[D_t,y]\left(\chi_{SU(2)\textbf{3}}[y]-D\right),
    \end{array}
    \right.
\end{equation}
where the factor that generates the spatial and  time derivative on the fields are
\begin{equation}
\begin{array}{lll}
    P_1[D,y]=\frac{1}{(1-y^2D)(1-D)(1-\frac{D}{y^2})},  \\
    \\
      P_0[D,z]=\frac{1}{1-D_t},
\end{array}   
\end{equation}
and the characters of each group are given by
\begin{equation}
\left\{
\begin{array}{lll}
         \chi_{U(1)}[q,x]=x^q,  \\
      \\\chi_{SU(2)\textbf{2}}[y]=y+\frac{1}{y},\\
      \\\chi_{SU(2)\textbf{3}}[y]=y^2+1+\frac{1}{y^2}.
\end{array}\right.
\end{equation}
Utilizing the above equations, and integrating the Eq.~\eqref{HPEThs} with the Haar measure 
\begin{equation}
\begin{array}{lll}
         \int d\mu_{U(1)}&=&\frac{1}{2\pi \mathbf{i}}\int_{|x|=1}\frac{1}{x}dx, \\
         \\
       \int d\mu_{SU(2)}&=&\frac{1}{2\pi \mathbf{i}}\int_{|y|=1}\frac{1}{2y}(1-y^2)(1-\frac{1}{y^2})dy,
\end{array}
\end{equation}
we obtain the two-fermion results as
\begin{eqnarray}
HS_{d=5}&=&BNN^{\dagger}+D^2NN^{\dagger}+ENN^{\dagger},\nonumber\\
HS_{d=6}&=&3BDNN^{\dagger}+D^3NN^{\dagger}+3EDNN^{\dagger},\nonumber\\
HS_{d=7}&=&B^2NN^{\dagger}+6BD^2NN^{\dagger}+D^4NN^{\dagger}+2BENN^{\dagger}+5ED^2NN^{\dagger}+E^2NN^{\dagger},\nonumber\\
HS_{d=8}&=&B^2D_tNN^{\dagger}+5B^2DNN^{\dagger}+10BD^3NN^{\dagger}+D^5NN^{\dagger}+2BED_tNN^{\dagger}\nonumber\\
&&+9BEDNN^{\dagger}+8ED^3NN^{\dagger}+E^2D_tNN^{\dagger}+4E^2DNN^{\dagger},
   \nonumber\\
HS_{d=9}&=&B^3NN^{\dagger}+B^2D_t^2NN^{\dagger}+5B^2D_tDNN^{\dagger}+16B^2D^2NN^{\dagger}+15BD^4NN^{\dagger}+D^6NN^{\dagger}\nonumber\\
&&+2B^2ENN^{\dagger}+2BED_t^2NN^{\dagger}+9BEDD_tNN^{\dagger}+26BED^2NN^{\dagger}+11ED^4NN^{\dagger}+2BE^2NN^{\dagger}\nonumber\\
&&+E^2D_t^2NN^{\dagger}+4E^2D_tDNN^{\dagger}+12E^2D^2NN^{\dagger}+E^3NN^{\dagger},
\end{eqnarray}
while the $P$ even and $T$ even results derived by methods in subsection \ref{cp+hs} are
\begin{eqnarray}\label{eq:HSHPETcp+}
    HS_{d=5}&=&BNN^{\dagger}+D^2NN^{\dagger},\nonumber\\
   HS_{d=6}&=&2EDNN^{\dagger},\nonumber\\
   HS_{d=7}&=&B^2NN^{\dagger}+4BD^2NN^{\dagger}+D^4NN^{\dagger}+E^2NN^{\dagger},\nonumber\\
HS_{d=8}&=&B^2D_tNN^{\dagger}+E^2D_tNN^{\dagger}+5BEDNN^{\dagger}+5ED^3NN^{\dagger},\nonumber\\
 HS_{d=9}&=&B^3NN^{\dagger}+B^2D_t^2NN^{\dagger}+10B^2D^2NN^{\dagger}+9BD^4NN^{\dagger}+D^6NN^{\dagger}\nonumber\\
 &&+4BEDD_tNN^{\dagger}+2BE^2NN^{\dagger}+E^2D_t^2NN^{\dagger}+8E^2D^2NN^{\dagger}.
\end{eqnarray}

\paragraph{Hilber Series for HQET}
For the HQET, the gauge group of the external gauge fields is $SU(3)$, so that the Hilbert series is given by
\begin{equation}
     H(D,\{N,E,B\} )=\int d\mu_{U(1)}\int d\mu_{SU(2)}\int d\mu_{SU(3)}\frac{1}{P_0P_1}PE[N\chi_N]PE[N^{\dagger}\chi_{N^{\dagger}}]PE[E\chi_E]PE[B\chi_B],
\end{equation}
and the characters are
\begin{equation} 
\left\{
\begin{array}{lll}
     \chi_{N}&=&P_1[D,y]\chi_{U(1)}[1,x]\chi_{SU(2)\textbf{2}}[y]\chi_{SU(3)\textbf{3}}[a,b],  \\
     \\
     \chi_{N^{\dagger}}&=&P_1[D,y]\chi_{U(1)}[-1,x]\chi_{SU(2)\textbf{2}}[y]\chi_{SU(3)\Bar{\mathbf{3}}}[a,b],
     \\
     \\
     \chi_{E}&=&P_1[D,y]P_0[D_t,y]\left(\chi_{SU(2)\textbf{3}}[y]-D\chi_{SU(2)\textbf{3}}[y]+D^2\right)\chi_{SU(3)\mathbf{8}}[a,b],
     \\
     \\
     \chi_{B}&=&P_1[D,y]P_0[D_t,y]\left(\chi_{SU(2)\textbf{3}}[y]-D\right)\chi_{SU(3)\textbf{8}}[a,b].
\end{array}\right.
\end{equation}
Here we keep the $U(1)$ to select invariants with the same number of $N$ and $N^{\dagger}$. The characters of fundamental and anti-fundamental representation of $SU(3)$ are
\begin{equation}
\begin{array}{lll}
      \chi_{SU(3)\textbf{3}}[a,b]=b+\frac{a}{b}+\frac{1}{a},  \\
     \\
     \chi_{SU(3)\Bar{\mathbf{3}}}[a,b]=a+\frac{b}{a}+\frac{1}{b},
\end{array}
\end{equation}
while the character of the adjoint representation is
\begin{equation}
    \chi_{SU(3)\textbf{8}}[a,b]=ab+\frac{b^2}{a}+\frac{a^2}{b}+2+\frac{a}{b^2}+\frac{b}{a^2}+\frac{1}{ab}.
\end{equation}
The Haar measures for $SU(3)$ is
\begin{equation}
    \int d\mu_{SU(3)}=\frac{1}{(2\pi \mathbf{i})^2}\int_{|a|=1}\int_{|b|=1}\frac{1}{6ab}(1-ab)(1-\frac{a^2}{b})(1-\frac{b^2}{a})(1-\frac{1}{ab})(1-\frac{a}{b^2})(1-\frac{b}{a^2}).
\end{equation}
According to section~\ref{cp+hs}, the $P$ and $T$ even results with fermion bilinear are
\begin{eqnarray}\label{eq:HQETHS}
HS_{d=5}&=&BNN^{\dagger}+D^2NN^{\dagger},\nonumber\\
HS_{d=6}&=&2EDNN^{\dagger},\nonumber\\
HS_{d=7}&=&3B^2NN^{\dagger}+4BD^2NN^{\dagger}+D^4NN^{\dagger}+3E^2NN^{\dagger},\nonumber\\
HS_{d=8}&=&3B^2D_tNN^{\dagger}+3E^2D_tNN^{\dagger}+14BEDNN^{\dagger}+5ED^3NN^{\dagger},\nonumber\\
HS_{d=9}&=&6B^3NN^{\dagger}+3B^2D_t^2NN^{\dagger}+29B^2D^2NN^{\dagger}+9BD^4NN^{\dagger}+D^6NN^{\dagger}\nonumber\\
&&+13BEDD_tNN^{\dagger}+13BE^2NN^{\dagger}+3E^2D_t^2NN^{\dagger}+22E^2D^2NN^{\dagger}.
\end{eqnarray}

\paragraph{Hilber Series for Pionless EFT}
For nucleon contact interaction, there are the spinor rotation group $SU(2)$, the $SU(2)$ isospin group as well as the $U(1)$ electric charge group. Notice that no gauge field is considered. The Hilbert series is given by
\begin{equation}
     H(\nabla,\{N\} )=\int d\mu_{st,SU(2)}\int d\mu_{U(1)}\int d\mu_{in,SU(2)}\frac{1}{P_1}PE[N\chi_N]PE[N^{\dagger}\chi_{N^{\dagger}}].
\end{equation}
The characters are
\begin{eqnarray}\label{chiN}
    \chi_{N}&=&P_1[\nabla,y]\chi_{U(1)}[1,x]\chi_{SU(2)\textbf{2}}[y]\chi_{SU(2)\textbf{2}}[z],\nonumber\\
    \chi_{N^{\dagger}}&=&P_1[\nabla,y]\chi_{U(1)}[-1,x]\chi_{SU(2)\textbf{2}}[y]\chi_{SU(2)\textbf{2}}[z],
\end{eqnarray}
where $\chi_{SU(2)\textbf{2}}[y]$, $\chi_{SU(2)\textbf{2}}[z]$ are characters for spinor symmetry group and isospin symmetry group, respectively. The results are
\begin{eqnarray}
HS_{d=6}&=&2N^2(N^{\dagger})^2,\nonumber\\
 HS_{d=7}&=&5\nabla N^2(N^{\dagger})^2,\nonumber\\
  HS_{d=8}&=&17\nabla^2N^2(N^{\dagger})^2,\nonumber\\
   HS_{d=9}&=&34\nabla^3N^2(N^{\dagger})^2+N^3(N^{\dagger})^3,\nonumber\\
   HS_{d=10}&=&74\nabla^4N^2(N^{\dagger})^2+5\nabla N^3(N^{\dagger})^3,\nonumber\\
  HS_{d=11}&=&128\nabla^5N^2(N^{\dagger})^2+28\nabla^2N^3(N^{\dagger})^3.
\end{eqnarray}
As it's discussed before, comparing to Ref.\cite{Kobach:2017xkw}, our results show that there is only one rotational $SU(2)$ 
symmetry group that needs to be considered in Eq.$~$(\ref{chiN}). The $SU(2)$ indices for spinor and $SO(3)$ indices for space vector turn out to be the fundamental and adjoint indices of the $SU(2)$ group. 

The  $P$ even and $T$ even Hilbert series are
\begin{itemize}
    \item N-N sector:
    \begin{align}
   HS_{d=6}&=2N^2(N^{\dagger})^2,\nonumber\\
    HS_{d=8}&=12\nabla^2N^2(N^{\dagger})^2,\nonumber\\
   HS_{d=10}&=45\nabla^4N^2(N^{\dagger})^2,
\end{align}
\item 3N sector:
\begin{align}
HS_{d=9}&=N^3(N^{\dagger})^3,\nonumber\\
HS_{d=11}&=18\nabla^2N^3(N^{\dagger})^3.
\end{align}
\end{itemize}

\paragraph{Hilber Series for DM-nucleon  Interaction}

The simplest NR spin-1/2 dark matter $\xi$ is considered to be no other internal group,  while the nucleons has a isospin $SU(2)$ symmetry. For this contact interaction, the Hilbert series is
\begin{equation}
\begin{array}{ll}
     H(\nabla,\{N,\xi\} )&=\int d\mu_{U(1)_N}\int d\mu_{U(1)_{\xi}}\int d\mu_{st,SU(2)}\int d\mu_{in,SU(2)}\\
     &
     \\
     &\times\frac{1}{P_1}PE[N\chi_N]PE[N^{\dagger}\chi_{N^{\dagger}}]PE[\xi\chi_{\xi}]PE[\xi^{\dagger}\chi_{\xi^{\dagger}}],
     \end{array}
\end{equation}
and the characters are given by
\begin{equation}\left\{
    \begin{array}{lll}
        \chi_{N}&=P_1[\nabla,y]\chi_{U(1)_N}[1,x]\chi_{SU(2)\textbf{2}}[y]\chi_{SU(2)\textbf{2}}[z],  \\
         \\
            \chi_{N^{\dagger}}&=P_1[\nabla,y]\chi_{U(1)_N}[-1,x]\chi_{SU(2)\textbf{2}}[y]\chi_{SU(2)\textbf{2}}[z],
            \\
            \\
            \chi_{\xi}&=P_1[\nabla,y]\chi_{U(1)_{\xi}}[1,x]\chi_{SU(2)\textbf{2}}[y],
            \\
            \\
            \chi_{\xi^{\dagger}}&=P_1[\nabla,y]\chi_{U(1)_{\xi}}[-1,x]\chi_{SU(2)\textbf{2}}[y].
    \end{array}\right.
\end{equation}
Here $U(1)_N, ~U(1)_{\xi}$ groups are related to their individual charge,  such that invariants are bilinears consisting of $N^{\dagger}N$ and $\xi^{\dagger}\xi$, and the particle numbers of nucleon and dark matter are separately conserved. According to the properties of the $P$ and $T$ transformation , the results are classified by 
\begin{itemize}
    \item $P$ even and $T$ even:
    \begin{align}
HS_{d=6}&=2NN^{\dagger}\xi\xi^{\dagger},\nonumber\\
HS_{d=8}&=17\nabla^2NN^{\dagger}\xi\xi^{\dagger},\nonumber\\
HS_{d=10}&=74\nabla^4NN^{\dagger}\xi\xi^{\dagger},
\end{align}
   \item $P$ even and $T$ odd:
    \begin{align}
HS_{d=8}&=10\nabla^2NN^{\dagger}\xi\xi^{\dagger},\nonumber\\
HS_{d=10}&=58\nabla^4NN^{\dagger}\xi\xi^{\dagger},
\end{align}

\item $P$ odd and $T$ even:
    \begin{align}
HS_{d=7}&=5\nabla NN^{\dagger}\xi\xi^{\dagger},\nonumber\\
HS_{d=9}&=34\nabla^3NN^{\dagger}\xi\xi^{\dagger},
\end{align}

\item $P$ odd and $T$ odd:
    \begin{align}
HS_{d=7}&=4\nabla NN^{\dagger}\xi\xi^{\dagger},\nonumber\\
HS_{d=9}&=30\nabla^3NN^{\dagger}\xi\xi^{\dagger}.
\end{align}

\end{itemize}

\section{Young Tensor Method for NR Theories}\label{ytm}

Having determined the number of independent operators via the Hilbert series, we proceed to explicitly construct the operator basis for the non-relativistic effective theory. Our approach employs the Young tensor method developed in~\cite{Li:2020gnx,Li:2022tec,Li:2020xlh,Ren:2022tvi}, and we extend it to the non-relativistic cases in this work. This method provides a systematic framework for deriving operator bases through the amplitude-operator correspondence. Since operators are built from derivatives, fields, and group invariant tensors, a complete and non-redundant basis can be obtained by classifying all possible invariant tensor structures for given field contents and derivative orders. These structures are in one-to-one correspondence with the semi-standard Young tableaux (SSYT).

Working within the fundamental representation of the $SU(2)_{\text{spin}}$ group significantly simplifies the construction procedure in the Young tensor method. Specifically, under the $SO(3)_{\text{spatial}}$ transformations, one encounters the complicated Fierz identities:
\begin{eqnarray}\label{eq:Fierz}
    \delta_i^j\delta_k^l&=&\frac{1}{2}\left[\delta_i^l\delta_k^j+(\sigma^I)_i^l(\sigma^I)_k^j\right],\\
    (\sigma^I)_i^j\delta_k^l&=&\frac{1}{2}\left[\delta_i^l(\sigma^I)_k^j+(\sigma^I)_i^l\delta_k^j-\mathbf{i}\epsilon^{IJK}(\sigma^J)_i^l(\sigma^K)_k^j\right],\label{eq:Fierz2}\\
    (\sigma^I)_i^j(\sigma^J)_k^l&=&\frac{1}{2}\left[\delta^{IJ}\delta_i^l\delta_k^j+\mathbf{i}\epsilon^{IKJ}\delta_i^l(\sigma^K)_k^j+\mathbf{i}\epsilon^{IJK}(\sigma^K)_i^l\delta_k^j\right.\nonumber\\
   &&\left.+(\sigma^J)_i^l(\sigma^I)_k^j-\delta^{IJ}(\sigma^K)_i^l(\sigma^K)_k^j+(\sigma^I)_i^l(\sigma^J)_k^j\right].\label{eq:Fierz3}
\end{eqnarray}
In contrast, when working with the $SU(2)_{\text{spin}}$ group, these redundancies reduce to the single Schouten identity:
\begin{equation}\label{eq:schouten0}
\epsilon^{rk}\epsilon^{ij} + \epsilon^{ri}\epsilon^{jk} + \epsilon^{rj}\epsilon^{ki} = 0,
\end{equation}
which is automatically eliminated when considering the SSYT.

This section is organized as follows. We begin in subsection~\ref{sec:building} by examining the building blocks for rotationally invariant theories, both with and without gauge interactions. Subsection~\ref{sec:redundancy} classifies potential redundancies among these building blocks. The core theoretical framework—the operator-Young tableau correspondence—is reviewed in subsection~\ref{sec:op-yt}. Finally, subsection~\ref{sec:basis} demonstrates the  procedure for constructing operator bases.

\subsection{Building Blocks}\label{sec:building}

We consider the rotational group $SU(2)_{\text{spin}}$ to be the homogeneous spacetime symmetry group for the NREFTs, including the Pionless EFT, the HPET and the HQET. In these NREFTs, the building blocks are fields and derivatives, which transform under irreducible representations of the $SU(2)_{\text{spin}}$ group. For the rotational symmetry, the invariant tensors  are 
\begin{equation}
	SU(2)_{\text{spin}}: \epsilon^{IJK}, \delta^{IJ}, ({\sigma}^{I})^{i}_j, \epsilon_{ij},\epsilon^{ij},
\end{equation}
where the $I,J,K$ and the (lower) $i,j,k$ are the adjoint and the fundamental indices, respectively\footnote{The rotational symmetry can be obtained from the Lorentz group which spontaneously breaks to the subgroup that keep the time-like reference vector $v^{\mu}$ invariant~\cite{Li:2025ejk}. This $SU(2)$ group is also the little group of $v^{\mu}$, and these fundamental indices $ijk$ are the little group indices of $v^{\mu}$. As a result, the corresponding on-shell scattering amplitude for these non-relativistic theories in the spinor-helicity formalism can be constructed.}. The generator of the $SU(2)_{\text{spin}}$ is  $\frac{1}{2}(\sigma^I)^i_j$, where $(\sigma^I)^i_j$ is the Pauli matrix. The (conjugate) heavy nucleon field ($N^{i\dagger}$) $N_i$ carries the (anti-) fundamental index $i$ of the $SU(2)_{\text{spin}}$, while the spatial derivative $\nabla^I$ carries the adjoint index of that group. As a result, in the $SU(2)_{\text{spin}}$, the building blocks can be written in terms of the fundamental indices as
\begin{equation}
N_i,\quad N^{\dagger}_j\equiv\epsilon_{ji}N^{\dagger i},\quad
\nabla_{ij}\equiv\epsilon_{ik}\nabla_{I}(\sigma^{I})^k_j.
\end{equation}
As in subsection~\ref{spingroup}, the spatial derivative $\nabla_{ij}$ is symmetric of $i, j$ since the Pauli matrix is traceless,
$
\epsilon^{ji}\nabla_{ij}=\nabla_I\text{tr}[\sigma^I]=0,
$
and for simplicity it can be denoted as $\nabla_{ii}$ (without the summation of the index $i$).

The time derivative $\partial_t$ and the spatial derivative $\nabla$ are denoted as the $(\mathbf{1})$ and $(\mathbf{3})$ representations, respectively. A spin-$j$ field  correspond to the $(2j+1)$-dimensional representations of the $SU(2)_{\text{spin}}$, denoted as $\mathbf{(2j+1)}$.  Specifically, for the heavy fermion $N$ and its conjugation as well as the derivatives, we have the following representations
\begin{eqnarray}
&&\partial_t\in (\mathbf{1}),\nonumber
\\
&&N_i\in(\mathbf{2}),\quad N^{\dagger}_{i}\equiv\epsilon_{ij}N^{j\dagger}\in(\mathbf{2})\nonumber
\\
&&\nabla_{ii}\in(\mathbf{3}).
\end{eqnarray}
If there is the gauge interaction, denoting  the spatial component of the electromagnetic field $\vec E$ and $\vec B$ as $E^I, B^I$, we have the additional vector representations of the $SU(2)_{\text{spin}}$,
\begin{eqnarray}\label{eq:EBdef}
    &&E_{ij}\equiv\epsilon_{ik}E^I(\sigma^I)^k_j\in(\mathbf{3}),\nonumber\\
    &&B_{ij}\equiv\epsilon_{ik}B^I(\sigma^I)^k_j\in(\mathbf{3}).
\end{eqnarray}

These building blocks transform as representations not only under the rotational $SU(2)_{\text{spin}}$ group but also under various internal symmetry groups. First, particle number conservation for the heavy fermions entails a global $U(1)$ symmetry, under which the fermions form bilinears like $N^\dagger N$. Furthermore, specific theories feature additional internal symmetries: nucleon contact interactions possess an $SU(2)_{\text{isospin}}$ isospin symmetry; the HPET has a $U(1)_{\text{EM}}$ gauge symmetry; and the HQET is endowed with an $SU(3)_{\text{color}}$ gauge symmetry. Here we investigate these symmetry respectively:
\begin{itemize}
    \item For the nucleon contact interaction,  the invariant tensors of the isospin symmetry are 
\begin{equation}
	SU(2)_{\text{isospin}}: \epsilon^{P_1P_2P_3}, \delta^{P_1P_2}, ({\tau}^{P_1})^{p_1}_{p_2}, \epsilon_{p_1p_2},\epsilon^{p_1p_2},
\end{equation}
where the $P_1,P_2,P_3$ and the (lower) $p_1,p_2,p_3$ are respectively the adjoint and the fundamental indices, thus the (conjugate) heavy nucleon field ($N^{{p_1}\dagger}$) $N_{p_1}$ carries the (anti-) fundamental index $p_1$ of the $SU(2)_{\text{isospin}}$.
Besides, since there is no field carries the adjoint indiex of the isospin,  we can  use the tensor $\delta^{p_1}_{p_2}=\epsilon^{p_1p_3}\epsilon_{p_3p_2}$  to contract the isospin indices on $N_{p_1}$ and $N^{\dagger p_2}$ without the presence of $(\tau^{P_1})^{p_1}_{p_2}$.

\item For the $U(1)_{\text{EM}}$ gauge interaction, gauge fields in Eq.~\eqref{eq:EBdef} are introduced. For the $SU(3)_{\text{color}}$ gauge interaction, the invariant tensors are
\begin{equation}\label{eq:su3colorit}
	SU(3)_{\text{color}}: f^{ABC}, d^{ABC},\delta^{AB}, (\lambda^{A})^{b}_a, \epsilon_{abc},\epsilon^{abc},
\end{equation}
where $(\lambda^{A})^{b}_a$ is the generator of the $SU(3)_{\text{color}}$ group. In this non-Abelian group, the heavy field $N_{a}(x)$ carries the fundamental index $a$ of the $SU(3)$ gauge group, while the Hermitian conjugate field $N^{\dagger a}$ carries the anti-fundamental index and it is rewritten by 
\begin{equation}
N^{\dagger}_{bc}\equiv\epsilon_{bca}N^{\dagger a}.
\end{equation}
In addition to Eq.~\eqref{eq:EBdef}, the non-Abelian electric and magnetic fields  $E^{A}$ and $B^A$ carry the adjoint index $A$ of the $SU(3)_{\text{color}}$ group. Due to the $SU(3)$ invariant tensors in Eq.~\eqref{eq:su3colorit}, the gauge indices of the non-Abelian electric and magnetic fields are rewritten in terms of the fundamental indices as
\begin{equation}
    (E)_{abc}\equiv  \epsilon_{acd}(\lambda^{A})^{d}_b E^A  ,\quad
(B)_{abc}\equiv  \epsilon_{acd}(\lambda^{A})^{d}_b B^A. 
\end{equation}
For simplicity, the gauge factors are extracted from the building blocks  in the following discussion. Note that when derivatives are replaced by the covariant derivatives $\partial_{\mu}\rightarrow D_{\mu}$, the gauge group indices are understood as the indices of the whole, i.e., $D E_{abc}=(D E)_{abc}$. Under the transformation of the gauge group, the (non-Abelian) electric field $\Vec{E}$ and magnetic field $\Vec{B}$ are invariant while the heavy field with the covariant derivatives  transforms as $D_{\mu}N\rightarrow e^{\mathbf{i}e\alpha(x)}D_{\mu}N$, just like $N$. However, during operator construction, we select those combinations with symmetric derivatives, e.g. $\nabla_I\nabla_JN\equiv\frac{1}{2}(D_ID_J+D_JD_I)N$ and $\partial_t\partial_tN\equiv D_tD_tN$ to remove the covariant derivatives commutators (CDC).

\end{itemize}

Here we summarize the building blocks in different NREFTs in Tab.~\ref{tab:fieldcontent} and as follows:
\begin{itemize}

\item For the nucleon contact interaction, in the  $SU(2)_{\text{spin}}\times SU(2)_{\text{isospin}}$ symmetry, the building blocks  are :
\begin{equation}\label{eq:NNbb}
    N_{i,p_1}\ ,\quad N_{i}^{p_1\dagger}\ ,\quad \nabla_{ii}.
\end{equation}
Note that the time derivative $\partial_t$ is exlcuded due to the EOM of $N$. 
For the spin-1/2 DM-nucleon contact interaction, there are the additional DM field
\begin{eqnarray}
    \xi_i,\quad\xi_i^{\dagger}.
\end{eqnarray}

\item For the HPET, in the  $SU(2)_{\text{spin}}\times U(1)_{\text{EM}}$ symmetry, the building blocks  are :
\begin{equation}
    N_{i}\ ,\quad N_{i}^{\dagger}\ ,\quad \nabla_{ii}\ ,\quad \partial_t\ ,\quad E_{ij},\quad B_{ij}.
\end{equation}
Note that the derivatives $\partial_t$ and $\nabla$ represent the symmetric derivatives.

\item 
For the HQET, in the  $SU(2)_{\text{spin}}\times SU(3)_{\text{color}}$ symmetry, the building blocks  are :
\begin{equation}
    N_{i,a}\ ,\quad N_{i,ab}^{\dagger}\ ,\quad \nabla_{ii}\ ,\quad \partial_t\ ,\quad E_{ij,abc}\ ,\quad B_{ij,abc}.
\end{equation}

\end{itemize}

\begin{table}[H]
    \centering
    \begin{tabular}{|c|c|c|cccc|}
    \hline
        &\text{Building Blocks} & $SU(2)_{\text{spin}}$ &$U(1)$&$SU(2)_{\text{isospin}}$&$U(1)_{\text{EM}}$&$SU(3)_{\text{color}}$ \\
        \hline
        Pionless EFT & $\begin{array}{c}
              N_{i,p_1}\\
              N_{i}^{p_1\dagger}\\
              \nabla_{ii} 
        \end{array}$&$\begin{array}{c}
             (\textbf{2})\\
              (\textbf{2})\\
              (\textbf{3}) 
        \end{array}$&$\begin{array}{c}
             1\\
              -1\\
              0 
        \end{array}$&$\begin{array}{c}
             (\textbf{2})\\
              (\textbf{2})\\
             (\textbf{1}) 
        \end{array}$&&\\
        DM&$\begin{array}{c}
             \xi_i  \\
             \xi_i^{\dagger}  
        \end{array}$&$\begin{array}{c}
             (\textbf{2})  \\
             (\textbf{2}) 
        \end{array}$&$\begin{array}{c}
             1'  \\
             -1' 
        \end{array}$&&&
        \\
        \hline
        HPET& $\begin{array}{c}
              N_{i}\\
              N_{i}^{\dagger}\\
              \nabla_{ii} \\
              \partial_t\\
              E_{ij}\\
              B_{ij}
        \end{array}$&$\begin{array}{c}
             (\textbf{2})\\
              (\textbf{2})\\
              (\textbf{3}) \\
             (\textbf{1})\\
              (\textbf{3})\\
              (\textbf{3}) 
        \end{array}$&$\begin{array}{c}
             1\\
              -1\\
             0 \\
             0\\
              0\\
             0 
        \end{array}$&&$\begin{array}{c}
             1\\
              -1\\
             0 \\
             0\\
              0\\
             0 
        \end{array}$&\\
        \hline
         HQET& $\begin{array}{c}
              N_{i,a}\\
              N_{i,ab}^{\dagger}\\
              \nabla_{ii} \\
              \partial_t\\
              E_{ij,abc}\\
              B_{ij,abc}
        \end{array}$&$\begin{array}{c}
             (\textbf{2})\\
              (\textbf{2})\\
              (\textbf{3}) \\
             (\textbf{1})\\
              (\textbf{3})\\
              (\textbf{3}) 
        \end{array}$&$\begin{array}{c}
             1\\
              -1\\
             0 \\
             0\\
              0\\
             0 
        \end{array}$&&&$\begin{array}{c}
            ( \mathbf{3})\\
              (\mathbf{\bar{3}})\\
            (\mathbf{1})\\
            (\mathbf{1})\\
             (\mathbf{8})\\
             (\mathbf{8}) 
        \end{array}$\\
        \hline
    \end{tabular}
    \caption{The building blocks of several NREFTs, as well as their representations under the rotation group and the internal symmetry. The dark matter and the nucleons each possess a distinct $U(1)$ symmetry, corresponding to the conservation of their respective particle numbers. To distinguish between these two groups, we label their charges in the table as $\pm1$ and $\pm1'$.}
    \label{tab:fieldcontent}
\end{table}

After determining the building blocks, in terms of the gauge factor $T$ from the internal symmetry, $\epsilon_{ij}$ from the $SU(2)_{\text{spin}}$, the derivatives and the fields $\phi$ including the heavy fields and the possible electromagnetic field, the operator can be formally written as
\begin{equation}\label{eq:opgeneral}
    O=T\times \epsilon\times \prod_{_l}\nabla^{\omega_{l}} \phi_{l}.
\end{equation}
Here $l$ labels the field and $\omega_{_{l}}$ is the number of the derivatives acting on that field $\phi_{_{l}}$. The operator given by Eq.~\eqref{eq:opgeneral} is non-redundant when the gauge factor $T$ and the invariant tensor $\epsilon$ are independent.

In the nucleon contact interaction, in terms of these building blocks in Eq.~\eqref{eq:NNbb}, and the invariant tensor $\epsilon^{ij}$ of the $SU(2)_{\text{spin}}$ as well as the isospin tensor factor $T^{p_1,...,p_{2m}}$ from the $SU(2)_{\text{isospin}}$, an operator with $m$ nucleon heavy field,  $\omega$ derivatives  from Eq.~\eqref{eq:opgeneral} reduces to
\begin{equation}\label{eq:oprt1}
O=T\times\epsilon^n\times\prod_{l=2}^{m}\times\prod_{l^{\prime}={1+m}}^{2m} N_1^{\dagger}\left(\nabla^{ \omega_{l}}N^{\dagger}_{l}\right)\left(\nabla^{\omega_{l^{\prime}}}N_{l^{\prime}}\right).
\end{equation} 
Here the number of the invariant tensor $\epsilon^{ij}$ is $n=\omega+m$. Besides, $l, l^{\prime}$ are the labels of the fields, and the number of the derivatives is $\omega=\sum_{l}\omega_{l}+\sum_{l^{\prime}}\omega_{l^{\prime}}$. Since the particle number is conserved, there is always a conjugation field $N_{l}^{\dagger}$ corresponds to a field $N_{l^{\prime}}$ with $l^{\prime}=l+m$. 
The labels $l, l^{\prime}$
are used to order every fields, making the identical fields become distinguishable temporarily, but note that only when we have symmetrized the repeated fields does Eq.$~$(\ref{eq:oprt1}) holds. For example, the four-fermion operator $ (N^{\dagger}N)(N^{\dagger}N)$ can be written as
\begin{eqnarray}\label{exop2}
        (N^{\dagger}N)(N^{\dagger}N)&\longrightarrow &(N_1^{\dagger}N_3)(N_2^{\dagger}N_4) \nonumber\\ &=&\delta_{p_1}^{p_3}\delta_{p_2}^{p_4}\epsilon^{i_1i_3}\epsilon^{i_2i_4}\left((N^{\dagger}_1)_{i_1}^{p_1}(N^{\dagger}_2)_{i_2}^{p_2}(N_3)_{i_3,p_3}(N_4)_{i_4,p_4}\right).
\end{eqnarray}
Here the labels are $l=1,2$, $l^{\prime}=3,4$, and the isospin tensor factor is $T=\delta_{p_1}^{p_3}\delta_{p_2}^{p_4}$. Another example is
\begin{eqnarray}\label{exop1}   (N^{\dagger}\sigma\cdot\overrightarrow{\nabla}N)(N^{\dagger}\sigma\cdot\overleftarrow{\nabla}N) &\longrightarrow&(N_1^{\dagger}\sigma\cdot\overrightarrow{\nabla}N_3)(N_2^{\dagger}\sigma\cdot\overleftarrow{\nabla}N_4)   \\
   &=&\delta_{p_1}^{p_3}\delta_{p_2}^{p_4}\epsilon^{i_1i'_3}\epsilon^{i'_3i_3}\epsilon^{i_2i'_2}\epsilon^{i'_2i_4}\left(\nabla_{i'_2i'_2}\nabla_{i'_3i'_3}(N^{\dagger}_1)_{i_1}^{p_1}(N^{\dagger}_2)_{i_2}^{p_2}(N_3)_{i_3,p_3}(N_4)_{i_4,p_4}\right).\nonumber
\end{eqnarray}
In this notation, $\nabla_{i'_2i'_2}$ is the derivative that acts on the field $N_2^{\dagger}$, while $\nabla_{i'_3i'_3}$ acts on the field $N_3$. Labeling these fields and its indices simplifies the procedure to obtain the basis without the redundancy.  However,  we must symmetrize the repeated fields in the above two examples.

Similarly, for the HPET, operators with total $\omega=\omega_{_{l}}+\sum_{l^{\prime}}\omega_{l^{\prime}}+\sum_{l^{\prime\prime}}\omega_{l^{\prime\prime}}$ derivatives, two-component spinor $N^{\dagger}$ and $N$, $n_E$ electric fields and $n_B$ magnetic fields from Eq.~\eqref{eq:opgeneral} reduces to
\begin{equation}\label{eq:oprt}
O=\epsilon^n\times\prod_{l^{\prime}}^{n_E}\times\prod_{l^{\prime\prime}}^{n_B}N^{\dagger}(\nabla^{\omega_{l^{\prime}}}E_{l^{\prime}})(\nabla^{\omega_{l^{\prime\prime}}}B_{l^{\prime\prime}})(\nabla^{\omega_{l}}N),
\end{equation} 
where $n=\omega+1+n_E+n_B$ is the number of invariant tensor $\epsilon$. The structures of the invariant tensor $\epsilon$ decide the operator for given fields and derivatives. For the HQET, additional  gauge factor $T$ is needed.

The representations $ r_E, r_B$ of $\nabla^{\omega}E$ and $\nabla^{\omega}B$ in Eq.~\eqref{eq:oprt} are determined as follows. Based on their $SU(2)_{\text{spin}}$ group characters given in Eq.~\eqref{chiHPET}, and taking into account the totally symmetric property of the derivatives acting on them, the $SU(2)_{\text{spin}}$ representations of the (non-Abelian) electric and magnetic fields can be expressed as:
\begin{eqnarray}\label{EOMEB}
  r_E&=& \frac{1}{1-\nabla^2} \left[ \sum_{\omega \geq 0} \nabla^\omega (\boldsymbol{\omega}+\mathbf{1}) + \nabla \right], \nonumber\\
r_B &= &\frac{1}{1-\nabla^2} \left[ \sum_{\omega \geq 0} \nabla^\omega \left( \boldsymbol{\omega} + \mathbf{\tfrac{1}{2}} \right) \times \mathbf{\tfrac{1}{2}} - \mathbf{0} \right],
\end{eqnarray}
where $\boldsymbol{\omega}$ is  the $(2\omega+1)$-dimensional representation of the $SU(2)$ group. The representation of $\nabla^{\omega}\Vec{E}$ decomposes into three parts: a totally symmetric part (e.g., $\nabla^I E^J + \nabla^J E^I$); terms involving $\nabla \cdot \Vec{E}$; and terms with $\nabla^2 \Vec{E}$. 
The representation of $\nabla^{\omega}\Vec{B}$ decomposes into two parts: the representation with $2\omega+1$ symmetric indices, formed by the $2\omega$ indices from $\omega$ derivatives and one of the two $SU(2)_{\text{spin}}$ indices of the $B_{ij}$ field; and terms with $\nabla^2 \Vec{B}$. As a result, after using there representations, the operator given by Eq.~\eqref{eq:oprt} is independent when the invariant tensors $\epsilon^n$ are independent.

\subsection{Redundancies}\label{sec:redundancy}
To derive the effective operators, the traditional method is to write down overcomplete operators and then remove the redundancies. In practice, the following redundancies are considered:
\begin{itemize}

\item \textbf{Integration by parts (IBP)}: In the effective action, total derivative operators correspond to boundary terms that typically vanish upon integration over all spacetime. To eliminate such operators, we can repeatedly apply IBP to ensure that no derivative acts on the first field, $(N_1)^\dagger$, after fixing the field ordering, as in Eq.~\eqref{eq:oprt1} and Eq.~\eqref{eq:oprt}.

\item \textbf{Equation of motion (EOM) for $N$}: The on-shell condition for the free heavy field is given by
\begin{equation}
\mathbf{i}\partial_t N(x) = \left(-\frac{\nabla^2}{2m} + \cdots \right) N(x),
\end{equation}
and this leading-order form remains valid even in the presence of interactions.
The EOM for $N$ implies that $\mathbf{i}\partial_t N$ need not be included as an independent building block in the operator basis, since it can be replaced via the EOM by higher-order spatial gradient terms. An analogous reasoning applies to $N^\dagger$. As a result, time derivatives $\partial_t$ do not appear in the operators for nucleon contact interactions. Similarly, in the HPET and the HQET, operators where $iD_t$ acts on the heavy field $N$ or its conjugate are eliminated from the basis. The same description applies to the spin-1/2 DM $\xi$ in this work.

\item \textbf{Fierz identity:} The Fierz identity captures the linear dependence between different spinor bilinears. After converting all the adjoint indices into the fundamental indices, the Fierz identity reduces to the Schouten identity for the $SU(2)_{\text{spin}}$:
\begin{equation}\label{eq:schouten}
\epsilon^{rk}\epsilon^{ij} + \epsilon^{ri}\epsilon^{jk} + \epsilon^{rj}\epsilon^{ki} = 0.
\end{equation}


\end{itemize}

For the nucleon contact interactions, the redundancies discussed above are sufficient to consider. However, in the presence of gauge interactions, as in the HPET and the HQET, the following additional redundancies are addressed:

\begin{itemize}

\item \textbf{Covariant derivative commutator (CDC):} The (non-Abelian) electric and magnetic fields, $E^I$ and $B^I$, carry the adjoint index of the $SU(2)_{\text{spin}}$ group. They are defined through the covariant derivative commutators:
\begin{equation}
E^I = \frac{\mathbf{i}}{g}\left[D_t, D^I\right], \quad B^I = \frac{\mathbf{i}}{2g} \left[D^J, D^K\right] \epsilon^{IJK}.
\end{equation}
To eliminate redundancies associated with these relations, all derivatives acting on the fields are treated as symmetric.

\item \textbf{Bianchi identity:} Expressed in terms of $\Vec{E}$ and $\Vec{B}$, the Bianchi identity takes the form:
\begin{equation}
\begin{aligned}
\nabla \times \Vec{E} &= -\partial_t \Vec{B}, \quad
\nabla \cdot \Vec{B} &= 0.
\end{aligned}
\end{equation}
The first redundancy is removed as follows: according to Eq.~\eqref{EOMEB}, the indices of $\vec{E}$ and $\nabla$ are symmetric. However, in the trace
\begin{equation}
\mathbf{i}(\nabla \times \Vec{E})^I = \text{tr}[\sigma^I \sigma^J \sigma^K] \nabla^J E^K = -\text{tr}[\sigma^I \sigma^K \sigma^J] \nabla^J E^K = -\text{tr}[\sigma^I \sigma^K \sigma^J] \nabla^K E^J,
\end{equation}
the anti-symmetry of $\sigma^J$ and $\sigma^K$ implies $\mathbf{i}(\nabla \times \Vec{E})^I = 0$ in our construction, leaving $\partial_t \vec{B}$ as the remaining term. The second redundancy is addressed by noting that, in Eq.~\eqref{EOMEB} the indices of $B_{kl}$ are symmetric with its derivative (e.g., in $\nabla_{ij}B_{kl}$, $i, j, k$ are symmetric), leading to the vanishing divergence:
\begin{equation}
\nabla \cdot \Vec{B} = \nabla_{ij} B_{kl} \epsilon^{jk} \epsilon^{li} = 0.
\end{equation}
Therefore, after using the corresponding representations of the $SU(2)_{\text{spin}}$, the EOM of $\vec E$ and $\vec B$ are removed.
\end{itemize}

\subsection{Operator-Young Tableau Correspondence}\label{sec:op-yt}

Unlike the conventional approaches,  we can directly write down the operators free of redundancies from the outset. As listed in the previous subsection, the IBP and the EOM (and the possible CDC and the Bianchi identity in the gauge interaction), are automatically removed in Eq.~\eqref{eq:oprt1} (and in Eq.~\eqref{eq:oprt}). The only remaining  redundancy arises from the Schouten identity applied to the $SU(2)_{\text{spin}}$ invariant tensor $\epsilon$.

In order to count the non-redundant   basis under the $SU(2)_{\text{spin}}$, we utilize the Young tableaux~\cite{Li:2020gnx}. The corresponding fundamental indices of the building blocks are ordered, then 
each anti-symmetric tensor corresponds to a two-columns Young tableau (denoted as $[1,1]$),

\begin{align}
   \epsilon^{ij}\sim  & ~\begin{ytableau}
i  \\
j
    \end{ytableau} \ , \notag 
\end{align}
with certain order, e.g., $i\leq j$.
For an operator, we suppose that the number of the invariant tensor $\epsilon$ is $n$, then there are total $2n$ indices. 
The singlet is derived from the $[n,n]$ SSYT,
\begin{equation} \label{eq:SSYT}
\epsilon^{i_1j_1}\epsilon^{i_2j_2}...\epsilon^{i_nj_n}\sim~
\begin{ytableau}
i_1 & i_2 & \cdots & i_n \\
j_1 & j_2 & \cdots & j_n
\end{ytableau}~,
\end{equation}
where $i_m\leq j_m$, $i_m\leq i_{m+1}$ and $j_m\leq j_{m+1}$ for all possible $m$.

In the SSYT, the Schouten identity is automatically removed. For Eq.~\eqref{eq:schouten} we have
\begin{equation}\label{eq:schouteny}
\epsilon^{rk}\epsilon^{ij}+\epsilon^{ri}\epsilon^{jk}+\epsilon^{rj}\epsilon^{ki}=0,
\end{equation}
and we can order the indices as $i\rightarrow1,j\rightarrow2,k\rightarrow3,r\rightarrow4$, then the above equation becomes
\begin{equation}\label{eq:schexample}
    \epsilon^{14}\epsilon^{23}=-\epsilon^{12}\epsilon^{34}+\epsilon^{13}\epsilon^{24}.
\end{equation}
In terms of the Young tableaux, this equals to 
\begin{equation}
\begin{ytableau}
1 & 2 \\
4 & 3
\end{ytableau}
~=~-~
\begin{ytableau}
1 & 3 \\
2 & 4
\end{ytableau}
~+~
\begin{ytableau}
1 & 2 \\
3 & 4
\end{ytableau}
~.
\end{equation}
The left-hand side of Eq.$~$($\ref{eq:schexample}$) can be treated as the redundant one and theen can be removed automatically by considering only the SSYT as the right-hand side.
As a result, the independent structures of the invariant tensor $\epsilon$ correspond to those SSYT.

Take the operator ($\ref{exop1}$) for example, it can be converted into the invariant tensors as
\begin{equation}\label{exop11}
     (N^{\dagger}\sigma\cdot\overrightarrow{\nabla}N)(N^{\dagger}\sigma\cdot\overleftarrow{\nabla}N)\longrightarrow (N_1^{\dagger}\sigma\cdot\overrightarrow{\nabla}N_3)(N_2^{\dagger}\sigma\cdot\overleftarrow{\nabla}N_4)\longrightarrow\epsilon^{i_1i_3'}\epsilon^{i_3'i_3}\epsilon^{i_2i_2'}\epsilon^{i_2'i_4},
\end{equation}
and this also corresponds to the Young tableau 
\begin{equation}
\epsilon^{i_1i_3'}\epsilon^{i_3'i_3}\epsilon^{i_2i_2'}\epsilon^{i_2'i_4}~\sim ~ 
\begin{ytableau}
i_1 & i_3' & i_2 & i_2' \\
i_3' & i_3 & i_2' & i_4
\end{ytableau}~.
\end{equation}
The operator-Young tableau correspondence converts the physical constraints of symmetry into a combinatorial problem of  semi-standard fillings, providing a clear  procedure for constructing a complete operator basis.

The non-redundant gauge factor $T$ can be determined in a similar procedure. Take the operator $N^{\dagger}\vec\sigma\cdot\vec BN$ in the HQET for another example. The rotation symmetry part is:
\begin{equation}
    N^{\dagger}\vec\sigma\cdot\vec BN=\epsilon^{i_1i_2}\epsilon^{j_2i_3} N_{i_1}^{\dagger}B_{i_2j_2}N_{i_3}\longrightarrow\epsilon^{i_1i_2}\epsilon^{j_2i_3}.
\end{equation}
Since the indices of the magnetic field $B$ are symmetric, we identifying $i_1<i_2=j_2<i_3$, and this corresponds to the only semi-standard Young tableau
\begin{equation}
\epsilon^{i_1i_2}\epsilon^{i_2i_3}~\sim~
\begin{ytableau}
i_1 & i_2 \\
i_2 & i_3
\end{ytableau} \ .
\end{equation}
For the $SU(3)_{\text{color}}$ gauge symmetry, the heavy field $N_{i,a}(x)$ carries the fundamental index $a$ of the $SU(3)$ gauge group, 
\begin{equation}
    N_a\sim\quad\begin{Young}
	a\cr
\end{Young}\ ,
\end{equation}
while the Hermitian conjugate field $N^{\dagger a}_{i}$ carries the anti-fundamental index can it is rewritten by
\begin{equation}
N^{\dagger}_{ab}\equiv\epsilon_{abc}N^{\dagger c}\sim \quad \begin{Young}
	a\cr
    b\cr
\end{Young}\ .
\end{equation}
The non-Abelian electric and magnetic fields  $E_{ij}^{A}$ and $B_{ij}^A$ carry the adjoint index $A$ of the $SU(3)$ group. The gauge indices of the non-Abelian electric and magnetic fields are rewritten in terms of the fundamental indices as
\begin{equation}
    (E)_{abc}\equiv  \epsilon_{acd}(\lambda^{A})^{d}_b E^A \sim \quad \begin{Young}
	a&b\cr
 c\cr
\end{Young},\quad
(B)_{abc}\equiv  \epsilon_{acd}(\lambda^{A})^{d}_b B^A \sim \quad \begin{Young}
	a&b\cr
 c\cr
\end{Young}.
\end{equation}
As a consequence, the gauge group part is
\begin{equation}
 (B)_{a_1b_1c_1}\equiv  \epsilon_{a_1c_1d_1}(\lambda^{A})^{d_1}_{b_1} B^A \sim \quad \begin{Young}
	$a_1$&$b_1$\cr
 $c_1$\cr
\end{Young}\ ,\quad  N^{\dagger}_{a_2b_2}\equiv\epsilon_{a_2b_2c_2}N^{\dagger c_2}\sim \quad \begin{Young}
	$a_2$\cr
    $b_2$\cr
\end{Young}\ ,\quad  N_{a_3}\sim\quad\begin{Young}
	$a_3$\cr
\end{Young}\ ,
\end{equation}
such that the only semi-standard Young tableau is 
\begin{equation}
   \begin{Young}
	$a_1$&$b_1$\cr
 $c_1$\cr
\end{Young}\quad\underrightarrow{\quad\begin{Young}
	$a_2$\cr
    $b_2$\cr
\end{Young}\quad}\quad \begin{Young}
	$a_1$&$b_1$\cr
    $c_1$&$a_2$\cr
    $b_2$\cr
\end{Young}\quad\underrightarrow{\quad\begin{Young}
	$a_3$\cr
\end{Young}\quad}\quad \begin{Young}
	$a_1$&$b_1$\cr
    $c_1$&$a_2$\cr
    $b_2$&$a_3$\cr
\end{Young}\ ,
\end{equation}
while the corresponding gauge factor is
\begin{equation}
 \epsilon^{a_1c_1b_2}\epsilon^{b_1a_2a_3}\epsilon_{a_1c_1d_1}(\lambda^{A})^{d_1}_{b_1} \epsilon_{a_2b_2c_2}=-2(\lambda^A)^{a_3}_{c_2}\equiv T_{SU(3)}^{(y)}.
\end{equation}

\subsection{Constructions of Operator Basis}\label{sec:basis}

Prior to basis construction, we establish three foundational definitions from the literature:

\begin{itemize}

\item \textbf{Type:} An operator type refers to a class of operators characterized by  given field content and derivatives. At this stage, one enumerates all possible combinations of building blocks that form singlets under both rotational and internal symmetry groups, while leaving the explicit contraction patterns implicit.

\item \textbf{Y-Basis:} The y-basis denotes a complete and independent operator set constructed through the operator--Young tableau correspondence. This framework automatically eliminates all conventional redundancies—including integration by parts (IBP) relations, equations of motion (EOM), Schouten identities, as well as the CDC (covariant derivatives commutator) and the Bianchi identity. Crucially, the y-basis does not enforce the permutation symmetry mandated by the spin-statistics theorem, thereby enabling separate treatment of rotational and internal symmetry structures. We denote the rotational y-basis as $\{\mathcal{B}^{(\text{y})}\}$ and the internal y-basis as $\{T^{(\text{y})}\}$, with the total y-basis defined by their tensor product: $\{\mathcal{T}^{(\text{y})}\}\equiv\{T^{(\text{y})}\}\otimes\{\mathcal{B}^{(\text{y})}\}$.

\item \textbf{P-Basis:} The p-basis constitutes a complete and independent operator basis obtained by imposing appropriate permutation symmetries on repeated fields within the y-basis. This basis carries definite permutation symmetry and directly corresponds to the physical effective operators, denoted as $\{\mathcal{T}^{(\text{p})}\}$.

\end{itemize}

We now turn to the systematic identification of all independent operators defined by Eq.~(\ref{eq:oprt1}).
For the rotational symmetry group $SU(2)_{\text{spin}}$, once the field content and the  derivatives are fixed, the only remaining freedom lies in the possible ways of contracting indices using the invariant  tensor $\epsilon_{ab}$. Each distinct contraction yields a unique operator; consequently, the problem reduces to classifying all linearly independent invariant tensor structures compatible with the given index configuration.

For a specified operator type, we first assign derivative indices according to the fields on which the derivatives act, thereby fixing the overall index structure. The independent invariant tensor structures for this index type are in one-to-one correspondence with semistandard Young tableaux (SSYTs), where each entry satisfies the condition that numbers are non-decreasing from left to right and  from top to bottom.  We denote the complete set of such tableaux by $\{\mathcal{A}\}$.

Since the CDC is removed, when multiple derivatives act on the same field, their associated $SU(2)_{\text{spin}}$ indices must be symmetrized. After performing this symmetrization for each tableau, we expand the resulting expressions in the SSYT basis $\{\mathcal{A}\}$. The linearly independent components obtained in this manner constitute the rotational y-basis, denoted $\{\mathcal{B}^{(\text{y})}\}$.

The construction of the y-basis assigns a distinct label to each field. To obtain physically admissible operators, however, the spin--statistics theorem imposes specific permutation-symmetry constraints on identical particles. Consequently, the y-basis must be further symmetrized so that the resulting operators are fully symmetric under the exchange of bosonic fields, or fully antisymmetric under the exchange of fermionic fields, in accordance with their quantum statistics.

For a permutation $\pi \in S_m$ acting on $m$ identical particles, the corresponding operators $\mathcal{T}^{(\mathrm{p})}$ in the resulting $\mathrm{p}$-basis must satisfy
\begin{equation}
    \mathcal{D}(\pi) \,\mathcal{T}^{(\mathrm{p})} = \pi \,\mathcal{T}^{(\mathrm{p})},
\end{equation}
where $\mathcal{D}(\pi)$ is the representation of $\pi$ on the $\mathrm{p}$-basis, determined by the particle statistics:
\begin{equation}\label{eq:permutationparity}
    \mathcal{D}(\pi) = 
    \begin{cases}
        \;1, & \text{for bosons}, \\[4pt]
        \;(-1)^{\operatorname{sgn}(\pi)}, & \text{for fermions}.
    \end{cases}
\end{equation}
Here $(-1)^{\operatorname{sgn}(\pi)}$ denotes the signature of $\pi$, which equals $+1$ for even permutations and $-1$ for odd permutations.

The $\mathrm{p}$-basis $\{\mathcal{T}^{(\mathrm{p})}\}$ is obtained from the y-basis $\{\mathcal{T}^{(\mathrm{y})}\}$ via the following procedure:
\begin{itemize}

\item First, we apply the Young operator $\mathcal{Y}_{i_1\cdots i_m}$ to enforce the appropriate permutation symmetry among identical particles $i_1\cdots i_m$ on each element of the y-basis:
\begin{equation}
   \mathcal{T}^{(\text{y})} \rightarrow \mathcal{Y}_{i_1\cdots i_m}\mathcal{T}^{(\text{y})}.
\end{equation}
We begin by deriving the generators of the symmetric group $S_m$, since every permutation of $m$ identical particles can be generated by the transposition $(12)$ and the cyclic permutation $(123\cdots m)$. The Young operators in the y-basis representation are subsequently obtained from these generators expressed in the same representation.

\item Second, we express the resulting symmetrized operators as linear combinations of the original y-basis elements
\begin{equation}\label{eq:yopptoy}
    (\mathcal{Y}_{i_1\cdots i_m}\mathcal{T}^{(\text{y})})_a = \sum_b x_{ab}^{(\text{p})} \mathcal{T}^{(\text{y})}_b,
\end{equation}
where the transformation matrix $x_{ab}^{(\text{p})}$ incorporates both the combinatorial factors from the Young operator and the appropriate statistical phases for bosonic or fermionic permutations.

\item Finally, we identify the linearly independent rows of the matrix $x_{ab}^{(\text{p})}$ to construct the p-basis $\{\mathcal{T}^{(\text{p})}\}$. This basis satisfies all permutation symmetry requirements dictated by the spin-statistics theorem and is termed the p-basis, reflecting its characterization as the polynomials of the SSYT tensors.
\end{itemize}

\paragraph{Coordinates and Hermitian Conjugation}

Any operator basis can be expanded in terms of the p-basis using the following decomposition:
\begin{equation}\label{eq:oonp}
    \mathcal{O}_m = \sum_{n=1}^{n_0} x_{mn}^{(\text{p})} \mathcal{T}_{n}^{(\text{p})},
\end{equation}
where the coefficients $x_{mn}^{(\text{p})}$ represent the unique coordinates of the operator $\mathcal{O}_m$ with respect to the p-basis operators $\mathcal{T}_n^{(\text{p})}$. Here, $n_0$ denotes the total number of operators in the p-basis. For a general operator basis $\{\mathcal{O}_m\}$ with unspecified $P$  and $T$  properties, the index $m$ ranges over $m = 1, 2, \dots, n_1$, where $n_1 = n_0$. In the restricted case of $P$-even and $T$-even operators, the range of $m$ remains $m = 1, 2, \dots, n_1$, but now $n_1 \leq n_0$. Consequently, any complete and independent operator basis $\{\mathcal{O}_m\}$ can be uniquely represented by its coordinates in the p-basis.

To expand a given operator $\widetilde{\mathcal{O}}$ in the basis $\{\mathcal{O}_m\}$, we write:
\begin{equation}
    \widetilde{\mathcal{O}} = \sum_{m=1}^{n_1} c_m \mathcal{O}_m.
\end{equation}
Direct reduction in this form may introduce computational complexity. However, this can be avoided by first expressing $\widetilde{\mathcal{O}}$ in terms of the p-basis and then employing Eq.~\eqref{eq:oonp}:
\begin{equation}\label{eq:otonp}
    \widetilde{\mathcal{O}} = \sum_{n=1}^{n_0} \left( \sum_{m=1}^{n_1} c_m x_{mn}^{(\text{p})} \right) \mathcal{T}_{n}^{(\text{p})} \equiv \sum_{n=1}^{n_0} c^{(\text{p})}_n \mathcal{T}_n^{(\text{p})}.
\end{equation}
The expansion coefficients $c_m$ in the $\{\mathcal{O}_m\}$ basis are subsequently determined by solving the linear equation:
\begin{equation}\label{eq:cord coef}
    \sum_{m=1}^{n_1} c_m x_{mn}^{(\text{p})} = c^{(\text{p})}_n.
\end{equation}

The Young tensor method significantly simplifies operator reduction. For any given operator basis, we first determine its unique coordinates relative to our y-basis. Subsequent symmetrization with Young operators then yields the coordinates in the p-basis, automatically satisfying the identical particle principle. This procedure ensures that each operator admits a unique expansion in terms of the p-basis. As concrete examples, the N-N operators presented in Refs.~\cite{Girlanda:2010ya, Filandri:2023qio} can be efficiently converted into our basis, with their independent coordinates readily determined. Similarly, the 3N operators from Ref.~\cite{Girlanda:2011fh} can be systematically mapped to our basis as well. Furthermore, by imposing constraints in the center-of-mass frame, we can construct the non-relativistic operator bases equivalent to those derived in Ref.~\cite{Xiao:2018jot}.

After the coordinates in the p‑basis are determined, we address Hermitian conjugation and the properties under $P$ and $T$.  
Consider an operator $\widetilde{\mathcal{O}}$ expanded in the p‑basis as in Eq.~\eqref{eq:otonp}.  
Its Hermitian conjugate is given by
\begin{equation}\label{eq:tildeodagger}
    \widetilde{\mathcal{O}}^{\dagger}
    = \sum_{n=1}^{n_0}  c_n^{(\mathrm{p})}\mathcal{T}_n^{(\mathrm{p})\dagger},
\end{equation} 
A Hermitian operator $\mathcal{O}^H$ can then be constructed via the linear combinations
\begin{equation}
       \mathcal{O}^H = \left\{\begin{array}{cc}
              \widetilde{\mathcal{O}} + \widetilde{\mathcal{O}}^{\dagger}&,  \\
            \\
            \mathbf{i}\left(\widetilde{\mathcal{O}} - \widetilde{\mathcal{O}}^{\dagger}\right)&. 
       \end{array}\right.
\end{equation}
The Hermitian conjugate of a p‑basis element $\mathcal{T}_n^{(\mathrm{p})}$ is obtained by exchanging the field indices with their conjugates, together with the derivatives acting on them, and subsequently multiplying the result by the permutation parity $\mathcal{D}(\pi)$ defined in Eq.~\eqref{eq:permutationparity}. Therefore, we obtain the Hermitian operators $\widetilde{\mathcal{O}} + \widetilde{\mathcal{O}}^{\dagger}$ and $\mathbf{i}\left(\widetilde{\mathcal{O}} - \widetilde{\mathcal{O}}^{\dagger}\right)$ which possess definite $P$ and $T$ transformation properties.

\subsubsection{HQET and HPET}
This subsubsection details the procedure for constructing the operator basis for both HQET and HPET using the Young tensor method. First, the y-basis is constructed directly from the relevant Young tableaux. This basis is then symmetrized, reducing it to the p-basis, which possesses the appropriate permutation symmetry. Finally, by applying the Hermitian conjugation, we derive specific linear combinations that exhibit definite behavior under the time-reversal transformations.

\paragraph{HQET}
As an illustration, consider the HQET operator type $E^{2}NN^{\dagger}$.  
Its building blocks and their index structures are
\begin{eqnarray}
(N^{\dagger}_{1})^{a}_{i_{1}}\,(E_{2})^{A}_{i_{2}i_{2}}\,(E_{3})^{B}_{i_{3}i_{3}}\,(N_{4})_{i_{4}b}\,,
\end{eqnarray}
where $i_{1},\dots ,i_{4}$ label the $SU(2)_{\text{spin}}$ indices, $a,b$ are the fundamental indices of the $SU(3)_{\text{color}}$, and $A,B$ denote the adjoint indices of the same color group.
In the absence of derivatives, the rotational y-basis is obtained directly from the semi-standard Young tableaux populated by the $SU(2)$ indices.  For the type $E^{2}NN^{\dagger}$, this construction yields exactly two independent rotational y-basis
\begin{eqnarray}\label{eq:E2rotationy}
&&\mathcal{B}^{(\text{y})}_1=\mathcal{A}_1=\epsilon^{i_1i_3}\epsilon^{i_2i_3}\epsilon^{i_2i_4}\sim\begin{ytableau} i_1 & i_2 & i_2  \\ i_3 & i_3 & i_4 \end{ytableau}~,\nonumber\\
&&\mathcal{B}^{(\text{y})}_2=\mathcal{A}_2=\epsilon^{i_1i_2}\epsilon^{i_2i_3}\epsilon^{i_3i_4}\sim\begin{ytableau} i_1 & i_2 & i_3  \\ i_2 & i_3 & i_4 \end{ytableau}~.
\end{eqnarray}
Similarly, the gauge y-basis for this type are
\begin{eqnarray}\label{eq:su3E2gauge}
   T_{SU(3),1}^{\text{(y)}}&=& d^{ABC}(\lambda^C)_a^b\ ,\nonumber
          \\ T_{SU(3),2}^{\text{(y)}}&=&\delta^{AB}\delta_a^b\ ,\nonumber
           \\
T_{SU(3),3}^{\text{(y)}}&=&\textbf{i}f^{ABC}(\lambda^C)_a^b.
\end{eqnarray}
Utilizing the tensor product of the rotational and gauge y-basis, we obtain the y-basis as
\begin{eqnarray}
    \{\mathcal{T}^{(\text{y})}_{\beta,a}\}&=&  \{\mathcal{B}^{(\text{y})}_a\}\otimes\{T_{SU(3),\beta}^{(\text{y})} \},\quad a=1,2; \beta=1,2,3,
\end{eqnarray}
where each component is defined by   
\begin{eqnarray}    \left(\begin{array}{c}
         \mathcal{T}^{(\text{y})}_{1}  \\
         \mathcal{T}^{(\text{y})}_{2} 
         \\
         \mathcal{T}^{(\text{y})}_{3}  \\
         \mathcal{T}^{(\text{y})}_{4} 
         \\
         \mathcal{T}^{(\text{y})}_{5}  \\
         \mathcal{T}^{(\text{y})}_{6} 
    \end{array}\right)&=&\left(\begin{array}{c}
         \mathcal{B}^{(\text{y})}_1T_{SU(3),1}^{(\text{y})}  \\
         \mathcal{B}^{(\text{y})}_1T_{SU(3),2}^{(\text{y})} 
         \\
         \mathcal{B}^{(\text{y})}_1T_{SU(3),3}^{(\text{y})} 
         \\
         \mathcal{B}^{(\text{y})}_2T_{SU(3),1}^{(\text{y})}
         \\
         \mathcal{B}^{(\text{y})}_2T_{SU(3),2}^{(\text{y})} 
         \\
         \mathcal{B}^{(\text{y})}_2T_{SU(3),3}^{(\text{y})} 
         
    \end{array}\right).
\end{eqnarray}

In the  type $E^{2}NN^{\dagger}$, the electromagnetic fields $E_2$ and $E_3$ constitute a pair of bonsonic repeated fields. The permutation symmetry associated with these identical fields is preserved in the p-basis. To construct this basis explicitly, we examine the transformation properties of the y-basis under the transposition $(23)$, which exchanges the labels of the two repeated fields,
\begin{eqnarray}
   (23)\left(\begin{array}{c}
         \mathcal{T}^{(\text{y})}_{1}  \\
         \mathcal{T}^{(\text{y})}_{2} 
         \\
         \mathcal{T}^{(\text{y})}_{3}  \\
         \mathcal{T}^{(\text{y})}_{4} 
         \\
         \mathcal{T}^{(\text{y})}_{5}  \\
         \mathcal{T}^{(\text{y})}_{6} 
    \end{array}\right)=\left(
\begin{array}{cccccc}
 0 & 0 & 0 & -1 & 0 & 0 \\
 0 & 0 & 0 & 0 & 1 & 0 \\
 0 & 0 & 0 & 0 & 0 & -1 \\
 -1 & 0 & 0 & 0 & 0 & 0 \\
 0 & 1 & 0 & 0 & 0 & 0 \\
 0 & 0 & -1 & 0 & 0 & 0 \\
\end{array}
\right)\left(\begin{array}{c}
         \mathcal{T}^{(\text{y})}_{1}  \\
         \mathcal{T}^{(\text{y})}_{2} 
         \\
         \mathcal{T}^{(\text{y})}_{3}  \\
         \mathcal{T}^{(\text{y})}_{4} 
         \\
         \mathcal{T}^{(\text{y})}_{5}  \\
         \mathcal{T}^{(\text{y})}_{6} 
    \end{array}\right), 
\end{eqnarray}
and then we obtain the representation of the transposition
\begin{eqnarray}
    D(23)=\left(
\begin{array}{cccccc}
 0 & 0 & 0 & -1 & 0 & 0 \\
 0 & 0 & 0 & 0 & 1 & 0 \\
 0 & 0 & 0 & 0 & 0 & -1 \\
 -1 & 0 & 0 & 0 & 0 & 0 \\
 0 & 1 & 0 & 0 & 0 & 0 \\
 0 & 0 & -1 & 0 & 0 & 0 \\
\end{array}
\right).
\end{eqnarray}
Utilizing the Young operator $\mathcal{Y}_{23}\equiv\frac{1}{2}\left[\mathbf{I}+(23)\right]$, where $\mathbf{I}$ is the identity, we can symmetrize the y-basis with the proper permutation symmetry
\begin{eqnarray}\label{eq:exampleHQETE2sym1}
    \mathcal{Y}_{23}\left(\begin{array}{c}
         \mathcal{T}^{(\text{y})}_{1}  \\
         \mathcal{T}^{(\text{y})}_{2} 
         \\
         \mathcal{T}^{(\text{y})}_{3}  \\
         \mathcal{T}^{(\text{y})}_{4} 
         \\
         \mathcal{T}^{(\text{y})}_{5}  \\
         \mathcal{T}^{(\text{y})}_{6} 
    \end{array}\right)
    =\frac{1}{2}\left[\mathbf{I}_{6\times6}+D(23)\right]\left(\begin{array}{c}
         \mathcal{T}^{(\text{y})}_{1}  \\
         \mathcal{T}^{(\text{y})}_{2} 
         \\
         \mathcal{T}^{(\text{y})}_{3}  \\
         \mathcal{T}^{(\text{y})}_{4} 
         \\
         \mathcal{T}^{(\text{y})}_{5}  \\
         \mathcal{T}^{(\text{y})}_{6} 
    \end{array}\right)
=\left(
\begin{array}{cccccc}
 \frac{1}{2} & 0 & 0 & -\frac{1}{2} & 0 & 0 \\
 0 & \frac{1}{2} & 0 & 0 & \frac{1}{2} & 0 \\
 0 & 0 & \frac{1}{2} & 0 & 0 & -\frac{1}{2} \\
 -\frac{1}{2} & 0 & 0 & \frac{1}{2} & 0 & 0 \\
 0 & \frac{1}{2} & 0 & 0 & \frac{1}{2} & 0 \\
 0 & 0 & -\frac{1}{2} & 0 & 0 & \frac{1}{2} \\
\end{array}
\right)\left(\begin{array}{c}
         \mathcal{T}^{(\text{y})}_{1}  \\
         \mathcal{T}^{(\text{y})}_{2} 
         \\
         \mathcal{T}^{(\text{y})}_{3}  \\
         \mathcal{T}^{(\text{y})}_{4} 
         \\
         \mathcal{T}^{(\text{y})}_{5}  \\
         \mathcal{T}^{(\text{y})}_{6} 
    \end{array}\right).
\end{eqnarray}
Of the six rows presented, three remain independent after the symmetrization. This independent set is identified as the p-basis:
\begin{eqnarray}\label{eq:exampleHQETE2sym2}
    \{\mathcal{T}^{(\text{p})}\}=\left\{\begin{array}{l}
        \mathcal{T}^{(\text{p})}_1=\dfrac{1}{2}\mathcal{T}^{(\text{y})}_1-\dfrac{1}{2}\mathcal{T}^{(\text{y})}_4,   \\
        \\[6pt]
        \mathcal{T}^{(\text{p})}_2=  \dfrac{1}{2}\mathcal{T}^{(\text{y})}_2+\dfrac{1}{2}\mathcal{T}^{(\text{y})}_5,
        \\[6pt]
        \\
        \mathcal{T}^{(\text{p})}_3=  \dfrac{1}{2}\mathcal{T}^{(\text{y})}_3-\dfrac{1}{2}\mathcal{T}^{(\text{y})}_6.
    \end{array}\right.
\end{eqnarray}
We can then find three independent operators
\begin{eqnarray}
    && N^{\dagger,a}(\vec\sigma\cdot\vec E^A)(\vec\sigma\cdot\vec E^B)N_bd^{ABC}(\lambda^C)_a^b+\text{h.c.}\ ,\nonumber
            \\
            &&N^{\dagger,a}(\vec\sigma\cdot\vec E^A)(\vec\sigma\cdot\vec E^B)N_b\delta^{AB}\delta_a^b+\text{h.c.}\nonumber\ ,
            \\
             &&N^{\dagger,a}(\vec\sigma\cdot\vec E^A)(\vec\sigma\cdot\vec E^B)N_b\textbf{i}f^{ABC}(\lambda^C)_a^b+\text{h.c.}\ .
\end{eqnarray}

As a second example, consider the HQET operator type $E^{2}D^{2}NN^{\dagger}$.  
the fundamental indices of the fields are assigned as $N^{\dagger}_{i_1}$, $E_{i_2i_2}$, $E_{i_3i_3}$, $N_{i_4}$.
Equation~\eqref{EOMEB} partitions the associated Young tableaux into three distinct parts:
\begin{enumerate}
\item[(i)]  tableaux in which the two derivatives are fully symmetrized with the electronic field indices they act upon;
\item[(ii)] reduced tableaux containing the rotational singlet $\vec{\nabla}\!\cdot\!\vec{E}^{A}\equiv [D^{I},E^{I,A}]$;
\item[(iii)] reduced tableaux containing the vector $\nabla^{2}E^{I,A}\equiv [D^{J},[D^{J},E^{I,A}]]$.
\end{enumerate}
Because the rotational structures (ii) and (iii) can be generated by the lower-dimensional types and by the Young tableaux in the same procedure, which are displayed in Eqs.~\eqref{eq:EDNN+DE1}, \eqref{eq:EDNN+DE2} and \eqref{eq:ENN+D2E}, we restrict the subsequent analysis to the first class (i) alone.

To construct the rotational y-basis, we consider covariant derivatives acting on the repeated electromagnetic field components $E_{i_2 i_2}$ and $E_{i_3 i_3}$. The relevant structure carrying rotational indices takes the form
\begin{eqnarray}\label{eq:indicestype}
N^{\dagger}_{i_1} \; \nabla_{i_2 i_2} E_{i_2 i_2} \, \nabla_{i_3 i_3} E_{i_3 i_3} \, N_{i_4},
\end{eqnarray}
where $\nabla_{ii}$ denotes a symmetric, traceless derivative operator appropriate for the $SU(2)_{\text{spin}}$ representations in this context.
A particular contraction of the indices in this structure corresponds to the following Young tableau:
\begin{eqnarray}
\epsilon^{i_1 i_3} \epsilon^{i_2 i_3} \epsilon^{i_2 i_3} \epsilon^{i_2 i_3} \epsilon^{i_2 i_4}
\sim
\begin{ytableau}
i_1 & i_2 & i_2 & i_2 & i_2 \\
i_3 & i_3 & i_3 & i_3 & i_4
\end{ytableau} \ .
\end{eqnarray}
This diagram make sure the corresponding structures are free from the redundancies including the Schouten identity and the EOM.
In an analogous manner, one can systematically generate all SSYT associated with this operator type; these are collected in Table~\ref{tab:E2D2ssyt}.
\begin{table}[H]
\centering
\scriptsize
\setlength{\tabcolsep}{1pt}
\renewcommand{\arraystretch}{1.8}
\begin{tabular}{@{}*{4}{c}@{}}
\toprule
$\mathcal{A}_1\sim\begin{ytableau} i_1 & i_2 & i_2 & i_2 & i_2 \\ i_3 & i_3 & i_3 & i_3 & i_4 \end{ytableau}$ &
$\mathcal{A}_2\sim\begin{ytableau} i_1 & i_2 & i_2 & i_2 & i_3 \\ i_2 & i_3 & i_3 & i_3 & i_4 \end{ytableau}$ &
$\mathcal{A}_3\sim\begin{ytableau} i_1 & i_2 & i_2 & i_2 & i_2 \\ i_3 & i_3 & i_4 & i_4' & i_4' \end{ytableau}$ &
$\mathcal{A}_4\sim\begin{ytableau} i_1 & i_2 & i_2 & i_2 & i_3 \\ i_2 & i_3 & i_4 & i_4' & i_4' \end{ytableau}$ \\
\midrule
$\mathcal{A}_5\sim\begin{ytableau} i_1 & i_2 & i_2 & i_2 & i_4 \\ i_2 & i_3 & i_3 & i_4' & i_4' \end{ytableau}$ &
$\mathcal{A}_6\sim\begin{ytableau} i_1 & i_2 & i_2 & i_3 & i_3 \\ i_3 & i_3 & i_4 & i_4' & i_4' \end{ytableau}$ &
$\mathcal{A}_7\sim\begin{ytableau} i_1 & i_2 & i_2 & i_3 & i_4 \\ i_3 & i_3 & i_3 & i_4' & i_4' \end{ytableau}$ &
$\mathcal{A}_8\sim\begin{ytableau} i_1 & i_2 & i_3 & i_3 & i_3 \\ i_2 & i_3 & i_4 & i_4' & i_4' \end{ytableau}$ \\
\midrule
$\mathcal{A}_9\sim\begin{ytableau} i_1 & i_2 & i_2 & i_3 & i_3 \\ i_4 & i_4' & i_4' & i_4'' & i_4'' \end{ytableau}$ &
$\mathcal{A}_{10}\sim\begin{ytableau} i_1 & i_2 & i_2 & i_3 & i_4 \\ i_3 & i_4' & i_4' & i_4'' & i_4'' \end{ytableau}$ &
$\mathcal{A}_{11}\sim\begin{ytableau} i_1 & i_2 & i_2 & i_3 & i_4' \\ i_3 & i_4 & i_4' & i_4'' & i_4'' \end{ytableau}$ &
$\mathcal{A}_{12}\sim\begin{ytableau} i_1 & i_2 & i_2 & i_4 & i_4' \\ i_3 & i_3 & i_4' & i_4'' & i_4'' \end{ytableau}$ \\
\midrule
$\mathcal{A}_{13}\sim\begin{ytableau} i_1 & i_2 & i_2 & i_4' & i_4' \\ i_3 & i_3 & i_4 & i_4'' & i_4'' \end{ytableau}$ &
$\mathcal{A}_{14}\sim\begin{ytableau} i_1 & i_2 & i_3 & i_3 & i_4 \\ i_2 & i_4' & i_4' & i_4'' & i_4'' \end{ytableau}$ &
$\mathcal{A}_{15}\sim\begin{ytableau} i_1 & i_2 & i_3 & i_3 & i_4' \\ i_2 & i_4 & i_4' & i_4'' & i_4'' \end{ytableau}$ &
$\mathcal{A}_{16}\sim\begin{ytableau} i_1 & i_2 & i_3 & i_4 & i_4' \\ i_2 & i_3 & i_4' & i_4'' & i_4'' \end{ytableau}$ \\
\midrule
$\mathcal{A}_{17}\sim\begin{ytableau} i_1 & i_2 & i_3 & i_4' & i_4' \\ i_2 & i_3 & i_4 & i_4'' & i_4'' \end{ytableau}$
\\
\bottomrule
\end{tabular}
\caption{The semi-standard Young tableaux of the type $E^2D^2NN^{\dagger}$. There are total 17 SSYT, and they are denoted as $\mathcal{A}_1,\cdots \mathcal{A}_{17}$.}
\label{tab:E2D2ssyt}
\end{table}

The rotational y-basis $\mathcal{B}^{(\text{y})}$ is subsequently constructed by symmetrizing the indices of derivatives that act the same field, i.e., by symmetrizing the indices  $i_4'$ and $i_4''$:
\begin{equation}\label{eq:E2D2yb}
\begin{array}{|rcl|}
\mathcal{B}^{\text{(y)}}_1 &=& \mathcal{A}_1, \\
\mathcal{B}^{\text{(y)}}_2 &=& \mathcal{A}_2, \\
\mathcal{B}^{\text{(y)}}_3 &=& \mathcal{A}_3, \\
\mathcal{B}^{\text{(y)}}_4 &=& \mathcal{A}_4, \\
\mathcal{B}^{\text{(y)}}_5 &=& \mathcal{A}_5, \\
\mathcal{B}^{\text{(y)}}_6 &=& \mathcal{A}_6, \\
\mathcal{B}^{\text{(y)}}_7 &=& \mathcal{A}_7,
\end{array}
\quad
\begin{array}{ccl|}
\mathcal{B}^{\text{(y)}}_8 &=& \mathcal{A}_8, \\
\mathcal{B}^{\text{(y)}}_9 &=& \mathcal{A}_9 - \mathcal{A}_{11} + \frac{1}{2}\mathcal{A}_{13} + \mathcal{A}_{15} - \mathcal{A}_{16} + \frac{1}{2}\mathcal{A}_{17}, \\
\mathcal{B}^{\text{(y)}}_{10} &=& \mathcal{A}_{10} - \frac{1}{2}\mathcal{A}_{11} - \frac{1}{2}\mathcal{A}_{12} + \frac{1}{2}\mathcal{A}_{13}, \\
\mathcal{B}^{\text{(y)}}_{11} &=& \frac{1}{2}\mathcal{A}_{13}, \\
\mathcal{B}^{\text{(y)}}_{12} &=& \mathcal{A}_{14} - \frac{1}{2}\mathcal{A}_{15}, \\
\mathcal{B}^{\text{(y)}}_{13} &=& \frac{1}{2}\mathcal{A}_{17}.
\end{array}\ .
\end{equation}
The gauge y-basis coincides with that already established for the type $E^{2}NN^{\dagger}$ and displayed in Eq.~\eqref{eq:su3E2gauge}.
\begin{eqnarray}\label{eq:E2D2gb}
   T_{SU(3),1}^{\text{(y)}}&=& d^{ABC}(\lambda^C)_a^b\ ,\nonumber
          \\ T_{SU(3),2}^{\text{(y)}}&=&\delta^{AB}\delta_a^b\ ,\nonumber
           \\
T_{SU(3),3}^{\text{(y)}}&=&\textbf{i}f^{ABC}(\lambda^C)_a^b.
\end{eqnarray}
Utilizing Eq.~\eqref{eq:E2D2yb} and \eqref{eq:E2D2gb}, we obtain the y-basis for the type $E^2D^2NN^{\dagger}$ by the tensor product
\begin{eqnarray}
    \{\mathcal{T}^{(\text{y})}_{\beta,a}\}&=&  \{\mathcal{B}^{(\text{y})}_a\}\otimes\{T_{SU(3),\beta}^{(\text{y})} \},\quad, a=1,\cdots,17;\beta=1,2,3.
\end{eqnarray}

To obtain the p-basis with proper permutation symmetry, following the similar procedure in Eq.~\eqref{eq:exampleHQETE2sym1} and \eqref{eq:exampleHQETE2sym2}, we symmetrize the y-basis and select the independent rows:
\begin{eqnarray}
   \underbrace{\mathcal{Y}_{23} \left\{\mathcal{T}^{(\text{y})}_{\beta,a}\right\}}_{\text{independent rows}}\longrightarrow\left\{\mathcal{T}^{\text{(p)}}_n\right\}.
\end{eqnarray}
Among the 51 elements of the y-basis $\left\{\mathcal{T}^{(\text{y})}_{\beta,a}\right\}$, application of the Young operator $\mathcal{Y}_{23}$ yields a set $\mathcal{Y}_{23} \left\{\mathcal{T}^{(\text{y})}_{\beta,a}\right\}$ that contains only 20 linearly independent rows. These independent combinations respect the required permutation symmetry under the transposition $(23)$ and constitute the p-basis. We denote the elements of this p-basis by $\mathcal{T}^{(\text{p})}_n$, where $n = 1, \dots, 20$,
\begin{equation}
\begin{array}{|ccl|}
\mathcal{T}^{(\text{p})}_1 &=& \frac{1}{2} \mathcal{T}^{(\text{y})}_1 - \frac{1}{2} \mathcal{T}^{(\text{y})}_4 \\
\mathcal{T}^{(\text{p})}_2 &=& \frac{1}{2} \mathcal{T}^{(\text{y})}_2 + \frac{1}{2} \mathcal{T}^{(\text{y})}_5 \\
\mathcal{T}^{(\text{p})}_3 &=& \frac{1}{2} \mathcal{T}^{(\text{y})}_3 - \frac{1}{2} \mathcal{T}^{(\text{y})}_6 \\
\mathcal{T}^{(\text{p})}_4 &=& \frac{1}{2} \mathcal{T}^{(\text{y})}_7 - \frac{1}{2} \mathcal{T}^{(\text{y})}_{22} \\
\mathcal{T}^{(\text{p})}_5 &=& \frac{1}{2} \mathcal{T}^{(\text{y})}_8 + \frac{1}{2} \mathcal{T}^{(\text{y})}_{23} \\
\mathcal{T}^{(\text{p})}_6 &=& \frac{1}{2} \mathcal{T}^{(\text{y})}_9 - \frac{1}{2} \mathcal{T}^{(\text{y})}_{24} \\
\mathcal{T}^{(\text{p})}_7 &=& \frac{1}{2} \mathcal{T}^{(\text{y})}_{10} - \frac{1}{2} \mathcal{T}^{(\text{y})}_{16} + \frac{1}{2} \mathcal{T}^{(\text{y})}_{19} \\
\mathcal{T}^{(\text{p})}_8 &=& \frac{1}{2} \mathcal{T}^{(\text{y})}_{11} + \frac{1}{2} \mathcal{T}^{(\text{y})}_{17} - \frac{1}{2} \mathcal{T}^{(\text{y})}_{20} \\
\mathcal{T}^{(\text{p})}_9 &=& \frac{1}{2} \mathcal{T}^{(\text{y})}_{12} - \frac{1}{2} \mathcal{T}^{(\text{y})}_{18} + \frac{1}{2} \mathcal{T}^{(\text{y})}_{21} \\
\mathcal{T}^{(\text{p})}_{10} &=& \frac{1}{2} \mathcal{T}^{(\text{y})}_{13} + \frac{1}{2} \mathcal{T}^{(\text{y})}_{19} 
\end{array}
\quad
\begin{array}{ccl|}
\mathcal{T}^{(\text{p})}_{11} &=& \frac{1}{2} \mathcal{T}^{(\text{y})}_{14} - \frac{1}{2} \mathcal{T}^{(\text{y})}_{20} \\
\mathcal{T}^{(\text{p})}_{12} &=& \frac{1}{2} \mathcal{T}^{(\text{y})}_{15} + \frac{1}{2} \mathcal{T}^{(\text{y})}_{21} \\
\mathcal{T}^{(\text{p})}_{13} &=& \mathcal{T}^{(\text{y})}_{25} \\
\mathcal{T}^{(\text{p})}_{14} &=& \mathcal{T}^{(\text{y})}_{27} \\
\mathcal{T}^{(\text{p})}_{15} &=& \frac{1}{2} \mathcal{T}^{(\text{y})}_{28} + \frac{1}{2} \mathcal{T}^{(\text{y})}_{34} - \frac{1}{2} \mathcal{T}^{(\text{y})}_{37} \\
\mathcal{T}^{(\text{p})}_{16} &=& \frac{1}{2} \mathcal{T}^{(\text{y})}_{29} - \frac{1}{2} \mathcal{T}^{(\text{y})}_{35} + \frac{1}{2} \mathcal{T}^{(\text{y})}_{38} \\
\mathcal{T}^{(\text{p})}_{17} &=& \frac{1}{2} \mathcal{T}^{(\text{y})}_{30} + \frac{1}{2} \mathcal{T}^{(\text{y})}_{36} - \frac{1}{2} \mathcal{T}^{(\text{y})}_{39} \\
\mathcal{T}^{(\text{p})}_{18} &=& \frac{1}{2} \mathcal{T}^{(\text{y})}_{31} - \frac{1}{2} \mathcal{T}^{(\text{y})}_{37} \\
\mathcal{T}^{(\text{p})}_{19} &=& \frac{1}{2} \mathcal{T}^{(\text{y})}_{32} + \frac{1}{2} \mathcal{T}^{(\text{y})}_{38} 
\\
\mathcal{T}^{(\text{p})}_{20} &=& \frac{1}{2} \mathcal{T}^{(\text{y})}_{33} - \frac{1}{2} \mathcal{T}^{(\text{y})}_{39}
\end{array}\ .
\end{equation}

However, not all of the 20 independent combinations in the p-basis $\left\{\mathcal{T}^{(\text{p})}_n\right\}$ are even under the time reversal. Since the operators in the HQET must be invariant under time reversal, we retain only those basis elements that are both linearly independent and $T$-even.
To identify such operators, we examine the transformation properties of the y-basis under Hermitian conjugation, 
\begin{equation}
    \left\{\mathcal{B}^{(\text{y})\dagger}_a\right\}\otimes \left\{T_{SU(3),\beta}^{\text{(y)}\dagger}\right\}=\left(D_{\text{dagB}}^{\text{(y)}} \otimes D_{\text{dagT}}^{\text{(y)}}\right) \left(\left\{\mathcal{B}^{(\text{y})}_a\right\}\otimes \left\{T_{SU(3),\beta}^{\text{(y)}}\right\}\right),
\end{equation}
\begin{eqnarray}
  \left( \begin{array}{c}
        \mathcal{B}^{(\text{y})\dagger}_1\\
        \mathcal{B}^{(\text{y})\dagger}_2\\
        \mathcal{B}^{(\text{y})\dagger}_3\\
        \mathcal{B}^{(\text{y})\dagger}_4\\
        \mathcal{B}^{(\text{y})\dagger}_5\\
        \mathcal{B}^{(\text{y})\dagger}_6\\
        \mathcal{B}^{(\text{y})\dagger}_7\\
        \mathcal{B}^{(\text{y})\dagger}_8\\
        \mathcal{B}^{(\text{y})\dagger}_9\\
        \mathcal{B}^{(\text{y})\dagger}_{10}\\
        \mathcal{B}^{(\text{y})\dagger}_{11}\\
        \mathcal{B}^{(\text{y})\dagger}_{12}\\
        \mathcal{B}^{(\text{y})\dagger}_{13}\\
   \end{array}\right)&=& \left(
\begin{array}{ccccccccccccc}
 0 & -1 & 0 & 0 & 0 & 0 & 0 & 0 & 0 & 0 & 0 & 0 & 0 \\
 -1 & 0 & 0 & 0 & 0 & 0 & 0 & 0 & 0 & 0 & 0 & 0 & 0 \\
 0 & 1 & 0 & 1 & -1 & 0 & 0 & 0 & 0 & 0 & 0 & 0 & 0 \\
 0 & 0 & 0 & 1 & 0 & 0 & 0 & 0 & 0 & 0 & 0 & 0 & 0 \\
 -1 & 0 & -1 & 1 & 0 & 0 & 0 & 0 & 0 & 0 & 0 & 0 & 0 \\
 0 & 1 & 0 & 0 & 0 & 0 & 0 & 1 & 0 & 0 & 0 & 0 & 0 \\
 0 & 1 & 0 & 0 & 0 & -1 & 1 & 1 & 0 & 0 & 0 & 0 & 0 \\
 1 & 0 & 0 & 0 & 0 & 1 & 0 & 0 & 0 & 0 & 0 & 0 & 0 \\
 1 & -1 & 1 & -1 & 1 & 1 & 0 & -1 & 1 & 0 & 0 & 0 & 0 \\
 0 & -1 & 0 & -1 & 1 & 1 & -1 & -1 & 1 & -1 & 0 & 0 & 0 \\
 0 & -1 & 0 & -1 & 1 & 0 & 0 & -1 & 0 & 0 & 0 & 0 & -1 \\
 0 & 0 & 0 & -1 & 0 & 0 & 0 & 0 & 1 & 0 & -1 & -1 & 1 \\
 -1 & 0 & -1 & 0 & 0 & -1 & 0 & 0 & 0 & 0 & -1 & 0 & 0 \\
\end{array}
\right)\left( \begin{array}{c}
        \mathcal{B}^{(\text{y})}_1\\
        \mathcal{B}^{(\text{y})}_2\\
        \mathcal{B}^{(\text{y})}_3\\
        \mathcal{B}^{(\text{y})}_4\\
        \mathcal{B}^{(\text{y})}_5\\
        \mathcal{B}^{(\text{y})}_6\\
        \mathcal{B}^{(\text{y})}_7\\
        \mathcal{B}^{(\text{y})}_8\\
        \mathcal{B}^{(\text{y})}_9\\
        \mathcal{B}^{(\text{y})}_{10}\\
        \mathcal{B}^{(\text{y})}_{11}\\
        \mathcal{B}^{(\text{y})}_{12}\\
        \mathcal{B}^{(\text{y})}_{13}\\
   \end{array}\right)\nonumber\\
   &=&D_{\text{dagB}}^{\text{(y)}} \left\{\mathcal{B}^{(\text{y})}_a\right\},
\end{eqnarray}

\begin{eqnarray}
    \left(\begin{array}{c}
         T_{SU(3),1}^{\text{(y)}\dagger}  \\
         T_{SU(3),1}^{\text{(y)}\dagger}  \\
         T_{SU(3),2}^{\text{(y)}\dagger}  
    \end{array}\right)&=&\left(
\begin{array}{ccc}
 1 & 0 & 0 \\
 0 & -1 & 0 \\
 0 & 0 & 1 \\
\end{array}
\right)\left(\begin{array}{c}
         T_{SU(3),1}^{\text{(y)}}  \\
         T_{SU(3),1}^{\text{(y)}}  \\
         T_{SU(3),2}^{\text{(y)}}  
    \end{array}\right)\nonumber\\
    &=&D_{\text{dagT}}^{\text{(y)}} \left\{T_{SU(3),\beta}^{\text{(y)}}\right\} ,
\end{eqnarray}
The matrices $D_{\text{dagB}}^{\text{(y)}}$ and $D_{\text{dagT}}^{\text{(y)}}$ act as representation matrices on the rotational and gauge y-basis, respectively. The representation matrix on the resulting p-basis, $D^{(\text{p})}_{\text{dag}}$, can therefore be obtained by the independent rows after the symmetrization,
\begin{equation}
\underbrace{\mathcal{Y}_{23}\left(D_{\text{dagB}}^{\text{(y)}} \otimes D_{\text{dagT}}^{\text{(y)}}\right) }_{\text{independent rows}}=D_{\text{dag}}^{(\text{p})},
\end{equation}
\begin{eqnarray}
   \mathcal{T}^{\text{(p)}\dagger}_n=\sum_{m=1}^{20}\left[D_{\text{dag}}^{(\text{p})}\right]_{nm}\mathcal{T}^{\text{(p)}}_m,\quad,n=1,\cdots,20,
\end{eqnarray}
where the representation matrix $D_{\text{dag}}^{\text{(p)}}$ on the p-basis read
\begin{eqnarray}
D_{\text{dag}}^{\text{(p)}}=\left(
\begin{array}{cccccccccccccccccccc}
 1 & 0 & 0 & 0 & 0 & 0 & 0 & 0 & 0 & 0 & 0 & 0 & 0 & 0 & 0 & 0 & 0 & 0 & 0 & 0 \\
 0 & 1 & 0 & 0 & 0 & 0 & 0 & 0 & 0 & 0 & 0 & 0 & 0 & 0 & 0 & 0 & 0 & 0 & 0 & 0 \\
 0 & 0 & 1 & 0 & 0 & 0 & 0 & 0 & 0 & 0 & 0 & 0 & 0 & 0 & 0 & 0 & 0 & 0 & 0 & 0 \\
 -1 & 0 & 0 & 0 & 0 & 0 & 1 & 0 & 0 & -1 & 0 & 0 & 0 & 0 & 0 & 0 & 0 & 0 & 0 & 0 \\
 0 & -1 & 0 & 0 & 0 & 0 & 0 & -1 & 0 & 0 & 1 & 0 & 0 & 0 & 0 & 0 & 0 & 0 & 0 & 0 \\
 0 & 0 & -1 & 0 & 0 & 0 & 0 & 0 & 1 & 0 & 0 & -1 & 0 & 0 & 0 & 0 & 0 & 0 & 0 & 0 \\
 0 & 0 & 0 & 0 & 0 & 0 & 1 & 0 & 0 & 0 & 0 & 0 & 0 & 0 & 0 & 0 & 0 & 0 & 0 & 0 \\
 0 & 0 & 0 & 0 & 0 & 0 & 0 & -1 & 0 & 0 & 0 & 0 & 0 & 0 & 0 & 0 & 0 & 0 & 0 & 0 \\
 0 & 0 & 0 & 0 & 0 & 0 & 0 & 0 & 1 & 0 & 0 & 0 & 0 & 0 & 0 & 0 & 0 & 0 & 0 & 0 \\
 -1 & 0 & 0 & -1 & 0 & 0 & 1 & 0 & 0 & 0 & 0 & 0 & 0 & 0 & 0 & 0 & 0 & 0 & 0 & 0 \\
 0 & 1 & 0 & 0 & 1 & 0 & 0 & -1 & 0 & 0 & 0 & 0 & 0 & 0 & 0 & 0 & 0 & 0 & 0 & 0 \\
 0 & 0 & -1 & 0 & 0 & -1 & 0 & 0 & 1 & 0 & 0 & 0 & 0 & 0 & 0 & 0 & 0 & 0 & 0 & 0 \\
 2 & 0 & 0 & 2 & 0 & 0 & -2 & 0 & 0 & 2 & 0 & 0 & 1 & 0 & 0 & 0 & 0 & 0 & 0 & 0 \\
 0 & 0 & 2 & 0 & 0 & 2 & 0 & 0 & -2 & 0 & 0 & 2 & 0 & 1 & 0 & 0 & 0 & 0 & 0 & 0 \\
 1 & 0 & 0 & 1 & 0 & 0 & -2 & 0 & 0 & 1 & 0 & 0 & 1 & 0 & -1 & 0 & 0 & 0 & 0 & 0 \\
 0 & 1 & 0 & 0 & 1 & 0 & 0 & 0 & 0 & 0 & -1 & 0 & 0 & 0 & 0 & 1 & 0 & 0 & 0 & 0 \\
 0 & 0 & 1 & 0 & 0 & 1 & 0 & 0 & -2 & 0 & 0 & 1 & 0 & 1 & 0 & 0 & -1 & 0 & 0 & 0 \\
 1 & 0 & 0 & 1 & 0 & 0 & -1 & 0 & 0 & 1 & 0 & 0 & 0 & 0 & 0 & 0 & 0 & 1 & 0 & 0 \\
 0 & 1 & 0 & 0 & 1 & 0 & 0 & 1 & 0 & 0 & -1 & 0 & 0 & 0 & 0 & 0 & 0 & 0 & 1 & 0 \\
 0 & 0 & 1 & 0 & 0 & 1 & 0 & 0 & -1 & 0 & 0 & 1 & 0 & 0 & 0 & 0 & 0 & 0 & 0 & 1 \\
\end{array}
\right).\nonumber\\
\end{eqnarray}
Consequently, from the p-basis we identify 14 independent combinations that satisfy both $T$-even and Hermitian:
\begin{eqnarray}
\left(\begin{array}{c}
     \mathcal{T}^{\text(p)\dagger}_1  \\
    \mathcal{T}^{\text(p)\dagger}_2 \\
    \mathcal{T}^{\text(p)\dagger}_3 \\
    \mathcal{T}^{\text(p)\dagger}_7 \\
    \mathcal{T}^{\text(p)\dagger}_9 \\
    -\mathcal{T}^{\text(p)\dagger}_4+\mathcal{T}^{\text(p)\dagger}_{10}\\
    
\mathcal{T}^{\text(p)\dagger}_5-\mathcal{T}^{\text(p)\dagger}_8+\mathcal{T}^{\text(p)\dagger}_{11} \\
-\mathcal{T}^{\text(p)\dagger}_6+\mathcal{T}^{\text(p)\dagger}_{12} \\
2\mathcal{T}^{\text(p)\dagger}_4+\mathcal{T}^{\text(p)\dagger}_{13} \\
2\mathcal{T}^{\text(p)\dagger}_6+\mathcal{T}^{\text(p)\dagger}_{14} \\
2\mathcal{T}^{\text(p)\dagger}_5-\mathcal{T}^{\text(p)\dagger}_{8} +2\mathcal{T}^{\text(p)\dagger}_{16}\\
\mathcal{T}^{\text(p)\dagger}_{4} +\mathcal{T}^{\text(p)\dagger}_{18}\\
\mathcal{T}^{\text(p)\dagger}_{5} +\mathcal{T}^{\text(p)\dagger}_{19}\\
\mathcal{T}^{\text(p)\dagger}_{6} +\mathcal{T}^{\text(p)\dagger}_{20}\\
\end{array}\right)=\left(\begin{array}{c}
     \mathcal{T}^{\text(p)}_1  \\
    \mathcal{T}^{\text(p)}_2 \\
    \mathcal{T}^{\text(p)}_3 \\
    \mathcal{T}^{\text(p)}_7 \\
    \mathcal{T}^{\text(p)}_9 \\
    -\mathcal{T}^{\text(p)}_4+\mathcal{T}^{\text(p)}_{10}\\
    
\mathcal{T}^{\text(p)}_5-\mathcal{T}^{\text(p)}_8+\mathcal{T}^{\text(p)}_{11} \\
-\mathcal{T}^{\text(p)}_6+\mathcal{T}^{\text(p)}_{12} \\
2\mathcal{T}^{\text(p)}_4+\mathcal{T}^{\text(p)}_{13} \\
2\mathcal{T}^{\text(p)}_6+\mathcal{T}^{\text(p)}_{14} \\
2\mathcal{T}^{\text(p)}_5-\mathcal{T}^{\text(p)}_{8} +2\mathcal{T}^{\text(p)}_{16}\\
\mathcal{T}^{\text(p)}_{4} +\mathcal{T}^{\text(p)}_{18}\\
\mathcal{T}^{\text(p)}_{5} +\mathcal{T}^{\text(p)}_{19}\\
\mathcal{T}^{\text(p)}_{6} +\mathcal{T}^{\text(p)}_{20}\\
\end{array}\right).
\end{eqnarray}
We can then obtain 14 $P$ even and $T$ even HQET operators of the type $E^2D^2NN^{\dagger}$: 
\begin{eqnarray}\label{eq:E2D20}
    &&N^{\dagger,a}[D^I,(\vec\sigma\cdot\vec E^B)][D^J,(\vec\sigma\cdot\vec E^A)]N_b\text{tr}\left[\sigma^I\sigma^J\right]d^{ABC}(\lambda^C)_a^b+\text{h.c.}\ ,\nonumber\\
     &&N^{\dagger,a}[D^I,(\vec\sigma\cdot\vec E^B)][D^J,(\vec\sigma\cdot\vec E^A)]N_b\text{tr}\left[\sigma^I\sigma^J\right]\delta^{AB}\delta_a^b+\text{h.c.}\ , \nonumber\\
     &&N^{\dagger,a}(\vec\sigma\cdot\vec E^B)[D^I,(\vec\sigma\cdot\vec E^A)]D^JN_b\text{tr}\left[\sigma^I\sigma^J\right]d^{ABC}(\lambda^C)_a^b+\text{h.c.}\ ,\nonumber\\
      &&N^{\dagger,a}(\vec\sigma\cdot\vec E^B)[D^I,(\vec\sigma\cdot\vec E^A)]D^JN_b\text{tr}\left[\sigma^I\sigma^J\right]\delta^{AB}\delta_a^b+\text{h.c.}\ , \nonumber\\
      &&N^{\dagger,a}[D^I,(\vec\sigma\cdot\vec E^A)](\vec\sigma\cdot\vec E^B)(\vec\sigma\cdot\vec D)\sigma^IN_bd^{ABC}(\lambda^C)_a^b+\text{h.c.}\ ,\nonumber\\
       &&N^{\dagger,a}[D^I,(\vec\sigma\cdot\vec E^A)](\vec\sigma\cdot\vec E^B)(\vec\sigma\cdot\vec D)\sigma^IN_b\delta^{AB}\delta_a^b+\text{h.c.}\ , \nonumber\\
        &&N^{\dagger,a}D^ID^JN_b\text{tr}\left[(\vec\sigma\cdot\vec E^A)\sigma^I\right]\text{tr}\left[(\vec\sigma\cdot\vec E^B)\sigma^J\right]d^{ABC}(\lambda^C)_a^b+\text{h.c.}\ ,\nonumber\\
        &&N^{\dagger,a}D^ID^JN_b\text{tr}\left[(\vec\sigma\cdot\vec E^A)\sigma^I\right]\text{tr}\left[(\vec\sigma\cdot\vec E^B)\sigma^J\right]\delta^{AB}\delta_a^b+\text{h.c.}\ , \nonumber\\
        &&N^{\dagger,a}(\vec\sigma\cdot\vec E^B)\sigma^I\sigma^J(\vec\sigma\cdot\vec E^A)D^ID^JN_b d^{ABC}(\lambda^C)_a^b+\text{h.c.}\ ,\nonumber\\
         &&N^{\dagger,a}(\vec\sigma\cdot\vec E^B)\sigma^I\sigma^J(\vec\sigma\cdot\vec E^A)D^ID^JN_b\delta^{AB}\delta_a^b+\text{h.c.}\ , \nonumber\\
          &&N^{\dagger,a}[D^I,(\vec\sigma\cdot\vec E^B)][D^J,(\vec\sigma\cdot\vec E^A)]N_b\text{tr}\left[\sigma^I\sigma^J\right]\textbf{i}f^{ABC}\delta_a^b+\text{h.c.}\ ,\nonumber\\
          &&N^{\dagger,a}(\vec\sigma\cdot\vec E^B)[D^I,(\vec\sigma\cdot\vec E^A)]D^JN_b\text{tr}\left[\sigma^I\sigma^J\right]\textbf{i}f^{ABC}\delta_a^b+\text{h.c.}\ ,\nonumber\\
          &&N^{\dagger,a}(\vec\sigma\cdot\vec E^B)(\vec\sigma\cdot\vec D)D^IN_b\text{tr}\left[(\vec\sigma\cdot\vec E^A)\sigma^I\right]\textbf{i}f^{ABC}\delta_a^b+\text{h.c.}\ ,\nonumber\\
           &&N^{\dagger,a}(\vec\sigma\cdot\vec E^B)\sigma^I\sigma^J(\vec\sigma\cdot\vec E^A)D^ID^JN_b \textbf{i}f^{ABC}\delta_a^b+\text{h.c.}\ ,
\end{eqnarray}

Furthermore, additional results are obtained by retaining the gauge y-basis $\{T_{\text{SU(3)},\beta}^{\text{(y)}}\}$ for operator  type $E^2D^2NN^{\dagger}$ and forming products with  the rotational y-basis $\{\mathcal{B}_a^{\text{(y)}}\}$ of the types $EDNN^{\dagger}$ and $E^2NN^{\dagger}$:
\begin{itemize}
    \item $ EDNN^{\dagger}\rightarrow E^2D^2NN^{\dagger}:$ 
\begin{eqnarray}\label{eq:EDNN+DE1}
     &&N^{\dagger,a}(\vec\sigma\cdot\vec E^A)(\vec\sigma\cdot\vec D)N_b[D^J,E^{J,B}]d^{ABC}(\lambda^C)_a^b+\text{h.c.}\ ,\nonumber\\
     &&N^{\dagger,a}(\vec\sigma\cdot\vec E^A)(\vec\sigma\cdot\vec D)N_b[D^J,E^{J,B}]\delta^{AB}\delta_a^c+\text{h.c.}\ ,\nonumber\\
     &&N^{\dagger,a}(\vec\sigma\cdot\vec E^A)(\vec\sigma\cdot\vec D)N_b[D^J,E^{J,B}]\textbf{i}f^{ABC}(\lambda^C)_a^b+\text{h.c.}\ .
\end{eqnarray}
\begin{eqnarray}\label{eq:EDNN+DE2}
     &&N^{\dagger,a}[D^I,E^{I,A}]N_b[D^J,E^{J,B}]d^{ABC}(\lambda^C)_a^b+\text{h.c.}\ ,\nonumber\\
     &&N^{\dagger,a}[D^I,E^{I,A}]N_b[D^J,E^{J,B}]\delta^{AB}\delta_a^c+\text{h.c.}\ ,
\end{eqnarray}
\item $E^2NN^{\dagger}\rightarrow E^2D^2NN^{\dagger}:$
\begin{eqnarray}\label{eq:ENN+D2E}
    &&N^{\dagger,a}(\vec\sigma\cdot\vec E^B)N_b[D^I,[D^I,(\vec\sigma\cdot\vec E^A)]]d^{ABC}(\lambda^C)_a^b+\text{h.c.}\ ,\nonumber\\
       &&N^{\dagger,a}(\vec\sigma\cdot\vec E^B)N_b[D^I,[D^I,(\vec\sigma\cdot\vec E^A)]]\delta^{AB}\delta_a^c+\text{h.c.}\ ,\nonumber\\
         &&N^{\dagger,a}(\vec\sigma\cdot\vec E^B)N_b[D^I,[D^I,(\vec\sigma\cdot\vec E^A)]]\textbf{i}f^{ABC}(\lambda^C)_a^b+\text{h.c.}\ .
\end{eqnarray}
\end{itemize}
Therefore, according to Eq.~\eqref{eq:E2D20}, \eqref{eq:EDNN+DE1}, \eqref{eq:EDNN+DE2}, and
\eqref{eq:ENN+D2E}, there are total 22 $P$ even $T$ even operators for the type $E^2D^2NN^{\dagger}$ in the HQET, which coincides with the results given by the Hilbert series Eq.~\eqref{eq:HQETHS}. The complete HQET operator bases up to dimension-9 are listed in appendix~\ref{ap:HQET}.

The systematical procedure to obtain the y-basis in this work  avoids potential omissions in constructing operators. While a method based on ordering the covariant derivatives is discussed in Ref.~\cite{Gunawardana:2017zix}, our framework offers a streamlined and redundancy-free construction.  As an illustration,  we compare the procedures in Ref.~\cite{Gunawardana:2017zix} with our method for the type $E^2NN^{\dagger}$ in the HQET: 
\begin{itemize}
    \item In that work, all possible  contractions of the spin-independent structure
    \begin{equation}\label{eq:gil1}
        \langle H|\bar h iD^{\mu}iD^{\nu}iD^{\rho}iD^{\sigma} h|H\rangle,
    \end{equation}
     as well as the spin-dependent structure
     \begin{equation}\label{eq:gil2}
         \langle H|\bar h iD^{\mu}iD^{\nu}iD^{\rho}iD^{\sigma} s^{\alpha} h|H\rangle,
     \end{equation}
     are considered explicitly.  Although these expressions are written with Lorentz indices, the contractions are effectively performed using $SO(3)$-invariant tensors due to the heavy-quark velocity $v^{\mu}$, and the resulting operators are $SO(3)$ singlets. Consequently, there are redundancies of the invariant tensor $\Pi^{\mu\nu}\epsilon^{\alpha\beta\lambda\rho}v_{\rho}$ originating from the Schouten identity. In contrast, our construction does not introduce the spin operator $s^{\alpha}$ explicitly. Instead,  we work directly with building blocks that transform  under the rotational and internal symmetries. The y-basis obtained via the semi-standard Young tableaux is inherently free from such redundancies.

\item The structures in Eq.~\eqref{eq:gil1} and Eq.~\eqref{eq:gil2}  are only related to the rotational structures, and does not systematically consider the independent gauge structures basis. In our approach, all independent gauge structures obtained by the semi-standard Young tableau are incorporated at the level of the y-basis, leaving the permutations of identical fields at the level of the p-basis,  as shown for example in Eq.~\eqref{eq:su3E2gauge}. For example, the gauge structure of the operator $N^{\dagger}E^A\lambda^AE^B\lambda^BN$ can be expanded by the gauge y-basis as
\begin{equation}
   \left(N^{\dagger}E^AE^BN\right)\lambda^A \lambda^B=\left(N^{\dagger}E^AE^BN\right)\left(\frac{1}{2}T_{SU(3),1}^{\text{(y)}}+\frac{1}{6}T_{SU(3),2}^{\text{(y)}}+\frac{1}{2}T_{SU(3),3}^{\text{(y)}}\right).
\end{equation}

Mixing the rotation and gauge y-basis while imposing  the permutation symmetry can lead to the omission of certain operators. For the type $E^2NN^{\dagger}$, the three independent operators are
\begin{eqnarray}\label{eq:opE^2}
     &&N^{\dagger,a}(\vec\sigma\cdot\vec E^A)(\vec\sigma\cdot\vec E^B)N_bd^{ABC}(\lambda^C)_a^b+\text{h.c.}\ ,\nonumber
            \\
 &&N^{\dagger,a}(\vec\sigma\cdot\vec E^A)(\vec\sigma\cdot\vec E^B)N_b\delta^{AB}\delta_a^b+\text{h.c.}\ , \nonumber
            \\           &&N^{\dagger,a}(\vec\sigma\cdot\vec E^A)(\vec\sigma\cdot\vec E^B)N_b\textbf{i}f^{ABC}(\lambda^C)_a^b+\text{h.c.}
\end{eqnarray}
Using the identity
\begin{equation}
    (\vec\sigma\cdot\vec E^A)(\vec\sigma\cdot\vec E^B)=E^{I,A}E^{J,B}\delta^{IJ}+\mathbf{i}\epsilon^{IJK}\sigma^KE^{I,A}E^{J,B},
\end{equation}
there are both symmetric and anti-symmetric part of the rotational  y-basis. 
When considering the symmetric part in the rotational structures in Eq.~\eqref{eq:gil1} and Eq.~\eqref{eq:gil2}, only the first two operators in Eq.~\eqref{eq:opE^2} are obtained. The anti-symmetric rotation y-basis and the anti-symmetric gauge y-basis give the last operator that can be incorrectly omitted.

\end{itemize}

\paragraph{HPET}

The same procedure applies directly to the HPET, where in this Abelian gauge theory we need only construct the y-basis with respect to the rotational $SU(2)_{\text{spin}}$ symmetry.  For the type $E^2NN^{\dagger}$ in the HPET, the rotational y-basis is the same as the Eq.~\eqref{eq:E2rotationy}
\begin{eqnarray}
&&\mathcal{B}^{(\text{y})}_1=\mathcal{A}_1=\epsilon^{i_1i_3}\epsilon^{i_2i_3}\epsilon^{i_2i_4}\sim\begin{ytableau} i_1 & i_2 & i_2  \\ i_3 & i_3 & i_4 \end{ytableau}~,\nonumber\\
&&\mathcal{B}^{(\text{y})}_2=\mathcal{A}_2=\epsilon^{i_1i_2}\epsilon^{i_2i_3}\epsilon^{i_3i_4}\sim\begin{ytableau} i_1 & i_2 & i_3  \\ i_2 & i_3 & i_4 \end{ytableau}~.
\end{eqnarray}
However, the transposition $(23)$ relates two elements,
\begin{eqnarray}
    (23)\left(\begin{array}{c}
         \mathcal{B}^{(\text{y})}_1  \\
         \mathcal{B}^{(\text{y})}_2
    \end{array}\right)=\left(\begin{array}{cc}
        0 & -1 \\
        -1 & 0
    \end{array}\right)\left(\begin{array}{c}
         \mathcal{B}^{(\text{y})}_1  \\
         \mathcal{B}^{(\text{y})}_2
    \end{array}\right),
\end{eqnarray}
such that the only independent p-basis element is 
\begin{eqnarray}
    \mathcal{T}^{(\text{p})}=\mathcal{B}^{(\text{y})}_1-\mathcal{B}^{(\text{y})}_2, 
\end{eqnarray}
which corresponds to the HPET operator
\begin{eqnarray}
    N^{\dagger}\vec E^2N.
\end{eqnarray}
For the type $E^2D^2NN^{\dagger}$ in the HPET, the y-basis is the same as Eq.~\eqref{eq:E2D2yb},
\begin{equation}
    \mathcal{T}^{(\text{y})}_a=\mathcal{B}^{\text{(y)}}_a,\quad a=1,\cdots,13,
\end{equation}
\begin{equation}
\begin{array}{|rcl|}
\mathcal{B}^{\text{(y)}}_1 &=& \mathcal{A}_1, \\
\mathcal{B}^{\text{(y)}}_2 &=& \mathcal{A}_2, \\
\mathcal{B}^{\text{(y)}}_3 &=& \mathcal{A}_3, \\
\mathcal{B}^{\text{(y)}}_4 &=& \mathcal{A}_4, \\
\mathcal{B}^{\text{(y)}}_5 &=& \mathcal{A}_5, \\
\mathcal{B}^{\text{(y)}}_6 &=& \mathcal{A}_6, \\
\mathcal{B}^{\text{(y)}}_7 &=& \mathcal{A}_7,
\end{array}
\quad
\begin{array}{ccl|}
\mathcal{B}^{\text{(y)}}_8 &=& \mathcal{A}_8, \\
\mathcal{B}^{\text{(y)}}_9 &=& \mathcal{A}_9 - \mathcal{A}_{11} + \frac{1}{2}\mathcal{A}_{13} + \mathcal{A}_{15} - \mathcal{A}_{16} + \frac{1}{2}\mathcal{A}_{17}, \\
\mathcal{B}^{\text{(y)}}_{10} &=& \mathcal{A}_{10} - \frac{1}{2}\mathcal{A}_{11} - \frac{1}{2}\mathcal{A}_{12} + \frac{1}{2}\mathcal{A}_{13}, \\
\mathcal{B}^{\text{(y)}}_{11} &=& \frac{1}{2}\mathcal{A}_{13}, \\
\mathcal{B}^{\text{(y)}}_{12} &=& \mathcal{A}_{14} - \frac{1}{2}\mathcal{A}_{15}, \\
\mathcal{B}^{\text{(y)}}_{13} &=& \frac{1}{2}\mathcal{A}_{17}.
\end{array}\ .
\end{equation}
After the symmetrization, we find that
\begin{eqnarray}
   \mathcal{Y}_{23}\left( \begin{array}{c}
        \mathcal{B}^{(\text{y})}_1\\
        \mathcal{B}^{(\text{y})}_2\\
        \mathcal{B}^{(\text{y})}_3\\
        \mathcal{B}^{(\text{y})}_4\\
        \mathcal{B}^{(\text{y})}_5\\
        \mathcal{B}^{(\text{y})}_6\\
        \mathcal{B}^{(\text{y})}_7\\
        \mathcal{B}^{(\text{y})}_8\\
        \mathcal{B}^{(\text{y})}_9\\
        \mathcal{B}^{(\text{y})}_{10}\\
        \mathcal{B}^{(\text{y})}_{11}\\
        \mathcal{B}^{(\text{y})}_{12}\\
        \mathcal{B}^{(\text{y})}_{13}\\
   \end{array}\right)= \left(
\begin{array}{ccccccccccccc}
 \frac{1}{2} & -\frac{1}{2} & 0 & 0 & 0 & 0 & 0 & 0 & 0 & 0 & 0 & 0 & 0 \\
 -\frac{1}{2} & \frac{1}{2} & 0 & 0 & 0 & 0 & 0 & 0 & 0 & 0 & 0 & 0 & 0 \\
 0 & 0 & \frac{1}{2} & 0 & 0 & 0 & 0 & -\frac{1}{2} & 0 & 0 & 0 & 0 & 0 \\
 0 & 0 & 0 & \frac{1}{2} & 0 & -\frac{1}{2} & \frac{1}{2} & 0 & 0 & 0 & 0 & 0 & 0 \\
 0 & 0 & 0 & 0 & \frac{1}{2} & 0 & \frac{1}{2} & 0 & 0 & 0 & 0 & 0 & 0 \\
 0 & 0 & 0 & -\frac{1}{2} & \frac{1}{2} & \frac{1}{2} & 0 & 0 & 0 & 0 & 0 & 0 & 0 \\
 0 & 0 & 0 & 0 & \frac{1}{2} & 0 & \frac{1}{2} & 0 & 0 & 0 & 0 & 0 & 0 \\
 0 & 0 & -\frac{1}{2} & 0 & 0 & 0 & 0 & \frac{1}{2} & 0 & 0 & 0 & 0 & 0 \\
 0 & 0 & 0 & 0 & 0 & 0 & 0 & 0 & 1 & 0 & 0 & 0 & 0 \\
 0 & 0 & 0 & 0 & 0 & 0 & 0 & 0 & 0 & \frac{1}{2} & 0 & \frac{1}{2} & -\frac{1}{2} \\
 0 & 0 & 0 & 0 & 0 & 0 & 0 & 0 & 0 & 0 & \frac{1}{2} & 0 & -\frac{1}{2} \\
 0 & 0 & 0 & 0 & 0 & 0 & 0 & 0 & 0 & \frac{1}{2} & -\frac{1}{2} & \frac{1}{2} & 0 \\
 0 & 0 & 0 & 0 & 0 & 0 & 0 & 0 & 0 & 0 & -\frac{1}{2} & 0 & \frac{1}{2} \\
\end{array}
\right)\left( \begin{array}{c}
        \mathcal{B}^{(\text{y})}_1\\
        \mathcal{B}^{(\text{y})}_2\\
        \mathcal{B}^{(\text{y})}_3\\
        \mathcal{B}^{(\text{y})}_4\\
        \mathcal{B}^{(\text{y})}_5\\
        \mathcal{B}^{(\text{y})}_6\\
        \mathcal{B}^{(\text{y})}_7\\
        \mathcal{B}^{(\text{y})}_8\\
        \mathcal{B}^{(\text{y})}_9\\
        \mathcal{B}^{(\text{y})}_{10}\\
        \mathcal{B}^{(\text{y})}_{11}\\
        \mathcal{B}^{(\text{y})}_{12}\\
        \mathcal{B}^{(\text{y})}_{13}\\
   \end{array}\right),
\end{eqnarray}
where the independent rows in the above equation are defined as the p-basis:
\begin{eqnarray}
    \left(\begin{array}{c}
         \mathcal{T}^{(\text{p})}_1  \\
         \mathcal{T}^{(\text{p})}_2  \\
         \mathcal{T}^{(\text{p})}_3  \\
         \mathcal{T}^{(\text{p})}_4  \\
         \mathcal{T}^{(\text{p})}_5  \\
         \mathcal{T}^{(\text{p})}_6  \\
         \mathcal{T}^{(\text{p})}_7  
    \end{array}\right)=\left(
\begin{array}{ccccccccccccc}
 \frac{1}{2} & -\frac{1}{2} & 0 & 0 & 0 & 0 & 0 & 0 & 0 & 0 & 0 & 0 & 0 \\
 0 & 0 & \frac{1}{2} & 0 & 0 & 0 & 0 & -\frac{1}{2} & 0 & 0 & 0 & 0 & 0 \\
 0 & 0 & 0 & \frac{1}{2} & 0 & -\frac{1}{2} & \frac{1}{2} & 0 & 0 & 0 & 0 & 0 & 0 \\
 0 & 0 & 0 & 0 & \frac{1}{2} & 0 & \frac{1}{2} & 0 & 0 & 0 & 0 & 0 & 0 \\
 0 & 0 & 0 & 0 & 0 & 0 & 0 & 0 & 1 & 0 & 0 & 0 & 0 \\
 0 & 0 & 0 & 0 & 0 & 0 & 0 & 0 & 0 & \frac{1}{2} & 0 & \frac{1}{2} & -\frac{1}{2} \\
 0 & 0 & 0 & 0 & 0 & 0 & 0 & 0 & 0 & 0 & \frac{1}{2} & 0 & -\frac{1}{2} \\
\end{array}
\right)\left( \begin{array}{c}
        \mathcal{B}^{(\text{y})}_1\\
        \mathcal{B}^{(\text{y})}_2\\
        \mathcal{B}^{(\text{y})}_3\\
        \mathcal{B}^{(\text{y})}_4\\
        \mathcal{B}^{(\text{y})}_5\\
        \mathcal{B}^{(\text{y})}_6\\
        \mathcal{B}^{(\text{y})}_7\\
        \mathcal{B}^{(\text{y})}_8\\
        \mathcal{B}^{(\text{y})}_9\\
        \mathcal{B}^{(\text{y})}_{10}\\
        \mathcal{B}^{(\text{y})}_{11}\\
        \mathcal{B}^{(\text{y})}_{12}\\
        \mathcal{B}^{(\text{y})}_{13}\\
   \end{array}\right).
\end{eqnarray}
Under the Hermitian conjugation, the p-basis transforms as
\begin{eqnarray}
   \mathcal{T}^{\text{(p)}}_n=\sum_{m=1}^{7}\left[D_{\text{dag}}^{(\text{p})}\right]_{nm}\mathcal{T}^{\text{(p)}}_m,\quad,n=1,\cdots,7,
\end{eqnarray}
where the representation matrix $D_{\text{dag}}^{\text{(p)}}$ on the p-basis read

\begin{eqnarray}
   D_{\text{dag}}^{\text{(p)}}= \left(
\begin{array}{ccccccc}
 1 & 0 & 0 & 0 & 0 & 0 & 0 \\
 -1 & 0 & 1 & -1 & 0 & 0 & 0 \\
 0 & 0 & 1 & 0 & 0 & 0 & 0 \\
 -1 & -1 & 1 & 0 & 0 & 0 & 0 \\
 2 & 2 & -2 & 2 & 1 & 0 & 0 \\
 1 & 1 & -2 & 1 & 1 & -1 & 0 \\
 1 & 1 & -1 & 1 & 0 & 0 & 1 \\
\end{array}
\right).
\end{eqnarray}
Consequently, we obtain the Hermitian combinations
\begin{eqnarray}
    \left(\begin{array}{c}
         \mathcal{T}^{\text{(p)}\dagger}_2+\mathcal{T}^{\text{(p)}\dagger}_7  \\
         2\mathcal{T}^{\text{(p)}\dagger}_2+\mathcal{T}^{\text{(p)}\dagger}_5\\
         -\mathcal{T}^{\text{(p)}\dagger}_2+\mathcal{T}^{\text{(p)}\dagger}_4 
         \\
         \mathcal{T}^{\text{(p)}\dagger}_3
         \\
         \mathcal{T}^{\text{(p)}\dagger}_1
    \end{array}\right)=\left(\begin{array}{c}
         \mathcal{T}^{\text{(p)}}_2+\mathcal{T}^{\text{(p)}}_7  \\
         2\mathcal{T}^{\text{(p)}}_2+\mathcal{T}^{\text{(p)}}_5\\
         -\mathcal{T}^{\text{(p)}}_2+\mathcal{T}^{\text{(p)}}_4 
         \\
         \mathcal{T}^{\text{(p)}}_3
         \\
         \mathcal{T}^{\text{(p)}}_1
    \end{array}\right),
\end{eqnarray}
and then we can find five independent Hermitian and $T$ even operators
\begin{eqnarray}
     &&N^{\dagger}[D^I,E^J][D^K,E^L](\delta^{IK}\delta^{JL}+\delta^{IL}\delta^{JK}N)+\text{h.c.}\ , \nonumber\\
            &&N^{\dagger}[D^I,E^J]E^KD^LN(\textbf{i}\epsilon^{JKL}\sigma^I+\textbf{i}\epsilon^{IKL}\sigma^L)N+\text{h.c.}\ , \nonumber\\
             &&N^{\dagger}[D^I,E^J]E^KD^L(\textbf{i}\epsilon^{IKM}\sigma^M\delta^{IL}+\textbf{i}\epsilon^{JKM}\sigma^M\delta^{IL})N+\text{h.c.}\ , \nonumber\\
             &&N^{\dagger}E^IE^JD^KD^L\delta^{IJ}\delta^{KL}N+\text{h.c.}\ , \nonumber\\
             &&N^{\dagger}E^IE^JD^KD^L\delta^{IK}\delta^{JL}N+\text{h.c.}\ .
\end{eqnarray}
Three additional operators arise from reduced Young tableaux associated with the operator types $EDNN^{\dagger}$ and $E^2NN^{\dagger}$. These are constructed using the same procedure outlined above; for brevity, we omit the intermediate steps and present only the resulting independent structures:
\begin{itemize}
\item $EDNN^{\dagger} \rightarrow E^2D^2NN^{\dagger}$:
\begin{eqnarray}
&& N^{\dagger} E^I D^J \sigma^K \, \mathbf{i} \epsilon^{IJK} N \, [D^M, E^N] \delta^{MN} + \text{h.c.}, \nonumber\\
&& N^{\dagger} [D^I, E^J] [D^K, E^L] \delta^{IJ} \delta^{KL} N,
\end{eqnarray}

\item  $E^2NN^{\dagger} \rightarrow E^2D^2NN^{\dagger}$:
\begin{eqnarray}
&& N^{\dagger} E^I [D^M, [D^N, E^J]] \delta^{IJ} \delta^{MN} N.
\end{eqnarray}
\end{itemize}
Together with the previously five operators, these contributions yield a total of eight linearly independent operators in the type $E^2D^2NN^{\dagger}$ that are simultaneously even under parity  and time reversal. This count is in exact agreement with the results from the Hilbert series given in Eq.~\eqref{eq:HSHPETcp+}.

Finally, the complete operator bases of the HPET  up to mass dimension nine is compiled in Appendix~\ref{app:HQETop}.

\subsubsection{Pionless EFT}

In this subsubsection, we present the construction of the N-N contact operators. For the type $N^2(N^{\dagger})^2$, there are precisely two SSYTs that generate the rotational y-basis:
\begin{equation}
    \mathcal{B}^{(\text{y})}_1=\epsilon^{i_1i_2}\epsilon^{i_3i_4}\sim~\begin{ytableau}
1 & 3 \\
2 & 4
\end{ytableau}
\ ,\quad \mathcal{B}^{(\text{y})}_2=\epsilon^{i_1i_3}\epsilon^{i_2i_4}\sim~
\begin{ytableau}
1 & 2 \\
3 & 4
\end{ytableau}\ .
\end{equation}
For the isospin $SU(2)_{\text{isospin}}$ symmetry, an analogous gauge y-basis is constructed using the Young tableaux, following the procedure detailed in Subsection~\ref{sec:op-yt}.
Denoting the isospin (anti-)fundamental indices of the fields as $(N_1^{\dagger})^{p_1}(N_2^{\dagger})^{p_2}$ $(N_{3}){p_3}(N_{4})_{p_4}$, the two independent isospin y-basis elements $T_{SU(2),\beta}^{(\text{y})}$ labeled by $\beta=1,2$ are 
\begin{eqnarray}\label{eq:4Ngauge}
    I_1\equiv T_{SU(2),1}^{(\text{y})}&=&\delta_{p_3}^{p_1}\delta_{p_4}^{p_2},\nonumber\\
    I_2\equiv T_{SU(2),2}^{(\text{y})}&=&\delta_{p_4}^{p_1}\delta_{p_3}^{p_2}.
\end{eqnarray}
The full y-basis for the operator type $N^2(N^{\dagger})^2$ is then obtained by taking the tensor product of 
the rotational and isospin y-basis
\begin{equation}
  \{\mathcal{T}^{(\text{y})}\}= \{\mathcal{B}^{(\text{y})}\}\otimes \{T_{SU(2)}^{(\text{y})}\}= \left\{\begin{array}{lll}
         \mathcal{T}^{(\text{y})}_1=\mathcal{B}_1^{(\text{y})}\otimes T_{SU(2),1}^{(\text{y})}=\epsilon^{i_1i_2}\epsilon^{i_3i_4}\delta_{p_3}^{p_1}\delta_{p_4}^{p_2}, \\
         \\
         \mathcal{T}^{(\text{y})}_2=\mathcal{B}_1^{(\text{y})}\otimes T_{SU(2),2}^{(\text{y})}=\epsilon^{i_1i_2}\epsilon^{i_3i_4}\delta_{p_4}^{p_1}\delta_{p_3}^{p_2},
         \\
         \\
         \mathcal{T}^{(\text{y})}_3=\mathcal{B}_2^{(\text{y})}\otimes T_{SU(2),1}^{(\text{y})}=\epsilon^{i_1i_3}\epsilon^{i_2i_4} \delta_{p_3}^{p_1}\delta_{p_4}^{p_2}, \\
         \\
         \mathcal{T}^{(\text{y})}_4=\mathcal{B}_2^{(\text{y})}\otimes T_{SU(2),2}^{(\text{y})}=\epsilon^{i_1i_3}\epsilon^{i_2i_4} \delta_{p_4}^{p_1}\delta_{p_3}^{p_2}.
    \end{array}\right.
\end{equation}

In the N–N contact interaction, the outgoing nucleon fields $N_1^{\dagger}$ and $N_2^{\dagger}$ form a pair of identical fermions, as do the incoming fields $N_3$ and $N_4$. Because nucleons are fermionic, the  operator must be anti-symmetric under the exchange of any two identical particles. Consequently, to properly impose the required anti-symmetry on the operator basis, we employ the Young operators that project onto the anti-symmetric representation for each pair. These Young operators are given by
\begin{eqnarray}\label{eq:yop1234}
\mathcal{Y}_{12} &=& \frac{1}{2}\left[E - (12)\right], \nonumber\\
\mathcal{Y}_{34} &=& \frac{1}{2}\left[E - (34)\right],
\end{eqnarray}
where $E$ denotes the identity permutation, and $(12)$ and $(34)$ represent the transpositions exchanging the outgoing and incoming nucleon labels, respectively. The minus signs in front of the transpositions encode the fermionic statistics: exchanging two identical fermions introduces a factor of $-1$, which is implemented here at the operator level through the Young operators. Applying the $\mathcal{Y}_{12}$ and $\mathcal{Y}_{34}$ to the y-basis ensures that the resulting p-basis respects the permutation symmetry.

Under the transposition $(12)$, the y-basis transforms as
\begin{eqnarray}
    (12)\left(\begin{array}{l}
        \mathcal{T}_1^{(\text{y})}  \\
         \mathcal{T}_2^{(\text{y})}\\
         \mathcal{T}_3^{(\text{y})}\\
         \mathcal{T}_4^{(\text{y})}
    \end{array}\right)=\left(\begin{array}{l}
              \epsilon^{i_2i_1}\epsilon^{i_3i_4}\delta_{p_3}^{p_2}\delta_{p_4}^{p_1}  \\
              \epsilon^{i_2i_1}\epsilon^{i_3i_4}\delta_{p_4}^{p_2}\delta_{p_3}^{p_1}
              \\
         \epsilon^{i_2i_3}\epsilon^{i_1i_4} \delta_{p_3}^{p_2}\delta_{p_4}^{p_1}
         \\
         \epsilon^{i_2i_3}\epsilon^{i_1i_4} \delta_{p_4}^{p_2}\delta_{p_3}^{p_1}
         \end{array}\right)=\left(\begin{array}{rrrr}
              0&-1&0&0  \\
              -1&0&0&0\\
              0&-1&0&1\\
              -1&0&1&0
         \end{array}\right)\left(\begin{array}{l}
        \mathcal{T}_1^{(\text{y})}  \\
         \mathcal{T}_2^{(\text{y})}\\
         \mathcal{T}_3^{(\text{y})}\\
         \mathcal{T}_4^{(\text{y})}
    \end{array}\right),
\end{eqnarray}
as a result, we obtain the representation of the transposition $(12)$ in the y-basis
\begin{equation}
    D(12)=\left(\begin{array}{rrrr}
              0&-1&0&0  \\
              -1&0&0&0\\
              0&-1&0&1\\
              -1&0&1&0
         \end{array}\right).
\end{equation}
Similarly, we obtain the representation of the transposition $(34)$ as
\begin{equation}
    D(34)=\left(\begin{array}{rrrr}
              0&-1&0&0  \\
              -1&0&0&0\\
              0&-1&0&1\\
              -1&0&1&0
         \end{array}\right).
\end{equation}
After that, we can then use the Young operators in Eq.~\eqref{eq:yop1234} to symmetrize the  y-basis as Eq.~\eqref{eq:yopptoy}:
\begin{eqnarray}
    \mathcal{Y}_{12}\mathcal{Y}_{34}\left(\begin{array}{l}
        \mathcal{T}_1^{(\text{y})}  \\
         \mathcal{T}_2^{(\text{y})}\\
         \mathcal{T}_3^{(\text{y})}\\
         \mathcal{T}_4^{(\text{y})}
    \end{array}\right)&=&\frac{1}{4}\left[\mathbf{I}_{4\times4}-D(12)-D(34)+D(12)D(34)\right]\left(\begin{array}{l}
        \mathcal{T}_1^{(\text{y})}  \\
         \mathcal{T}_2^{(\text{y})}\\
         \mathcal{T}_3^{(\text{y})}\\
         \mathcal{T}_4^{(\text{y})}
    \end{array}\right)\\
    &=&\left(\begin{array}{rrrr}
              \frac{1}{2}&\frac{1}{2}&0&0  \\
              \frac{1}{2}&\frac{1}{2}&0&0\\
              0&\frac{1}{2}&\frac{1}{2}&-\frac{1}{2}\\
              \frac{1}{2}&0&-\frac{1}{2}&\frac{1}{2}
         \end{array}\right)\left(\begin{array}{l}
        \mathcal{T}_1^{(\text{y})}  \\
         \mathcal{T}_2^{(\text{y})}\\
         \mathcal{T}_3^{(\text{y})}\\
         \mathcal{T}_4^{(\text{y})}
    \end{array}\right).
\end{eqnarray}
Since only the first and the third rows are independent, we obtain the p-basis for the type $N^2(N^{\dagger})^2$ as
\begin{equation}
    \{\mathcal{T}^{(\text{p})}\}=\left\{\begin{array}{l}
        \mathcal{T}^{(\text{p})}_1=\frac{1}{2}\mathcal{T}^{(\text{y})}_1+\frac{1}{2}\mathcal{T}^{(\text{y})}_2,   \\
       \\
       \mathcal{T}^{(\text{p})}_2=  \frac{1}{2}\mathcal{T}^{(\text{y})}_2+\frac{1}{2}\mathcal{T}^{(\text{y})}_3-\frac{1}{2}\mathcal{T}^{(\text{y})}_4.
    \end{array}\right.
\end{equation}
Having obtained the p-basis for the type $N^2(N^{\dagger})^2$, two familiar  operators in Eq.~\eqref{eq:csct} can be converted in to this p-basis:  
\begin{eqnarray}
(N^{\dagger}N)(N^{\dagger}N)&=&\epsilon^{i_1i_3}\epsilon^{i_2i_4}(N^{\dagger})^{p_1}_{i_1}(N^{\dagger})^{p_2}_{i_2}N_{i_3,p_1}N_{i_4,p_2}
\nonumber\\
&\leftrightarrow&\mathcal{Y}_{12}\mathcal{Y}_{34}\epsilon^{i_1i_3}\epsilon^{i_2i_4}\delta_{p_3}^{p_1}\delta_{p_4}^{p_2}\nonumber\\
  &  =&\mathcal{T}_2^{(\text{p})},
\end{eqnarray}
\begin{eqnarray}
    (N^{\dagger}\vec\sigma N)\cdot(N^{\dagger}\vec \sigma N)&=&\left(-2\epsilon^{i_1i_4}\epsilon^{i_2i_3}-\epsilon^{i_1i_3}\epsilon^{i_2i_4}\right)(N^{\dagger})^{p_1}_{i_1}(N^{\dagger})^{p_2}_{i_2}N_{i_3,p_1}N_{i_4,p_2}\nonumber\\
    &\leftrightarrow&\mathcal{Y}_{12}\mathcal{Y}_{34}\left(-2\epsilon^{i_1i_4}\epsilon^{i_2i_3}-\epsilon^{i_1i_3}\epsilon^{i_2i_4}\right)\delta_{p_3}^{p_1}\delta_{p_4}^{p_2}\nonumber\\
    &=&2\mathcal{T}_1^{(\text{p})}-3\mathcal{T}_2^{(\text{p})}.
\end{eqnarray}
In the first line of the above equation, we utilize the Fierz identity in Eq.~\eqref{eq:Fierz} to reduce the Pauli matrices.

As a further illustration we analyze the  type $N^{2}(N^{\dagger})^{2}\nabla^{2}$.  
For this operator type, the rotational $SU(2)_{\text{spin}}$  alone produces 36 independent tensors $\mathcal A_{a}$ ($a=1,\dots ,36$); each tensor is in one-to-one correspondence with a semi-standard Young tableau listed in Tab.~\ref{examplessyt1}.  
The indices $i_{1},i_{2},i_{3},i_{4}$ carried by the $\epsilon$-tensors are the fundamental $SU(2)_{\text{spin}}$ indices of the nucleon fields $N^{\dagger}_1$, $N^{\dagger}_2$, $N_3$ and $N_4$.  
The primed pair $i'_{2},i''_{2}$ (with the ordering convention $i'_{2}<i''_{2}$) labels the fundamental indices of the first and second derivatives, where $\nabla_{i'_{2}i'_{2}}$ and $\nabla_{i''_{2}i''_{2}}$, both acting on the  field $N^{\dagger}_{2}$.

\begin{table}[htbp]
\centering
\scriptsize
\setlength{\tabcolsep}{1pt}
\renewcommand{\arraystretch}{1.8}
\begin{tabular}{@{}*{6}{c}@{}}
\toprule
$\mathcal{A}_1\sim\begin{ytableau} i_1 & i_2 & i'_2 & i'_2 \\ i''_2 & i''_2 & i_3 & i_4 \end{ytableau}$ &
$\mathcal{A}_2\sim\begin{ytableau} i_1 & i_2 & i'_2 & i''_2 \\ i'_2 & i''_2 & i_3 & i_4 \end{ytableau}$ &
$\mathcal{A}_3\sim\begin{ytableau} i_1 & i_2 & i'_2 & i_3 \\ i'_2 & i''_2 & i''_2 & i_4 \end{ytableau}$ &
$\mathcal{A}_4\sim\begin{ytableau} i_1 & i_2 & i''_2 & i''_2 \\ i'_2 & i'_2 & i_3 & i_4 \end{ytableau}$ &
$\mathcal{A}_5\sim\begin{ytableau} i_1 & i'_2 & i'_2 & i''_2 \\ i_2 & i''_2 & i_3 & i_4 \end{ytableau}$ &
$\mathcal{A}_6\sim\begin{ytableau} i_1 & i'_2 & i'_2 & i_3 \\ i_2 & i''_2 & i''_2 & i_4 \end{ytableau}$ \\
\midrule
$\mathcal{A}_7\sim\begin{ytableau} i_1 & i_2 & i'_2 & i'_2 \\ i_3 & i'_3 & i'_3 & i_4 \end{ytableau}$ &
$\mathcal{A}_8\sim\begin{ytableau} i_1 & i_2 & i'_2 & i_3 \\ i'_2 & i'_3 & i'_3 & i_4 \end{ytableau}$ &
$\mathcal{A}_9\sim\begin{ytableau} i_1 & i_2 & i'_2 & i'_3 \\ i'_2 & i_3 & i'_3 & i_4 \end{ytableau}$ &
$\mathcal{A}_{10}\sim\begin{ytableau} i_1 & i_2 & i_3 & i'_3 \\ i'_2 & i'_2 & i'_3 & i_4 \end{ytableau}$ &
$\mathcal{A}_{11}\sim\begin{ytableau} i_1 & i'_2 & i'_2 & i_3 \\ i_2 & i'_3 & i'_3 & i_4 \end{ytableau}$ &
$\mathcal{A}_{12}\sim\begin{ytableau} i_1 & i'_2 & i'_2 & i'_3 \\ i_2 & i_3 & i'_3 & i_4 \end{ytableau}$ \\
\midrule
$\mathcal{A}_{13}\sim\begin{ytableau} i_1 & i_2 & i'_2 & i'_2 \\ i_3 & i_4 & i'_4 & i'_4 \end{ytableau}$ &
$\mathcal{A}_{14}\sim\begin{ytableau} i_1 & i_2 & i'_2 & i_3 \\ i'_2 & i_4 & i'_4 & i'_4 \end{ytableau}$ &
$\mathcal{A}_{15}\sim\begin{ytableau} i_1 & i_2 & i'_2 & i_4 \\ i'_2 & i_3 & i'_4 & i'_4 \end{ytableau}$ &
$\mathcal{A}_{16}\sim\begin{ytableau} i_1 & i_2 & i_3 & i_4 \\ i'_2 & i'_2 & i'_4 & i'_4 \end{ytableau}$ &
$\mathcal{A}_{17}\sim\begin{ytableau} i_1 & i'_2 & i'_2 & i_3 \\ i_2 & i_4 & i'_4 & i'
_4 \end{ytableau}$ &
$\mathcal{A}_{18}\sim\begin{ytableau} i_1 & i'_2 & i'_2 & i_4 \\ i_2 & i_3 & i'_4 & i'_4 \end{ytableau}$ \\
\midrule
$\mathcal{A}_{19}\sim\begin{ytableau} i_1 & i_2 & i_3 & i'_3 \\ i'_3 & i''_3 & i''_3 & i_4 \end{ytableau}$ &
$\mathcal{A}_{20}\sim\begin{ytableau} i_1 & i_2 & i_3 & i''_3 \\ i'_3 & i'_3 & i''_3 & i_4 \end{ytableau}$ &
$\mathcal{A}_{21}\sim\begin{ytableau} i_1 & i_2 & i'_3 & i'_3 \\ i_3 & i''_3 & i''_3 & i_4 \end{ytableau}$ &
$\mathcal{A}_{22}\sim\begin{ytableau} i_1 & i_2 & i'_3 & i''_3 \\ i_3 & i'_3 & i''_3 & i_4 \end{ytableau}$ &
$\mathcal{A}_{23}\sim\begin{ytableau} i_1 & i_3 & i'_3 & i'_3 \\ i_2 & i''_3 & i''_3 & i_4 \end{ytableau}$ &
$\mathcal{A}_{24}\sim\begin{ytableau} i_1 & i_3 & i'_3 & i''_3 \\ i_2 & i'_3 & i''_3 & i_4 \end{ytableau}$ \\
\midrule
$\mathcal{A}_{25}\sim\begin{ytableau} i_1 & i_2 & i_3 & i'_3 \\ i'_3 & i_4 & i'_4 & i'_4 \end{ytableau}$ &
$\mathcal{A}_{26}\sim\begin{ytableau} i_1 & i_2 & i_3 & i_4 \\ i'_3 & i'_3 & i'_4 & i'_4 \end{ytableau}$ &
$\mathcal{A}_{27}\sim\begin{ytableau} i_1 & i_2 & i'_3 & i'_3 \\ i_3 & i_4 & i'_4 & i'_4 \end{ytableau}$ &
$\mathcal{A}_{28}\sim\begin{ytableau} i_1 & i_2 & i'_3 & i_4 \\ i_3 & i'_3 & i'_4 & i'_4 \end{ytableau}$ &
$\mathcal{A}_{29}\sim\begin{ytableau} i_1 & i_3 & i'_3 & i'_3 \\ i_2 & i_4 & i'_4 & i'_4 \end{ytableau}$ &
$\mathcal{A}_{30}\sim\begin{ytableau} i_1 & i_3 & i'_3 & i_4 \\ i_2 & i'_3 & i'_4 & i'_4 \end{ytableau}$ \\
\midrule
$\mathcal{A}_{31}\sim\begin{ytableau} i_1 & i_2 & i_3 & i_4 \\ i'_4 & i'_4 & i''_4 & i''_4 \end{ytableau}$ &
$\mathcal{A}_{32}\sim\begin{ytableau} i_1 & i_2 & i_3 & i'_4 \\ i_4 & i'_4 & i''_4 & i''_4 \end{ytableau}$ &
$\mathcal{A}_{33}\sim\begin{ytableau} i_1 & i_2 & i_4 & i'_4 \\ i_3 & i'_4 & i''_4 & i''_4 \end{ytableau}$ &
$\mathcal{A}_{34}\sim\begin{ytableau} i_1 & i_2 & i'_4 & i'_4 \\ i_3 & i_4 & i''_4 & i''_4 \end{ytableau}$ &
$\mathcal{A}_{35}\sim\begin{ytableau} i_1 & i_3 & i_4 & i'_4 \\ i_2 & i'_4 & i''_4 & i''_4 \end{ytableau}$ &
$\mathcal{A}_{36}\sim\begin{ytableau} i_1 & i_3 & i'_4 & i'_4 \\ i_2 & i_4 & i''_4 & i''_4 \end{ytableau}$ \\
\bottomrule
\end{tabular}
\caption{ Complete set of $\mathcal{A}$ for operator type $N^2(N^{\dagger})^2\nabla^2$. The table is organized in reading order: the first row contains $\mathcal{A}_1$ through $\mathcal{A}_6$, the second row contains $\mathcal{A}_7$ through $\mathcal{A}_{12}$, and subsequent rows continue accordingly.\label{examplessyt1}}
\end{table}

Upon symmetrizing the derivatives, each tensor expands into a polynomial expression, and some resulting terms no longer correspond to the SSYT. As a concrete example, consider the symmetrization of the derivative indices in the tensor $\mathcal{A}_2$:
\begin{align}\label{eq:mathcalA_2symd}
\mathcal{A}_2 \equiv \epsilon^{i_1 i'_2} \epsilon^{i_2 i''_2} \epsilon^{i'_2 i_3} \epsilon^{i''_2 i_4} \longrightarrow \mathcal{B}^{(\text{y})}_2 &\equiv \frac{1}{2} \epsilon^{i_1 i'_2} \epsilon^{i_2 i''_2} \epsilon^{i'_2 i_3} \epsilon^{i''_2 i_4} + \frac{1}{2} \epsilon^{i_1 i''_2} \epsilon^{i_2 i'_2} \epsilon^{i''_2 i_3} \epsilon^{i'_2 i_4}, \nonumber \\
 &= \frac{1}{2}\mathcal{A}_2 + \frac{1}{2} \epsilon^{i_1 i''_2} \epsilon^{i_2 i'_2} \epsilon^{i''_2 i_3} \epsilon^{i'_2 i_4}.
\end{align}
When the derivative indices are ordered as $i_2' < i_2''$, the second term on the right-hand side does not correspond to a valid SSYT, as seen from its Young tableau representation:
\begin{equation}
  \epsilon^{i_1 i''_2} \epsilon^{i_2 i'_2} \epsilon^{i''_2 i_3} \epsilon^{i'_2 i_4} \sim
\begin{ytableau}
i_1 & i_2 & i''_2 & i'_2 \\
i''_2 & i'_2 & i_3 & i_4
\end{ytableau}\ .
\end{equation}
Consequently, this term is not linearly independent, which can be shown using the Schouten identity in Eq.~\eqref{eq:schouteny}:
\begin{eqnarray}\label{eq:exampleyschouten}
     \epsilon^{i_1 i''_2} \epsilon^{i_2 i'_2} \epsilon^{i''_2 i_3} \epsilon^{i'_2 i_4} & = & \epsilon^{i_1 i'_2} \epsilon^{i_2 i''_2} \epsilon^{i'_2 i_3} \epsilon^{i''_2 i_4} - \epsilon^{i_1 i'_2} \epsilon^{i_2 i''_2} \epsilon^{i'_2 i''_2} \epsilon^{i_3 i_4} \nonumber \\
     & & - \epsilon^{i_1 i_2} \epsilon^{i'_2 i''_2} \epsilon^{i'_2 i_3} \epsilon^{i''_2 i_4} + \epsilon^{i_1 i_2} \epsilon^{i'_2 i''_2} \epsilon^{i'_2 i''_2} \epsilon^{i_3 i_4} \nonumber \\
     & = & \mathcal{A}_2 - \mathcal{A}_3 - \mathcal{A}_5 + \mathcal{A}_6.
\end{eqnarray}
Substituting Eq.~\eqref{eq:exampleyschouten} into Eq.~\eqref{eq:mathcalA_2symd} yields the derivative-symmetrized result:
\begin{equation}\label{eq:exampleybasisnn}
    \mathcal{A}_2 \rightarrow \mathcal{B}^{(\text{y})}_2 \equiv \mathcal{A}_2 - \frac{1}{2}\mathcal{A}_3 - \frac{1}{2}\mathcal{A}_5 + \frac{1}{2}\mathcal{A}_6.
\end{equation}

In general, a derivative-symmetrized tensor can be expanded in the original SSYT basis $\mathcal{A}_b$  as:
\begin{equation}
    \mathcal{B}^{(\text{y})}_a = \sum_{b=1}^{36} x^{(\text{y})}_{ab} \mathcal{A}_b,
\end{equation}
where $x^{(\text{y})}_{ab}$ are expansion coefficients. The set of linearly independent vectors $\{\mathcal{B}^{(\text{y})}_a\}$ constitutes the rotational y-basis, which corresponds precisely to the linearly independent rows of the coefficient matrix $x^{(\text{y})}_{ab}$. For the type $\nabla^2N^2(N^{\dagger})^2$, the complete set of the rotational y-basis is
\begin{equation}\label{eq:4N2Db}
\begin{array}{|ccl|}
\mathcal{B}^{(\text{y})}_1 &=& \frac{1}{2} \mathcal{A}_1 + \frac{1}{2} \mathcal{A}_4 \\
\mathcal{B}^{(\text{y})}_2 &=& \mathcal{A}_2 - \frac{1}{2} \mathcal{A}_3 - \frac{1}{2} \mathcal{A}_5 + \frac{1}{2} \mathcal{A}_6 \\
\mathcal{B}^{(\text{y})}_3 &=& \frac{1}{2} \mathcal{A}_6 \\
\mathcal{B}^{(\text{y})}_4 &=& \mathcal{A}_7 \\
\mathcal{B}^{(\text{y})}_5 &=& \mathcal{A}_8 \\
\mathcal{B}^{(\text{y})}_6 &=& \mathcal{A}_9 \\
\mathcal{B}^{(\text{y})}_7 &=& \mathcal{A}_{10} \\
\mathcal{B}^{(\text{y})}_8 &=& \mathcal{A}_{11} \\
\mathcal{B}^{(\text{y})}_9 &=& \mathcal{A}_{12} \\
\mathcal{B}^{(\text{y})}_{10} &=& \mathcal{A}_{13} \\
\mathcal{B}^{(\text{y})}_{11} &=& \mathcal{A}_{14} \\
\mathcal{B}^{(\text{y})}_{12} &=& \mathcal{A}_{15} \\
\mathcal{B}^{(\text{y})}_{13} &=& \mathcal{A}_{16}
\end{array}\quad
\begin{array}{ccl|}
\mathcal{B}^{(\text{y})}_{14} &=& \mathcal{A}_{17} \\
\mathcal{B}^{(\text{y})}_{15} &=& \mathcal{A}_{18} \\
\mathcal{B}^{(\text{y})}_{16} &=& \frac{1}{2} \mathcal{A}_{19} + \frac{1}{2} \mathcal{A}_{20} - \frac{1}{2} \mathcal{A}_{22} \\
\mathcal{B}^{(\text{y})}_{17} &=& \frac{1}{2} \mathcal{A}_{19} + \frac{1}{2} \mathcal{A}_{20} - \frac{1}{2} \mathcal{A}_{21} \\
\mathcal{B}^{(\text{y})}_{18} &=& \frac{1}{2} \mathcal{A}_{23} - \frac{1}{2} \mathcal{A}_{24} \\
\mathcal{B}^{(\text{y})}_{19} &=& \mathcal{A}_{25} \\
\mathcal{B}^{(\text{y})}_{20} &=& \mathcal{A}_{26} \\
\mathcal{B}^{(\text{y})}_{21} &=& \mathcal{A}_{27} \\
\mathcal{B}^{(\text{y})}_{22} &=& \mathcal{A}_{28} \\
\mathcal{B}^{(\text{y})}_{23} &=& \mathcal{A}_{29} \\
\mathcal{B}^{(\text{y})}_{24} &=& \mathcal{A}_{30} \\
\mathcal{B}^{(\text{y})}_{25} &=& \mathcal{A}_{31} - \frac{1}{2} \mathcal{A}_{32} - \frac{1}{2} \mathcal{A}_{33} + \frac{1}{2} \mathcal{A}_{34} + \frac{1}{2} \mathcal{A}_{35} - \frac{1}{2} \mathcal{A}_{36} \\
\mathcal{B}^{(\text{y})}_{26} &=& \frac{1}{2} \mathcal{A}_{34} - \frac{1}{2} \mathcal{A}_{36} \\
\mathcal{B}^{(\text{y})}_{27} &=& \frac{1}{2} \mathcal{A}_{34}
\end{array}
\end{equation}
Taking the tensor product of the rotational y-basis $\{{\mathcal{B}^{(\text{y})}}\}$ with the isospin y-basis $\{{T_{SU(2)}^{(\text{y})}}\}$ in Eq.~\eqref{eq:4Ngauge} yields the complete y-basis $\{{\mathcal{T}^{(\text{y})}}\}$:
\begin{equation}
 \{\mathcal{T}^{(\text{y})}_{\beta,a}\}= \{\mathcal{B}^{(\text{y})}_a\}\otimes\{T_{SU(2),\beta}^{(\text{y})} \},\quad a=1,\cdots,27;\beta=1,2.
\end{equation}

To restore the permutation symmetry, we apply the appropriate Young operators to each monomial within the tensor product. As a specific example, consider the y-basis element $\mathcal{B}^{(\text{y})}_2 $  given in Eq.~\eqref{eq:exampleybasisnn}, combined with the isospin tensor $T_{\text{SU(2)},1}^{(\text{y})}=\delta_{p_3}^{p_1}\delta_{p_4}^{p_2}$. The action of the Young operators is as follows:
\begin{equation}
   \mathcal{Y}_{12}\mathcal{Y}_{34} \left(\mathcal{B}^{(\text{y})}_2 \otimes T_{\text{SU(2)},1}^{(\text{y})}\right) = \mathcal{Y}_{12}\mathcal{Y}_{34} \left[ \left( \mathcal{A}_2 - \frac{1}{2}\mathcal{A}_3 - \frac{1}{2}\mathcal{A}_5 + \frac{1}{2}\mathcal{A}_6 \right) \otimes T_{\text{SU(2)},1}^{(\text{y})} \right].
\end{equation}
Using the explicit form of the Young operator from Eq.~\eqref{eq:yop1234}, the first term on the right-hand side becomes
\begin{eqnarray}
    && \mathcal{Y}_{12}\mathcal{Y}_{34} \left( \mathcal{A}_2 \otimes T_{\text{SU(2)},1}^{(\text{y})} \right) \nonumber\\
    =&& \mathcal{Y}_{12}\mathcal{Y}_{34} \left( \epsilon ^{i_1 i'_2} \epsilon ^{i_2 i''_2} \epsilon ^{i'_2 i_3} \epsilon ^{i''_2 i_4} \delta_{p_3}^{p_1}\delta_{p_4}^{p_2} \right) \nonumber\\
    =&& \frac{1}{4} \left[ E - (12) - (34) + (12)(34) \right] \left( \epsilon ^{i_1 i'_2} \epsilon ^{i_2 i''_2} \epsilon ^{i'_2 i_3} \epsilon ^{i''_2 i_4} \delta_{p_3}^{p_1}\delta_{p_4}^{p_2} \right) \nonumber\\
    =&& \frac{1}{4} \left( \epsilon ^{i_1 i'_2} \epsilon ^{i_2 i''_2} \epsilon ^{i'_2 i_3} \epsilon ^{i''_2 i_4} \delta_{p_3}^{p_1}\delta_{p_4}^{p_2}
          - \epsilon ^{i_2 i'_1} \epsilon ^{i_1 i''_1} \epsilon ^{i'_1 i_3} \epsilon ^{i''_1 i_4} \delta_{p_3}^{p_2}\delta_{p_4}^{p_1} \right. \nonumber\\
    && \left. - \epsilon ^{i_1 i'_2} \epsilon ^{i_2 i''_2} \epsilon ^{i'_2 i_4} \epsilon ^{i''_2 i_3} \delta_{p_4}^{p_1}\delta_{p_3}^{p_2}
          + \epsilon ^{i_2 i'_1} \epsilon ^{i_1 i''_1} \epsilon ^{i'_1 i_4} \epsilon ^{i''_1 i_3} \delta_{p_4}^{p_2}\delta_{p_3}^{p_1} \right).
\end{eqnarray}
The symmetrization of the remaining terms $\mathcal{A}_3$, $\mathcal{A}_5$, and $\mathcal{A}_6$ proceeds similarly. This procedure is repeated for all elements of the y-basis. Finally, by selecting the linearly independent rows from the reduced form in Eq.~\eqref{eq:yopptoy}, we obtain the p-basis for operators of this type.

Following the method outlined above, the complete p-basis is:
\begin{small}

\begin{equation}
\begin{array}{rcl}
\mathcal{T}^{(\text{p})}_1 & = & \frac{1}{2} \mathcal{T}^{(\text{y})}_1 + \frac{1}{4} \mathcal{T}^{(\text{y})}_4 - \frac{1}{4} \mathcal{T}^{(\text{y})}_5 + \frac{1}{4} \mathcal{T}^{(\text{y})}_6 - \frac{1}{2} \mathcal{T}^{(\text{y})}_7 + \frac{1}{4} \mathcal{T}^{(\text{y})}_{10} - \frac{1}{4} \mathcal{T}^{(\text{y})}_{11} - \frac{1}{4} \mathcal{T}^{(\text{y})}_{12}+ \frac{1}{2} \mathcal{T}^{(\text{y})}_{13} + \frac{1}{4} \mathcal{T}^{(\text{y})}_{15} - \frac{1}{4} \mathcal{T}^{(\text{y})}_{17}\\
&&- \frac{1}{4} \mathcal{T}^{(\text{y})}_{19} + \frac{1}{2} \mathcal{T}^{(\text{y})}_{20} + \frac{1}{4} \mathcal{T}^{(\text{y})}_{21} - \frac{1}{4} \mathcal{T}^{(\text{y})}_{22} + \frac{1}{4} \mathcal{T}^{(\text{y})}_{25} -\frac{1}{2} \mathcal{T}^{(\text{y})}_{28} - \frac{1}{4} \mathcal{T}^{(\text{y})}_{31} + \frac{1}{4} \mathcal{T}^{(\text{y})}_{32} - \frac{1}{4} \mathcal{T}^{(\text{y})}_{33} + \frac{1}{2} \mathcal{T}^{(\text{y})}_{34} - \frac{1}{4} \mathcal{T}^{(\text{y})}_{37}\\
&&+ \frac{1}{4} \mathcal{T}^{(\text{y})}_{38} + \frac{1}{4} \mathcal{T}^{(\text{y})}_{39}- \frac{1}{2} \mathcal{T}^{(\text{y})}_{40} - \frac{1}{4} \mathcal{T}^{(\text{y})}_{42} + \frac{1}{4} \mathcal{T}^{(\text{y})}_{44} + \frac{1}{4} \mathcal{T}^{(\text{y})}_{46} - \frac{1}{2} \mathcal{T}^{(\text{y})}_{47} - \frac{1}{4} \mathcal{T}^{(\text{y})}_{48} + \frac{1}{4} \mathcal{T}^{(\text{y})}_{49} - \frac{1}{4} \mathcal{T}^{(\text{y})}_{53}, \\
\mathcal{T}^{(\text{p})}_2 & = & \frac{1}{2} \mathcal{T}^{(\text{y})}_2 - \frac{1}{4} \mathcal{T}^{(\text{y})}_5 + \frac{1}{2} \mathcal{T}^{(\text{y})}_6 - \frac{1}{2} \mathcal{T}^{(\text{y})}_7 + \frac{1}{4} \mathcal{T}^{(\text{y})}_8 - \frac{1}{4} \mathcal{T}^{(\text{y})}_9 - \frac{1}{4} \mathcal{T}^{(\text{y})}_{11} - \frac{1}{4} \mathcal{T}^{(\text{y})}_{12}+ \frac{1}{2} \mathcal{T}^{(\text{y})}_{13} + \frac{1}{4} \mathcal{T}^{(\text{y})}_{14} - \frac{1}{4} \mathcal{T}^{(\text{y})}_{16}\\
&&+ \frac{1}{4} \mathcal{T}^{(\text{y})}_{18} - \frac{1}{4} \mathcal{T}^{(\text{y})}_{19} + \frac{1}{2} \mathcal{T}^{(\text{y})}_{20} - \frac{1}{4} \mathcal{T}^{(\text{y})}_{22} + \frac{1}{4} \mathcal{T}^{(\text{y})}_{23} + \frac{1}{4} \mathcal{T}^{(\text{y})}_{25} - \frac{1}{4} \mathcal{T}^{(\text{y})}_{26} - \frac{1}{2} \mathcal{T}^{(\text{y})}_{29} + \frac{1}{2} \mathcal{T}^{(\text{y})}_{30} + \frac{1}{4} \mathcal{T}^{(\text{y})}_{32} - \frac{1}{2} \mathcal{T}^{(\text{y})}_{33}\\
&&+ \frac{1}{2} \mathcal{T}^{(\text{y})}_{34} + \frac{1}{4} \mathcal{T}^{(\text{y})}_{36} + \frac{1}{4} \mathcal{T}^{(\text{y})}_{38} + \frac{1}{4} \mathcal{T}^{(\text{y})}_{39} - \frac{1}{2} \mathcal{T}^{(\text{y})}_{40} - \frac{1}{4} \mathcal{T}^{(\text{y})}_{42} + \frac{1}{4} \mathcal{T}^{(\text{y})}_{43} - \frac{1}{4} \mathcal{T}^{(\text{y})}_{46} + \frac{1}{4} \mathcal{T}^{(\text{y})}_{48} - \frac{1}{2} \mathcal{T}^{(\text{y})}_{51} + \frac{1}{4} \mathcal{T}^{(\text{y})}_{54}, \\
\mathcal{T}^{(\text{p})}_3 & = & \frac{1}{2} \mathcal{T}^{(\text{y})}_3 + \frac{1}{4} \mathcal{T}^{(\text{y})}_8 + \frac{1}{4} \mathcal{T}^{(\text{y})}_{14} - \frac{1}{4} \mathcal{T}^{(\text{y})}_{15} + \frac{1}{4} \mathcal{T}^{(\text{y})}_{18} + \frac{1}{4} \mathcal{T}^{(\text{y})}_{23} - \frac{1}{4} \mathcal{T}^{(\text{y})}_{26} + \frac{1}{4} \mathcal{T}^{(\text{y})}_{27} + \frac{1}{2} \mathcal{T}^{(\text{y})}_{30} + \frac{1}{4} \mathcal{T}^{(\text{y})}_{35} + \frac{1}{4} \mathcal{T}^{(\text{y})}_{41} \\
&&- \frac{1}{4} \mathcal{T}^{(\text{y})}_{42} + \frac{1}{4} \mathcal{T}^{(\text{y})}_{44} + \frac{1}{4} \mathcal{T}^{(\text{y})}_{46} + \frac{1}{4} \mathcal{T}^{(\text{y})}_{50} - \frac{1}{4} \mathcal{T}^{(\text{y})}_{53} + \frac{1}{4} \mathcal{T}^{(\text{y})}_{54},\\
\mathcal{T}^{(\text{p})}_4 & = & \frac{1}{4} \mathcal{T}^{(\text{y})}_4 - \frac{1}{4} \mathcal{T}^{(\text{y})}_{10} + \frac{1}{4} \mathcal{T}^{(\text{y})}_{11} + \frac{1}{4} \mathcal{T}^{(\text{y})}_{19} - \frac{1}{4} \mathcal{T}^{(\text{y})}_{21} - \frac{1}{4} \mathcal{T}^{(\text{y})}_{27} + \frac{1}{4} \mathcal{T}^{(\text{y})}_{31} - \frac{1}{4} \mathcal{T}^{(\text{y})}_{37}+ \frac{1}{4} \mathcal{T}^{(\text{y})}_{38} - \frac{1}{4} \mathcal{T}^{(\text{y})}_{39} + \frac{1}{4} \mathcal{T}^{(\text{y})}_{46} \\
&&- \frac{1}{4} \mathcal{T}^{(\text{y})}_{48} + \frac{1}{4} \mathcal{T}^{(\text{y})}_{51} - \frac{1}{4} \mathcal{T}^{(\text{y})}_{53}, \\
\mathcal{T}^{(\text{p})}_5 & = & \frac{1}{4} \mathcal{T}^{(\text{y})}_5 + \frac{1}{4} \mathcal{T}^{(\text{y})}_{11} - \frac{1}{4} \mathcal{T}^{(\text{y})}_{12} - \frac{1}{4} \mathcal{T}^{(\text{y})}_{14} + \frac{1}{4} \mathcal{T}^{(\text{y})}_{15} + \frac{1}{4} \mathcal{T}^{(\text{y})}_{19} - \frac{1}{4} \mathcal{T}^{(\text{y})}_{22} - \frac{1}{4} \mathcal{T}^{(\text{y})}_{23}+ \frac{1}{4} \mathcal{T}^{(\text{y})}_{24} + \frac{1}{4} \mathcal{T}^{(\text{y})}_{26} - \frac{1}{4} \mathcal{T}^{(\text{y})}_{27} \\
&&+ \frac{1}{4} \mathcal{T}^{(\text{y})}_{32} - \frac{1}{4} \mathcal{T}^{(\text{y})}_{37} + \frac{1}{4} \mathcal{T}^{(\text{y})}_{38} - \frac{1}{4} \mathcal{T}^{(\text{y})}_{39} - \frac{1}{4} \mathcal{T}^{(\text{y})}_{43} - \frac{1}{4} \mathcal{T}^{(\text{y})}_{46} + \frac{1}{4} \mathcal{T}^{(\text{y})}_{48} - \frac{1}{4} \mathcal{T}^{(\text{y})}_{51}, \\
\mathcal{T}^{(\text{p})}_6 & = & \frac{1}{4} \mathcal{T}^{(\text{y})}_6 + \frac{1}{4} \mathcal{T}^{(\text{y})}_{11} - \frac{1}{4} \mathcal{T}^{(\text{y})}_{14} + \frac{1}{4} \mathcal{T}^{(\text{y})}_{19} - \frac{1}{4} \mathcal{T}^{(\text{y})}_{23} + \frac{1}{4} \mathcal{T}^{(\text{y})}_{24} + \frac{1}{4} \mathcal{T}^{(\text{y})}_{26} + \frac{1}{4} \mathcal{T}^{(\text{y})}_{34} - \frac{1}{4} \mathcal{T}^{(\text{y})}_{37} + \frac{1}{4} \mathcal{T}^{(\text{y})}_{38} - \frac{1}{4} \mathcal{T}^{(\text{y})}_{42}\\
&&+ \frac{1}{4} \mathcal{T}^{(\text{y})}_{44} - \frac{1}{4} \mathcal{T}^{(\text{y})}_{48} + \frac{1}{4} \mathcal{T}^{(\text{y})}_{49}, \\
\mathcal{T}^{(\text{p})}_7 & = & \displaystyle\frac{1}{4} \mathcal{T}^{(\text{y})}_7 + \frac{1}{4} \mathcal{T}^{(\text{y})}_{13} + \frac{1}{4} \mathcal{T}^{(\text{y})}_{20} + \frac{1}{4} \mathcal{T}^{(\text{y})}_{25} + \frac{1}{4} \mathcal{T}^{(\text{y})}_{34} + \frac{1}{4} \mathcal{T}^{(\text{y})}_{40} + \frac{1}{4} \mathcal{T}^{(\text{y})}_{43} + \frac{1}{4} \mathcal{T}^{(\text{y})}_{45} - \frac{1}{4} \mathcal{T}^{(\text{y})}_{47} - \frac{1}{4} \mathcal{T}^{(\text{y})}_{48} + \frac{1}{4} \mathcal{T}^{(\text{y})}_{49}, \\
\mathcal{T}^{(\text{p})}_8 & = & \frac{1}{4} \mathcal{T}^{(\text{y})}_8 - \frac{1}{4} \mathcal{T}^{(\text{y})}_{14} + \frac{1}{4} \mathcal{T}^{(\text{y})}_{15} - \frac{1}{4} \mathcal{T}^{(\text{y})}_{23} + \frac{1}{2} \mathcal{T}^{(\text{y})}_{26} - \frac{1}{2} \mathcal{T}^{(\text{y})}_{27} - \frac{1}{4} \mathcal{T}^{(\text{y})}_{35} + \frac{1}{4} \mathcal{T}^{(\text{y})}_{41} - \frac{1}{4} \mathcal{T}^{(\text{y})}_{42} - \frac{1}{2} \mathcal{T}^{(\text{y})}_{45} - \frac{1}{4} \mathcal{T}^{(\text{y})}_{50}, \\
\mathcal{T}^{(\text{p})}_9 & = & \frac{1}{4} \mathcal{T}^{(\text{y})}_9 - \frac{1}{4} \mathcal{T}^{(\text{y})}_{14} - \frac{1}{4} \mathcal{T}^{(\text{y})}_{23} + \frac{1}{4} \mathcal{T}^{(\text{y})}_{24} + \frac{1}{4} \mathcal{T}^{(\text{y})}_{26} - \frac{1}{4} \mathcal{T}^{(\text{y})}_{27} - \frac{1}{4} \mathcal{T}^{(\text{y})}_{35} - \frac{1}{4} \mathcal{T}^{(\text{y})}_{41} - \frac{1}{4} \mathcal{T}^{(\text{y})}_{50} + \frac{1}{4} \mathcal{T}^{(\text{y})}_{51},\\
\mathcal{T}^{(\text{p})}_{10} & = & -\frac{1}{4} \mathcal{T}^{(\text{y})}_4 + \frac{1}{4} \mathcal{T}^{(\text{y})}_6 - \frac{1}{4} \mathcal{T}^{(\text{y})}_9 + \frac{1}{4} \mathcal{T}^{(\text{y})}_{10} - \frac{1}{2} \mathcal{T}^{(\text{y})}_{16} + \frac{1}{2} \mathcal{T}^{(\text{y})}_{17} - \frac{1}{4} \mathcal{T}^{(\text{y})}_{21} - \frac{1}{4} \mathcal{T}^{(\text{y})}_{31} + \frac{1}{4} \mathcal{T}^{(\text{y})}_{34} + \frac{1}{4} \mathcal{T}^{(\text{y})}_{37} - \frac{1}{4} \mathcal{T}^{(\text{y})}_{38}\\
&&+ \frac{1}{4} \mathcal{T}^{(\text{y})}_{39} - \frac{1}{4} \mathcal{T}^{(\text{y})}_{42} + \frac{1}{4} \mathcal{T}^{(\text{y})}_{43} + \frac{1}{4} \mathcal{T}^{(\text{y})}_{48} - \frac{1}{4} \mathcal{T}^{(\text{y})}_{50} + \frac{1}{2} \mathcal{T}^{(\text{y})}_{53}, \\
\mathcal{T}^{(\text{p})}_{11} & = & \frac{1}{4} \mathcal{T}^{(\text{y})}_6 - \frac{1}{4} \mathcal{T}^{(\text{y})}_9 + \frac{1}{4} \mathcal{T}^{(\text{y})}_{11} - \frac{1}{4} \mathcal{T}^{(\text{y})}_{16} + \frac{1}{4} \mathcal{T}^{(\text{y})}_{17} - \frac{1}{4} \mathcal{T}^{(\text{y})}_{21} + \frac{1}{4} \mathcal{T}^{(\text{y})}_{22} + \frac{1}{4} \mathcal{T}^{(\text{y})}_{34} - \frac{1}{4} \mathcal{T}^{(\text{y})}_{38} + \frac{1}{4} \mathcal{T}^{(\text{y})}_{39} - \frac{1}{4} \mathcal{T}^{(\text{y})}_{42}\\
&&+ \frac{1}{4} \mathcal{T}^{(\text{y})}_{43} + \frac{1}{4} \mathcal{T}^{(\text{y})}_{48} + \frac{1}{4} \mathcal{T}^{(\text{y})}_{53}, \\
\mathcal{T}^{(\text{p})}_{12} & = & -\frac{1}{4} \mathcal{T}^{(\text{y})}_5 + \frac{1}{4} \mathcal{T}^{(\text{y})}_6 + \frac{1}{4} \mathcal{T}^{(\text{y})}_8 - \frac{1}{4} \mathcal{T}^{(\text{y})}_9 + \frac{1}{4} \mathcal{T}^{(\text{y})}_{12} - \frac{1}{4} \mathcal{T}^{(\text{y})}_{16} + \frac{1}{4} \mathcal{T}^{(\text{y})}_{17} + \frac{1}{4} \mathcal{T}^{(\text{y})}_{18} + \frac{1}{4} \mathcal{T}^{(\text{y})}_{19} - \frac{1}{4} \mathcal{T}^{(\text{y})}_{21} + \frac{1}{4} \mathcal{T}^{(\text{y})}_{24}\\
&&+ \frac{1}{4} \mathcal{T}^{(\text{y})}_{32} - \frac{1}{4} \mathcal{T}^{(\text{y})}_{37} + \frac{1}{4} \mathcal{T}^{(\text{y})}_{39} - \frac{1}{4} \mathcal{T}^{(\text{y})}_{42} + \frac{1}{4} \mathcal{T}^{(\text{y})}_{48} + \frac{1}{4} \mathcal{T}^{(\text{y})}_{53}, \\
\mathcal{T}^{(\text{p})}_{13} & = & \frac{1}{4} \mathcal{T}^{(\text{y})}_7 + \frac{1}{4} \mathcal{T}^{(\text{y})}_{13} + \frac{1}{4} \mathcal{T}^{(\text{y})}_{17} + \frac{1}{4} \mathcal{T}^{(\text{y})}_{19} - \frac{1}{4} \mathcal{T}^{(\text{y})}_{20} - \frac{1}{4} \mathcal{T}^{(\text{y})}_{21} + \frac{1}{4} \mathcal{T}^{(\text{y})}_{22} + \frac{1}{4} \mathcal{T}^{(\text{y})}_{34} + \frac{1}{4} \mathcal{T}^{(\text{y})}_{40} + \frac{1}{4} \mathcal{T}^{(\text{y})}_{47} + \frac{1}{4} \mathcal{T}^{(\text{y})}_{52}, \\
\mathcal{T}^{(\text{p})}_{14} & = & -\frac{1}{4} \mathcal{T}^{(\text{y})}_9 + \frac{1}{4} \mathcal{T}^{(\text{y})}_{14} - \frac{1}{4} \mathcal{T}^{(\text{y})}_{18} - \frac{1}{4} \mathcal{T}^{(\text{y})}_{23} + \frac{1}{4} \mathcal{T}^{(\text{y})}_{24} + \frac{1}{4} \mathcal{T}^{(\text{y})}_{35} - \frac{1}{4} \mathcal{T}^{(\text{y})}_{41} - \frac{1}{4} \mathcal{T}^{(\text{y})}_{50} + \frac{1}{4} \mathcal{T}^{(\text{y})}_{51} + \frac{1}{4} \mathcal{T}^{(\text{y})}_{54}, \\
\mathcal{T}^{(\text{p})}_{15} & = & \frac{1}{4} \mathcal{T}^{(\text{y})}_{16} - \frac{1}{4} \mathcal{T}^{(\text{y})}_{25} + \frac{1}{4} \mathcal{T}^{(\text{y})}_{27} - \frac{1}{4} \mathcal{T}^{(\text{y})}_{43} + \frac{1}{4} \mathcal{T}^{(\text{y})}_{45} + \frac{1}{4} \mathcal{T}^{(\text{y})}_{52} - \frac{1}{4} \mathcal{T}^{(\text{y})}_{53}, \\
\mathcal{T}^{(\text{p})}_{16} & = & \frac{1}{4} \mathcal{T}^{(\text{y})}_{17} - \frac{1}{4} \mathcal{T}^{(\text{y})}_{25} - \frac{1}{4} \mathcal{T}^{(\text{y})}_{43} + \frac{1}{4} \mathcal{T}^{(\text{y})}_{52}, \\
\mathcal{T}^{(\text{p})}_{17} & = & \frac{1}{4} \mathcal{T}^{(\text{y})}_{18} - \frac{1}{4} \mathcal{T}^{(\text{y})}_{26} + \frac{1}{4} \mathcal{T}^{(\text{y})}_{27} + \frac{1}{4} \mathcal{T}^{(\text{y})}_{44} - \frac{1}{4} \mathcal{T}^{(\text{y})}_{53} + \frac{1}{4} \mathcal{T}^{(\text{y})}_{54}.
\end{array}
\end{equation}
\end{small}

The nuclear interaction is invariant under the space inversion and the time reversal. All the operators in the type $\nabla^2N^2(N^{\dagger})^2$ are $P$ even. To find the $T$ even combinations, notice that the nucleon field $N_3$ has conjugation $N_1^{\dagger}$, while the nucleon field $N_4$ has conjugation $N_2^{\dagger}$. 
Then the Hermitian conjugation of the p-basis is obtained by
\begin{equation}
    \mathcal{T}^{(\text{p})\dagger}=\pi\mathcal{T}^{(\text{p})}=\mathcal{D}(\pi)\left[\mathcal{T}^{(\text{p})}(1\leftrightarrow3, 2\leftrightarrow4)\right],
\end{equation}
where $\mathcal{D}(\pi)$ defined in Eq.~\eqref{eq:permutationparity} is the signature of the permutation $\pi$.
the Hermitian conjugations are reduced to the original p-basis to derive the unique expression with their coordinates. We find that
the p-basis transforms under the Hermitian conjugation as

The nuclear interaction is invariant under both parity  and time reversal. All operators of the type $\nabla^2 N^2 (N^\dagger)^2$ are inherently even under parity. To identify the subset that is also even under time reversal, we examine their behavior under Hermitian conjugation.
In this context, the outgoing nucleon fields $N_1^\dagger$ and $N_2^\dagger$ are Hermitian conjugates of the incoming fields $N_3$ and $N_4$, respectively. Consequently, Hermitian conjugation acts on the p-basis operators by inverting all the indices and  exchanging the incoming and outgoing labels: $1 \leftrightarrow 3$ and $2 \leftrightarrow 4$, including the labels of the derivatives. 
To determine the $T$-even combinations, each Hermitian conjugate $\mathcal{T}^{(\text{p})\dagger}$ is reduced back onto the original p-basis. This reduction yields a unique expression in terms of the original coordinates, allowing us to read off the transformation law under Hermitian conjugation. We find that the p-basis transforms as follows:
\begin{eqnarray}
    \mathcal{T}_n^{(\text{p}) \dagger}=\sum_{m=1}^{17}\left[D_{\text{dag}}^{\text{(p)}}\right]_{nm}\mathcal{T}^{(\text{p})}_m,\quad n=1,\cdots,17,
\end{eqnarray}
where the representation matrix is
\begin{eqnarray}
 D_{\text{dag}}^{\text{(p)}}=   \left(
\begin{array}{ccccccccccccccccc}
 0 & 0 & 0 & 0 & 0 & 0 & 0 & 0 & 0 & 0 & 0 & 0 & 0 & 0 & 0 & -1 & 0 \\
 0 & 0 & 0 & 0 & 0 & 0 & 0 & 0 & 0 & 0 & 0 & 0 & 0 & 0 & -1 & 0 & 1 \\
 0 & 0 & 0 & 0 & 0 & 0 & 0 & 0 & 0 & 0 & 0 & 0 & 0 & 0 & 0 & 0 & 1 \\
 0 & 0 & 0 & 1 & -1 & 0 & 0 & 0 & 1 & 0 & 0 & 0 & 0 & 0 & 0 & 0 & 0 \\
 0 & 0 & 0 & 0 & 0 & 0 & 0 & 0 & 1 & 0 & 0 & 0 & 0 & 0 & 0 & 0 & 0 \\
 0 & 0 & 0 & 0 & 0 & 1 & 0 & 0 & 0 & 0 & 0 & 0 & 0 & 0 & 0 & 0 & 0 \\
 0 & 0 & 0 & 0 & 0 & 0 & 1 & 0 & 0 & 0 & 0 & 0 & 0 & 0 & 0 & 0 & 0 \\
 0 & 0 & 0 & 0 & 0 & 0 & 0 & 1 & 0 & 0 & 0 & 0 & 0 & 0 & 0 & 0 & 0 \\
 0 & 0 & 0 & 0 & 1 & 0 & 0 & 0 & 0 & 0 & 0 & 0 & 0 & 0 & 0 & 0 & 0 \\
 0 & 0 & 0 & 0 & 0 & 0 & 0 & 0 & 0 & 1 & 0 & 0 & 0 & 0 & 0 & 0 & 0 \\
 0 & 0 & 0 & 0 & 0 & 0 & 0 & 0 & 0 & 0 & 1 & 0 & 0 & 0 & 0 & 0 & 0 \\
 0 & 0 & 0 & 0 & 0 & 0 & 0 & 0 & 0 & 0 & 1 & 0 & 0 & -1 & 0 & 0 & 0 \\
 0 & 0 & 0 & 0 & 0 & 0 & 0 & 0 & 0 & 0 & 0 & 0 & 1 & 0 & 0 & 0 & 0 \\
 0 & 0 & 0 & 0 & 0 & 0 & 0 & 0 & 0 & 0 & 1 & -1 & 0 & 0 & 0 & 0 & 0 \\
 0 & -1 & 1 & 0 & 0 & 0 & 0 & 0 & 0 & 0 & 0 & 0 & 0 & 0 & 0 & 0 & 0 \\
 -1 & 0 & 0 & 0 & 0 & 0 & 0 & 0 & 0 & 0 & 0 & 0 & 0 & 0 & 0 & 0 & 0 \\
 0 & 0 & 1 & 0 & 0 & 0 & 0 & 0 & 0 & 0 & 0 & 0 & 0 & 0 & 0 & 0 & 0 \\
\end{array}
\right).
\end{eqnarray}
Thus, of the 17 p-basis elements there are 12 $P, T$ even Hermitian combinations of p-basis:
\begin{eqnarray}
 \left(\begin{array}{c}
      \mathcal{T}^{(\text{p}) \dagger}_{3}+\mathcal{T}^{(\text{p}) \dagger}_{17}  \\
      -\mathcal{T}^{(\text{p}) \dagger}_{1}+\mathcal{T}^{(\text{p}) \dagger}_{16} 
      \\
       -\mathcal{T}^{(\text{p}) \dagger}_{2}+\mathcal{T}^{(\text{p}) \dagger}_{3} +\mathcal{T}^{(\text{p}) \dagger}_{15} \\
        -\mathcal{T}^{(\text{p}) \dagger}_{12}+\mathcal{T}^{(\text{p}) \dagger}_{14}
        \\
        \mathcal{T}^{(\text{p}) \dagger}_{13}\\
        \mathcal{T}^{(\text{p}) \dagger}_{11}
        \\
        \mathcal{T}^{(\text{p}) \dagger}_{10}
        \\
        \mathcal{T}^{(\text{p}) \dagger}_{4}+\mathcal{T}^{(\text{p}) \dagger}_{9}\\
        \mathcal{T}^{(\text{p}) \dagger}_{8}\\
        \mathcal{T}^{(\text{p}) \dagger}_{7}\\
        \mathcal{T}^{(\text{p}) \dagger}_{6}\\
        -\mathcal{T}^{(\text{p}) \dagger}_{4}+\mathcal{T}^{(\text{p}) \dagger}_{5}\\
 \end{array}\right)
 =\left(\begin{array}{c}
      \mathcal{T}^{(\text{p})  }_{3}+\mathcal{T}^{(\text{p})  }_{17}  \\
      -\mathcal{T}^{(\text{p})  }_{1}+\mathcal{T}^{(\text{p})  }_{16} 
      \\
       -\mathcal{T}^{(\text{p})  }_{2}+\mathcal{T}^{(\text{p})  }_{3} +\mathcal{T}^{(\text{p})  }_{15} \\
        -\mathcal{T}^{(\text{p})  }_{12}+\mathcal{T}^{(\text{p})  }_{14}
        \\
        \mathcal{T}^{(\text{p})  }_{13}\\
        \mathcal{T}^{(\text{p})  }_{11}
        \\
        \mathcal{T}^{(\text{p})  }_{10}
        \\
        \mathcal{T}^{(\text{p})  }_{4}+\mathcal{T}^{(\text{p})  }_{9}\\
        \mathcal{T}^{(\text{p})  }_{8}\\
        \mathcal{T}^{(\text{p})  }_{7}\\
        \mathcal{T}^{(\text{p})  }_{6}\\
        -\mathcal{T}^{(\text{p})  }_{4}+\mathcal{T}^{(\text{p})  }_{5}\\
 \end{array}\right).
\end{eqnarray}
And then we can obtain 12 independent $P, T$ even Hermitian operators for the type $\nabla^2N^2N^{\dagger2}$ in the nucleon-nucleon interaction:
\begin{eqnarray}
\mathcal{O}^{p^{\prime}(2)}_1 &=& (N^{\dagger}\Vec{\sigma}\cdot\overrightarrow{\nabla}\overrightarrow{\nabla}^iN^{\dagger})(N\sigma^iN)I_1+\text{h.c.} \nonumber\\
\mathcal{O}^{p^{\prime}(2)}_2 &=& (N^{\dagger}\sigma^iN)(N^{\dagger}\overleftarrow{\nabla}^i\Vec{\sigma}\cdot\overleftarrow{\nabla}N)I_1+\text{h.c.}\nonumber\\
\mathcal{O}^{p^{\prime}(2)}_3 &=& -(N^{\dagger}\overrightarrow{\nabla}^2N^{\dagger})(NN)I_1+\text{h.c.}\nonumber\\
\mathcal{O}^{p^{\prime}(2)}_4 &=& -(N^{\dagger}\overrightarrow{\nabla}^iN)(N^{\dagger}\sigma^i\Vec{\sigma}\cdot\overleftarrow{\nabla}N)I_1+\text{h.c.}\nonumber\\
\mathcal{O}^{p^{\prime}(2)}_5 &=& -(N^{\dagger}\Vec{\sigma}\cdot\overrightarrow{\nabla}\sigma^iN^{\dagger})(N\overleftarrow{\nabla}^iN)I_1+\text{h.c.}\nonumber\\
\mathcal{O}^{p^{\prime}(2)}_6 &=& -(N^{\dagger}\sigma^i\Vec{\sigma}\cdot\overrightarrow{\nabla}N)(N^{\dagger}\overleftarrow{\nabla}^iN)I_2+h.c.\nonumber \\
\mathcal{O}^{p^{\prime}(2)}_7 &=& -(N^{\dagger}\Vec{\sigma}\cdot\overrightarrow{\nabla}N^{\dagger})(N\Vec{\sigma}\cdot\overleftarrow{\nabla}N)I_1+\text{h.c.}\nonumber\\
\mathcal{O}^{p^{\prime}(2)}_8 &=& 2(N^{\dagger}\overrightarrow{\nabla}^iN^{\dagger})(N\overleftarrow{\nabla}^iN)I_1+\text{h.c.}\nonumber \\
\mathcal{O}^{p^{\prime}(2)}_9 &=& -2(N^{\dagger}N)(N^{\dagger}\overleftarrow{\nabla}\cdot\overrightarrow{\nabla}N)I_1+\text{h.c.}\nonumber\\
\mathcal{O}^{p^{\prime}(2)}_{10} &=& -(N^{\dagger}\sigma^i\sigma^jN)(N^{\dagger}\overleftarrow{\nabla}^i\overrightarrow{\nabla}^jN)I_1+\text{h.c.}\nonumber\\
\mathcal{O}^{p^{\prime}(2)}_{11} &=& (N^{\dagger}\sigma^i\sigma^jN)(N^{\dagger}\overleftarrow{\nabla}^i\overrightarrow{\nabla}^jN)I_2+\text{h.c.}\nonumber\\
\mathcal{O}^{p^{\prime}(2)}_{12} &=& (N^{\dagger}\vec{\sigma}\cdot\overrightarrow{\nabla}N^{\dagger})(N\Vec{\sigma}\cdot\overrightarrow{\nabla}N)I_1+\text{h.c.}\ ,
\end{eqnarray}
where  $I_1$ and $I_2$ are the isospin structures defined in Eq.~\eqref{eq:4Ngauge}.

Following the same procedure, the operator bases for the nucleon-nucleon interaction up to $\mathcal{O}(Q^4)$ are listed in appendix~\ref{ap:NN}, and the operator bases for the three-nucleon interaction up to $\mathcal{O}(Q^2)$ are listed in appendix~\ref{ap:3N}.

\subsubsection{DM-nucleon contact interaction}

Having discussed the background and the field content for the non-relativistic DM and the nucleons in subsection~\ref{eftreview} and \ref{sec:building}, in this subsubsection we explicitly show how to construct the operator basis for the DM-nucleon contact interaction.

The rotational y-basis for the DM--nucleon contact operators coincides with that of the nucleon--nucleon operators.  
The Dark matter in this work possesses no internal symmetry other than a $U(1)$ charge, which stems from the particle-number conservation in the non-relativistic scattering.
Although nucleons obey the isospin symmetry, the only admissible isospin structure for the nucleon bilinear in  a DM--nucleon four‑fermion operator is the isoscalar combination $N^{\dagger,a}N_a$; moreover, there is no repeated field, i.e., no field appears more than once in the operator.  
Consequently, the p-basis and the y-basis are obtained directly from the rotational y-basis, and the three sets are identical:
\begin{eqnarray}
    \{\mathcal{T}^{(\text{p})}\}=    \{\mathcal{T}^{(\text{y})}\}    =\{\mathcal{B}^{(\text{y})}\}.
\end{eqnarray}
Thus we can construct the DM-nucleon contact operator basis directly from the rotational y-basis $\{\mathcal{B}^{(\text{y})}\}$.

Explicitly, the leading-order $\mathcal{O}(v)$ interaction is described by the type $NN^{\dagger}\xi\xi^{\dagger}$, and the rotational y-basis in this type correspond to the only two possible SSYT 
\begin{equation}
    \mathcal{B}^{(\text{y})}_1=\epsilon^{i_1i_2}\epsilon^{i_3i_4}\sim~\begin{ytableau}
1 & 3 \\
2 & 4
\end{ytableau}
\ ,\quad \mathcal{B}^{(\text{y})}_2=\epsilon^{i_1i_3}\epsilon^{i_2i_4}\sim~
\begin{ytableau}
1 & 2 \\
3 & 4
\end{ytableau}\ ,
\end{equation}
which correspond to two operators 
\begin{eqnarray}
   O^{(0)}_1 &=&(N_1^{\dagger}N_2)(\xi^{\dagger}_3\xi_4),\nonumber\\
    O^{(0)}_2 &=&(N_1^{\dagger}\xi^{\dagger}_3)(N_2\xi_4).
\end{eqnarray}
At $\mathcal{O}(v^2)$, this interaction is described by the type $\nabla^2NN^{\dagger}\xi\xi^{\dagger}$, and the rotational y-basis is the same as the Eq.~\eqref{eq:4N2Db} in the nucleons four-fermion operators due to the similar rotational structures,
\begin{equation}\label{eq:DMNyb}
\begin{array}{|ccl|}
\mathcal{B}^{(\text{y})}_1 &=& \frac{1}{2} \mathcal{A}_1 + \frac{1}{2} \mathcal{A}_4 \\
\mathcal{B}^{(\text{y})}_2 &=& \mathcal{A}_2 - \frac{1}{2} \mathcal{A}_3 - \frac{1}{2} \mathcal{A}_5 + \frac{1}{2} \mathcal{A}_6 \\
\mathcal{B}^{(\text{y})}_3 &=& \frac{1}{2} \mathcal{A}_6 \\
\mathcal{B}^{(\text{y})}_4 &=& \mathcal{A}_7 \\
\mathcal{B}^{(\text{y})}_5 &=& \mathcal{A}_8 \\
\mathcal{B}^{(\text{y})}_6 &=& \mathcal{A}_9 \\
\mathcal{B}^{(\text{y})}_7 &=& \mathcal{A}_{10} \\
\mathcal{B}^{(\text{y})}_8 &=& \mathcal{A}_{11} \\
\mathcal{B}^{(\text{y})}_9 &=& \mathcal{A}_{12} \\
\mathcal{B}^{(\text{y})}_{10} &=& \mathcal{A}_{13} \\
\mathcal{B}^{(\text{y})}_{11} &=& \mathcal{A}_{14} \\
\mathcal{B}^{(\text{y})}_{12} &=& \mathcal{A}_{15} \\
\mathcal{B}^{(\text{y})}_{13} &=& \mathcal{A}_{16}
\end{array}\quad
\begin{array}{ccl|}
\mathcal{B}^{(\text{y})}_{14} &=& \mathcal{A}_{17} \\
\mathcal{B}^{(\text{y})}_{15} &=& \mathcal{A}_{18} \\
\mathcal{B}^{(\text{y})}_{16} &=& \frac{1}{2} \mathcal{A}_{19} + \frac{1}{2} \mathcal{A}_{20} - \frac{1}{2} \mathcal{A}_{22} \\
\mathcal{B}^{(\text{y})}_{17} &=& \frac{1}{2} \mathcal{A}_{19} + \frac{1}{2} \mathcal{A}_{20} - \frac{1}{2} \mathcal{A}_{21} \\
\mathcal{B}^{(\text{y})}_{18} &=& \frac{1}{2} \mathcal{A}_{23} - \frac{1}{2} \mathcal{A}_{24} \\
\mathcal{B}^{(\text{y})}_{19} &=& \mathcal{A}_{25} \\
\mathcal{B}^{(\text{y})}_{20} &=& \mathcal{A}_{26} \\
\mathcal{B}^{(\text{y})}_{21} &=& \mathcal{A}_{27} \\
\mathcal{B}^{(\text{y})}_{22} &=& \mathcal{A}_{28} \\
\mathcal{B}^{(\text{y})}_{23} &=& \mathcal{A}_{29} \\
\mathcal{B}^{(\text{y})}_{24} &=& \mathcal{A}_{30} \\
\mathcal{B}^{(\text{y})}_{25} &=& \mathcal{A}_{31} - \frac{1}{2} \mathcal{A}_{32} - \frac{1}{2} \mathcal{A}_{33} + \frac{1}{2} \mathcal{A}_{34} + \frac{1}{2} \mathcal{A}_{35} - \frac{1}{2} \mathcal{A}_{36} \\
\mathcal{B}^{(\text{y})}_{26} &=& \frac{1}{2} \mathcal{A}_{34} - \frac{1}{2} \mathcal{A}_{36} \\
\mathcal{B}^{(\text{y})}_{27} &=& \frac{1}{2} \mathcal{A}_{34}
\end{array}\ .
\end{equation}

Furthermore, to derive the operators with definite properties under the $P$ and $T$ transformations, we  investigate the Hermitian conjugation of the y-basis in Eq.~\eqref{eq:DMNyb}.
Under the Hermitian conjugation, the transformation matrix $D_{\text{dag}}$ of the y-basis $\{\mathcal{B}^{(\text{y})}\}$ for the type $\nabla^2NN^{\dagger}\xi\xi^{\dagger}$ reads
\begin{equation}
\mathcal{B}^{(\text{y})\dagger}_a=\sum_{b=1}^{27}\left[D_{\text{dag}}\right]_{ab}\mathcal{B}^{(\text{y})}_b,\quad,a=1,\cdots,27,
\end{equation}
\begin{tiny}
\begin{equation}
\begin{array}{ll}
     &  D_{\text{dag}}=   \\
     & 
 \left(
\begin{array}{ccccccccccccccccccccccccccc}
 1 & 0 & 0 & 1 & -1 & 1 & -2 & 0 & 0 & 1 & -1 & -1 & 2 & 0 & 1 & 0 & -1 & 0 & -1 & 2 & 1 & -1 & 0 & 0 & 1 & 0 & 0 \\
 0 & 1 & 0 & 0 & -1 & 2 & -2 & 1 & -1 & 0 & -1 & -1 & 2 & 1 & 0 & -1 & 0 & 1 & -1 & 2 & 0 & -1 & 1 & 0 & 1 & -1 & 0 \\
 0 & 0 & 1 & 0 & 0 & 0 & 0 & 1 & 0 & 0 & 0 & 0 & 0 & 1 & -1 & 0 & 0 & 1 & 0 & 0 & 0 & 0 & 1 & 0 & 0 & -1 & 1 \\
 0 & 0 & 0 & 0 & 0 & 0 & 0 & 0 & 0 & -1 & 1 & 0 & 0 & 0 & 0 & 0 & 0 & 0 & 1 & 0 & -1 & 0 & 0 & 0 & 0 & 0 & -1 \\
 0 & 0 & 0 & 0 & 0 & 0 & 0 & 0 & 0 & 0 & 1 & -1 & 0 & -1 & 1 & 0 & 0 & 0 & 1 & 0 & 0 & -1 & -1 & 1 & 0 & 1 & -1 \\
 0 & 0 & 0 & 0 & 0 & 0 & 0 & 0 & 0 & 0 & 1 & 0 & 0 & -1 & 0 & 0 & 0 & 0 & 1 & 0 & 0 & 0 & -1 & 1 & 0 & 1 & 0 \\
 0 & 0 & 0 & 0 & 0 & 0 & 0 & 0 & 0 & 0 & 0 & 0 & 1 & 0 & 0 & 0 & 0 & 0 & 0 & 1 & 0 & 0 & 0 & 0 & 1 & 0 & 0 \\
 0 & 0 & 0 & 0 & 0 & 0 & 0 & 0 & 0 & 0 & 0 & 0 & 0 & -1 & 1 & 0 & 0 & 0 & 0 & 0 & 0 & 0 & -1 & 0 & 0 & 2 & -2 \\
 0 & 0 & 0 & 0 & 0 & 0 & 0 & 0 & 0 & 0 & 0 & 0 & 0 & -1 & 0 & 0 & 0 & 0 & 0 & 0 & 0 & 0 & -1 & 1 & 0 & 1 & -1 \\
 0 & 0 & 0 & -1 & 0 & 1 & 0 & 0 & -1 & 0 & 0 & 0 & 0 & 0 & 0 & -2 & 2 & 0 & 0 & 0 & -1 & 0 & 0 & 0 & 0 & 0 & 0 \\
 0 & 0 & 0 & 0 & 0 & 1 & 0 & 0 & -1 & 0 & 0 & 0 & 0 & 0 & 0 & -1 & 1 & 0 & 0 & 0 & -1 & 1 & 0 & 0 & 0 & 0 & 0 \\
 0 & 0 & 0 & 0 & -1 & 1 & 0 & 1 & -1 & 0 & 0 & 0 & 0 & 0 & 0 & -1 & 1 & 1 & 1 & 0 & -1 & 0 & 0 & 1 & 0 & 0 & 0 \\
 0 & 0 & 0 & 0 & 0 & 0 & 1 & 0 & 0 & 0 & 0 & 0 & 0 & 0 & 0 & 0 & 1 & 0 & 1 & -1 & -1 & 1 & 0 & 0 & 0 & 0 & 0 \\
 0 & 0 & 0 & 0 & 0 & 0 & 0 & 0 & -1 & 0 & 0 & 0 & 0 & 0 & 0 & 0 & 0 & -1 & 0 & 0 & 0 & 0 & -1 & 1 & 0 & 0 & 0 \\
 0 & 0 & 0 & 0 & 0 & 0 & 0 & 1 & -1 & 0 & 0 & 0 & 0 & 0 & 0 & 0 & 0 & 1 & 0 & 0 & 0 & 0 & 0 & 1 & 0 & 0 & 0 \\
 0 & 0 & 0 & 0 & 0 & 0 & 0 & 0 & 0 & 0 & 0 & 0 & 0 & 0 & 0 & 0 & 0 & 0 & 0 & 0 & 0 & 0 & 0 & 0 & -1 & 0 & 1 \\
 0 & 0 & 0 & 0 & 0 & 0 & 0 & 0 & 0 & 0 & 0 & 0 & 0 & 0 & 0 & 0 & 0 & 0 & 0 & 0 & 0 & 0 & 0 & 0 & -1 & 0 & 0 \\
 0 & 0 & 0 & 0 & 0 & 0 & 0 & 0 & 0 & 0 & 0 & 0 & 0 & 0 & 0 & 0 & 0 & 0 & 0 & 0 & 0 & 0 & 0 & 0 & 0 & -1 & 1 \\
 0 & 0 & 0 & 0 & 0 & 0 & 0 & 0 & 0 & 0 & 0 & 0 & 0 & 0 & 0 & 0 & 0 & 0 & 0 & 0 & 1 & -1 & 0 & 0 & 0 & 0 & 0 \\
 0 & 0 & 0 & 0 & 0 & 0 & 0 & 0 & 0 & 0 & 0 & 0 & 0 & 0 & 0 & 0 & 0 & 0 & -1 & 1 & 1 & -1 & 0 & 0 & 0 & 0 & 0 \\
 0 & 0 & 0 & 0 & 0 & 0 & 0 & 0 & 0 & 0 & 0 & 0 & 0 & 0 & 0 & 0 & 0 & 0 & 0 & 0 & 1 & 0 & 0 & 0 & 0 & 0 & 0 \\
 0 & 0 & 0 & 0 & 0 & 0 & 0 & 0 & 0 & 0 & 0 & 0 & 0 & 0 & 0 & 0 & 0 & 0 & -1 & 0 & 1 & 0 & 0 & 0 & 0 & 0 & 0 \\
 0 & 0 & 0 & 0 & 0 & 0 & 0 & 0 & 0 & 0 & 0 & 0 & 0 & 0 & 0 & 0 & 0 & 0 & 0 & 0 & 0 & 0 & 1 & 0 & 0 & 0 & 0 \\
 0 & 0 & 0 & 0 & 0 & 0 & 0 & 0 & 0 & 0 & 0 & 0 & 0 & 0 & 0 & 0 & 0 & 0 & 0 & 0 & 0 & 0 & 0 & 1 & 0 & 0 & 0 \\
 0 & 0 & 0 & 0 & 0 & 0 & 0 & 0 & 0 & 0 & 0 & 0 & 0 & 0 & 0 & 0 & -1 & 0 & 0 & 0 & 0 & 0 & 0 & 0 & 0 & 0 & 0 \\
 0 & 0 & 0 & 0 & 0 & 0 & 0 & 0 & 0 & 0 & 0 & 0 & 0 & 0 & 0 & 1 & -1 & -1 & 0 & 0 & 0 & 0 & 0 & 0 & 0 & 0 & 0 \\
 0 & 0 & 0 & 0 & 0 & 0 & 0 & 0 & 0 & 0 & 0 & 0 & 0 & 0 & 0 & 1 & -1 & 0 & 0 & 0 & 0 & 0 & 0 & 0 & 0 & 0 & 0 \\
\end{array}
\right)
\end{array},
\end{equation}
\end{tiny}
Among these 27 y-basis elements $\mathcal{B}^{(\text{y})}_a$, we find 17 Hermitian combinations,

\begin{equation}
\begin{array}{rcl}
(O_{H})_1 & = & -\mathcal{B}^{(\text{y})}_1 + \mathcal{B}^{(\text{y})}_2 - \mathcal{B}^{(\text{y})}_3 - \mathcal{B}^{(\text{y})}_4 + \mathcal{B}^{(\text{y})}_6 - \mathcal{B}^{(\text{y})}_9 + \mathcal{B}^{(\text{y})}_{27}, \\
(O_{H})_2 & = & -\mathcal{B}^{(\text{y})}_1 + \mathcal{B}^{(\text{y})}_2 - \mathcal{B}^{(\text{y})}_4 + \mathcal{B}^{(\text{y})}_6 + \mathcal{B}^{(\text{y})}_8 - \mathcal{B}^{(\text{y})}_9 + \mathcal{B}^{(\text{y})}_{26}, \\
(O_{H})_3 & = & -\mathcal{B}^{(\text{y})}_1 - \mathcal{B}^{(\text{y})}_4 + \mathcal{B}^{(\text{y})}_5 - \mathcal{B}^{(\text{y})}_6 + 2\mathcal{B}^{(\text{y})}_7 + \mathcal{B}^{(\text{y})}_{25}, \\
(O_{H})_4 & = & \mathcal{B}^{(\text{y})}_{24} \\
(O_{H})_5 & = & \mathcal{B}^{(\text{y})}_{23}, \\
(O_{H})_6 & = & \mathcal{B}^{(\text{y})}_1 - \mathcal{B}^{(\text{y})}_2 + \mathcal{B}^{(\text{y})}_3 + \mathcal{B}^{(\text{y})}_4 + \mathcal{B}^{(\text{y})}_{11} + \mathcal{B}^{(\text{y})}_{22}, \\
(O_{H})_7 & = & \mathcal{B}^{(\text{y})}_{21}, \\
(O_{H})_8 & = & \mathcal{B}^{(\text{y})}_1 - \mathcal{B}^{(\text{y})}_2 + \mathcal{B}^{(\text{y})}_3 + \mathcal{B}^{(\text{y})}_4 + \mathcal{B}^{(\text{y})}_{11} + \mathcal{B}^{(\text{y})}_{20}, \\
(O_{H})_9 & = & \mathcal{B}^{(\text{y})}_1 - \mathcal{B}^{(\text{y})}_2 + \mathcal{B}^{(\text{y})}_3 + \mathcal{B}^{(\text{y})}_4 + \mathcal{B}^{(\text{y})}_{11} + \mathcal{B}^{(\text{y})}_{19}, \\
(O_{H})_{10} & = & \mathcal{B}^{(\text{y})}_3 + \mathcal{B}^{(\text{y})}_8 + \mathcal{B}^{(\text{y})}_{18}, \\
(O_{H})_{11} & = & -\mathcal{B}^{(\text{y})}_1 - \mathcal{B}^{(\text{y})}_4 + \mathcal{B}^{(\text{y})}_5 - \mathcal{B}^{(\text{y})}_6 + 2\mathcal{B}^{(\text{y})}_7 + \mathcal{B}^{(\text{y})}_{17}, \\
(O_{H})_{12} & = & -\mathcal{B}^{(\text{y})}_2 + \mathcal{B}^{(\text{y})}_3 + \mathcal{B}^{(\text{y})}_5 - 2\mathcal{B}^{(\text{y})}_6 + 2\mathcal{B}^{(\text{y})}_7 + \mathcal{B}^{(\text{y})}_9 + \mathcal{B}^{(\text{y})}_{16}, \\
(O_{H})_{13} & = & -\mathcal{B}^{(\text{y})}_3 - \mathcal{B}^{(\text{y})}_9 + \mathcal{B}^{(\text{y})}_{15}, \\
(O_{H})_{14} & = & \mathcal{B}^{(\text{y})}_3 + \mathcal{B}^{(\text{y})}_8 - \mathcal{B}^{(\text{y})}_9 + \mathcal{B}^{(\text{y})}_{14}, \\
(O_{H})_{15} & = & \mathcal{B}^{(\text{y})}_2 - \mathcal{B}^{(\text{y})}_3 - \mathcal{B}^{(\text{y})}_5 + \mathcal{B}^{(\text{y})}_6 - \mathcal{B}^{(\text{y})}_7 - \mathcal{B}^{(\text{y})}_{11} + \mathcal{B}^{(\text{y})}_{13}, \\
(O_{H})_{16} & = & -\mathcal{B}^{(\text{y})}_3 - \mathcal{B}^{(\text{y})}_5 - \mathcal{B}^{(\text{y})}_{11} + \mathcal{B}^{(\text{y})}_{12}, \\
(O_{H})_{17} & = & 2\mathcal{B}^{(\text{y})}_1 - 2\mathcal{B}^{(\text{y})}_2 + 2\mathcal{B}^{(\text{y})}_3 + \mathcal{B}^{(\text{y})}_4 - \mathcal{B}^{(\text{y})}_6 + \mathcal{B}^{(\text{y})}_9 + \mathcal{B}^{(\text{y})}_{10}.
\end{array}
\end{equation}
In terms of these combinations, we can thus find 17 $P, T$ even operators for the DM-nucleon contact interaction as
\begin{equation}\label{eq:opd2DMNT+}
    \begin{array}{l|c}
 O^{(2)}{}_1 & \left(N^{\dagger }{}_1 \vec{\sigma }\cdot \nabla _2 N_2\right) \left(\xi ^{\dagger }{}_3 \vec{\sigma }\cdot \nabla _2 \xi _4\right) +\text{h.c.},\\
 O^{(2)}{}_2 & \left(N_2 \vec{\sigma }\cdot \nabla _2 \xi _4\right) \left(N^{\dagger }{}_1 \vec{\sigma }\cdot \nabla _2 \xi ^{\dagger }{}_3\right) +\text{h.c.}, \\
 O^{(2)}{}_3 & -\left(N^{\dagger }{}_1 \vec{\sigma }\cdot \nabla _2 \vec{\sigma }\cdot \nabla _2 N_2\right) \left(\xi ^{\dagger }{}_3 \xi _4\right) +\text{h.c.}, \\
 O^{(2)}{}_4 & -\left(N_2 \vec{\sigma }\cdot \nabla _3 \vec{\sigma }\cdot \nabla _2 \xi _4\right) \left(N^{\dagger }{}_1 \xi ^{\dagger }{}_3\right) +\text{h.c.}, \\
 O^{(2)}{}_5 & -\left(N^{\dagger }{}_1 \vec{\sigma }\cdot \nabla _2 \vec{\sigma }\cdot \nabla _3 N_2\right) \left(\xi ^{\dagger }{}_3 \xi _4\right)  +\text{h.c.},\\
 O^{(2)}{}_6 & \left(N^{\dagger }{}_1 \vec{\sigma }\cdot \nabla _2 \vec{\sigma }\cdot \nabla _3 \xi _4\right) \left(N_2 \xi ^{\dagger }{}_3\right)  +\text{h.c.},\\
 O^{(2)}{}_7 & -\left(N^{\dagger }{}_1 \vec{\sigma }\cdot \nabla _2 N_2\right) \left(\xi ^{\dagger }{}_3 \vec{\sigma }\cdot \nabla _3 \xi _4\right)  +\text{h.c.},\\
 O^{(2)}{}_8 & -\left(N^{\dagger }{}_1 N_2\right) \left(\xi ^{\dagger }{}_3 \xi _4\right) \text{tr}\left(\vec{\sigma }\cdot \nabla _2 \vec{\sigma }\cdot \nabla _3\right)  +\text{h.c.},\\
 O^{(2)}{}_9 & -\left(N^{\dagger }{}_1 N_2\right) \left(\xi ^{\dagger }{}_3 \vec{\sigma }\cdot \nabla _2 \vec{\sigma }\cdot \nabla _3 \xi _4\right) +\text{h.c.}, \\
 O^{(2)}{}_{10} & -\left(N_2 \xi _4\right) \left(N^{\dagger }{}_1 \xi ^{\dagger }{}_3\right) \text{tr}\left(\vec{\sigma }\cdot \nabla _2 \vec{\sigma }\cdot \nabla _4\right) +\text{h.c.}, \\
 O^{(2)}{}_{11} & -\left(N_2 \xi _4\right) \left(N^{\dagger }{}_1 \vec{\sigma }\cdot \nabla _2 \vec{\sigma }\cdot \nabla _4 \xi ^{\dagger }{}_3\right) +\text{h.c.}, \\
 O^{(2)}{}_{12} & -\left(N^{\dagger }{}_1 \vec{\sigma }\cdot \nabla _2 \vec{\sigma }\cdot \nabla _4 \xi _4\right) \left(N_2 \xi ^{\dagger }{}_3\right) +\text{h.c.}, \\
 O^{(2)}{}_{13} & \left(N^{\dagger }{}_1 \vec{\sigma }\cdot \nabla _2 N_2\right) \left(\xi ^{\dagger }{}_3 \vec{\sigma }\cdot \nabla _4 \xi _4\right)  +\text{h.c.},\\
 O^{(2)}{}_{14} & -\left(N^{\dagger }{}_1 N_2\right) \left(\xi ^{\dagger }{}_3 \vec{\sigma }\cdot \nabla _4 \vec{\sigma }\cdot \nabla _2 \xi _4\right)  +\text{h.c.},\\
 O^{(2)}{}_{15} & -\left(N^{\dagger }{}_1 \vec{\sigma }\cdot \nabla _3 \xi _4\right) \left(N_2 \vec{\sigma }\cdot \nabla _3 \xi ^{\dagger }{}_3\right) +\text{h.c.}, \\
 O^{(2)}{}_{16} & -\left(N^{\dagger }{}_1 \vec{\sigma }\cdot \nabla _3 N_2\right) \left(\xi ^{\dagger }{}_3 \vec{\sigma }\cdot \nabla _3 \xi _4\right)  +\text{h.c.},\\
 O^{(2)}{}_{17} & -\left(N^{\dagger }{}_1 N_2\right) \left(\xi ^{\dagger }{}_3 \vec{\sigma }\cdot \nabla _3 \vec{\sigma }\cdot \nabla _3 \xi _4\right)  +\text{h.c.}\ .\\
\end{array}
\end{equation}
Besides, there are  10 anti-Hermitian combinations
\begin{equation}
\begin{array}{lcl}
(O_{aH})_1 & = & -\mathcal{B}^{(\text{y})}_{16} + \mathcal{B}^{(\text{y})}_{17} + \mathcal{B}^{(\text{y})}_{27}, \\
(O_{aH})_2 & = & -\mathcal{B}^{(\text{y})}_{16} + \mathcal{B}^{(\text{y})}_{17} + \mathcal{B}^{(\text{y})}_{18} + \mathcal{B}^{(\text{y})}_{26}, \\
(O_{aH})_3 & = & \mathcal{B}^{(\text{y})}_{17} + \mathcal{B}^{(\text{y})}_{25}, \\
(O_{aH})_4 & = & \mathcal{B}^{(\text{y})}_4 - \mathcal{B}^{(\text{y})}_5 + \mathcal{B}^{(\text{y})}_6 + \mathcal{B}^{(\text{y})}_8 - \mathcal{B}^{(\text{y})}_9 + \mathcal{B}^{(\text{y})}_{10} - \mathcal{B}^{(\text{y})}_{11} - \mathcal{B}^{(\text{y})}_{12} + \mathcal{B}^{(\text{y})}_{18} + \mathcal{B}^{(\text{y})}_{24}, \\
(O_{aH})_5 & = & \mathcal{B}^{(\text{y})}_4 - \mathcal{B}^{(\text{y})}_5 + \mathcal{B}^{(\text{y})}_6 + \mathcal{B}^{(\text{y})}_8 + \mathcal{B}^{(\text{y})}_{10} - \mathcal{B}^{(\text{y})}_{11} - \mathcal{B}^{(\text{y})}_{12} + \mathcal{B}^{(\text{y})}_{14} + 2\mathcal{B}^{(\text{y})}_{18} + \mathcal{B}^{(\text{y})}_{23}, \\
(O_{aH})_6 & = & \mathcal{B}^{(\text{y})}_4 + \mathcal{B}^{(\text{y})}_{10} - \mathcal{B}^{(\text{y})}_{11} + \mathcal{B}^{(\text{y})}_{16} - \mathcal{B}^{(\text{y})}_{17} + \mathcal{B}^{(\text{y})}_{22}, \\
(O_{aH})_7 & = & \mathcal{B}^{(\text{y})}_4 - \mathcal{B}^{(\text{y})}_6 + \mathcal{B}^{(\text{y})}_9 + \mathcal{B}^{(\text{y})}_{10} + 2\mathcal{B}^{(\text{y})}_{16} - 2\mathcal{B}^{(\text{y})}_{17} + \mathcal{B}^{(\text{y})}_{21}, \\
(O_{aH})_8 & = & -\mathcal{B}^{(\text{y})}_7 + \mathcal{B}^{(\text{y})}_{13} - \mathcal{B}^{(\text{y})}_{17} + \mathcal{B}^{(\text{y})}_{20}, \\
(O_{aH})_9 & = & -\mathcal{B}^{(\text{y})}_6 + \mathcal{B}^{(\text{y})}_9 + \mathcal{B}^{(\text{y})}_{11} + \mathcal{B}^{(\text{y})}_{16} - \mathcal{B}^{(\text{y})}_{17} + \mathcal{B}^{(\text{y})}_{19}, \\
(O_{aH})_{10} & = & \mathcal{B}^{(\text{y})}_4 - \mathcal{B}^{(\text{y})}_5 + \mathcal{B}^{(\text{y})}_6 + \mathcal{B}^{(\text{y})}_{10} - \mathcal{B}^{(\text{y})}_{11} - \mathcal{B}^{(\text{y})}_{12} + \mathcal{B}^{(\text{y})}_{15},
\end{array}
\end{equation}
where we can then find 10 $P$ even $T$ odd operators for the DM-nucleon contact interaction
\begin{equation}\label{eq:opd2DMNT-}
    \begin{array}{l|c}
 \tilde{O}^{(2)}{}_1 & \mathbf{i}\left(N^{\dagger }{}_1 \vec{\sigma }\cdot \nabla _2 N_2\right) \left(\xi ^{\dagger }{}_3 \vec{\sigma }\cdot \nabla _2 \xi _4\right)+\text{h.c.}, \\
 \tilde{O}^{(2)}{}_2 &\mathbf{i} \left(N_2 \vec{\sigma }\cdot \nabla _2 \xi _4\right) \left(N^{\dagger }{}_1 \vec{\sigma }\cdot \nabla _2 \xi ^{\dagger }{}_3\right) +\text{h.c.}\\
 \tilde{O}^{(2)}{}_3 & -\mathbf{i}\left(N^{\dagger }{}_1 \vec{\sigma }\cdot \nabla _2 \vec{\sigma }\cdot \nabla _2 N_2\right) \left(\xi ^{\dagger }{}_3 \xi _4\right)+\text{h.c.}, \\
 \tilde{O}^{(2)}{}_4 & -\mathbf{i}\left(N_2 \vec{\sigma }\cdot \nabla _3 \vec{\sigma }\cdot \nabla _2 \xi _4\right) \left(N^{\dagger }{}_1 \xi ^{\dagger }{}_3\right) +\text{h.c.}\\
 \tilde{O}^{(2)}{}_5 & -\mathbf{i}\left(N^{\dagger }{}_1 \vec{\sigma }\cdot \nabla _2 \vec{\sigma }\cdot \nabla _3 N_2\right) \left(\xi ^{\dagger }{}_3 \xi _4\right) +\text{h.c.}\\
 \tilde{O}^{(2)}{}_6 & \mathbf{i}\left(N^{\dagger }{}_1 \vec{\sigma }\cdot \nabla _2 \vec{\sigma }\cdot \nabla _3 \xi _4\right) \left(N_2 \xi ^{\dagger }{}_3\right)+\text{h.c.}, \\
 \tilde{O}^{(2)}{}_7 & -\mathbf{i}\left(N^{\dagger }{}_1 \vec{\sigma }\cdot \nabla _2 N_2\right) \left(\xi ^{\dagger }{}_3 \vec{\sigma }\cdot \nabla _3 \xi _4\right) +\text{h.c.}\\
 \tilde{O}^{(2)}{}_8 & -\mathbf{i}\left(N^{\dagger }{}_1 N_2\right) \left(\xi ^{\dagger }{}_3 \xi _4\right) \text{tr}\left(\vec{\sigma }\cdot \nabla _2 \vec{\sigma }\cdot \nabla _3\right)+\text{h.c.}, \\
 \tilde{O}^{(2)}{}_9 & -\mathbf{i}\left(N^{\dagger }{}_1 N_2\right) \left(\xi ^{\dagger }{}_3 \vec{\sigma }\cdot \nabla _2 \vec{\sigma }\cdot \nabla _3 \xi _4\right)+\text{h.c.}, \\
 \tilde{O}^{(2)}{}_{10} & -\mathbf{i}\left(N_2 \xi _4\right) \left(N^{\dagger }{}_1 \vec{\sigma }\cdot \nabla _2 \vec{\sigma }\cdot \nabla _4 \xi ^{\dagger }{}_3\right)+\text{h.c.}\ . 
\end{array}
\end{equation}
Therefore, we explicitly obtain the operator basis for the type $\nabla^2NN^{\dagger}\xi\xi^{\dagger}$, where the $P$ even $T$ even results are in Eq.~\eqref{eq:opd2DMNT+}, and the $P$ even $T$ odd results are in \eqref{eq:opd2DMNT-}. Following the same procedure, we list the operator basis with definite $P$ and $T$ transformation properties up to $\mathcal{O}(v^4)$ in appendix~\ref{ap:DMN}.

\section{Conclusion}
\label{sec:sum}

\begin{figure}[htbp]
    \centering
    \includegraphics[width=11cm, height=9cm]{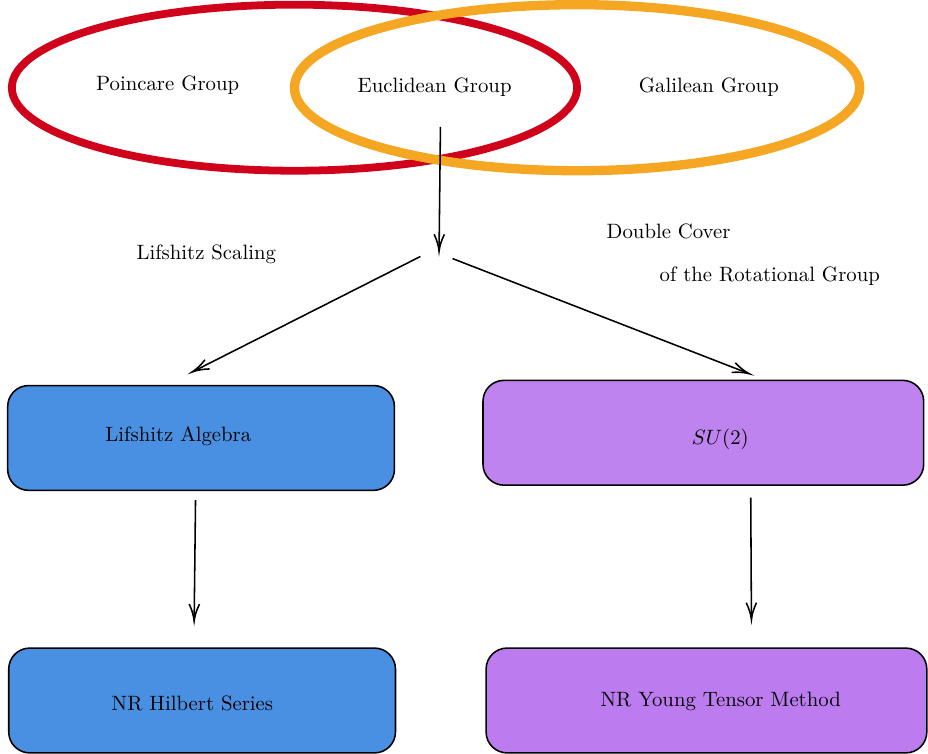}
    \caption{Schematic flowchart of the procedure to derive the non-relativistic operator bases in this work. The Euclidean group is the common subgroup of both the Poincar\'{e} group and the Galilean group. The Lifshitz algebra is used to obtain the NR Hilbert series, while the semi-standard Young tableaux of the $SU(2)$ group are utilized to explicitly construct the NR operators.}
    \label{fig:method_flowchart}
\end{figure}

In this work, we have systematically developed a framework for constructing operator bases in non-relativistic effective field theories. For the NR EFTs, the building blocks transform linearly under the Euclidean group. The additional boost transformations determine the underlying Galilean symmetry or Poincar\'{e} symmetry.
The non-relativistic operator bases  is built on $SO(3)$ rotational symmetry, with fields and covariant derivatives transforming under its double cover $Spin(3) \simeq SU(2)$. The simplification procedure, summarized in Fig.~\ref{fig:method_flowchart}, is combined with a consistent power‑counting scheme to organize the effective Lagrangian order by order.

Technically, we improve the Hilbert series~\cite{Kobach:2017xkw,Kobach:2018nmt} by eliminating redundancies associated with spin operators in the counting of multi-fermion non-relativistic operators. This is accomplished by working with the double cover of the rotation group. Furthermore, we extend the Young tensor method to the non-relativistic systems by employing the $SU(2)$ semi‑standard Young tableaux. The operators obtained are automatically free of redundancies.
The two methods provide a powerful cross check, ensuring both completeness and independence of the resulting operator bases.

Key improvements of our approach include:
\begin{itemize}
    \item 
    \textbf{Treatment of the repeated fields.}  The Young tensor method provides a systematic way of eliminating redundancies that arise from the presence of identical fields. Through the operator–Young tableau correspondence, operators can be symmetrized directly by the Young operators, thereby yielding a set of independent non-relativistic operator bases that possess the correct permutation symmetry.

    \item \textbf{Avoidance of the redundant spin operator.} Unlike Ref.~\cite{Kobach:2017xkw}, which introduces a spin operator and encounters redundancies in multi‑fermion sectors, we work directly with the $SU(2)$ rotational building blocks. This allows us to apply the Hilbert‑series method systematically to four‑ and six‑fermion operators without such ambiguities.
    
    \item \textbf{Automatic removal of the redundancies.} The equations of motion, integration‑by‑parts, and the Schouten identities are removed automatically, whereas previous constructions~\cite{Gunawardana:2017zix,Filandri:2023qio,Girlanda:2011fh,Nasoni:2023adf} required manual elimination.

    \item\textbf{Treatment of the discrete symmetries.} While the spatial inversion ($P$) and the time reversal ($T$) properties were handled manually in the earlier work~\cite{Kobach:2017xkw}, we compute the Hilbert‑series and the p‑basis from the Young tensor method with specified $P$ and $T$ properties, thereby automatically selecting only operators that are invariant under the required discrete symmetries.
\end{itemize}

We have applied this framework to several physically relevant theories and obtained complete sets of  operators listed in appendix:
\begin{itemize}
    \item The operator bases for the HPET and the HQET at dimension 9  presented here for the first time in appendix~\ref{app:HQETop} and~\ref{ap:HQET}. As a cross check, we find our results at dimension 5–8  agree  with the existing results~\cite{Gunawardana:2017zix,Kobach:2017xkw,Mannel:1994kv,Manohar:1997qy,Caswell:1985ui,Nio:1997fg,Hill:2012rh}.
    \item Nucleon‑nucleon  contact operators up to $\mathcal{O}(p^4)$ in appendix~\ref{ap:NN} and three‑nucleon contact operators up to $\mathcal{O}(p^2)$ in appendix~\ref{ap:3N},  which confirm earlier work in the general frame~\cite{Filandri:2023qio,Girlanda:2011fh,Nasoni:2023adf} and in the center-of-mass frame~\cite{Epelbaum:2004fk,Xiao:2018jot}.
    \item The rotational (and translation) invariant  spin‑$1/2$ dark‑matter–nucleon contact operators with specified $P$ and $T$ properties up to $\mathcal{O}(v^4)$ in appendix~\ref{ap:DMN}, derived here for the first time.  Here we don't impose the Galilean boost invariance or the center-of-mass condition.
\end{itemize}
Following the same technique, the NRQCD operators can be written similarly, with additional internal color symmetry imposed.

The procedure developed here is general and can be directly adapted to other NREFTs, such as the heavy black hole effective theory~\cite{Bern:2019nnu,Bern:2019crd}. Looking forward, our results open several promising directions. The operator bases, expressed in terms of semi‑standard Young tableaux, provide a natural starting point for studying the non‑trivial relations among Wilson coefficients that arise when Lorentz symmetry is restored through top‑down matching to a relativistic ultraviolet theory~\cite{Filandri:2023qio,Nasoni:2023adf,Xiao:2018jot,Bishara:2016hek,Bishara:2017pfq,Cohen:2019btp}. Moreover, the explicit operator–amplitude correspondence inherent in our construction offers a clear path toward deriving the non‑relativistic heavy black hole scattering amplitudes~\cite{Bern:2019nnu,Bern:2019crd,Damgaard:2019lfh,Aoude:2020onz}, a compelling topic for future investigation.

\section*{Acknowledgements}
We would like to express our gratitude to Zhe Ren and Hao Sun for the valuable assistance provided on the Young tensor method. 
This work is supported by the National Science Foundation of China under Grants No. 12347105, No. 12375099 and No. 12447101, and the National Key Research and Development Program of China Grant No. 2020YFC2201501, No. 2021YFA0718304.
\appendix
\section{Complete Operator Basis for HPET}\label{app:HQETop}

Using the Young tensor method described above, we derive the complete and independent operator bases up to dimension 9, which agree with the Hilbert series results presented in section~\ref{HSforHQET}. The bases up to dimension 8, given in Tab.~\ref{tab:HPET8}, are consistent with the results in Refs.~\cite{Manohar:1997qy,Caswell:1985ui,Nio:1997fg,Kobach:2017xkw,Hill:2012rh}. The $P$-even and $T$-even operators at dimension 9 are listed in Tab.~\ref{tab:HQETbasis}.

All derivatives acting on fields are understood to be symmetrized, thereby eliminating redundancies arising from the covariant derivative commutators (CDC). For instance, $D^I D^J N$ is interpreted as $\frac{1}{2}\{D^I,D^J\}N$, and $[D^I,[D^I,B^J]]$ is equivalent to $[D^2,B^J]$. Moreover, contractions involving the spinor $SU(2)$ adjoint indices—which coincide with the fundamental indices of the rotation group $SO(3)$—are expressed as $\sigma^I B^I \equiv \vec\sigma\cdot\vec B$. Explicit forms of the operators are provided in the subsequent subsections.

\subsection{Dimension 5-8 }

\begin{table}[H]\scriptsize
    \centering
    \begin{tabular}{|c|c|c|c|}
    \hline
         Dimension&Type& Hilbert Series  &Operators  \\
         \hline
         Dim-5&$BNN^{\dagger}$&$BNN^{\dagger}$ & $N^{\dagger}\vec\sigma\cdot\vec B N$\\
         \cline{2-4}
         &$D^2NN^{\dagger}$&$D^2NN^{\dagger}$&$N^{\dagger}\vec D^2N$\\
        \hline
        Dim-6&$EDNN^{\dagger}$&$2EDNN^{\dagger}$&$\begin{array}{c}
             \mathbf{i}\epsilon^{IJK}N^{\dagger}E^ID^J\sigma^KN+\text{h.c.}  \\
             N^{\dagger}[D^I,E^I]N 
        \end{array}$
        \\
        \hline
        Dim-7&$B^2NN^{\dagger}$&$B^2NN^{\dagger}$&$N^{\dagger}\vec B^2N$\\
        \cline{2-4}
        &$BD^2NN^{\dagger}$&$4BD^2NN^{\dagger}$&$\begin{array}{c}
           \mathbf{i}\epsilon^{IJP} N^{\dagger}[D^I,B^J]D^K\sigma^L\sigma^P\delta^{KL}N+\text{h.c.} \\
             N^{\dagger}B^ID^JD^K\sigma^L\delta^{IL}\delta^{JK}N+\text{h.c.}\\
             N^{\dagger}B^ID^JD^K\sigma^L\delta^{IJ}\delta^{KL}N+\text{h.c.}\\
             N^{\dagger}[D^M,[D^N,B^I]]\sigma^J\delta^{IJ}\delta^{MN}N
        \end{array}$
        \\
        \cline{2-4}
        &$D^4NN^{\dagger}$&$D^4NN^{\dagger}$&$N^{\dagger}D^ID^JD^KD^L\delta^{IJ}\delta^{KL}N$
        \\
        \cline{2-4}
        &$E^2NN^{\dagger}$&$E^2NN^{\dagger}$&$N^{\dagger}\vec E^2N$
        \\
        \hline
        
        Dim-8 &  $B^2D_tNN^{\dagger}$&$B^2D_tNN^{\dagger}$&$N^{\dagger}B^I[D_t,B^J]\sigma^K\epsilon^{IJK}N$  \\
          \cline{2-4}
         &$E^2D_tNN^{\dagger}$&$E^2D_tNN^{\dagger}$&$N^{\dagger}E^I[D_t,E^J]\sigma^K\epsilon^{IJK}N$\\
         \cline{2-4}
        & $BEDNN^{\dagger}$& $5BEDNN^{\dagger}$&
         $\begin{array}{c}
              N^{\dagger}[D^I,E^J]B^K\sigma^L\delta^{IK}\delta^{JL}N+\text{h.c.} \\
              N^{\dagger}E^I[D^J,B^K]\sigma^L\delta^{IJ}\delta^{KL}N+\text{h.c.}\\
              N^{\dagger}E^I[D^J,B^K]\sigma^L\delta^{IK}\delta^{JL}N+\text{h.c.}\\
              N^{\dagger}E^IB^JD^K\mathbf{i}\epsilon^{IJK}N+\text{h.c.}\\
              N^{\dagger}[D^I,E^J]B^K\sigma^L\delta^{IJ}\delta^{KL}N
         \end{array}$
         \\
         \cline{2-4}
       
         &$ED^3NN^{\dagger}$&$5ED^3NN^{\dagger}$&$\begin{array}{c}
              N^{\dagger}[D^I,E^J]D^KD^L\delta^{IK}\delta^{JL}N+\text{h.c.}  \\
              N^{\dagger}E^ID^JD^KD^L\delta^{KL}\mathbf{i}\sigma^M\epsilon^{IJM}N+\text{h.c.}\\
              N^{\dagger}[D^I,E^J]D^KD^L\delta^{IJ}\delta^{KL}N\\
              N^{\dagger}[D^M,[D^N,E^I]]\delta^{MN}D^J\sigma^K\mathbf{i}\epsilon^{IJK}N+\text{h.c.}\\
              N^{\dagger}[D^I,[D^M,[D^N,E^J]]]\delta^{MN}\delta^{IJ}N
               
         \end{array}
         $
         \\
         \hline
    \end{tabular}
    \caption{The HPET operators at dimension 5 to dimension 8.}\label{tab:HPET8}
\end{table}

       
               
         

\subsection{Dimension 9 }

\begin{table}[H]\scriptsize
    \centering
    \begin{tabular}{|c|c|c|}
    \hline
         Type & Hilbert Series & Operators  \\
         \hline
         $B^3NN^{\dagger}$&$B^3NN^{\dagger}$&$N^{\dagger}B^IB^JB^K\sigma^L\delta^{IJ}\delta^{KL} N$\\
         \hline
        $B^2D_t^2NN^{\dagger}$& $B^2D_t^2NN^{\dagger}$&$N^{\dagger}B^I[D_t,[D_t,B^J]]\delta^{IJ} N$\\
        \hline
       $B^2D^2NN^{\dagger}$& $10B^2D^2NN^{\dagger}$&$\begin{array}{c}
       N^{\dagger}[D^I,B^J][D^K,B^L]\delta^{IK}\delta^{JL}N+\text{h.c.}\\
       N^{\dagger}[D^I,B^J][D^K,B^L]\delta^{IL}\delta^{JK}N+\text{h.c.}\\
       N^{\dagger}B^I[D^J,B^K]D^L\mathbf{i}\epsilon^{IJM}\sigma^M\delta^{KL}N+\text{h.c.}\\
        N^{\dagger}B^I[D^J,B^K]D^L\mathbf{i}\epsilon^{IKM}\sigma^M\delta^{JL}N+\text{h.c.}\\
        N^{\dagger}B^I[D^J,B^K]D^L\mathbf{i}\epsilon^{IJL}\sigma^K+\text{h.c.}\\
        N^{\dagger}B^I[D^J,B^K]D^L\mathbf{i}\epsilon^{IKL}\sigma^J+\text{h.c.}\\
        N^{\dagger}B^I[D^J,B^K]D^L\mathbf{i}\epsilon^{JKL}\sigma^I+\text{h.c.}\\
        N^{\dagger}B^IB^JD^KD^L\delta^{IJ}\delta^{KL}N+\text{h.c.}\\
         N^{\dagger}B^IB^JD^KD^L\delta^{IK}\delta^{JL}N+\text{h.c.}\\
         N^{\dagger}[D^I,[D^J,B^K]]B^L\delta^{IJ}\delta^{KL}N
       \end{array}$\\
       \hline
       $BD^4NN^{\dagger}$&$9BD^4NN^{\dagger}$&$\begin{array}{c}
           N^{\dagger}[D^I,[D^J,B^K]]D^LD^M\sigma^K\delta^{IL}\delta^{JM}N+\text{h.c.}\\
             N^{\dagger}[D^I,[D^J,B^K]]D^LD^M\sigma^J\delta^{KL}\delta^{IM}N+\text{h.c.}\\
             N^{\dagger}[D^I,B^J]D^KD^L\sigma^M\mathbf{i}\epsilon^{IJK}\delta^{LM}N+\text{h.c.}\\
             N^{\dagger}B^ID^JD^KD^LD^M\sigma^I\delta^{JK}\delta^{LM}N+\text{h.c.}\\
             N^{\dagger}B^ID^JD^KD^LD^M\sigma^J\delta^{IK}\delta^{LM}N+\text{h.c.}\\
             N^{\dagger}[D^M,[D^N,[D^I,B^J]]]D^K\sigma^L\mathbf{i}\epsilon^{IJP}\sigma^P\delta^{KL}\delta^{MN}N+\text{h.c.} \\
             N^{\dagger}[D^M,[D^N,B^I]]D^JD^K\sigma^L\delta^{IL}\delta^{JK}\delta^{MN}N+\text{h.c.}\\
             N^{\dagger}[D^M,[D^N,B^I]]D^JD^K\sigma^L\delta^{IJ}\delta^{KL}\delta^{MN}N+\text{h.c.}\\
             N^{\dagger}[D^M,[D^N,[D^P,[D^Q,B^I]]]\sigma^J\delta^{IJ}\delta^{MN}\delta^{PQ}N
       \end{array}$\\
       \hline
        $D^6NN^{\dagger}$&$D^6NN^{\dagger}$&$N^{\dagger}D^ID^JD^KD^LD^MD^N\delta^{IJ}\delta^{KL}\delta^{MN} N$\\
        \hline
        $BEDD_tNN^{\dagger}$&$4BEDD_tNN^{\dagger}$&$\begin{array}{c}
             N^{\dagger}[D_t,E^I][D^J,B^K]\epsilon^{IJK}N\\
             N^{\dagger}\mathbf{i}[D_t,E^I]B^JD^K\sigma^L\delta^{IJ}\delta^{KL}N+\text{h.c.}
             \\ N^{\dagger}\mathbf{i}[D_t,E^I]B^JD^K\sigma^L\delta^{IK}\delta^{JL}N+\text{h.c.}
             \\
              N^{\dagger}\mathbf{i}[D_t,E^I]B^JD^K\sigma^L\delta^{IL}\delta^{JK}N+\text{h.c.}
        \end{array}$\\
        \hline
        $BE^2NN^{\dagger}$&$2BE^2NN^{\dagger}$&$\begin{array}{c}
             N^{\dagger}B^IE^JE^K\sigma^L\delta^{IL}\delta^{JK}N \\
             N^{\dagger}B^IE^JE^K\sigma^L\delta^{IJ}\delta^{KL}N
        \end{array}$\\
        \hline
        $E^2D_t^2NN^{\dagger}$& $E^2D_t^2NN^{\dagger}$& $N^{\dagger}E^I[D_t,[D_t,E^J]]\delta^{IJ}N$\\
        \hline
        $E^2D^2NN^{\dagger}$&$8E^2D^2NN^{\dagger}$&$\begin{array}{c}
            N^{\dagger}[D^I,E^J][D^K,E^L](\delta^{IK}\delta^{JL}+\delta^{IL}\delta^{JK})N+\text{h.c.}\\
            N^{\dagger}[D^I,E^J]E^KD^L(\mathbf{i}\epsilon^{JKL}\sigma^I+\mathbf{i}\epsilon^{IKL}\sigma^J)N+\text{h.c.}\\
             N^{\dagger}[D^I,E^J]E^KD^L(\mathbf{i}\epsilon^{IKM}\sigma^M\delta^{JL}+\mathbf{i}\epsilon^{JKM}\sigma^M\delta^{IL})N+\text{h.c.}\\
             N^{\dagger}E^IE^JD^KD^L\delta^{IJ}\delta^{KL}N+\text{h.c.}\\
             N^{\dagger}E^IE^JD^KD^L\delta^{IK}\delta^{JL}N+\text{h.c.}\\
             N^{\dagger}E^I[D^M,[D^N,E^J]]\delta^{IJ}\delta^{MN}N\\
               N^{\dagger}E^ID^J\sigma^K\mathbf{i}\epsilon^{IJK}N[D^M,E^N]\delta^{MN}+\text{h.c.}  \\
             N^{\dagger}[D^I,E^J][D^K,E^L]\delta^{IJ}\delta^{KL}N 
             
        \end{array}$\\
\hline
    \end{tabular}
      \caption{The HPET operator Basis at dimension 9.}
    \label{tab:HQETbasis}
\end{table}
\section{Complete Operator Basis for HQET}\label{ap:HQET}
The HQET bases up to dimension 8, given in Tab.~\ref{tab:HQET5-7} and Tab.~\ref{tab:HQET8}, are consistent with the results given by the Hilbert series in Eq.~\eqref{eq:HQETHS} and the earlier findings in Refs.~\cite{Mannel:1994kv,Manohar:1997qy,Kobach:2017xkw,Gunawardana:2017zix}. The complete and independent HQET operators at dimension 9 are listed in subsection~\ref{sec:HQETopb9}.

\subsection{Dimension 5-8 }
\begin{table}[H]
\scriptsize
    \centering
    \begin{tabular}{|c|c|c|c|}
    \hline
         Dimension&Type& Hilbert Series  &Operators  \\
         \hline
         Dim-5&$BNN^{\dagger}$&$BNN^{\dagger}$ & $N^{\dagger,a}(\vec\sigma\cdot\vec B^A)N_b(\lambda^A)_a^b$\\
         \cline{2-4}
         &$D^2NN^{\dagger}$&$D^2NN^{\dagger}$&$N^{\dagger}\vec D^2N$\\
        \hline
        Dim-6&$EDNN^{\dagger}$&$2EDNN^{\dagger}$&$\begin{array}{l}
             N^{\dagger,a}(\vec\sigma\cdot\vec E^A)(\vec\sigma\cdot\vec D)N_b(\lambda^A)_a^b+\text{h.c.}  \\
             N^{\dagger,a}[D^I,E^{I,A}]N_b(\lambda^A)_a^b+\text{h.c.} 
        \end{array}$
        \\
        \hline
        Dim-7&$B^2NN^{\dagger}$&$3B^2NN^{\dagger}$&
        $
        \begin{array}{l}
            N^{\dagger,a}(\vec\sigma\cdot\vec B^A)(\vec\sigma\cdot\vec B^B)N_bd^{ABC}(\lambda^C)_a^b+\text{h.c.}
            \\
            N^{\dagger,a}(\vec\sigma\cdot\vec B^A)(\vec\sigma\cdot\vec B^B)N_b\delta^{AB}\delta_a^b+\text{h.c.}
            \\
             N^{\dagger,a}(\vec\sigma\cdot\vec B^A)(\vec\sigma\cdot\vec B^B)N_b\textbf{i}f^{ABC}(\lambda^C)_a^b+\text{h.c.}
        \end{array}$\\
        \cline{2-4}
        &$BD^2NN^{\dagger}$&$4BD^2NN^{\dagger}$&$\begin{array}{l}
            N^{\dagger,a}[D^I,(\vec\sigma\cdot\vec B^A)](\vec\sigma\cdot\vec D)\sigma^IN_b(\lambda^A)_a^b+\text{h.c.} \\
             N^{\dagger,a}\sigma^I(\vec\sigma\cdot\vec B^A)(\vec\sigma\cdot\vec D)D^IN_b(\lambda^A)_a^b+\text{h.c.}\\
             N^{\dagger,a}(\vec\sigma\cdot\vec B^A)(\vec\sigma\cdot\vec D)(\vec\sigma\cdot\vec D)N_b(\lambda^A)_a^b+\text{h.c.}\\
             N^{\dagger,a}[D^I,[D^I,(\vec\sigma\cdot\vec B^A)]]N_b(\lambda^A)_a^b+\text{h.c.}
        \end{array}$
        \\
        \cline{2-4}
        &$D^4NN^{\dagger}$&$D^4NN^{\dagger}$&$N^{\dagger}\vec D^4N$
        \\
        \cline{2-4}
        &$E^2NN^{\dagger}$&$3E^2NN^{\dagger}$&$ \begin{array}{l}
            N^{\dagger,a}(\vec\sigma\cdot\vec E^A)(\vec\sigma\cdot\vec E^B)N_bd^{ABC}(\lambda^C)_a^b+\text{h.c.}
            \\
            N^{\dagger,a}(\vec\sigma\cdot\vec E^A)(\vec\sigma\cdot\vec E^B)N_b\delta^{AB}\delta_a^b+\text{h.c.}
            \\
             N^{\dagger,a}(\vec\sigma\cdot\vec E^A)(\vec\sigma\cdot\vec E^B)N_b\textbf{i}f^{ABC}(\lambda^C)_a^b+\text{h.c.}
        \end{array}$
        \\
        \hline
    
    \end{tabular}
    \caption{The HQET operators at dimension 5 to dimension 7.}\label{tab:HQET5-7}
    
\end{table}

\begin{table}[H]
\scriptsize
    \centering
    \begin{tabular}{|c|c|c|c|}
    \hline
         Dimension&Type& Hilbert Series  &Operators  \\
      \hline
        Dim-8 &  $B^2D_tNN^{\dagger}$&$3B^2D_tNN^{\dagger}$& $\begin{array}{l}
            N^{\dagger,a}(\vec\sigma\cdot\vec B^A)(\vec\sigma\cdot [\textbf{i}D_t,\vec B^B])N_bd^{ABC}(\lambda^C)_a^b+\text{h.c.}
            \\
            N^{\dagger,a}(\vec\sigma\cdot\vec B^A)(\vec\sigma\cdot [\textbf{i}D_t,\vec B^B])N_b\delta^{AB}\delta_a^b+\text{h.c.}
            \\
             N^{\dagger,a}(\vec\sigma\cdot\vec B^A)(\vec\sigma\cdot[\textbf{i}D_t,\vec B^B])N_b\textbf{i}f^{ABC}(\lambda^C)_a^b+\text{h.c.}
        \end{array}$\\
         \cline{2-4}
         &  $E^2D_tNN^{\dagger}$&$3E^2D_tNN^{\dagger}$& $\begin{array}{l}
            N^{\dagger,a}(\vec\sigma\cdot\vec E^A)(\vec\sigma\cdot [\textbf{i}D_t,\vec E^B])N_bd^{ABC}(\lambda^C)_a^b+\text{h.c.}
            \\
            N^{\dagger,a}(\vec\sigma\cdot\vec E^A)(\vec\sigma\cdot [\textbf{i}D_t,\vec E^B])N_b\delta^{AB}\delta_a^b+\text{h.c.}
            \\
             N^{\dagger,a}(\vec\sigma\cdot\vec E^A)(\vec\sigma\cdot[\textbf{i}D_t,\vec E^B])N_b\textbf{i}f^{ABC}(\lambda^C)_a^b+\text{h.c.}
        \end{array}$\\
         \cline{2-4}
        & $BEDNN^{\dagger}$& $14BEDNN^{\dagger}$&
         $\begin{array}{l}
              N^{\dagger,a}[D^I,(\vec\sigma\cdot\vec E^A)](\vec\sigma\cdot\vec B^B)\sigma^IN_bd^{ABC}(\lambda^C)_a^b+\text{h.c.} \\
              N^{\dagger,a}[D^I,(\vec\sigma\cdot\vec E^A)](\vec\sigma\cdot\vec B^B)\sigma^IN_b\delta^{AB}\delta_a^b+\text{h.c.}\\
              N^{\dagger,a}\sigma^IN_b\text{tr}\left[
            (\vec\sigma\cdot\vec E^A)[D^I,(\vec\sigma\cdot\vec B^B)]
              \right]d^{ABC}(\lambda^C)_a^b+\text{h.c.}\\
              N^{\dagger,a}\sigma^IN_b\text{tr}\left[
            (\vec\sigma\cdot\vec E^A)[D^I,(\vec\sigma\cdot\vec B^B)]
              \right]\delta^{AB}\delta_a^b+\text{h.c.}\\
              N^{\dagger,a}\sigma^I(\vec\sigma\cdot\vec E^A)[D^I,(\vec\sigma\cdot\vec B^B)]N_bd^{ABC}(\lambda^C)_a^b+\text{h.c.}\\
              N^{\dagger,a}\sigma^I(\vec\sigma\cdot\vec E^A)[D^I,(\vec\sigma\cdot\vec B^B)]N_b\delta^{AB}\delta_a^b+\text{h.c.}\\

              N^{\dagger,a}(\vec\sigma\cdot\vec B^A)\sigma^I(\vec\sigma\cdot\vec E^B)D^IN_bd^{ABC}(\lambda^C)_a^b+\text{h.c.}\\
              N^{\dagger,a}(\vec\sigma\cdot\vec B^A)\sigma^I(\vec\sigma\cdot\vec E^B)D^IN_b\delta^{AB}\delta_a^b+\text{h.c.}\\

              N^{\dagger,a}\sigma^I(\vec\sigma\cdot\vec E^A)[D^I,(\vec\sigma\cdot\vec B^B)]N_b\textbf{i}f^{ABC}(\lambda^C)_a^b+\text{h.c.}
              \\

              N^{\dagger,a}(\vec\sigma\cdot\vec B^A)\sigma^I(\vec\sigma\cdot\vec E^B)D^IN_b\textbf{i}f^{ABC}(\lambda^C)_a^b+\text{h.c.}\\

              N^{\dagger,a}(\vec\sigma\cdot\vec B^A)(\vec\sigma\cdot\vec E^B)(\vec\sigma\cdot\vec D)N_b\textbf{i}f^{ABC}(\lambda^C)_a^b+\text{h.c.}
              \\

              N^{\dagger,a}(\vec\sigma\cdot E^A)D^IN_b\text{tr}\left[(\vec\sigma\cdot\vec B^B)\sigma^I\right]\textbf{i}f^{ABC}(\lambda^C)_a^b+\text{h.c.}
              \\
              
              N^{\dagger,a}(\vec\sigma\cdot\vec B^A)N_b[D^I,E^{I,B}]d^{ABC}(\lambda^C)_a^b+\text{h.c.}
              \\
              N^{\dagger,a}(\vec\sigma\cdot\vec B^A)N_b[D^I,E^{I,B}]\delta^{AB}\delta_a^b+\text{h.c.}
             
         \end{array}$
         \\
         \cline{2-4}
       
         &$ED^3NN^{\dagger}$&$5ED^3NN^{\dagger}$&$\begin{array}{l}
              N^{\dagger,a}D^ID^JN_b\text{tr}\left[[D^K,(\vec\sigma\cdot\vec E^A)]\sigma^I\right]\text{tr}\left[\sigma^J\sigma^K\right](\lambda^A)_a^b+\text{h.c.}  \\

           N^{\dagger,a}\sigma^I(\vec\sigma\cdot\vec E^A)(\vec\sigma\cdot\vec D)(\vec\sigma\cdot\vec D)D^IN_b(\lambda^A)_a^b+\text{h.c.}  \\
           
N^{\dagger,a}\vec D^2N_b[D^I,E^{I,A}](\lambda^A)_a^b+\text{h.c.} \\ 

N^{\dagger,a}[D^I,[D^I,(\vec\sigma\cdot\vec E^A)]](\vec\sigma\cdot\vec D)N_b(\lambda^A)_a^b +\text{h.c.}   \\
                N^{\dagger,a}N_b[D^J,[D^I,[D^I,E^{J,A}]]](\lambda^A)_a^b+\text{h.c.}
               
         \end{array}
         $
         \\
         \hline
    \end{tabular}
    \caption{The HQET operators at dimension 8.}\label{tab:HQET8}
    
\end{table}

\subsection{Dimension 9}\label{sec:HQETopb9}

\paragraph{Type $B^3NN^{\dagger}$}  In this type, the Hilbert series counts: 
\begin{equation}
    6B^3NN^{\dagger}.
\end{equation} Utilizing the Young tensor method we obtain the operators:
\begin{eqnarray}
    &&N^{\dagger,a}(\vec\sigma\cdot\vec B^B)N_b\text{tr}\left[(\vec\sigma\cdot\vec B^A)(\vec\sigma\cdot\vec B^C)\right]\delta^{AC}(\lambda^B)_a^b+\text{h.c.}\ ,\nonumber\\
    &&N^{\dagger,a}(\vec\sigma\cdot\vec B^B)N_b\text{tr}\left[(\vec\sigma\cdot\vec B^A)(\vec\sigma\cdot\vec B^C)\right]\delta^{BC}(\lambda^A)_a^b+\text{h.c.}\ ,\nonumber\\
    &&N^{\dagger,a}(\vec\sigma\cdot\vec B^B)N_b\text{tr}\left[(\vec\sigma\cdot\vec B^A)(\vec\sigma\cdot\vec B^C)\right]d^{ABD}d^{CDE}(\lambda^E)_a^b+\text{h.c.}\ ,\nonumber\\
     &&N^{\dagger,a}(\vec\sigma\cdot\vec B^B)N_b\text{tr}\left[(\vec\sigma\cdot\vec B^A)(\vec\sigma\cdot\vec B^C)\right]d^{ABC}\delta_a^b+\text{h.c.}\ ,\nonumber\\
      &&N^{\dagger,a}(\vec\sigma\cdot\vec B^B)(\vec\sigma\cdot\vec B^A)(\vec\sigma\cdot\vec B^C)N_bd^{BDE}\textbf{i}f^{ACD}(\lambda^E)_a^b+\text{h.c.}\ , \nonumber\\
       &&N^{\dagger,a}(\vec\sigma\cdot\vec B^B)(\vec\sigma\cdot\vec B^A)(\vec\sigma\cdot\vec B^C)N_b\textbf{i}f^{ABC}\delta_a^b+\text{h.c.}\ .
\end{eqnarray}

\paragraph{Type $B^2D_t^2NN^{\dagger}$} In this type, the Hilbert series counts: 
\begin{equation}
    3B^2D_t^2NN^{\dagger}.
\end{equation} Utilizing the Young tensor method we obtain the operators:
\begin{eqnarray}
           && N^{\dagger,a}(\vec\sigma\cdot\vec B^A)(\vec\sigma\cdot [D_t,[D_t,\vec B^B]])N_bd^{ABC}(\lambda^C)_a^b+\text{h.c.}\ ,\nonumber
            \\
            &&  N^{\dagger,a}(\vec\sigma\cdot\vec B^A)(\vec\sigma\cdot [D_t,[D_t,\vec B^B]])N_b\delta^{AB}\delta_a^b+\text{h.c.}\ ,\nonumber
            \\
              && N^{\dagger,a}(\vec\sigma\cdot\vec B^A)(\vec\sigma\cdot[D_t,[D_t,\vec B^B]])N_b\textbf{i}f^{ABC}(\lambda^C)_a^b+\text{h.c.} \ .
\end{eqnarray}

\paragraph{Type $B^2D^2NN^{\dagger}$} In this type, the Hilbert series counts: 
\begin{equation}
    29B^2D^2NN^{\dagger}.
\end{equation} Utilizing the Young tensor method we obtain the operators:
\begin{eqnarray}
&&N^{\dagger,a}\sigma^I\sigma^J[D^I,(\vec\sigma\cdot\vec B^B)][D^J,(\vec\sigma\cdot\vec B^A)]N_bd^{ABC}
    (\lambda^C)_a^b+\text{h.c.} \ , \nonumber\\
&&N^{\dagger,a}\sigma^I\sigma^J[D^I,(\vec\sigma\cdot\vec B^B)][D^J,(\vec\sigma\cdot\vec B^A)]N_b\delta^{AB}\delta_a^b+\text{h.c.} \ , \nonumber\\
&&N^{\dagger,a}[D^I,(\vec\sigma\cdot\vec B^A)][D^J,(\vec\sigma\cdot\vec B^B)]N_b\text{tr}\left[\sigma^I\sigma^J\right]d^{ABC}
    (\lambda^C)_a^b+\text{h.c.} \ , \nonumber\\
    &&N^{\dagger,a}[D^I,(\vec\sigma\cdot\vec B^A)][D^J,(\vec\sigma\cdot\vec B^B)]N_b\text{tr}\left[\sigma^I\sigma^J\right]\delta^{AB}\delta_a^b+\text{h.c.} \ , \nonumber\\
    &&N^{\dagger,a}(\vec\sigma\cdot\vec B^B)\sigma^ID^JN_b\text{tr}\left[[D^I,(\vec\sigma\cdot\vec B^A)]\sigma^J\right]d^{ABC}
    (\lambda^C)_a^b+\text{h.c.} \ , \nonumber\\
     &&N^{\dagger,a}(\vec\sigma\cdot\vec B^B)\sigma^ID^JN_b\text{tr}\left[[D^I,(\vec\sigma\cdot\vec B^A)]\sigma^J\right]\delta^{AB}\delta_a^b+\text{h.c.} \ , \nonumber\\
     &&N^{\dagger,a}[D^J,(\vec\sigma\cdot\vec B^A)]\sigma^I(\vec\sigma\cdot\vec B^B)\sigma^JD^IN_bd^{ABC}
    (\lambda^C)_a^b+\text{h.c.} \ , \nonumber\\
    &&N^{\dagger,a}[D^J,(\vec\sigma\cdot\vec B^A)]\sigma^I(\vec\sigma\cdot\vec B^B)\sigma^JD^IN_b\delta^{AB}\delta_a^b+\text{h.c.} \ , \nonumber\\
    &&N^{\dagger,a}[D^I,(\vec\sigma\cdot\vec B^A)](\vec\sigma\cdot\vec D)N_b\text{tr}\left[\sigma^I(\vec\sigma\cdot\vec B^B)\right]d^{ABC}
    (\lambda^C)_a^b+\text{h.c.} \ , \nonumber\\
      &&N^{\dagger,a}[D^I,(\vec\sigma\cdot\vec B^A)](\vec\sigma\cdot\vec D)N_b\text{tr}\left[\sigma^I(\vec\sigma\cdot\vec B^B)\right]\delta^{AB}\delta_a^b+\text{h.c.} \ , \nonumber\\
      &&N^{\dagger,a}[D^I,(\vec\sigma\cdot\vec B^A)]\sigma^ID^JN_b\text{tr}[(\vec\sigma\cdot\vec B^B)\sigma^J]d^{ABC}
    (\lambda^C)_a^b+\text{h.c.} \ , \nonumber\\
       &&N^{\dagger,a}[D^I,(\vec\sigma\cdot\vec B^A)]\sigma^ID^JN_b\text{tr}[(\vec\sigma\cdot\vec B^B)\sigma^J]\delta^{AB}\delta_a^b+\text{h.c.} \ , \nonumber\\
       &&N^{\dagger,a}[D^I,(\vec\sigma\cdot\vec B^A)]\sigma^I(\vec\sigma\cdot\vec B^B)(\vec\sigma\cdot\vec D)N_bd^{ABC}
    (\lambda^C)_a^b+\text{h.c.} \ , \nonumber\\
       &&N^{\dagger,a}[D^I,(\vec\sigma\cdot\vec B^A)]\sigma^I(\vec\sigma\cdot\vec B^B)(\vec\sigma\cdot\vec D)N_b\delta^{AB}\delta_a^b+\text{h.c.} \ , \nonumber\\
       &&N^{\dagger,a}D^ID^JN_b\text{tr}\left[(\vec\sigma\cdot\vec B^A)\sigma^I\right]\text{tr}\left[(\vec\sigma\cdot\vec B^B)\sigma^J\right]d^{ABC}
    (\lambda^C)_a^b+\text{h.c.} \ , \nonumber\\
       &&N^{\dagger,a}D^ID^JN_b\text{tr}\left[(\vec\sigma\cdot\vec B^A)\sigma^I\right]\text{tr}\left[(\vec\sigma\cdot\vec B^B)\sigma^J\right]\delta^{AB}\delta_a^b+\text{h.c.} \ , \nonumber\\
       &&N^{\dagger,a}(\vec\sigma\cdot\vec B^B)\sigma^I\sigma^J(\vec\sigma\cdot\vec B^A)D^ID^JN_bd^{ABC}
    (\lambda^C)_a^b+\text{h.c.} \ , \nonumber\\
         &&N^{\dagger,a}(\vec\sigma\cdot\vec B^B)\sigma^I\sigma^J(\vec\sigma\cdot\vec B^A)D^ID^JN_b\delta^{AB}\delta_a^b+\text{h.c.} \ , 
\end{eqnarray}
\begin{eqnarray}
    &&N^{\dagger,a}[D^I,[D^J,(\vec\sigma\cdot\vec B^A)]]\sigma^I(\vec\sigma\cdot\vec B^B)\sigma^JN_b\textbf{i}f^{ABC}(\lambda^C)_a^b+\text{h.c.}\ , \nonumber\\
    &&N^{\dagger,a}\sigma^I\sigma^J[D^I,(\vec\sigma\cdot\vec B^B)][D^J,(\vec\sigma\cdot\vec B^A)]N_b\textbf{i}f^{ABC}(\lambda^C)_a^b+\text{h.c.}\ , \nonumber\\
  &&N^{\dagger,a}[D^I,(\vec\sigma\cdot\vec B^A)][D^J,(\vec\sigma\cdot\vec B^B)]N_b\text{tr}\left[\sigma^I\sigma^J\right]\textbf{i}f^{ABC}(\lambda^C)_a^b+\text{h.c.}\ , \nonumber\\
  &&N^{\dagger,a}[D^I,(\vec\sigma\cdot\vec B^A)]\sigma^I[D^J,(\vec\sigma\cdot\vec B^B)]\sigma^JN_b\textbf{i}f^{ABC}(\lambda^C)_a^b+\text{h.c.}\ , \nonumber\\
  &&N^{\dagger,a}(\vec\sigma\cdot\vec B^B)\sigma^ID^JN_b\text{tr}\left[[D^I,(\vec\sigma\cdot\vec B^A)]\sigma^J\right]\textbf{i}f^{ABC}(\lambda^C)_a^b+\text{h.c.}\ , \nonumber\\
    &&N^{\dagger,a}[D^J,(\vec\sigma\cdot\vec B^A)]\sigma^I(\vec\sigma\cdot\vec B^B)\sigma^JD^IN_b\textbf{i}f^{ABC}(\lambda^C)_a^b+\text{h.c.}\ , \nonumber\\
    &&N^{\dagger,a}(\vec\sigma\cdot\vec B^B)\sigma^JD^ID^JN_b\text{tr}\left[(\vec\sigma\cdot\vec B^A)\sigma^I\right]\textbf{i}f^{ABC}(\lambda^C)_a^b+\text{h.c.}\ , \nonumber\\
    &&N^{\dagger,a}(\vec\sigma\cdot\vec B^B)\sigma^I\sigma^J(\vec\sigma\cdot\vec B^A)D^ID^JN_b\textbf{i}f^{ABC}(\lambda^C)_a^b+\text{h.c.}\ , 
\end{eqnarray}
\begin{eqnarray}
    &&N^{\dagger,a}(\vec\sigma\cdot\vec B^B)[D^I,[D^I,(\vec\sigma\cdot\vec B^A)]]d^{ABC}(\lambda^C)_a^b+\text{h.c.}\ ,\nonumber\\
       &&N^{\dagger,a}(\vec\sigma\cdot\vec B^B)[D^I,[D^I,(\vec\sigma\cdot\vec B^A)]]\delta^{AB}\delta_a^c+\text{h.c.}\ ,\nonumber\\
         &&N^{\dagger,a}(\vec\sigma\cdot\vec B^B)[D^I,[D^I,(\vec\sigma\cdot\vec B^A)]]\textbf{i}f^{ABC}(\lambda^C)_a^b+\text{h.c.}\ .
\end{eqnarray}

\paragraph{Type $BD^4NN^{\dagger}$}
In this type, the Hilbert series counts: 
\begin{equation}
    9BD^4NN^{\dagger}.
\end{equation} Utilizing the Young tensor method we obtain the operators:
\begin{eqnarray}
    &&N^{\dagger,a}[D^I,[D^J(\vec\sigma\cdot\vec B^A)]](\vec\sigma\cdot\vec D)\sigma^ID^KN_b\text{tr}\left[\sigma^J\sigma^K\right](\lambda^A)_a^b+\text{h.c.}\ ,\nonumber\\
    &&N^{\dagger,a}[D^K,[D^I,(\vec\sigma\cdot\vec B^A)]]\sigma^I(\vec\sigma\cdot\vec D)D^JN_b\text{tr}[\sigma^K\sigma^J](\lambda^A)_a^b+\text{h.c.}\ ,\nonumber\\
    &&N^{\dagger,a}\sigma^I\sigma^J\sigma^K[D^J,(\vec\sigma\cdot\vec B^A)](\vec\sigma\cdot\vec D)D^ID^KN_b(\lambda^A)_a^b+\text{h.c.}\ ,\nonumber\\
    &&N^{\dagger,a}\sigma^K(\vec\sigma\cdot\vec B^A)(\vec\sigma\cdot\vec D)D^ID^JD^KN_b\text{tr}\left[\sigma^I\sigma^J\right](\lambda^A)_a^b+\text{h.c.}\ ,\nonumber\\
    &&N^{\dagger,a}(\vec\sigma\cdot\vec D)(\vec\sigma\cdot\vec D)(\vec\sigma\cdot\vec D)D^IN_b\text{tr}\left[(\vec\sigma\cdot\vec B^A)\sigma^I\right](\lambda^A)_a^b+\text{h.c.}\ ,\nonumber\\
    && N^{\dagger,a}[D^I,[D^J,[D^J,(\vec\sigma\cdot\vec B^A)]]](\vec\sigma\cdot\vec D)\sigma^IN_b(\lambda^A)_a^b+\text{h.c.} \ ,\nonumber\\
             &&N^{\dagger,a}\sigma^I[D^J,[D^J,(\vec\sigma\cdot\vec B^A)]](\vec\sigma\cdot\vec D)D^IN_b(\lambda^A)_a^b+\text{h.c.}\ ,\nonumber\\
             &&N^{\dagger,a}[D^J,[D^J,(\vec\sigma\cdot\vec B^A)]](\vec\sigma\cdot\vec D)(\vec\sigma\cdot\vec D)N_b(\lambda^A)_a^b+\text{h.c.}\ ,\nonumber\\
             &&N^{\dagger,a}[D^I,[D^I,[D^J,[D^J,(\vec\sigma\cdot\vec B^A)]]]]N_b(\lambda^A)_a^b+\text{h.c.}\ .
\end{eqnarray}

\paragraph{Type $D^6NN^{\dagger}$}
In this type, the only operator is
\begin{equation}
    N^{\dagger}D^6N.
\end{equation}

\paragraph{Type $BEDD_tNN^{\dagger}$}
In this type, the Hilbert series counts: 
\begin{equation}
    13BEDD_tNN^{\dagger}.
\end{equation} Utilizing the Young tensor method we obtain the operators:
\begin{eqnarray}
    &&N^{\dagger,a}[D^I,(\vec\sigma\cdot\vec E^A)][\textbf{i}D_t,(\vec\sigma\cdot\vec B^B)]\sigma^IN_b\textbf{i}f^{ABC}(\lambda^C)_a^b+\text{h.c.}\ ,\nonumber\\
    &&N^{\dagger,a}\sigma^IN_b\text{tr}\left[(\vec\sigma\cdot\vec E^A)[D^I,[\textbf{i}D_t,(\vec\sigma\cdot\vec B^B)]]\right]\textbf{i}f^{ABC}(\lambda^C)_a^b+\text{h.c.}\ ,\nonumber\\
    &&N^{\dagger,a}\sigma^I(\vec\sigma\cdot\vec E^A)[D^I,[\textbf{i}D_t,(\vec\sigma\cdot\vec B^B)]]N_b    \textbf{i}f^{ABC}(\lambda^C)_a^b+\text{h.c.}\ ,\nonumber\\
    &&N^{\dagger,a}[\textbf{i}D_t,(\vec\sigma\cdot\vec B^B)]\sigma^I(\vec\sigma\cdot\vec E^A)D^IN_b\textbf{i}f^{ABC}(\lambda^C)_a^b+\text{h.c.}\ ,\nonumber\\
     &&N^{\dagger,a}\sigma^I(\vec\sigma\cdot\vec E^A)[D^I,[\textbf{i}D_t,(\vec\sigma\cdot\vec B^B)]]N_bd^{ABC}(\lambda^C)_a^b+\text{h.c.}\ ,\nonumber\\
     &&N^{\dagger,a}\sigma^I(\vec\sigma\cdot\vec E^A)[D^I,[\textbf{i}D_t,(\vec\sigma\cdot\vec B^B)]]N_b\delta^{AB}\delta_a^c+\text{h.c.}\ ,\nonumber\\
      &&N^{\dagger,a}[\textbf{i}D_t,(\vec\sigma\cdot\vec B^B)]\sigma^I(\vec\sigma\cdot\vec E^A)D^IN_bd^{ABC}(\lambda^C)_a^b+\text{h.c.}\ ,\nonumber\\
      &&N^{\dagger,a}[\textbf{i}D_t,(\vec\sigma\cdot\vec B^B)]\sigma^I(\vec\sigma\cdot\vec E^A)D^IN_b\delta^{AB}\delta_a^c+\text{h.c.}\ ,\nonumber \\
      &&N^{\dagger,a}[\textbf{i}D_t,(\vec\sigma\cdot\vec B^B)](\vec\sigma\cdot\vec E^A)(\vec\sigma\cdot\vec D)N_bd^{ABC}(\lambda^C)_a^b+\text{h.c.}\ ,\nonumber\\
       &&N^{\dagger,a}[\textbf{i}D_t,(\vec\sigma\cdot\vec B^B)](\vec\sigma\cdot\vec E^A)(\vec\sigma\cdot\vec D)N_b\delta^{AB}\delta_a^c+\text{h.c.}\ ,\nonumber \\
        &&N^{\dagger,a}(\vec\sigma\cdot\vec E^A)D^IN_b\text{tr}\left[[\textbf{i}D_t,(\vec\sigma\cdot\vec B^B)]\sigma^I\right]d^{ABC}(\lambda^C)_a^b+\text{h.c.}\ ,\nonumber\\
        &&N^{\dagger,a}(\vec\sigma\cdot\vec E^A)D^IN_b\text{tr}\left[[\textbf{i}D_t,(\vec\sigma\cdot\vec B^B)]\sigma^I\right]\delta^{AB}\delta_a^c+\text{h.c.}\ ,\nonumber \\
        &&N^{\dagger,a}[\textbf{i}D_t,(\vec\sigma\cdot\vec B^B)]N_b[D^I,E^{I,A}]\textbf{i}f^{ABC}(\lambda^C)_a^b+\text{h.c.}\ .
\end{eqnarray}

\paragraph{Type $BE^2NN^{\dagger}$}
In this type, the Hilbert series counts: 
\begin{equation}
    13BE^2NN^{\dagger}.
\end{equation} Utilizing the Young tensor method we obtain the operators:
\begin{eqnarray}
    &&N^{\dagger,a}(\vec\sigma\cdot\vec E^B)N_b\text{tr}\left[(\vec\sigma\cdot\vec E^A)(\vec\sigma\cdot\vec B^C)\right]\delta^{AC}(\lambda^B)_a^b+\text{h.c.} \ , \nonumber\\
     &&N^{\dagger,a}(\vec\sigma\cdot\vec E^B)N_b\text{tr}\left[(\vec\sigma\cdot\vec E^A)(\vec\sigma\cdot\vec B^C)\right]\delta^{BC}(\lambda^A)_a^b+\text{h.c.} \ , \nonumber\\
      &&N^{\dagger,a}(\vec\sigma\cdot\vec E^B)N_b\text{tr}\left[(\vec\sigma\cdot\vec E^A)(\vec\sigma\cdot\vec B^C)\right]d^{ABD}d^{CDE}(\lambda^E )_a^b+\text{h.c.} \ , \nonumber\\
      &&N^{\dagger,a}(\vec\sigma\cdot\vec E^B)N_b\text{tr}\left[(\vec\sigma\cdot\vec E^A)(\vec\sigma\cdot\vec B^C)\right]f^{ABD}f^{CDE}(\lambda^E)_a^b+\text{h.c.} \ , \nonumber\\
      &&N^{\dagger,a}(\vec\sigma\cdot\vec E^B)N_b\text{tr}\left[(\vec\sigma\cdot\vec E^A)(\vec\sigma\cdot\vec B^C)\right]d^{ABC}\delta_a^b+\text{h.c.} \ , \nonumber\\
      &&N^{\dagger,a}(\vec\sigma\cdot\vec E^B)N_b\text{tr}\left[(\vec\sigma\cdot\vec E^A)(\vec\sigma\cdot\vec B^C)\right]\delta^{AB}(\lambda^C)_a^b+\text{h.c.} \ , \nonumber\\
       &&N^{\dagger,a}(\vec\sigma\cdot\vec E^B)(\vec\sigma\cdot\vec E^A)(\vec\sigma\cdot\vec B^C)N_b\delta^{AC}(\lambda^B)_a^b+\text{h.c.} \ , \nonumber\\
       &&N^{\dagger,a}(\vec\sigma\cdot\vec E^B)(\vec\sigma\cdot\vec E^A)(\vec\sigma\cdot\vec B^C)N_bd^{ABD}d^{CDE}(\lambda^E)_a^b+\text{h.c.} \ , \nonumber\\
       &&N^{\dagger,a}(\vec\sigma\cdot\vec E^B)(\vec\sigma\cdot\vec E^A)(\vec\sigma\cdot\vec B^C)N_bd^{ABC}\delta_a^b+\text{h.c.} \ , \nonumber\\
       &&N^{\dagger,a}(\vec\sigma\cdot\vec E^B)(\vec\sigma\cdot\vec E^A)(\vec\sigma\cdot\vec B^C)N_b\delta^{AB}(\lambda^C)_a^b+\text{h.c.} \ , \nonumber\\
       &&N^{\dagger,a}(\vec\sigma\cdot\vec E^B)(\vec\sigma\cdot\vec E^A)(\vec\sigma\cdot\vec B^C)N_bd^{BDE}\textbf{i}f^{ACD}(\lambda^E)_a^b+\text{h.c.} \ , \nonumber\\
        &&N^{\dagger,a}(\vec\sigma\cdot\vec E^B)(\vec\sigma\cdot\vec E^A)(\vec\sigma\cdot\vec B^C)N_bd^{ACD}\textbf{i}f^{BDE}(\lambda^E)_a^b+\text{h.c.} \ , \nonumber\\
        &&N^{\dagger,a}(\vec\sigma\cdot\vec E^B)(\vec\sigma\cdot\vec E^A)(\vec\sigma\cdot\vec B^C)N_b\textbf{i}f^{ABC}\delta_a^b+\text{h.c.} \ .
\end{eqnarray}

\paragraph{Type $E^2D_t^2NN^{\dagger}$}
In this type, the Hilbert series counts: 
\begin{equation}
    3E^2D_t^2NN^{\dagger}.
\end{equation} Utilizing the Young tensor method we obtain the operators:
\begin{eqnarray}
    &&N^{\dagger,a}(\vec\sigma\cdot\vec E^A)[D_t,[D_t,(\vec\sigma\cdot\vec E^B)]]N_bd^{ABC}(\lambda^C)_a^b+\text{h.c.}\ ,\nonumber
            \\
           && N^{\dagger,a}(\vec\sigma\cdot\vec E^A)[D_t,[D_t,(\vec\sigma\cdot\vec E^B)]]N_b\delta^{AB}\delta_a^b+\text{h.c.}\ ,\nonumber
            \\
             &&N^{\dagger,a}(\vec\sigma\cdot\vec E^A)[D_t,[D_t,(\vec\sigma\cdot\vec E^B)]]N_b\textbf{i}f^{ABC}(\lambda^C)_a^b+\text{h.c.}\ .
\end{eqnarray}

\paragraph{Type $E^2D^2NN^{\dagger}$}
In this type, the Hilbert series counts: 
\begin{equation}
    22E^2D^2NN^{\dagger}.
\end{equation} Utilizing the Young tensor method we obtain the operators:
\begin{eqnarray}
    &&N^{\dagger,a}[D^I,(\vec\sigma\cdot\vec E^B)][D^J,(\vec\sigma\cdot\vec E^A)]N_b\text{tr}\left[\sigma^I\sigma^J\right]d^{ABC}(\lambda^C)_a^b+\text{h.c.}\ ,\nonumber\\
     &&N^{\dagger,a}[D^I,(\vec\sigma\cdot\vec E^B)][D^J,(\vec\sigma\cdot\vec E^A)]N_b\text{tr}\left[\sigma^I\sigma^J\right]\delta^{AB}\delta_a^b+\text{h.c.}\ , \nonumber\\
     &&N^{\dagger,a}(\vec\sigma\cdot\vec E^B)[D^I,(\vec\sigma\cdot\vec E^A)]D^JN_b\text{tr}\left[\sigma^I\sigma^J\right]d^{ABC}(\lambda^C)_a^b+\text{h.c.}\ ,\nonumber\\
      &&N^{\dagger,a}(\vec\sigma\cdot\vec E^B)[D^I,(\vec\sigma\cdot\vec E^A)]D^JN_b\text{tr}\left[\sigma^I\sigma^J\right]\delta^{AB}\delta_a^b+\text{h.c.}\ , \nonumber\\
      &&N^{\dagger,a}[D^I,(\vec\sigma\cdot\vec E^A)](\vec\sigma\cdot\vec E^B)(\vec\sigma\cdot\vec D)\sigma^IN_bd^{ABC}(\lambda^C)_a^b+\text{h.c.}\ ,\nonumber\\
       &&N^{\dagger,a}[D^I,(\vec\sigma\cdot\vec E^A)](\vec\sigma\cdot\vec E^B)(\vec\sigma\cdot\vec D)\sigma^IN_b\delta^{AB}\delta_a^b+\text{h.c.}\ , \nonumber\\
        &&N^{\dagger,a}D^ID^JN_b\text{tr}\left[(\vec\sigma\cdot\vec E^A)\sigma^I\right]\text{tr}\left[(\vec\sigma\cdot\vec E^B)\sigma^J\right]d^{ABC}(\lambda^C)_a^b+\text{h.c.}\ ,\nonumber\\
        &&N^{\dagger,a}D^ID^JN_b\text{tr}\left[(\vec\sigma\cdot\vec E^A)\sigma^I\right]\text{tr}\left[(\vec\sigma\cdot\vec E^B)\sigma^J\right]\delta^{AB}\delta_a^b+\text{h.c.}\ , \nonumber\\
        &&N^{\dagger,a}(\vec\sigma\cdot\vec E^B)\sigma^I\sigma^J(\vec\sigma\cdot\vec E^A)D^ID^JN_b d^{ABC}(\lambda^C)_a^b+\text{h.c.}\ ,\nonumber\\
         &&N^{\dagger,a}(\vec\sigma\cdot\vec E^B)\sigma^I\sigma^J(\vec\sigma\cdot\vec E^A)D^ID^JN_b\delta^{AB}\delta_a^b+\text{h.c.}\ , \nonumber\\
          &&N^{\dagger,a}[D^I,(\vec\sigma\cdot\vec E^B)][D^J,(\vec\sigma\cdot\vec E^A)]N_b\text{tr}\left[\sigma^I\sigma^J\right]\textbf{i}f^{ABC}\delta_a^b+\text{h.c.}\ ,\nonumber\\
          &&N^{\dagger,a}(\vec\sigma\cdot\vec E^B)[D^I,(\vec\sigma\cdot\vec E^A)]D^JN_b\text{tr}\left[\sigma^I\sigma^J\right]\textbf{i}f^{ABC}\delta_a^b+\text{h.c.}\ ,\nonumber\\
          &&N^{\dagger,a}(\vec\sigma\cdot\vec E^B)(\vec\sigma\cdot\vec D)D^IN_b\text{tr}\left[(\vec\sigma\cdot\vec E^A)\sigma^I\right]\textbf{i}f^{ABC}\delta_a^b+\text{h.c.}\ ,\nonumber\\
           &&N^{\dagger,a}(\vec\sigma\cdot\vec E^B)\sigma^I\sigma^J(\vec\sigma\cdot\vec E^A)D^ID^JN_b \textbf{i}f^{ABC}\delta_a^b+\text{h.c.}\ ,
\end{eqnarray}

\begin{eqnarray}
    &&N^{\dagger,a}(\vec\sigma\cdot\vec E^B)N_b[D^I,[D^I,(\vec\sigma\cdot\vec E^A)]]d^{ABC}(\lambda^C)_a^b+\text{h.c.}\ ,\nonumber\\
       &&N^{\dagger,a}(\vec\sigma\cdot\vec E^B)N_b[D^I,[D^I,(\vec\sigma\cdot\vec E^A)]]\delta^{AB}\delta_a^c+\text{h.c.}\ ,\nonumber\\
         &&N^{\dagger,a}(\vec\sigma\cdot\vec E^B)N_b[D^I,[D^I,(\vec\sigma\cdot\vec E^A)]]\textbf{i}f^{ABC}(\lambda^C)_a^b+\text{h.c.}\ ,
\end{eqnarray}

\begin{eqnarray}
     &&N^{\dagger,a}N_b[D^I,E^{I,A}][D^J,E^{J,B}]d^{ABC}(\lambda^C)_a^b+\text{h.c.}\ ,\nonumber\\
     &&N^{\dagger,a}N_b[D^I,E^{I,A}][D^J,E^{J,B}]\delta^{AB}\delta_a^c+\text{h.c.}\ ,
\end{eqnarray}

\begin{eqnarray}
     &&N^{\dagger,a}(\vec\sigma\cdot\vec E^A)(\vec\sigma\cdot\vec D)N_b[D^J,E^{J,B}]d^{ABC}(\lambda^C)_a^b+\text{h.c.}\ ,\nonumber\\
     &&N^{\dagger,a}(\vec\sigma\cdot\vec E^A)(\vec\sigma\cdot\vec D)N_b[D^J,E^{J,B}]\delta^{AB}\delta_a^c+\text{h.c.}\ ,\nonumber\\
     &&N^{\dagger,a}(\vec\sigma\cdot\vec E^A)(\vec\sigma\cdot\vec D)N_b[D^J,E^{J,B}]\textbf{i}f^{ABC}(\lambda^C)_a^b+\text{h.c.}\ .
\end{eqnarray}
\section{Complete Operator Basis for N-N Contact Interaction}\label{ap:NN}

The numbers of the nucleon--nucleon contact interaction operators, categorized by their types, are presented in Tab.~\ref{number2d}. The counts derived from the Hilbert series and the p-basis are in complete agreement. Specifically, for the $P$-even and $T$-even sector, there are 2 operators at $\mathcal{O}(Q^0)$, 12 at $\mathcal{O}(Q^2)$, and 45 at $\mathcal{O}(Q^4)$.

\begin{table}[H]
\small	
\centering
\begin{tabular}{|rr|c|c|c|}
	\hline
	&& Type $N^2(N^{\dagger})^2$ & Type $N^2(N^{\dagger})^2\nabla^2$ & Type $N^2(N^{\dagger})^2\nabla^4$ \\
	\hline
	& Hilbert series & $2N^2(N^{\dagger})^2$ & $17N^2(N^{\dagger})^2\nabla^2$ & $74N^2(N^{\dagger})^2\nabla^4$ \\
	\hline
	& p-basis & $2N^2(N^{\dagger})^2$ & $17N^2(N^{\dagger})^2\nabla^2$ & $74N^2(N^{\dagger})^2\nabla^4$ \\
	\hline
	$P$ even, $T$ even & Hilbert series & $2N^2(N^{\dagger})^2$ & $12N^2(N^{\dagger})^2\nabla^2$ & $45N^2(N^{\dagger})^2\nabla^4$ \\
	\hline
	$P$ even, $T$ even & p-basis & $2N^2(N^{\dagger})^2$ & $12N^2(N^{\dagger})^2\nabla^2$ & $45N^2(N^{\dagger})^2\nabla^4$ \\
	\hline
\end{tabular}
\caption{Operator counts for the nucleon--nucleon contact interactions, comparing results from the Hilbert series and the p-basis. The upper rows show the total number of operators, while the lower rows list the counts for the $P$-even and $T$-even subset.}
\label{number2d}
\end{table}

In the general framework, at the $\mathcal{O}(Q^0)$, the basis contains two operators:
\begin{equation}
    O_S = (N^{\dagger}N)(N^{\dagger}N), \qquad 
O_T = (N^{\dagger}\vec{\sigma}N) \cdot (N^{\dagger}\vec{\sigma}N).
\end{equation}
The basis of $\mathcal{O}(Q^2)$ operators is provided in Tab.~\ref{4f2d}, and the $\mathcal{O}(Q^4)$ basis is given in Tab.~\ref{4f4d}. In our convention, $N_3$ and $N_4$ represent identical incoming particles, while $N^{\dagger}_1$ and $N^{\dagger}_2$ represent identical outgoing particles. Consequently, the derivatives $-\mathbf{i}\nabla_3$, $-\mathbf{i}\nabla_4$ correspond to the initial-state momenta, and $\mathbf{i}\nabla_1$, $\mathbf{i}\nabla_2$ to the final-state momenta in the configuration space.
The isospin index structures are defined as
\begin{equation}
\begin{aligned}
I_1 &= \delta_{p_3}^{p_1} \delta_{p_4}^{p_2}, \\
I_2 &= \delta_{p_3}^{p_2} \delta_{p_4}^{p_1}.
\end{aligned}
\end{equation}

In the following subsections, we present the explicit forms of the nucleon--nucleon operators with two and four derivatives, which are consistent with the results in Refs.~\cite{Pastore:2009is,Girlanda:2010ya, Filandri:2023qio}.

 

\subsection{2 Derivatives}
\begin{table}[H]\small	
\caption{N-N Operator basis with 2 derivatives.}
\label{4f2d}
\begin{center}
 \begin{tabular}{|l|l|}
	\hline
	$\mathcal{O}^{p^{\prime}(2)}_1=(N^{\dagger}\Vec{\sigma}\cdot\overrightarrow{\nabla}\overrightarrow{\nabla}^iN^{\dagger})(N\sigma^iN)I_1+\text{h.c.}$
 &
 $\mathcal{O}^{p^{\prime}(2)}_2=(N^{\dagger}\sigma^iN)(N^{\dagger}\overleftarrow{\nabla}^i\Vec{\sigma}\cdot\overleftarrow{\nabla}N)I_1+\text{h.c.}$\\
 
 $\mathcal{O}^{p^{\prime}(2)}_3=-(N^{\dagger}\overrightarrow{\nabla}^2N^{\dagger})(NN)I_1+\text{h.c.}$
 &
 $\mathcal{O}^{p^{\prime}(2)}_4=-(N^{\dagger}\overrightarrow{\nabla}^iN)(N^{\dagger}\sigma^i\Vec{\sigma}\cdot\overleftarrow{\nabla}N)I_1+\text{h.c.}$\\
 $\mathcal{O}^{p^{\prime}(2)}_5=-(N^{\dagger}\Vec{\sigma}\cdot\overrightarrow{\nabla}\sigma^iN^{\dagger})(N\overleftarrow{\nabla}^iN)I_1+\text{h.c.}$
 &
 $\mathcal{O}^{p^{\prime}(2)}_6=-(N^{\dagger}\sigma^i\Vec{\sigma}\cdot\overrightarrow{\nabla}N)(N^{\dagger}\overleftarrow{\nabla}^iN)I_2+\text{h.c.}$
 \\
 $\mathcal{O}^{p^{\prime}(2)}_7=-(N^{\dagger}\Vec{\sigma}\cdot\overrightarrow{\nabla}N^{\dagger})(N\Vec{\sigma}\cdot\overleftarrow{\nabla}N)I_1+\text{h.c.}$
 &
 $\mathcal{O}^{p^{\prime}(2)}_8=2(N^{\dagger}\overrightarrow{\nabla}^iN^{\dagger})(N\overleftarrow{\nabla}^iN)I_1+\text{h.c.}$\\
 $\mathcal{O}^{p^{\prime}(2)}_9=-2(N^{\dagger}N)(N^{\dagger}\overleftarrow{\nabla}\cdot\overrightarrow{\nabla}N)I_1+\text{h.c.}$
 &
$\mathcal{O}^{p^{\prime}(2)}_{10}=-(N^{\dagger}\sigma^i\sigma^jN)(N^{\dagger}\overleftarrow{\nabla}^i\overrightarrow{\nabla}^jN)I_1+\text{h.c.}$\\
$\mathcal{O}^{p^{\prime}(2)}_{11}=(N^{\dagger}\sigma^i\sigma^jN)(N^{\dagger}\overleftarrow{\nabla}^i\overrightarrow{\nabla}^jN)I_2+\text{h.c.}$
&
$\mathcal{O}^{p^{\prime}(2)}_{12}=(N^{\dagger}\vec{\sigma}\cdot\overrightarrow{\nabla}N^{\dagger})(N\Vec{\sigma}\cdot\overrightarrow{\nabla}N)I_1+\text{h.c.}$\\
\hline

\end{tabular}
\end{center}	
\end{table}
\subsection{4 Derivatives}
\begin{table}[H]\scriptsize
\caption{N-N Operator basis with 4 derivatives.}
\label{4f4d}
\begin{center}
 \begin{tabular}{|l|l|}
	\hline
	$\mathcal{O}^{p^{\prime}(4)}_1=-2(N^{\dagger}\Vec{\sigma}\cdot\overrightarrow{\nabla}\overrightarrow{\nabla}^i\overrightarrow{\nabla}^2N^{\dagger})(N\sigma^iN)I_1+\text{h.c.}$
 &
 $\mathcal{O}^{p^{\prime}(4)}_2=-2(N^{\dagger}\sigma^iN)(N^{\dagger}\Vec{\sigma}\cdot\overleftarrow{\nabla}\overleftarrow{\nabla}^i\overleftarrow{\nabla}^2N)I_1+\text{h.c.}$\\
 $\mathcal{O}^{p^{\prime}(4)}_3=(N^{\dagger}(\vec{\sigma}\cdot\overrightarrow{\nabla})^4N^{\dagger})(NN)I_1+\text{h.c.}$
 &
 $\mathcal{O}^{p^{\prime}(4)}_4=-\dd{\rsd\rd{i}\rd{j}}\nn{\sigma^i\lsd\sigma^j}I_1+\text{h.c.}$\\
  $\mathcal{O}^{p^{\prime}(4)}_{5}=-\dn{\sigma^i\rsd\sigma^j}\dn{\ld{i}\ld{j}\lsd}I_1+\text{h.c.}$
  &
   $\mathcal{O}^{p^{\prime}(4)}_{6}=\dd{\rsd\sigma^i(\rsd)^2}\nn{\ld{i}}I_1+\text{h.c.}$\\
      
$\mathcal{O}^{p^{\prime}(4)}_{7}=\dn{\sigma^i\sigma^j}\dn{\ld{i}\lsd\rd{j}}I_2+\text{h.c.}$
&
$\mathcal{O}^{p^{\prime}(4)}_{8}=\dd{\rsd\rd{i}\rd{j}}\nn{\sigma^i\sigma^j\lsd}I_1+\text{h.c.}$\\
$\mathcal{O}^{p^{\prime}(4)}_{9}=-2\dn{\sigma^i\rd{j}}\dn{\ld{j}\lsd\ld{i}}I_1+\text{h.c.}$
&
$\mathcal{O}^{p^{\prime}(4)}_{10}=2\dd{(\rsd)^2\rd{i}}\nn{\ld{i}}I_1+\text{h.c.}$\\
$\mathcal{O}^{p^{\prime}(4)}_{11}=-2\dd{\rsd\rd{i}\rd{j}}\nn{\sigma^i\rd{j}}I_1+\text{h.c.}$
&
$\mathcal{O}^{p^{\prime}(4)}_{12}=-\dd{\rsd\sigma^i\rsd\rd{j}}\nn{\sigma^j\rd{i}}I_1+\text{h.c.}$\\
$\mathcal{O}^{p^{\prime}(4)}_{13}=\dn{\sigma^i\sigma^j}\dn{(\lsd)^2\ld{i}\rd{j}}I_1+\text{h.c.}$
&
$\mathcal{O}^{p^{\prime}(4)}_{14}=-\dn{\sigma^i\rsd}\dn{(\lsd)^2\ld{i}}I_2+\text{h.c.}$\\
$\mathcal{O}^{p^{\prime}(4)}_{15}=-\dd{\rsd\rd{i}\rd{j}}\nn{\rsd\sigma^i\sigma^j}I_1+\text{h.c.}$
&
$\mathcal{O}^{p^{\prime}(4)}_{16}=-\dn{\sigma^i}\dn{\lsd\rsd\lsd\ld{i}}I_1+\text{h.c.}$\\
$\mathcal{O}^{p^{\prime}(4)}_{17}=-\dd{(\rsd)^2\rd{i}}\nn{\sigma^i\rsd}I_1+\text{h.c.}$
&
$\mathcal{O}^{p^{\prime}(4)}_{18}=\dn{\rd{i}\rd{j}}\dn{\sigma^i\lsd\sigma^j\lsd}I_1+\text{h.c.}$\\
$\mathcal{O}^{p^{\prime}(4)}_{19}=\dd{\rsd\sigma^i\rsd\sigma^j}\nn{\ld{i}\ld{j}}I_1+\text{h.c.}$
&
$\mathcal{O}^{p^{\prime}(4)}_{20}=\dn{\sigma^i\sigma^j\sigma^k\sigma^l}\dn{\ld{i}\ld{k}\rd{j}\rd{l}}I_2+\text{h.c.}$\\
$\mathcal{O}^{p^{\prime}(4)}_{21}=2\dn{\sigma^i\sigma^j}\dn{\ld{i}\ld{k}\rd{k}\rd{j}}I_2+\text{h.c.}$
&
$\mathcal{O}^{p^{\prime}(4)}_{22}=\dd{\rsd\rd{i}}\nn{\lsd\sigma^i\lsd}I_1+\text{h.c.}$\\
$\mathcal{O}^{p^{\prime}(4)}_{23}=2\dd{\rsd\rd{i}}\nn{\lsd\ld{i}}I_1+\text{h.c.}$
&
$\mathcal{O}^{p^{\prime}(4)}_{24}=2\dd{\rsd\sigma^i\rd{j}}\nn{\ld{j}\ld{i}}I_1+\text{h.c.}$\\
$\mathcal{O}^{p^{\prime}(4)}_{25}=4\dd{\rd{i}\rd{j}}\nn{\ld{i}\ld{j}}I_1+\text{h.c.}$
&
$\mathcal{O}^{p^{\prime}(4)}_{26}=\dn{\rd{i}}\dn{\sigma^i\lsd\ld{j}\rd{j}}I_1+\text{h.c.}$\\
$\mathcal{O}^{p^{\prime}(4)}_{27}=\dn{\sigma^i\sigma^j\rd{k}}\dn{\ld{i}\rd{j}\sigma^k\lsd}I_1+\text{h.c.}$
&
$\mathcal{O}^{p^{\prime}(4)}_{28}=-\dn{\sigma^i\rsd\sigma^j\sigma^k}\dn{\ld{i}\rd{j}\ld{k}}I_1+\text{h.c.}$\\
$\mathcal{O}^{p^{\prime}(4)}_{29}=-2\dn{\sigma^i\rsd}\dn{\ld{i}\ld{j}\rd{j}}I_2+\text{h.c.}$
&$\mathcal{O}^{p^{\prime}(4)}_{30}=\dd{\rsd\rd{i}}\nn{\rsd\lsd\sigma^i}I_1+\text{h.c.}$\\
$\mathcal{O}^{p^{\prime}(4)}_{31}=\dd{\rsd\rd{i}}\nn{\rsd\ld{i}}I_1+\text{h.c.}$
&
$\mathcal{O}^{p^{\prime}(4)}_{32}=-\dd{\rsd\rd{i}}\nn{\sigma^i\ld{j}\rd{j}}I_1+\text{h.c.}$\\
$\mathcal{O}^{p^{\prime}(4)}_{33}=-\dd{\rsd\rd{i}}\nn{\sigma^i\lsd\rsd}I_1+\text{h.c.}$
&
$\mathcal{O}^{p^{\prime}(4)}_{34}=\dd{\rsd\rd{i}\sigma^j}\nn{\ld{j}\rsd\sigma^i}I_1+\text{h.c.}$\\
$\mathcal{O}^{p^{\prime}(4)}_{35}=\dn{\sigma^i\rsd\sigma^j}\dn{\ld{i}\ld{j}}I_2+\text{h.c.}$
&
$\mathcal{O}^{p^{\prime}(4)}_{36}=-\dn{\sigma^i\sigma^j\sigma^k\rsd}\dn{\ld{i}\ld{k}\rd{j}}I_2+\text{h.c.}$\\
$\mathcal{O}^{p^{\prime}(4)}_{37}=2\dd{(\rsd)^2}\nn{\ld{i}\rd{i}}I_1+\text{h.c.}$
&
$\mathcal{O}^{p^{\prime}(4)}_{38}=\dd{(\rsd)^2}\nn{\lsd\rsd}I_1+\text{h.c.}$\\
$\mathcal{O}^{p^{\prime}(4)}_{39}=2\dn{}\dn{(\ld{i}\rd{i})^2}I_1+\text{h.c.}$
&
$\mathcal{O}^{p^{\prime}(4)}_{40}=2\dn{\sigma^i\sigma^j}\dn{\ld{i}\rd{j}\ld{k}\rd{k}}I_1+\text{h.c.}$\\
$\mathcal{O}^{p^{\prime}(4)}_{41}=-2\dn{\sigma^i\rsd\rd{j}}\dn{\ld{j}\ld{i}}I_2+\text{h.c.}$
&
$\mathcal{O}^{p^{\prime}(4)}_{42}=-\dn{\sigma^i(\rsd)^2\sigma^j}\dn{\ld{i}\ld{j}}I_2+\text{h.c.}$\\
$\mathcal{O}^{p^{\prime}(4)}_{43}=-2\dd{\rsd\rd{i}}\nn{\rsd\rd{i}}I_1+\text{h.c.}$
&
$\mathcal{O}^{p^{\prime}(4)}_{44}=-\dd{\rsd\rd{i}}\nn{(\rsd)^2\sigma^i}I_1+\text{h.c.}$\\
$\mathcal{O}^{p^{\prime}(4)}_{45}=\dn{\sigma^i\sigma^j\sigma^k\sigma^l}\dn{\ld{i}\ld{k}\rd{j}\rd{l}}I_1+\text{h.c.}$&\\
\hline

\end{tabular}
\end{center}	
\end{table}

\section{Complete Operator 
Basis for 3N Contact Interaction}\label{ap:3N}

In Tab.~\ref{table:6ft}, the numbers of the three-nucleon contact interaction operators are given. There is only one independent operator at the leading order $\mathcal{O}(Q^0)$: $O^{(0)}=(N^{\dagger}N )(N^{\dagger}N)(N^{\dagger}N) $, in the general frame, which is in agreement with Refs.~\cite{Epelbaum:2002vt,Girlanda:2011fh}. There are 18 $P$-even and $T$-even operators at $\mathcal{O}(Q^2)$. 
\begin{table}[H]\small	
\begin{center}
 \begin{tabular}{|rr|c|c|}
	\hline
	&&Type $N^3(N^{\dagger})^3$& Type $N^3(N^{\dagger})^3\nabla^2$\\
  \hline
  &Hilbert series&$1N^3(N^{\dagger})^3$&$28N^3(N^{\dagger})^3\nabla^2$\\
  \hline
 & p-basis&$1N^3(N^{\dagger})^3$&$28N^3(N^{\dagger})^3\nabla^2$\\
	\hline
 $P$ even, $T$ even & Hilbert series&$1N^3(N^{\dagger})^3$&$18N^3(N^{\dagger})^3\nabla^2$\\
 \hline
 $P$ even, $T$ even &p-basis&$1N^3(N^{\dagger})^3$&$18N^3(N^{\dagger})^3\nabla^2$\\
 \hline
\end{tabular}
\end{center}	
\caption{The number of operators in the type $N^3(N^{\dagger})^3$ and $N^3(N^{\dagger})^3\nabla^2$.}\label{table:6ft}
\end{table}
 
\subsection{2 Derivatives}

In the present notation, $N_4$, $N_5$ and $N_6$ denote identical incoming particles, while $N^{\dagger}_1$, $N^{\dagger}_2$ and $N^{\dagger}_3$ denote identical outgoing particles. Consequently, the derivatives $-\mathbf{i}\nabla_4$, $-\mathbf{i}\nabla_5$, $-\mathbf{i}\nabla_6$ correspond to initial-state momenta, and $\mathbf{i}\nabla_1$, $\mathbf{i}\nabla_2$, $\mathbf{i}\nabla_3$ to final-state momenta in configuration space.

For the isospin structure, the fields are taken with explicit isospin indices: $(N_1^{\dagger})^{p_1}$, $(N_2^{\dagger})^{p_2}$, $(N_3^{\dagger})^{p_3}$, $(N_4)_{p_4}$, $(N_5)_{p_5}$, $(N_6)_{p_6}$. The five independent isospin y-basis tensors are
\begin{eqnarray}
    I_1 \equiv T_{SU(2),1}^{(y)} &=& \delta^{p_4}_{p_1}\,\delta_{p_2}^{p_5}\,\delta_{p_3}^{p_6},\nonumber\\
    I_2 \equiv T_{SU(2),2}^{(y)} &=& \delta^{p_6}_{p_1}\,\delta_{p_2}^{p_5}\,\delta_{p_3}^{p_4},\nonumber\\
    I_3 \equiv T_{SU(2),3}^{(y)} &=& \delta^{p_5}_{p_1}\,\delta_{p_2}^{p_4}\,\delta_{p_3}^{p_6},\nonumber\\
    I_4 \equiv T_{SU(2),4}^{(y)} &=& \delta^{p_6}_{p_1}\,\delta_{p_2}^{p_4}\,\delta_{p_3}^{p_5},\nonumber\\
    I_5 \equiv T_{SU(2),5}^{(y)} &=& \delta^{p_5}_{p_1}\,\delta_{p_2}^{p_6}\,\delta_{p_3}^{p_4}.
\end{eqnarray}
The familiar spin-isospin structures $\vec{\tau}_1\!\cdot\!\vec{\tau}_2$, $\vec{\tau}_1\!\cdot\!\vec{\tau}_3$, $\vec{\tau}_2\!\cdot\!\vec{\tau}_3$, and $\vec{\tau}_1\!\cdot\!\vec{\tau}_2\!\times\!\vec{\tau}_3$ can be expanded in this basis via Fierz and Schouten identities:
\begin{eqnarray}
    (\tau_1^I)_{p_1}^{p_4}(\tau_2^I)_{p_2}^{p_5}\delta_{p_3}^{p_6}
        &=& \bigl(2\delta_{p_1}^{p_5}\delta_{p_2}^{p_4} - \delta_{p_1}^{p_4}\delta_{p_2}^{p_5}\bigr)\delta_{p_3}^{p_6}
        = 2I_3 - I_1,
\end{eqnarray}
\begin{eqnarray}
    (\tau_1^I)_{p_1}^{p_4}(\tau_3^I)_{p_3}^{p_6}\delta_{p_2}^{p_5}
        &=& \bigl(2\delta_{p_1}^{p_6}\delta_{p_3}^{p_4} - \delta_{p_1}^{p_4}\delta_{p_3}^{p_6}\bigr)\delta_{p_2}^{p_5}
        = 2I_2 - I_1,
\end{eqnarray}
\begin{eqnarray}
    (\tau_2^I)_{p_2}^{p_5}(\tau_3^I)_{p_3}^{p_6}\delta_{p_1}^{p_4}
        &=& \bigl(2\delta_{p_2}^{p_6}\delta_{p_3}^{p_5} - \delta_{p_2}^{p_5}\delta_{p_3}^{p_6}\bigr)\delta_{p_1}^{p_4}
        = I_1 - 2I_2 - 2I_3 + 2I_4 + 2I_5,
\end{eqnarray}
\begin{eqnarray}
    \epsilon^{IJK}(\tau_1^I)_{p_1}^{p_4}(\tau_2^J)_{p_2}^{p_5}(\tau_3^K)_{p_3}^{p_6}
        &=& 2\mathbf{i}\,(I_4 - I_5).
\end{eqnarray}
For later convenience, we also define the combination
\begin{eqnarray}
    I_6 \equiv I_1 - I_2 - I_3 + I_4 + I_5
        &=& \delta^{p_4}_{p_1}\,\delta_{p_2}^{p_6}\,\delta_{p_3}^{p_5}.
\end{eqnarray}

The complete set of three‑nucleon contact operators at $\mathcal{O}(Q^2)$ is listed in Table~\ref{6f2dop}, which is consistent with the results in Ref.~\cite{Girlanda:2011fh}.

\begin{table}[H]\scriptsize
\caption{Non-relativistic 3N basis with two derivatives.}
\label{6f2dop}
\begin{center}
$\begin{array}{|l|l|}
\hline
O^{(2)}_1= (N^{\dagger}\sigma^iN)(N^{\dagger}\overleftarrow{\nabla}^i\overleftarrow{\nabla}^jN^{\dagger})(N\sigma^jN)(I_2-I_4)+\text{h.c.} & O^{(2)}_2= (N^{\dagger}\sigma^i\overrightarrow{\nabla}^{i}\overrightarrow{\nabla}^{j}N^{\dagger})(N^{\dagger}N)(N\sigma^jN)(I_1-I_3)+\text{h.c.} \\
O^{(2)}_3= (N^{\dagger}\sigma^i\overrightarrow{\nabla}^i\overrightarrow{\nabla}^jN^{\dagger})(N^{\dagger}\sigma^jN)(NN)(I_4-I_6)+\text{h.c.} & O^{(2)}_4=(N^{\dagger}\sigma^j\overrightarrow{\nabla}^jN^{\dagger})(N\sigma^jN)(N^{\dagger}\overleftarrow{\nabla}^iN)(I_1-I_3)+\text{h.c.} \\
 O^{(2)}_5=(N^{\dagger}\overrightarrow{\nabla}^iN^{\dagger})(N^{\dagger}\overleftarrow{\nabla}^j\sigma^iN)(N\sigma^jN)(I_5-I_6)+\text{h.c.}& O^{(2)}_6=(N^{\dagger}\overrightarrow{\nabla}^iN^{\dagger})(N^{\dagger}\overleftarrow{\nabla}^j\sigma^i\sigma^jN)(NN)(I_1-I_2)+\text{h.c.} \\
 O^{(2)}_7=(N^{\dagger}\sigma^iN)(N^{\dagger}\overleftarrow{\nabla}^i\sigma^jN)(N^{\dagger}\overleftarrow{\nabla}^jN)(I_6)+\text{h.c.}& O^{(2)}_8= (N^{\dagger}\sigma^iN)(N^{\dagger}\overleftarrow{\nabla}^i\overrightarrow{\nabla}^jN^{\dagger})(N\sigma^jN)(I_1-I_6)+\text{h.c.} \\
 O^{(2)}_9=(N^{\dagger}\sigma^i\sigma^jN)(N^{\dagger}\overleftarrow{\nabla}^i\overrightarrow{\nabla}^jN^{\dagger})(NN)(I_3-I_5)+\text{h.c.} & O^{(2)}_{10}=(N^{\dagger}\overrightarrow{\nabla}^iN)(N^{\dagger}\overleftarrow{\nabla}^j\sigma^i\sigma^jN)(N^{\dagger}N))(I_1)+\text{h.c.}\\
 O^{(2)}_{11}=(N^{\dagger}N^{\dagger})(N^{\dagger}\overleftarrow{\nabla}^j\sigma^i\sigma^jN)(N\overleftarrow{\nabla}^iN)(I_1-I_2)+\text{h.c.} & O^{(2)}_{12}=(N^{\dagger}N^{\dagger})(N^{\dagger}\overleftarrow{\nabla}^i\overrightarrow{\nabla}^jN)(N\sigma^i\sigma^jN)(I_3-I_4) +\text{h.c.}\\
 O^{(2)}_{13}=(N^{\dagger}N^{\dagger})(N^{\dagger}\overleftarrow{\nabla}^i\overrightarrow{\nabla}^iN)(NN)(I_3-I_4) +\text{h.c.}&  O^{(2)}_{14}=(N^{\dagger}\sigma^i\sigma^j\overrightarrow{\nabla}^iN^{\dagger})(N^{\dagger}N)(N\overleftarrow{\nabla}^jN)(I_4-I_6) +\text{h.c.}\\
  O^{(2)}_{15}=(N^{\dagger}\sigma^i\sigma^jN)(N^{\dagger}\overleftarrow{\nabla}^i\overrightarrow{\nabla}^jN)(N^{\dagger}N)(I_4)+\text{h.c.}&  O^{(2)}_{16}=(N^{\dagger}\sigma^i\sigma^jN)(N^{\dagger}\overleftarrow{\nabla}^iN^{\dagger})(N\overleftarrow{\nabla}^jN)(I_2-I_4)+\text{h.c.}\\
  O^{(2)}_{17}=(N^{\dagger}\sigma^i\overrightarrow{\nabla}^jN)((N^{\dagger}\overleftarrow{\nabla}^iN^{\dagger})(N\sigma^jN)(I_1-I_6)+\text{h.c.}& O^{(2)}_{18}= (N^{\dagger}\sigma^i\overrightarrow{\nabla}^iN^{\dagger})(N^{\dagger}\overrightarrow{\nabla}^jN)(N\sigma^jN)(I_2-I_5)+\text{h.c.}\\
   \hline
\end{array}$
\end{center}
\end{table}

\section{Complete Operator Basis for spin-1/2 DM-N Interaction}\label{ap:DMN}

Tab.~\ref{DMNnumber} lists the number of operators of each type, categorized by their $P$ and $T$ transformation properties. The results obtained from the Hilbert series and the Young tensor method are in agreement.

At $\mathcal{O}(v^0)$, there are two operators 
\begin{eqnarray}
   O^{(0)}_1 &=&(N_1^{\dagger}N_2)(\xi^{\dagger}_3\xi_4)\nonumber\\
    O^{(0)}_2 &=&(N_1^{\dagger}\xi^{\dagger}_3)(N_2\xi_4),
\end{eqnarray}
Since there is no repeated field, the field subscripts can be retained unambiguously. 
The isospin indices of the nucleons are contracted with each other, rendering the isospin structure trivial. Consequently, only the spinor indices require consideration.

To illustrate the notation, consider the operator $(N^{\dagger}_1\vec\sigma\cdot\nabla_2\xi_4)(N_2\xi^{\dagger}_3)$. It can be rewriten as
\begin{equation}
    (N^{\dagger}_1\vec\sigma\cdot\nabla_2\xi_4)(N_2\xi^{\dagger}_3)=\left(N^{\dagger,i,p}_1~(\sigma^I)_i^j~\xi_{4,j}\right)\left((\nabla_2^I N_{2,p}^k)~\xi^{\dagger}_{3,k}\right),
\end{equation}
where $i,j,k$ denote fundamental indices of the spinor $SU(2)$ group, $p$ is the fundamental index of the isospin $SU(2)$ group, and $I$ is the adjoint index of the spinor $SU(2)$ group. The subscript on $\nabla_2$ indicates the field on which the derivative acts.
As a further example, the trace $\text{tr}(\vec{\sigma}\cdot\nabla_2\vec\sigma\cdot\nabla_3)$ evaluates to
\begin{eqnarray}
\text{tr}(\vec{\sigma}\cdot\nabla_2\vec\sigma\cdot\nabla_3) = (\sigma^I)_i^j(\sigma^J)_j^i \nabla_2^I\nabla_3^J = 2\delta^{IJ}\nabla_2^I\nabla_3^J.
\end{eqnarray}

In the following parts we list the explicit forms of the operators.

\begin{table}[H]
    \centering
\begin{tabular}{|c|c|c|c|}
    \hline
      Type& $P$~$T$&Hilbert series&Young tensor method\\
      \hline
      $NN^{\dagger}\xi\xi^{\dagger}$&$+~ +$&$2NN^{\dagger}\xi\xi^{\dagger}$&$2NN^{\dagger}\xi\xi^{\dagger}$\\
      \hline
      $\nabla NN^{\dagger}\xi\xi^{\dagger}$&$-~+$&   $5\nabla NN^{\dagger}\xi\xi^{\dagger}$&   $5\nabla NN^{\dagger}\xi\xi^{\dagger}$\\
      \cline{2-4}
      &$-~-$&   $4\nabla NN^{\dagger}\xi\xi^{\dagger}$&   $4\nabla NN^{\dagger}\xi\xi^{\dagger}$\\
      \hline
      $\nabla^2NN^{\dagger}\xi\xi^{\dagger}$&$+~+$&$17\nabla^2NN^{\dagger}\xi\xi^{\dagger}$&$17\nabla^2NN^{\dagger}\xi\xi^{\dagger}$\\
      \cline{2-4}
        &$+~-$&$10\nabla^2NN^{\dagger}\xi\xi^{\dagger}$&$10\nabla^2NN^{\dagger}\xi\xi^{\dagger}$\\
        \hline
         $\nabla^3NN^{\dagger}\xi\xi^{\dagger}$&$-~+$&$34\nabla^2NN^{\dagger}\xi\xi^{\dagger}$&$34\nabla^2NN^{\dagger}\xi\xi^{\dagger}$\\
      \cline{2-4}
        &$-~-$&$30\nabla^2NN^{\dagger}\xi\xi^{\dagger}$&$30\nabla^2NN^{\dagger}\xi\xi^{\dagger}$\\
        \hline
         $\nabla^4NN^{\dagger}\xi\xi^{\dagger}$&$+~+$&$74\nabla^2NN^{\dagger}\xi\xi^{\dagger}$&$74\nabla^2NN^{\dagger}\xi\xi^{\dagger}$\\
      \cline{2-4}
        &$+~-$&$58\nabla^2NN^{\dagger}\xi\xi^{\dagger}$&$58\nabla^2NN^{\dagger}\xi\xi^{\dagger}$\\
        \hline
    \end{tabular}
    \caption{The number of operators in the type $NN^{\dagger}\xi\xi^{\dagger}$, $\nabla NN^{\dagger}\xi\xi^{\dagger}$, $\nabla^2NN^{\dagger}\xi\xi^{\dagger}$, $\nabla^3NN^{\dagger}\xi\xi^{\dagger}$ and $\nabla^4NN^{\dagger}\xi\xi^{\dagger}$. With the specified transformation properties of the $P$ and $T$ transformation, we list the number of operators respectively. The Hilbert series and the Young tensor method give the consistent results.}
    \label{DMNnumber}
 
\end{table}

\subsection{1 Derivative}
\begin{table}[H]\scriptsize
    \centering
 $$
\begin{array}{|cc|}
 O^{(1)}{}_1 & \left(N^{\dagger }{}_1 \vec{\sigma }\cdot \nabla _2 \xi _4\right) \left(N_2 \xi ^{\dagger }{}_3\right) +\text{h.c.}\\
 O^{(1)}{}_2 & -\left(N^{\dagger }{}_1 \vec{\sigma }\cdot \nabla _2 N_2\right) \left(\xi ^{\dagger }{}_3 \xi _4\right) +\text{h.c.}\\
 O^{(1)}{}_3 & -\left(N^{\dagger }{}_1 N_2\right) \left(\xi ^{\dagger }{}_3 \vec{\sigma }\cdot \nabla _2 \xi _4\right)+\text{h.c.} \\
 O^{(1)}{}_4 & -\left(N^{\dagger }{}_1 \vec{\sigma }\cdot \nabla _3 N_2\right) \left(\xi ^{\dagger }{}_3 \xi _4\right)+\text{h.c.} \\
 O^{(1)}{}_5 & \left(N_2 \vec{\sigma }\cdot \nabla _3 \xi _4\right) \left(N^{\dagger }{}_1 \xi ^{\dagger }{}_3\right)+\text{h.c.} \\
\end{array}
$$
    \caption{P-odd T-odd Operators at $\mathcal{O}(v^1)$.}

\end{table}

\begin{table}[H]\scriptsize
 $$
\begin{array}{|cc|}
 \tilde{O}^{(1)}{}_1 & i\left(N^{\dagger }{}_1 \vec{\sigma }\cdot \nabla _2 \xi _4\right) \left(N_2 \xi ^{\dagger }{}_3\right) +\text{h.c.}\\
 \tilde{O}^{(1)}{}_2 & -i\left(N^{\dagger }{}_1 \vec{\sigma }\cdot \nabla _2 N_2\right) \left(\xi ^{\dagger }{}_3 \xi _4\right)+\text{h.c.} \\
 \tilde{O}^{(1)}{}_3 & -i\left(N^{\dagger }{}_1 N_2\right) \left(\xi ^{\dagger }{}_3 \vec{\sigma }\cdot \nabla _2 \xi _4\right) +\text{h.c.}\\
 \tilde{O}^{(1)}{}_4 & i\left(N_2 \vec{\sigma }\cdot \nabla _3 \xi _4\right) \left(N^{\dagger }{}_1 \xi ^{\dagger }{}_3\right)+\text{h.c.} \\
\end{array}
$$
\caption{P-odd T-even Operators at $\mathcal{O}(v^1)$.}
\end{table}

\subsection{2 Derivatives}
\begin{table}[H]\scriptsize
$$
\begin{array}{|cc|}
 O^{(2)}{}_1 & \left(N^{\dagger }{}_1 \vec{\sigma }\cdot \nabla _2 N_2\right) \left(\xi ^{\dagger }{}_3 \vec{\sigma }\cdot \nabla _2 \xi _4\right) +\text{h.c.}\\
 O^{(2)}{}_2 & \left(N_2 \vec{\sigma }\cdot \nabla _2 \xi _4\right) \left(N^{\dagger }{}_1 \vec{\sigma }\cdot \nabla _2 \xi ^{\dagger }{}_3\right) +\text{h.c.} \\
 O^{(2)}{}_3 & -\left(N^{\dagger }{}_1 \vec{\sigma }\cdot \nabla _2 \vec{\sigma }\cdot \nabla _2 N_2\right) \left(\xi ^{\dagger }{}_3 \xi _4\right) +\text{h.c.} \\
 O^{(2)}{}_4 & -\left(N_2 \vec{\sigma }\cdot \nabla _3 \vec{\sigma }\cdot \nabla _2 \xi _4\right) \left(N^{\dagger }{}_1 \xi ^{\dagger }{}_3\right) +\text{h.c.} \\
 O^{(2)}{}_5 & -\left(N^{\dagger }{}_1 \vec{\sigma }\cdot \nabla _2 \vec{\sigma }\cdot \nabla _3 N_2\right) \left(\xi ^{\dagger }{}_3 \xi _4\right)  +\text{h.c.}\\
 O^{(2)}{}_6 & \left(N^{\dagger }{}_1 \vec{\sigma }\cdot \nabla _2 \vec{\sigma }\cdot \nabla _3 \xi _4\right) \left(N_2 \xi ^{\dagger }{}_3\right)  +\text{h.c.}\\
 O^{(2)}{}_7 & -\left(N^{\dagger }{}_1 \vec{\sigma }\cdot \nabla _2 N_2\right) \left(\xi ^{\dagger }{}_3 \vec{\sigma }\cdot \nabla _3 \xi _4\right)  +\text{h.c.}\\
 O^{(2)}{}_8 & -\left(N^{\dagger }{}_1 N_2\right) \left(\xi ^{\dagger }{}_3 \xi _4\right) \text{tr}\left(\vec{\sigma }\cdot \nabla _2 \vec{\sigma }\cdot \nabla _3\right)  +\text{h.c.}\\
\end{array}
$$
    \caption{P-even T-even Operators at $\mathcal{O}(v^2)$.}
 
\end{table}

\begin{table}[H]\scriptsize
$$
\begin{array}{|cc|}
 \tilde{O}^{(2)}{}_1 & i\left(N^{\dagger }{}_1 \vec{\sigma }\cdot \nabla _2 N_2\right) \left(\xi ^{\dagger }{}_3 \vec{\sigma }\cdot \nabla _2 \xi _4\right)+\text{h.c.} \\
 \tilde{O}^{(2)}{}_2 &i \left(N_2 \vec{\sigma }\cdot \nabla _2 \xi _4\right) \left(N^{\dagger }{}_1 \vec{\sigma }\cdot \nabla _2 \xi ^{\dagger }{}_3\right) +\text{h.c.}\\
 \tilde{O}^{(2)}{}_3 & -i\left(N^{\dagger }{}_1 \vec{\sigma }\cdot \nabla _2 \vec{\sigma }\cdot \nabla _2 N_2\right) \left(\xi ^{\dagger }{}_3 \xi _4\right)+\text{h.c.} \\
 \tilde{O}^{(2)}{}_4 & -i\left(N_2 \vec{\sigma }\cdot \nabla _3 \vec{\sigma }\cdot \nabla _2 \xi _4\right) \left(N^{\dagger }{}_1 \xi ^{\dagger }{}_3\right) +\text{h.c.}\\
 \tilde{O}^{(2)}{}_5 & -i\left(N^{\dagger }{}_1 \vec{\sigma }\cdot \nabla _2 \vec{\sigma }\cdot \nabla _3 N_2\right) \left(\xi ^{\dagger }{}_3 \xi _4\right) +\text{h.c.}\\
 \tilde{O}^{(2)}{}_6 & i\left(N^{\dagger }{}_1 \vec{\sigma }\cdot \nabla _2 \vec{\sigma }\cdot \nabla _3 \xi _4\right) \left(N_2 \xi ^{\dagger }{}_3\right)+\text{h.c.} \\
 \tilde{O}^{(2)}{}_7 & -i\left(N^{\dagger }{}_1 \vec{\sigma }\cdot \nabla _2 N_2\right) \left(\xi ^{\dagger }{}_3 \vec{\sigma }\cdot \nabla _3 \xi _4\right) +\text{h.c.}\\
 \tilde{O}^{(2)}{}_8 & -i\left(N^{\dagger }{}_1 N_2\right) \left(\xi ^{\dagger }{}_3 \xi _4\right) \text{tr}\left(\vec{\sigma }\cdot \nabla _2 \vec{\sigma }\cdot \nabla _3\right)+\text{h.c.} \\
 \tilde{O}^{(2)}{}_9 & -i\left(N^{\dagger }{}_1 N_2\right) \left(\xi ^{\dagger }{}_3 \vec{\sigma }\cdot \nabla _2 \vec{\sigma }\cdot \nabla _3 \xi _4\right)+\text{h.c.} \\
 \tilde{O}^{(2)}{}_{10} & -i\left(N_2 \xi _4\right) \left(N^{\dagger }{}_1 \vec{\sigma }\cdot \nabla _2 \vec{\sigma }\cdot \nabla _4 \xi ^{\dagger }{}_3\right)+\text{h.c.} \\
\end{array}
$$
    \caption{P-even T-odd Operators at $\mathcal{O}(v^2)$.}

\end{table}

\subsection{3 Derivatives}
\begin{table}[H]\scriptsize
  $$
\begin{array}{|cc|cc|}
 O^{(3)}{}_1 & -\left(N^{\dagger }{}_1 \vec{\sigma }\cdot \nabla _2 \xi _4\right) \left(N_2 \vec{\sigma }\cdot \nabla _2 \vec{\sigma }\cdot \nabla _2 \xi ^{\dagger }{}_3\right)+\text{h.c.} & O^{(3)}{}_{18} & \left(N^{\dagger }{}_1 \vec{\sigma }\cdot \nabla _3 N_2\right) \left(\xi ^{\dagger }{}_3 \xi _4\right) \text{tr}\left(\vec{\sigma }\cdot \nabla _2 \vec{\sigma }\cdot \nabla _3\right)+\text{h.c.} \\
 O^{(3)}{}_2 & \left(N^{\dagger }{}_1 \vec{\sigma }\cdot \nabla _2 N_2\right) \left(\xi ^{\dagger }{}_3 \vec{\sigma }\cdot \nabla _2 \vec{\sigma }\cdot \nabla _2 \xi _4\right) +\text{h.c.}& O^{(3)}{}_{19} & -\left(N_2 \vec{\sigma }\cdot \nabla _3 \xi _4\right) \left(N^{\dagger }{}_1 \xi ^{\dagger }{}_3\right) \text{tr}\left(\vec{\sigma }\cdot \nabla _2 \vec{\sigma }\cdot \nabla _3\right) +\text{h.c.}\\
 O^{(3)}{}_3 & \left(N^{\dagger }{}_1 \vec{\sigma }\cdot \nabla _2 \vec{\sigma }\cdot \nabla _2 N_2\right) \left(\xi ^{\dagger }{}_3 \vec{\sigma }\cdot \nabla _2 \xi _4\right) +\text{h.c.}& O^{(3)}{}_{20} & -\left(N_2 \vec{\sigma }\cdot \nabla _3 \vec{\sigma }\cdot \nabla _2 \vec{\sigma }\cdot \nabla _3 \xi _4\right) \left(N^{\dagger }{}_1 \xi ^{\dagger }{}_3\right) +\text{h.c.}\\
 O^{(3)}{}_4 & -\left(N^{\dagger }{}_1 \vec{\sigma }\cdot \nabla _2 \xi _4\right) \left(N_2 \xi ^{\dagger }{}_3\right) \text{tr}\left(\vec{\sigma }\cdot \nabla _2 \vec{\sigma }\cdot \nabla _3\right) +\text{h.c.}& O^{(3)}{}_{21} & -\left(N_2 \vec{\sigma }\cdot \nabla _3 \xi _4\right) \left(N^{\dagger }{}_1 \vec{\sigma }\cdot \nabla _2 \vec{\sigma }\cdot \nabla _3 \xi ^{\dagger }{}_3\right)+\text{h.c.} \\
\end{array}
$$
    \caption{P-odd T-odd Operators at $\mathcal{O}(v^3)$.}

\end{table}

\begin{table}[H]\scriptsize
$$
\begin{array}{|cc|cc|}
 \tilde{O}^{(3)}{}_1 & -i\left(N^{\dagger }{}_1 \vec{\sigma }\cdot \nabla _2 \xi _4\right) \left(N_2 \vec{\sigma }\cdot \nabla _2 \vec{\sigma }\cdot \nabla _2 \xi ^{\dagger }{}_3\right) +\text{h.c.}& \tilde{O}^{(3)}{}_{16} & -i\left(N^{\dagger }{}_1 \vec{\sigma }\cdot \nabla _2 \xi _4\right) \left(N_2 \vec{\sigma }\cdot \nabla _2 \vec{\sigma }\cdot \nabla _2 \xi ^{\dagger }{}_3\right)+\text{h.c.} \\
 \tilde{O}^{(3)}{}_2 & i\left(N^{\dagger }{}_1 \vec{\sigma }\cdot \nabla _2 N_2\right) \left(\xi ^{\dagger }{}_3 \vec{\sigma }\cdot \nabla _2 \vec{\sigma }\cdot \nabla _2 \xi _4\right) +\text{h.c.}& \tilde{O}^{(3)}{}_{17} &i \left(N^{\dagger }{}_1 \vec{\sigma }\cdot \nabla _2 N_2\right) \left(\xi ^{\dagger }{}_3 \vec{\sigma }\cdot \nabla _2 \vec{\sigma }\cdot \nabla _2 \xi _4\right) +\text{h.c.}\\
 \tilde{O}^{(3)}{}_3 & i\left(N^{\dagger }{}_1 \vec{\sigma }\cdot \nabla _2 \vec{\sigma }\cdot \nabla _2 N_2\right) \left(\xi ^{\dagger }{}_3 \vec{\sigma }\cdot \nabla _2 \xi _4\right) +\text{h.c.}& \tilde{O}^{(3)}{}_{18} & i\left(N^{\dagger }{}_1 \vec{\sigma }\cdot \nabla _2 \vec{\sigma }\cdot \nabla _2 N_2\right) \left(\xi ^{\dagger }{}_3 \vec{\sigma }\cdot \nabla _2 \xi _4\right) +\text{h.c.}\\
 \tilde{O}^{(3)}{}_4 & -i\left(N^{\dagger }{}_1 \vec{\sigma }\cdot \nabla _2 \xi _4\right) \left(N_2 \xi ^{\dagger }{}_3\right) \text{tr}\left(\vec{\sigma }\cdot \nabla _2 \vec{\sigma }\cdot \nabla _3\right) +\text{h.c.}& \tilde{O}^{(3)}{}_{19} & -i\left(N^{\dagger }{}_1 \vec{\sigma }\cdot \nabla _2 \xi _4\right) \left(N_2 \xi ^{\dagger }{}_3\right) \text{tr}\left(\vec{\sigma }\cdot \nabla _2 \vec{\sigma }\cdot \nabla _3\right) +\text{h.c.}\\
\end{array}
$$
    \caption{P-odd T-even Operators at $\mathcal{O}(v^3)$.}

\end{table}

\subsection{4 Derivatives}
\begin{table}[H]\tiny
$$
\begin{array}{|ccc|ccc|}
 O^{(4)}{}_1 & -\left(N^{\dagger }{}_1 \vec{\sigma }\cdot \nabla _2 N_2\right) \left(\xi ^{\dagger }{}_3 \vec{\sigma }\cdot \nabla _2 \xi _4\right) \text{tr}\left(\vec{\sigma }\cdot \nabla _2 \vec{\sigma }\cdot \nabla _2\right) & +\text{h.c.} & O^{(4)}{}_{38} & \left(N_2 \vec{\sigma }\cdot \nabla _3 \vec{\sigma }\cdot \nabla _2 \vec{\sigma }\cdot \nabla _3 \vec{\sigma }\cdot \nabla _2 \xi _4\right) \left(N^{\dagger }{}_1 \xi ^{\dagger }{}_3\right) & +\text{h.c.} \\
 O^{(4)}{}_2 & -\left(N_2 \vec{\sigma }\cdot \nabla _2 \xi _4\right) \left(N^{\dagger }{}_1 \vec{\sigma }\cdot \nabla _2 \xi ^{\dagger }{}_3\right) \text{tr}\left(\vec{\sigma }\cdot \nabla _2 \vec{\sigma }\cdot \nabla _2\right) & +\text{h.c.} & O^{(4)}{}_{39} & \left(N^{\dagger }{}_1 \vec{\sigma }\cdot \nabla _2 \vec{\sigma }\cdot \nabla _3 \vec{\sigma }\cdot \nabla _2 \vec{\sigma }\cdot \nabla _3 N_2\right) \left(\xi ^{\dagger }{}_3 \xi _4\right) & +\text{h.c.} \\
 O^{(4)}{}_3 & \left(N^{\dagger }{}_1 \vec{\sigma }\cdot \nabla _2 \vec{\sigma }\cdot \nabla _2 \vec{\sigma }\cdot \nabla _2 \vec{\sigma }\cdot \nabla _2 N_2\right) \left(\xi ^{\dagger }{}_3 \xi _4\right) & +\text{h.c.} & O^{(4)}{}_{40} & -\left(N^{\dagger }{}_1 \vec{\sigma }\cdot \nabla _2 \vec{\sigma }\cdot \nabla _3 \vec{\sigma }\cdot \nabla _2 \vec{\sigma }\cdot \nabla _3 \xi _4\right) \left(N_2 \xi ^{\dagger }{}_3\right) & +\text{h.c.} \\
 O^{(4)}{}_4 & -\left(N^{\dagger }{}_1 \vec{\sigma }\cdot \nabla _2 N_2\right) \left(\xi ^{\dagger }{}_3 \vec{\sigma }\cdot \nabla _2 \vec{\sigma }\cdot \nabla _3 \vec{\sigma }\cdot \nabla _2 \xi _4\right) & +\text{h.c.} & O^{(4)}{}_{41} & -\left(N^{\dagger }{}_1 \vec{\sigma }\cdot \nabla _2 \vec{\sigma }\cdot \nabla _3 \xi _4\right) \left(N_2 \xi ^{\dagger }{}_3\right) \text{tr}\left(\vec{\sigma }\cdot \nabla _2 \vec{\sigma }\cdot \nabla _3\right) & +\text{h.c.} \\
\end{array}
$$
    \caption{P-even T-even Operators at $\mathcal{O}(v^4)$.}
 
\end{table}

\begin{table}[H]\tiny
  $$
\begin{array}{|ccc|ccc|}
 \tilde{O}^{(4)}{}_1 & -i \left(N^{\dagger }{}_1 \vec{\sigma }\cdot \nabla _2 N_2\right) \left(\xi ^{\dagger }{}_3 \vec{\sigma }\cdot \nabla _2 \xi _4\right) \text{tr}\left(\vec{\sigma }\cdot \nabla _2 \vec{\sigma }\cdot \nabla _2\right) & +\text{h.c.} & \tilde{O}^{(4)}{}_{30} & -i \left(N^{\dagger }{}_1 \vec{\sigma }\cdot \nabla _2 N_2\right) \left(\xi ^{\dagger }{}_3 \vec{\sigma }\cdot \nabla _2 \xi _4\right) \text{tr}\left(\vec{\sigma }\cdot \nabla _2 \vec{\sigma }\cdot \nabla _2\right) & +\text{h.c.} \\
 \tilde{O}^{(4)}{}_2 & -i \left(N_2 \vec{\sigma }\cdot \nabla _2 \xi _4\right) \left(N^{\dagger }{}_1 \vec{\sigma }\cdot \nabla _2 \xi ^{\dagger }{}_3\right) \text{tr}\left(\vec{\sigma }\cdot \nabla _2 \vec{\sigma }\cdot \nabla _2\right) & +\text{h.c.} & \tilde{O}^{(4)}{}_{31} & -i \left(N_2 \vec{\sigma }\cdot \nabla _2 \xi _4\right) \left(N^{\dagger }{}_1 \vec{\sigma }\cdot \nabla _2 \xi ^{\dagger }{}_3\right) \text{tr}\left(\vec{\sigma }\cdot \nabla _2 \vec{\sigma }\cdot \nabla _2\right) & +\text{h.c.} \\
 \tilde{O}^{(4)}{}_3 & i \left(N^{\dagger }{}_1 \vec{\sigma }\cdot \nabla _2 \vec{\sigma }\cdot \nabla _2 \vec{\sigma }\cdot \nabla _2 \vec{\sigma }\cdot \nabla _2 N_2\right) \left(\xi ^{\dagger }{}_3 \xi _4\right) & +\text{h.c.} & \tilde{O}^{(4)}{}_{32} & i \left(N^{\dagger }{}_1 \vec{\sigma }\cdot \nabla _2 \vec{\sigma }\cdot \nabla _2 \vec{\sigma }\cdot \nabla _2 \vec{\sigma }\cdot \nabla _2 N_2\right) \left(\xi ^{\dagger }{}_3 \xi _4\right) & +\text{h.c.} \\
 \tilde{O}^{(4)}{}_4 & -i \left(N^{\dagger }{}_1 \vec{\sigma }\cdot \nabla _2 N_2\right) \left(\xi ^{\dagger }{}_3 \vec{\sigma }\cdot \nabla _2 \vec{\sigma }\cdot \nabla _3 \vec{\sigma }\cdot \nabla _2 \xi _4\right) & +\text{h.c.} & \tilde{O}^{(4)}{}_{33} & -i \left(N^{\dagger }{}_1 \vec{\sigma }\cdot \nabla _2 N_2\right) \left(\xi ^{\dagger }{}_3 \vec{\sigma }\cdot \nabla _2 \vec{\sigma }\cdot \nabla _3 \vec{\sigma }\cdot \nabla _2 \xi _4\right) & +\text{h.c.} \\
\end{array}
$$
    \caption{P-even T-odd Operators at $\mathcal{O}(v^4)$.}
 
\end{table}

\bibliography{ref}
\end{document}